\documentclass[a4paper,fleqn,10pt]{article}
\pdfoutput=1
\usepackage{amsmath}
\usepackage{mathtools}
\usepackage{amssymb}
\usepackage{bbm}
\usepackage{array}
\usepackage{calc}
\usepackage{longtable}
\usepackage{multirow}
\usepackage{slashed}
\usepackage{pstricks}
\usepackage{graphicx}
\usepackage{xspace}
\usepackage{units}
\usepackage{tikz}
\usepackage{textcomp}

\numberwithin{equation}{section}
\usepackage{cite}
\usepackage[pdfborder={0 0 0}]{hyperref}
\usepackage[format=hang,labelfont=bf,hypcap=true]{caption}
\usepackage{subcaption}
\usepackage{sectsty}
\usepackage{enumitem}
\allsectionsfont{\sffamily}
\subsubsectionfont{\mdseries\itshape\large}
\makeatletter
\DeclareRobustCommand*{\bfseries}{%
  \not@math@alphabet\bfseries\mathbf
  \fontseries\bfdefault\selectfont
  \boldmath
}
\makeatother
\let\spreprint\empty
\newcommand{\preprint}[1]{\def\spreprint{\protect#1}}
\let\sinstitute\empty
\newcommand{\institute}[1]{\def\sinstitute{\protect#1}}
\makeatletter
\renewcommand{\maketitle}{\begingroup
  \null\thispagestyle{empty}%
    \ifx\spreprint\empty
      \vskip 5ex
    \else
      \flushright\large\spreprint\vskip 10ex
    \fi
    \vskip 5ex
    \flushleft
      {\sffamily\bfseries\huge\@title}\vskip 6ex
      \@author\vskip 2ex
      \ifx\sinstitute\empty
      \else
        {\small\sinstitute}
      \fi
    \vskip 5ex
  \endgroup
}
\makeatother
\renewenvironment{abstract}{\begin{center}
  {\large\sffamily\bfseries Abstract: }
  \begin{minipage}[t]{0.75\textwidth}
}{\end{minipage}\end{center}\vskip 10ex}

\numberwithin{equation}{section}
\allowdisplaybreaks[2]

\newcommand{\LHAPDF}{L\protect\scalebox{0.8}{HAPDF}\xspace}

\newcommand{\MSbar}{\ensuremath{\overline{\text{MS}}}\xspace}

\newcommand{\Rivet}{R\protect\scalebox{0.8}{IVET}\xspace}

\newcommand{\Recola}{R\protect\scalebox{0.8}{ECOLA}\xspace}

\newcommand{\Sherpa}{S\protect\scalebox{0.8}{HERPA}\xspace}

\newcommand{\Amegic}{A\protect\scalebox{0.8}{MEGIC}\xspace}

\long\def\symbolfootnote[#1]#2{\begingroup%
\def\thefootnote{\fnsymbol{footnote}}\footnote[#1]{#2}\endgroup}

\newcommand{\done}{{\rm d}}
\newcommand{\order}{\mathcal{O}}

\newcommand{\nnb}{\nonumber}
\newcommand{\bea}{\begin{eqnarray}}
\newcommand{\eea}{\end{eqnarray}}
\newcommand{\bi}{\begin{itemize}}
\newcommand{\ei}{\end{itemize}}
\newcommand{\hl}{\vphantom{$\int_A^B$}}

\newcommand{\tbar}{{\ensuremath{\bar{t}}}}
\newcommand{\ttbar}{{\ensuremath{t\tbar}}}
\newcommand{\Mttbar}{\ensuremath{M_\ttbar}}
\newcommand{\MTttbar}{\ensuremath{M_{\mathrm{T}}^\ttbar}}
\newcommand{\MTttbart}{\ensuremath{M_{\mathrm{T},\tau}^\ttbar}}
\newcommand{\bttbar}{\ensuremath{\beta_\ttbar}}
\newcommand{\Yttbar}{\ensuremath{Y_\ttbar}}
\newcommand{\dphittbar}{\ensuremath{\Delta\phi_{\ttbar}}}
\newcommand{\dPhittbar}{\ensuremath{\Delta\Phi_{\ttbar}}}
\newcommand{\mTt}{\ensuremath{m_{\mathrm{T}}^{t}}}
\newcommand{\mTtt}{\ensuremath{m_{\mathrm{T},\tau}^{t}}}
\newcommand{\mTtbar}{\ensuremath{m_{\mathrm{T}}^{\tbar}}}
\newcommand{\mTtbart}{\ensuremath{m_{\mathrm{T},\tau}^{\tbar}}}
\newcommand{\lamtau}{\ensuremath{\lambda_\tau}}
\newcommand{\nbar}{{\ensuremath{\bar{n}}}}
\newcommand{\pp}{{\ensuremath{\phantom{\prime}}}}
\newcommand{\hn}{{\ensuremath{{h_n^\pp}}}}
\newcommand{\hnp}{{\ensuremath{{h_n^{\prime}}}}}
\newcommand{\hnbar}{{\ensuremath{{h_\nbar^\pp}}}}
\newcommand{\hnbarp}{{\ensuremath{{h_\nbar^{\prime}}}}}

\newcommand{\qT}{\ensuremath{q_{\mathrm{T}}}}
\newcommand{\qTvec}{\ensuremath{\vec{q}_{\mathrm{T}}}}

\newcommand{\bT}{\ensuremath{b_\mathrm{T}}}
\newcommand{\bTvec}{\ensuremath{\vec{b}_{\mathrm{T}}}}

\newcommand{\kTvec}{\ensuremath{\vec{k}_{\mathrm{T}}}}

\newcommand{\NLO}{\ensuremath{\text{NLO}}\xspace}
\newcommand{\NNLO}{\ensuremath{\text{N$^2$LO}}\xspace}
\newcommand{\NNNLO}{\ensuremath{\text{N$^3$LO}}\xspace}
\newcommand{\NLOs}{\ensuremath{\text{NLO$_\text{s}$}}\xspace}
\newcommand{\NNLOs}{\ensuremath{\text{N$^2$LO$_\text{s}$}}\xspace}

\newcommand{\NLL}{\ensuremath{\text{NLL}}\xspace}
\newcommand{\NLLp}{\ensuremath{\text{NLL$'$}}\xspace}
\newcommand{\NNLL}{\ensuremath{\text{N$^2$LL}}\xspace}
\newcommand{\NNLLp}{\ensuremath{\text{N$^2$LL$'$}}\xspace}

\newcommand{\aNNLLp}{\ensuremath{\text{aN$^2$LL$'$}}\xspace}
\newcommand{\NLLNLO}{\ensuremath{\text{NLL+NLO}}\xspace}

\newcommand{\NNLLNNLO}{\ensuremath{\text{N$^2$LL+N$^2$LO}}\xspace}
\newcommand{\NNLLpNNLO}{\ensuremath{\text{N$^2$LL$'$+N$^2$LO}}\xspace}

\newcommand{\aNNLLpNNLO}{\ensuremath{\text{aN$^2$LL$'$+N$^2$LO}}\xspace}
\newcommand{\SCET}{\ensuremath{\text{SCET}}\xspace}

\newcommand{\SCETII}{\ensuremath{\text{SCET$_{\mathrm{II}}$}}\xspace}

\newlist{myitemize}{itemize}{3}
\setlist[myitemize]{leftmargin=14em}

\newcolumntype{C}{>{\centering\arraybackslash}p{0.14\textwidth}}

\newlength{\unitcharwidth}
\hypersetup{
  pdfauthor={Wan-Li Ju, Marek Schoenherr},
  pdftitle={Projected transverse momentum resummation in top-antitop pair production at LHC}
}
\preprint{IPPP/22/73\\MCnet-22-20}
\author{Wan-Li Ju${}^{(a,b)}$, Marek Sch{\"o}nherr${}^{(a)}$}
\title{Projected transverse momentum resummation in top-antitop pair production at LHC}
\institute{${}^{(a)}$~Institute for Particle Physics Phenomenology, Durham University, Durham DH1 3LE, United Kingdom\\
${}^{(b)}$~INFN, Sezione di Milano, Via Celoria 16, 20133 Milano, Italy}
\newcommand{\changed}[1]{#1}
\begin{document}
\vspace*{10mm}
\maketitle
\vspace*{20mm}
\begin{abstract} 
   The transverse momentum distribution of the $\ttbar$ system is of
   both  experimental and theoretical \changed{interest}. 
   In the presence of azimuthally asymmetric divergences, pursuing
   resummation at high logarithmic precision is rather
   demanding in general.
   In this paper, we propose the projected transverse momentum spectrum
   $\done\sigma_{\ttbar}/\done q_{\tau}$, which is derived from the
   classical $\qTvec$ spectrum by integrating out the rejection component
   $q_{\tau_{\perp}}$ with respect to a reference unit vector $\vec{\tau}$,
   to serve as an alternative solution to remove these asymmetric
   divergences, in addition to the azimuthally averaged case
   $\done\sigma_{\ttbar}/\done |\qTvec|$.
   In the context of the effective field theories, SCET$_{\mathrm{II}}$
   and HQET, we will demonstrate that in spite of the $q_{\tau_{\perp}}$
   integrations, the leading asymptotic terms of
   $\done\sigma_{\ttbar}/\done q_{\tau}$ still observe the factorisation
   pattern in terms of the hard, beam, and soft functions in the vicinity
   of $ q_{\tau}=0$~GeV.    
   \changed{
   Then, with the help of the renormalisation group equation techniques,
   we carry out the resummation at \NLLNLO, \NNLLNNLO,
   and approximate \NNLLpNNLO accuracy on three observables of interest,
   $\done\sigma_{\ttbar}/\done q_{\mathrm{T,in}}$,
   $\done\sigma_{\ttbar}/\done q_{\mathrm{T,out}}$,
   and $\done\sigma_{\ttbar}/\done \dphittbar$, within the domain $M_{t\bar{t}}\ge400~$GeV.
   The first two cases are obtained by choosing $\vec{\tau}$ parallel
   and perpendicular to the top quark transverse momentum, respectively.
   The azimuthal de-correlation $\dphittbar$ of the $t\bar{t}$ pair is
   evaluated through its kinematical connection to $q_{\mathrm{T,out}}$.}
   This is the first time the azimuthal
   spectrum $\dphittbar$ is appraised at or beyond the
   \NNLL level including a consistent treatment of both beam collinear
   and soft radiation.
\end{abstract}
\newpage
\tableofcontents
\section{Introduction}
\label{sec:intro}

Hadroproduction of top-antitop pairs plays a pivotal part
in the physics programme of the LHC experiments due to its role
in the precise extractions of fundamental parameters
of the Standard Model (SM).
It has thus drawn plenty of theoretical and experimental
attention in the recent years.
On the experimental side, the inclusive top-pair production
cross section has been measured at the colliding energies \changed{$\sqrt{s}=5.02~$TeV~\cite{ATLAS:2021xhc,ATLAS:2022jbj,CMS:2017zpm,CMS:2021gwv}, $7~$TeV~\cite{CMS:2013yjt,CMS:2016yys,ATLAS:2022aof}, $8~$TeV~\cite{CMS:2015auz,CMS:2016csa,CMS:2016yys,ATLAS:2022aof}, $13~$TeV~\cite{ATLAS:2019hau,ATLAS:2020aln,ATLAS:2020ccu,CMS:2019snc,CMS:2016rtp,CMS:2018fks,CMS:2021vhb}
and $13.6~$TeV~\cite{CMS:2022elr}},
respectively, whilst a large number of the differential
spectra have been published in the latest analyses
\cite{CMS:2019esx,ATLAS:2020ccu,CMS:2021vhb,ATLAS:2022xfj,
  CMS:2020tvq,CMS:2022uae}, including the transverse momentum
of the $\ttbar$ system $q_{\mathrm{T}}$, the invariant
mass of the top quark pair $\Mttbar$, and the azimuthal
opening angle of the top and antitop quarks $\dPhittbar$.
Theoretical calculations of these spectra also have since long
attracted a lot of interest in the community.
While NLO QCD corrections to top-pair production were determined
already some time ago~\cite{Nason:1987xz,Beenakker:1988bq,
  Beenakker:1990maa,Mangano:1991jk}, recent advances have
reached the NNLO accuracy \cite{Czakon:2013goa,Czakon:2015owf,
  Czakon:2016ckf,Czakon:2017dip,Czakon:2017wor,Catani:2019iny,
  Catani:2020tko}.
Top-quark decay effects were considered in \cite{Gao:2012ja,
  Brucherseifer:2013iv,Catani:2019hip,Behring:2019iiv,
  Czakon:2020qbd} and the electroweak (EW) corrections in
\cite{Bernreuther:2010ny,Kuhn:2006vh,Bernreuther:2006vg,
  Kuhn:2013zoa,Hollik:2011ps,Pagani:2016caq,Gutschow:2018tuk,
  Denner:2016jyo,Czakon:2017wor}.
Along with the progress made in fixed-order calculations, in
a bid to improve the perturbative convergence and in turn the
predictivity of the theoretical results, resummed calculations
have also been carried out within a variety of frameworks and
the kinematical limits.
Examples include the mechanic threshold \cite{Kidonakis:2014pja,
  Kidonakis:2010dk,Kidonakis:2009ev,Kidonakis:2014isa,
  Kidonakis:2019yji,Ahrens:2010zv,Ferroglia:2012ku,
  Ferroglia:2013awa,Pecjak:2016nee,Czakon:2018nun,
  Hinderer:2014qta}, the top-quark pair production threshold
\cite{Beneke:2009ye,Beneke:2010da,Beneke:2011mq,Cacciari:2011hy,
  Piclum:2018ndt,Ju:2020otc,Ju:2019mqc}, the low transverse
momentum domain~\cite{Zhu:2012ts,Li:2013mia,Catani:2014qha,
  Catani:2017tuc,Catani:2018mei}, and the narrow jettiness
regime~\cite{Alioli:2021ggd}.
Very recently, the combination of the fixed-order results with
a parton shower has been discussed in \cite{Mazzitelli:2020jio,
  Mazzitelli:2021mmm}.

This work will investigate the projection of the $\ttbar$
system's transverse momentum with respect to a reference
unit vector $\vec{\tau}$ on the azimuthal plane,
more explicitly,
$q_{\tau}\equiv|\vec{q}_{\tau_{\|}}|\equiv|\qTvec\cdot \vec{\tau}|$.
In contrast to the traditional transverse momentum spectrum
$\done\sigma_{\ttbar}/\done \qT$, $\qT=|\qTvec|$, where both components of
$\qTvec$ are measured and thereby constrained, the present
observable $\done\sigma_{\ttbar}/\done q_{\tau}$ only concerns
the projected piece $\vec{q}_{\tau_{\|}}$, leaving the
perpendicular part $\vec{q}_{\tau_{\perp}}$ unresolved and,
hence, it should be integrated out.
As will be demonstrated in this paper, the act of integrating
out the perpendicular components will introduce new and
distinguishing features to the $q_{\tau}$ spectrum,  particularly in 
regards to the treatment on the azimuthal asymmetric contributions
\cite{Nadolsky:2007ba,Catani:2010pd,Catani:2014qha,Catani:2017tuc}.
 
Induced by the soft and collinear radiation, the fixed-order
calculation of the $q_{\tau}$ distribution exhibits substantial
higher-order corrections in the small $q_{\tau}$ region.
This, thus, necessitates a resummation of the dominant contributions
in this regime to all orders to stabilise the perturbative predictions.
In order to accomplish this target, one of the prerequisite
conditions is to determine the dynamic regions driving the
asymptotic behaviour in the limit $q_{\tau}\to0$.
For the classic transverse momentum resummation, this
analysis was first presented for Drell-Yan production in
\cite{Collins:1980ui} by means of inspecting the power laws
of a generic configuration on the pinch singularity surface
\cite{Sterman:1978bi,Landau:1959fi,Coleman:1965xm}.
It was proven that the leading singular behaviour in the small $\qT$
domain was well captured by the beam-collinear, soft, and hard
regions.
However, this conclusion cannot be straightforwardly   applied
onto the $q_{\tau}$ resummation in top-pair production, as the
deep recoil configuration
$|\vec{q}_{\tau_{\perp}}|\sim M_{\ttbar}\gg q_{\tau}$, which stems
from the integral over the perpendicular component, was kinematically
excluded in \cite{Collins:1980ui}.
Therefore, for delivering an honest and self-consistent study on
$\done\sigma_{\ttbar}/\done q_{\tau}$, this work will reappraise
the scalings of the relevant configurations, comprising both the
$\qT$-like configuration $\Mttbar\gg|\vec{q}_{\tau_{\perp}}|\sim q_{\tau}$
and the asymmetric one $|\vec{q}_{\tau_{\perp}}|\sim\Mttbar\gg q_{\tau}$.

To this end, we will exploit the strategy of expansion by regions
\cite{Beneke:1997zp,Smirnov:2002pj,Smirnov:2012gma,Jantzen:2011nz}
to motivate the momentum modes governing the low $q_{\tau}$ regime,
which will cover the beam-collinear, soft, central-jet, and hard
regions.
Then, the soft-collinear effective theory (SCET) \cite{Bauer:2001yt,
  Bauer:2001ct,Bauer:2000yr,Bauer:2000ew,Bauer:2002nz,Beneke:2002ph,
  Beneke:2002ni,Bauer:2002aj,Lange:2003pk,Beneke:2003pa} and the
heavy quark effective theory (HQET)~\cite{Eichten:1989zv,Georgi:1990um,
  Grinstein:1990mj,Neubert:1993mb} are used to embody those dynamic
modes, thereby calculating the effective amplitudes and the respective
differential cross sections.
After carrying out a multipole expansion, the results constructed
by those dynamic regions all reflect the unambiguous scaling behaviors,
which can be determined from the power prescriptions of the relevant
effective fields.
From the outcome, we point out that the leading asymptotic behavior
of $\done\sigma_{\ttbar}/\done q_{\tau}$ is still resultant of the
symmetric configuration $\Mttbar\gg |\vec{q}_{\tau_{\perp}}|\sim q_{\tau}$,
which is in practice dictated by the beam-collinear, soft, and hard
momenta, akin to the case of $\done\sigma_{\ttbar}/\done\qT$,
whereas the contributions from the
$|\vec{q}_{\tau_{\perp}}|\sim\Mttbar\gg q_{\tau}$ pattern are
suppressed by at least one power of
$\lambda_{\tau}\equiv q_{\tau}/\Mttbar$.
 
Upon the identification of the leading regions, we make use of
the decoupling properties of the soft modes~\cite{Bauer:2001yt,
  Beneke:2010da} to derive the factorisation formula for
$\done\sigma_{\ttbar}/\done q_{\tau}$.
Owing to the integration over $\vec{q}_{\tau_{\perp}}$, the
impact space integrals herein are all reduced from 2D to 1D.
Thus, the azimuthal asymmetric contributions, which in principle
contain divergent terms in the asymptotic regime in the general
$\done\sigma_{\ttbar}/\done \qTvec$ cross section after completing
the inverse Fourier transformations, do not contribute any
divergences to the $q_{\tau}$ spectrum.
This is the second $\qTvec$-based observable free of asymmetric
singularities in addition to the azimuthally averaged spectra
$\done\sigma_{\ttbar}/\done \qT$~\cite{Zhu:2012ts,Li:2013mia,Catani:2018mei}.

To implement the resummation, we employ the renormalisation group
equations (RGE) and the rapidity renormalisation group equations (RaGE)
to evolve the intrinsic scales in the respective ingredients and
in turn accomplish the logarithmic exponentiations~\cite{Chiu:2011qc,
  Chiu:2012ir,Li:2016axz,Li:2016ctv}.
Alternative approaches can also be found in \cite{Collins:1984kg,
  Catani:2000vq,Bozzi:2005wk,Bozzi:2007pn,Ebert:2016gcn,Monni:2016ktx,
  Bizon:2017rah,Becher:2010tm,GarciaEchevarria:2011rb,Becher:2011dz}.
For assessing the resummation accuracy, we take the logarithmic
counting rule $L_{\mathrm{M}}\sim\alpha_s^{-1}\sim\lambda_L^{-1}$
throughout and organise the perturbative corrections to the relevant
sectors in line with the following prescription,
 \begin{equation}\label{eq:intro:def_res_order}
  \begin{split}
 \frac{   \done{\sigma}_{\ttbar}} {\done q_{\tau}  }
    \sim&\;
    \sigma^\text{Born}_{\ttbar}
    \,
      \exp\bigg[\;
              \underbrace{L_{\mathrm{M}}\, f_0(\alpha_sL_{\mathrm{M}})}_{(\text{LL})}
            +\underbrace{f_1(\alpha_sL_{\mathrm{M}})}_{(\text{NLL,NLL$'$})}
            +\underbrace{\alpha_s\, f_2(\alpha_sL_{\mathrm{M}})}_{(\text{N$^2$LL,N$^2$LL$'$})}
            +\underbrace{\alpha^2_s\, f_3(\alpha_sL_{\mathrm{M}})}_{(\text{N$^3$LL,N$^3$LL$'$})}
            +\ldots\;
          \bigg] \\
        &\quad\times
      \bigg\{
        1          (\text{LL,NLL})+
        \alpha_s   (\text{NLL$'$,N$^2$LL})+
        \alpha_s^2 (\text{N$^2$LL$'$,N$^3$LL})+
        \alpha_s^3 (\text{N$^3$LL$'$,N$^4$LL})+
        \ldots
      \bigg\} \, .
  \end{split}
\end{equation} 
Therein, the desired precisions of the anomalous dimensions are
specified between the square brackets in the exponent as for a
given logarithmic accuracy, while the according requirements on
the fixed-order elements are presented within the curly brackets.
In this work, we will evaluate and compare the resummed $q_{\tau}$
spectra on the next-to-leading-logarithmic (NLL), \NNLL, and
approximate \NNLLp (\aNNLLp) levels.

The paper is structured as follows.
In Sec.~\ref{sec:fac}, we will utilise the strategy of expansion
by dynamic regions \cite{Beneke:1997zp,Smirnov:2002pj,Smirnov:2012gma,
  Jantzen:2011nz} and effective field theories, i.e.\
\SCETII~\cite{Bauer:2002aj,Lange:2003pk,Beneke:2003pa}
and HQET~\cite{Eichten:1989zv,Georgi:1990um,Grinstein:1990mj,Neubert:1993mb},
to derive the factorisation formula governing the leading asymptotic
behaviour of ${\done{\sigma}_{\ttbar}} / {\done q_{\tau}  }$ in the
limit $q_{\tau}\to0$.
Then, the (rapidity) renormalisation group equations will be solved
in Sec.~\ref{sec:res} for the respective sectors participating into
the factorisation formula, from which we exponentiate the characteristic
logarithmic constituents in the impact space and thereby accomplish
the resummation of the singular terms in the momentum space.
Sec.~\ref{sec:results} will be devoted to the numeric evaluations on
the spectra $q_{\tau}$.
Therein, we will at first validate the approximations of our
factorisation formula up to \NNLO, and then present the resummation
improved differential distributions for three particular observables,
$\done\sigma_{\ttbar}/q_{\mathrm{T},\mathrm{in}}$,
$\done\sigma_{\ttbar}/q_{\mathrm{T},\mathrm{out}}$, and
$\done\sigma_{\ttbar}/\dphittbar$. 
\changed{
$q_{\mathrm{T},\mathrm{in(out)}}$ is a special case of $q_{\tau}$ on the choice of $\vec{\tau}$ parallel (perpendicular) to the top quark transverse momentum, while  $\dphittbar$ represents   the azimuthal de-correlation  of the $t\bar{t}$ pair  and can be extracted  through its kinematical connection to $q_{\mathrm{T,out}}$.} Finally, we will offer some concluding remarks in Sec.~\ref{sec:conclusions}.

\section{Factorisation}
\label{sec:fac}

\subsection{Kinematics and the factorised cross section}
\label{sec:fac:kin}

\begin{figure}[t!]
  \centering
  \includegraphics[width=0.45\textwidth]{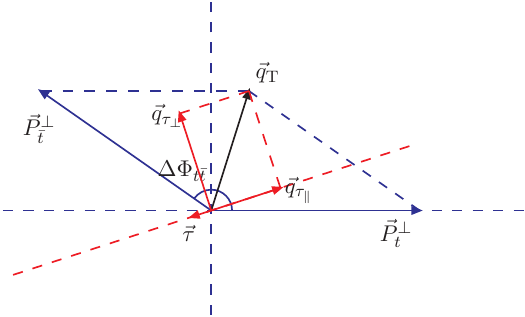}
  \caption{
    The kinematics on the transverse plane in the laboratory reference frame.
    $\vec{P}_{t(\tbar)}^{\perp}$ stands for the transverse momentum
    of the (anti-)top quark.
    $\dPhittbar$ is the azimuthal opening angle between
    the top and anti-top quarks.
    $\vec{\tau}$ is a unit reference vector in the transverse plane.
    \label{fig:KinConfig}
  }
\end{figure}

We start this section with the elaboration on the kinematics. 
As illustrated in Fig.~\ref{fig:KinConfig}, the main concern of
this work is on the interplay between a reference unit vector
$\vec{\tau}$ and the transverse momentum $\qTvec$ of the
$\ttbar$ system.
By means of the reference vector $\vec{\tau}$,  $\qTvec$ can be
decomposed into two parts,  the projection component
$\vec{q}_{\tau_{\|}}$ and  the rejection one
$\vec{q}_{\tau_{\perp}}$, i.e.,
\begin{align} 
\qTvec= \vec{q}_{\tau_{\perp}}+ \vec{q}_{\tau_{\|}}= q_{\tau_{\perp}}\vec{\tau}\times\vec{n}+ q_{\tau_{\|}}\vec{\tau}\,,
\end{align} 
where $\vec{n}$ is another unit vector pointing to one of beam
directions in the laboratory reference frame.
In the numeric implementation presented in this paper, the
magnitude of the projection $\vec{q}_{\tau_{\|}}$ is of
primary interest, which will hereafter be dubbed
$q_{\tau}\equiv |\vec{q}_{\tau_{\|}}|$.

The fixed-order calculation on the $q_{\tau}$ spectrum can be
realised using the QCD factorisation theorem of \cite{Collins:1989gx},
that is,
\begin{equation}\label{eq:QCDF}
  \begin{split}
    \frac{\done^5{\sigma_{\ttbar}}}
         {\done \Mttbar^2\,\done^2\vec{P}_{t}^{\perp}\,
          \done \Yttbar\,\done q_{\tau}}
    \,=\;&
      \sum_{\mathrm{sign}[P_{t}^{z}]}\frac{1}{16s(2\pi)^6 }
      \int \done^2 \qTvec\,\Theta_{\mathrm{kin}}\,
      \delta\Big[q_{\tau}-|\qTvec \cdot \vec{\tau}|\Big]\,
      \frac{\Sigma_{\ttbar}}{\MTttbar\,|P_{t}^{z}|}\,,
  \end{split}
\end{equation}
where $\Mttbar$ denotes the invariant mass of the $\ttbar$ system,
and $s$ is the colliding energy.
In this work we will concentrate on $\sqrt{s}=13$~TeV throughout.
$\Yttbar$ and $\MTttbar$ are for the pseudorapidity
and the transverse mass of the $\ttbar$ pair in the laboratory
frame (LF), respectively.
$P_{t}^{z}$ represents the longitudinal components of the top quark
momentum measured from the $z$-direction rest frame ($z$RF) of the
$\ttbar$ pair.
The $z$RF can be obtained through boosting the LF along one of the
beam directions until the longitudinal momentum of the $\ttbar$
pair has been eliminated.

To perform the integral of $\qTvec$ in Eq.~\eqref{eq:QCDF}, it is
of essence to establish suitable kinematical boundaries to fulfill
energy-momentum conservation condition.
To this end, we introduce the function $\Theta_{\mathrm{kin}}$ to
impose the following constraints,
\begin{equation}\label{eq:thetakin:ft}
  \begin{split}
    \Theta_{\mathrm{kin}}
    \,=\,&
      \Theta\Big[\sqrt{s}-\MTttbar-|\qTvec|\Big]\,
      \Theta\Big[\MTttbar-\mTt-\mTtbar\Big]\,
      \Theta\left[\sinh^{-1}
                  \left(\sqrt{\frac{(\Mttbar^2+s)^2}
                                   {4\,s\,{\MTttbar}^2}-1}
                  \right)
                  -|\Yttbar|\right]\,,
  \end{split}
\end{equation}
where $\Theta[\dots]$ is the usual Heaviside function.
Therein, $\mTt$ and $\mTtbar$ are the transverse
masses of the top and anti-top quarks in the LF, respectively.
Finally, Eq.~\eqref{eq:QCDF} also entails ${\Sigma_{\ttbar}}$,
encoding the contributions from all partonic processes, defined as,
\begin{equation}
  \begin{split}\label{eq:def:QCD:parton}
    {\Sigma_{\ttbar}}
    \,=&\;
      \sum_{i,j} \int^1_0 \frac{\done x_n}{x_n}\,
                          \frac{\done x_{\nbar}}{x_{\nbar}}\,
      f_{i/N}(x_n)\,f_{j/\bar{N}}(x_{\nbar})\,
      \sum_{ r}\,\int\, \prod_{m}^r\done \Phi_{k_m}\;
      (2\pi)^4\delta^4\left(p_i+p_{j}-p_{t}-p_{\bar{t}}-\sum_{m} k_m\right)\\
    &\;
      \times\overline{\sum_{\mathrm{hel, col}}}
            \left|\mathcal{M}(i+j\to t+\bar{t}+X ) \right|^2 \,.
  \end{split}
\end{equation}
Here $f_{i/N}(x)$ is the parton distribution function (PDF) for the
parton $i$ within the proton $N$ carrying the momentum fraction $x$.
$\done \Phi_{k_m}$ characterises
the phase space integral of the $m$-th emitted parton, that is, 
\begin{equation}
  \begin{split}\label{eq:def:QCD:phaseInt}
    \done \Phi_{k_m}
    \,\equiv\,
      \frac{\done^4 k_{m}}{(2\pi)^3}\,
      \delta(k_m^2)\,\Theta(k_m^0)
    \,=\,
      \frac{1}{2}\;\frac{\done y_{m}}{2\pi}\;
      \frac{\done^2 \vec{k}_{m}^{\perp}}{(2\pi)^2}\,,
  \end{split}
\end{equation}
where $y_{m}$ and $\vec{k}_{m}^{\perp}$ indicate the rapidity and
transverse momentum of the occurring emission, respectively.
$|\mathcal{M}|^2$ is the squared transition amplitude
for the partonic processes of the indices indicated.
Substituting Eq.~\eqref{eq:def:QCD:parton} into Eq.~\eqref{eq:QCDF},
it is ready to perform the fixed-order calculations of the spectra
of $q_{\tau}$.
In the vicinity of $q_{\tau}=0$~GeV, however, this perturbative
expansion fails to converge, and
an asymptotic expansion of $\done\sigma_{\ttbar}/\done q_{\tau}$
can be carried out in the small parameter
$\lamtau\equiv q_{\tau}/\Mttbar$,
\begin{align}\label{eq:AsyExp}
\frac{\done  {\sigma_{\ttbar}} }{\done q_{\tau} }
 \;\sim\;
 \sigma^{\ttbar}_{\mathrm{B}}
 \,
 \sum_{m,n}
  \,
   \left[
   \frac{\alpha_s(\Mttbar)}{4\pi}\right]^m
    \,
    \left[
    \underbrace{c^{(0)}_{m,n}\,\frac{\ln^{n}(\lamtau)}{\lamtau}}_{\mathrm{LP}}
    +
     \,
     \underbrace{c^{(1)}_{m,n} \, \ln^{n}(\lamtau)}_{\mathrm{NLP}}
     +  
     \,
     \underbrace{c^{(2)}_{m,n} \, \lamtau\ln^{n}(\lamtau)}_{\mathrm{N}^2\mathrm{LP}}
     + \dots \right]\;,
\end{align}
indicating the leading, next-to-leading and next-to-next-to-leading power
terms in $\lambda_\tau$, labelled LP, NLP, and N$^2$LP, respectively.
Therein, $\sigma^{\ttbar}_{\mathrm{B}}$ is the Born level total
cross section of the process $pp\to \ttbar+X$,
$\alpha_s$ denotes the strong coupling constant, and
$c^{(k)}_{m,n}$ represents the coefficient for the asymptotic constituent with the superscript $k$ specifying the occurring power.
Thus, conventionally, the leading power terms
$c^{(0)}_{m,n}\ln^n(\lamtau)/\lamtau$
are associated with the most singular behaviors
in the low $q_{\tau}$ domain and also the main concern of this work.
It is important to note, however, that also the next-to-leading power
terms, $c^{(1)}_{m,n}\ln^{n}(\lamtau)$, are divergent as
$\lambda_\tau\to 0$.

\subsection{Dynamic regions}
\label{sec:fac:dynreg}

Based on the strategy of expansion of dynamic regions
\cite{Beneke:1997zp,Smirnov:2002pj,Smirnov:2012gma,Jantzen:2011nz},
the asymptotic series of Eq.~\eqref{eq:AsyExp} can be interpreted
with the aid of a set of regions from the phase space and loop integrals.
This work, in particular, will choose the formalism of
\cite{Jantzen:2011nz}.
We base the definition of our regions on the criterion of
domain completeness, i.e.\ the existence of a set of non-intersecting
dynamic regions that cover the whole integration domain.
This criterion plays an essential role in consistently extrapolating
the expanded integrands from their own convergent domains to the
entire integration ranges.
Other constraints are also imposed therein, including the
regularisation of the expanded integrands and the (at least partial)
commutativity amongst the asymptotic expansions.
The former case can be fulfilled by introducing the rapidity
regulator~\cite{Manohar:2006nz,Becher:2010tm,Becher:2011dz,Chiu:2011qc,Chiu:2012ir,Li:2016axz,Li:2016ctv} in implementing the
\SCETII formalism \cite{Bauer:2002aj,Lange:2003pk,Beneke:2003pa}.
However, for the latter criterion, we assume that all the
non-commutative dynamic regions, such as the collinear-plane
modes~\cite{Jantzen:2011nz}, will cancel out in the eventual
$q_{\tau}$ spectra.
It merits noting that, this ansatz, together with the proposal
of \cite{Jantzen:2011nz}, has been scrutinised only within the
one-loop integrals in the various kinematical limits.
We regard their effectiveness on $\ttbar$ hadroproduction as
the primary hypothesis in this work.
Recent developments on the criteria to implement the region
analysis can be found in
\cite{Ananthanarayan:2018tog,Plenter:2020lop,Heinrich:2021dbf}.

As the first application of the domain completeness, we explore the
relevant modes for the $\vec{q}_{\tau_{\perp}}$ integral here.
From Eq.~\eqref{eq:QCDF}, two types of scales will be
involved, one of $\order(q_{\tau})$ and the other of
$\order(\Mttbar)$.
In order to disentangle the influences of those two scales and to
also fulfil the constraints of \cite{Jantzen:2011nz}, we identify
the dynamic regions for the rejection component as follows
\begin{align}
  \mathbf{Isotropic}\textrm{-}\mathbf{recoil}:
  &\;
    \Mttbar\gg |\vec{q}_{\tau_{\perp}}|\sim q_{\tau}\,,
    \label{eq:def:region:tt:sym}\\
  \mathbf{Asymmetric}\textrm{-}\mathbf{recoil}:
  &\;
    |\vec{q}_{\tau_{\perp}}|\sim \Mttbar\gg q_{\tau}\,.
    \label{eq:def:region:tt:asy}
\end{align}
To precisely separate both regimes and still cover the complete
integration range, we introduce the auxiliary boundary $\Lambda_{\tau}$
satisfying $M_\ttbar\gg\Lambda_{\tau}\gg q_{\tau}$, from which the two
non-intersecting intervals read
$|\vec{q}_{\tau_{\perp}}|\lesssim\Lambda_{\tau}$ and
$\Lambda_{\tau} \lesssim|\vec{q}_{\tau_{\perp}}|$.
However, as demonstrated in \cite{Jantzen:2011nz}, those auxiliary
boundary dependences will all drop out after assembling all
relevant domains.
Their dependence is thus dropped in the following.
An analogous analysis is also applied to the phase space integral
for the real emissions in Eq.~\eqref{eq:def:QCD:phaseInt},
which leads to,
\begin{align}
  \mathbf{Soft}:\;
  &\;
    {y_m} \sim 0\,,\quad\quad\quad\quad~ |\vec{k}_m^{\perp}|\sim q_{\tau}\,,
    \label{eq:def:region:soft:parton}\\
  \mathbf{Beam}\textrm{-}\mathbf{collinear}\textrm{-}n:\,
  &\;
    {y_m} \sim +\ln\lamtau\,,\quad\quad |\vec{k}_m^{\perp}|\sim q_{\tau}\,,
    \label{eq:def:region:beamN:parton}\\
  \mathbf{Beam}\textrm{-}\mathbf{collinear}\textrm{-}\nbar:\;
  &\;
    {y_m} \sim -\ln\lamtau\,,\quad\quad |\vec{k}_m^{\perp}|\sim q_{\tau}\,,
    \label{eq:def:region:beamNbar:parton}\\
  \mathbf{Jet}\textrm{-}\mathbf{collinear}\textrm{-}{n_{\mathrm{J}}}:\;
  &\;
    {y_m} \sim 0\,,\quad\quad\quad\quad~ |\vec{k}_m^{\perp}|\sim \Mttbar\,.\label{eq:def:region:jet:parton}
\end{align}
Therein, the decomposition of the $\vec{k}_{m}^{\perp}$ integration
range respects a similar pattern as the $\vec{q}_{\tau_{\perp}}$ case.
The rapidity integrals are broken down with respect to the reference
points $\pm \Lambda_{y_{m}}$ with the relationship
$0\ll\Lambda_{y_{m}}\ll\ln\lamtau$, from which three
non-overlapping regions emerge, ${y_m}\lesssim -\Lambda_{y_{m}}$,
$-\Lambda_{y_{m}}\lesssim{y_m}\lesssim  \Lambda_{y_{m}}$, and
$\Lambda_{y_{m}}\lesssim {y_m}$.
As before, the dependence on the auxiliary boundary $\Lambda_{y_{m}}$
cancels, and we omit it in the following.
In addition, please note that in deriving
Eq.~\eqref{eq:def:region:jet:parton} the super-hard-collinear
domains ${y_m} \sim\pm \ln\lamtau$ and
$|\vec{k}_m^{\perp}|\sim \Mttbar$ are ignored as they
explicitly contradict the energy-momentum conservation condition
in Eq.~\eqref{eq:def:QCD:parton}.


Apart from the above momentum modes, the amplitudes in
Eq.~\eqref{eq:def:QCD:parton} also contain the loop integrals.
A region analysis of this case can proceed in principle in a
similar way as for the real emission corrections.
However, due to the variability in the offshellness and
the varying multiplicity of the external particles,
exhausting all relevant circumstances herein is usually
more challenging.
To this end, this work will utilise momentum
regions that can lead to pinched singularities (PS)
in a hadron collider process~\cite{Landau:1959fi,Coleman:1965xm,Sterman:1978bi,Collins:1980ui,Collins:1981ta,Collins:1985ue},
which in general consists of hard, collinear, soft,
Coulomb, and Glauber regions.
During our calculation, to circumvent the complexity induced
by Coulomb singularities, we introduce the lower cutoff
upon $\Mttbar$ in the phase space integral to stay clear of
the $\ttbar$ production threshold.
In order to cope with the remaining hard, collinear, and soft
components, we apply effective field theories, i.e.\
\SCETII~\cite{Bauer:2002aj,Lange:2003pk,Beneke:2003pa}
and HQET~\cite{Eichten:1989zv,Georgi:1990um,Grinstein:1990mj,Neubert:1993mb},
onto the massless and heavy partons, respectively.
Regarding Glauber gluon exchanges, even though their cancellation
for inclusive observables has been demonstrated in
colour-singlet production in hadronic collisions using a variety of
approaches~\cite{Collins:1997sr,Collins:2004nx,Gaunt:2014ska,Schwartz:2018obd},
a systematic discussion of their effects on the production of
coloured systems, like hadronic $\ttbar$ pair production, is
still absent.
While leftover soft contributions were observed in $\ttbar$
production in the context of light-cone ordered perturbation
theory~\cite{Chang:1968bh,Sterman:1978bj,Mitov:2012gt}, it
nevertheless deserves further confirmation from
perturbative QCD, in particular as to whether those soft
remnants are eikonalisable or not
\cite{Rothstein:2016bsq,Rogers:2010dm}.
In this work, we will follow the approach of
\cite{Catani:2014qha,Catani:2017tuc}, and assume the
irrelevance of the Glauber contributions.
Recent developments in generalizing \SCET to encode the Glauber
interactions can be found in \cite{Donoghue:2009cq,Bauer:2010cc,Fleming:2014rea,Rothstein:2016bsq,Moult:2022lfy}.

We are now ready to summarise the dynamic modes presiding over
the virtual and real corrections to the $q_{\tau}$ spectra,
\begin{align}
  \label{eq:regions:hard}
  \mathbf{Hard}:
  &
    \quad k_{h} ~\sim\;\Mttbar\,\big[\order(1),\order(1),\order(1)\big]_n\,,
    \\
  \label{eq:regions:soft}
  \mathbf{Soft}:
  &
    \quad k_{s}~ \sim\;\Mttbar\,\big[\order(\lamtau),\order(\lamtau),\order(\lamtau)\big]_n\,,
    \\
  \label{eq:regions:beam:n}
  \mathbf{Beam}\textrm{-}\mathbf{collinear}\textrm{-}{ n}:
  &
    \quad k_{c} ~\sim\;\Mttbar\,\big[\order(1),\order(\lamtau^2),\order(\lamtau )\big]_n\,,
    \\
  \label{eq:regions:beam:nbar}
  \mathbf{Beam}\textrm{-}\mathbf{collinear}\textrm{-}{ \nbar}:
  &
  \quad k_{\bar{c}} ~\sim\;\Mttbar\,\big[\order(\lamtau^2),\order(1),\order(\lamtau )\big]_n\,,
  \\
  \label{eq:regions:jets}
    \mathbf{Jet}\textrm{-}\mathbf{collinear}\textrm{-}{n_{\mathrm{J}}}:
  &
  \quad k_{\mathrm{J}} ~\sim\;\Mttbar\,\big[\order(1),\order(\lamtau^2),\order(\lamtau )\big]_{n_{\mathrm{J}}}\,.
\end{align} 
In writing those momentum modes,  the light-cone coordinate system has been applied, from which an arbitrary momentum $p^{\mu}$ are  decomposed and reforged as
\begin{equation}\label{eq:def:lc}
  p^{\mu}
  =
    \frac{p\cdot \nbar_k}{2}n_k^{\mu}+\frac{p\cdot n_k}{2}\nbar_k^{\mu}+ {p}^{ \perp,\mu}_{n_k}
  \equiv
    \frac{p_{\nbar_k}}{2}n_k^{\mu}+\frac{p_{ n_k}}{2}\nbar_k^{\mu}+ {p}^{ \perp,\mu}_{n_k}
  \equiv
    \big[p_{\nbar_k},p_{n_k},{p}_{n_k}^{\perp} \big]_{n_k}.
\end{equation}
Therein, $n_k$ and $\nbar_k$ are two light-like  vectors with the
normalisation conditions $n_k\cdot \nbar_k=2$ and $n^0_k=\nbar^0_k=1$.
Throughout this paper, we will make use of the symbols $n$ ($\nbar$)
and $n_{\mathrm{J}} $ to characterise the positive (negative) beam and
jet directions, respectively.
In accordance with Eq.~\eqref{eq:def:region:jet:parton}, at least one of
the transverse components of the vector $n^{\mu}_{\mathrm{J}}$ should be
non-vanishing.
In establishing Eqs.~\eqref{eq:regions:soft}-\eqref{eq:regions:jets},
the offshellnesses of the collinear and soft fluctuations are assigned
to be of $\order(\lambda^2_{\tau})$ so as to accommodate the projected
transverse momentum.
The hard-collinear degree of freedom, e.g.,\
$k_{hc}\sim\Mttbar\,[\order(\lamtau),\order(1),\order(\sqrt{\lamtau})]_n$, is neglected in the
analysis as it always results in a scaleless integral for the leading
power accuracy investigated in this paper.
 
By comparison with the calculations on $\done\sigma_{\ttbar}/\done\qT$
of \cite{Zhu:2012ts,Li:2013mia,Catani:2014qha,Bonciani:2015sha,Catani:2017tuc,Catani:2018mei}, in addition to the common pattern of
the hard, beam-collinear, and soft regions, this work will also take
into account the central jet mode as shown in Eq.~\eqref{eq:regions:jets}.
For the low $\qT$ domain, the subleading nature of the jet region has
been demonstrated in \cite{Collins:1980ui,Collins:1981uk} by analyzing
the powers of the relevant PS surfaces.
This conclusion is also expected to hold for our symmetric configuration
in Eq.~\eqref{eq:def:region:tt:sym}.
\changed{
Nevertheless, aside from the isotropic recoil, the region expansion
strategy in \cite{Jantzen:2011nz} also leads to the asymmetric
configuration of Eq.~\eqref{eq:def:region:tt:asy} participating
in the $q_{\tau}$ spectra.
In light of the different scaling behaviors in both regions, it is
\emph{a priori} not clear whether the conclusion in
\cite{Collins:1980ui,Collins:1981uk} derived for the isotropic
pattern of Eq.~\eqref{eq:def:region:tt:sym} is still applicable
to the asymmetric configuration of Eq.~\eqref{eq:def:region:tt:asy}.
To this end, we will
revisit the power rules of all the relevant configurations induced by
Eqs.~\eqref{eq:def:region:tt:sym}-\eqref{eq:def:region:tt:asy} and
Eqs.~\eqref{eq:regions:hard}-\eqref{eq:regions:jets} in the following
subsections.
}

\changed{
To facilitate our discussion, we categorise the $q_{\tau}$
spectrum according to the number of the embedded jet modes
$N_{\mathrm{J}}$, more explicitly,
\begin{equation}\label{eq:def:NJ0:NJm}
  \frac{\done^5{\sigma_{\ttbar}}}
       {\done \Mttbar^2\,\done^2\vec{P}_{t}^{\perp}\,\done\Yttbar\,\done q_{\tau}}
  \longrightarrow\,
    \sum_{m=0}^{\infty}\,
    \frac{\done^5{\sigma_{\ttbar}}}
         {\done \Mttbar^2\,\done^2\vec{P}_{t}^{\perp}\,\done\Yttbar\,\done q_{\tau}}
    \Bigg|_{N_{\mathrm{J}}=m}
    \,.
\end{equation}
In Sec.~\ref{sec:fac:NJ0}, we will concentrate on the $N_{\mathrm{J}}=0$
configuration, elaborating on the factorisation properties of the
occurring constituents and utilising SCET$_{\mathrm{II}}$ and HQET
to determine the power accuracy.
Sec.~\ref{sec:fac:NJm} will then be devoted to all configurations
comprising at least one jet.
This discussion will be subdivided into Sec.~\ref{sec:fac:NJ1},
examining the $N_{\mathrm{J}}=1$ configuration as an example to
present the generic scaling feature in presence of the jet mode,
and Sec.~\ref{sec:fac:NJ2}, where we move on to the more
general situation, the $N_{\mathrm{J}}\ge2$ contributions,
enumerating all the possible scaling behaviors brought about by
the various jet momenta present.
At last, we will summarise our observations and compare the power
prescriptions derived in EFT with those established in
\cite{Collins:1980ui,Collins:1981uk}.
}

\subsection{The case of \texorpdfstring{$N_{\mathrm{J}}=0$}{NJ=0}}
\label{sec:fac:NJ0}

\begin{figure} [h!]
  \centering
  \includegraphics[width=0.5\textwidth]{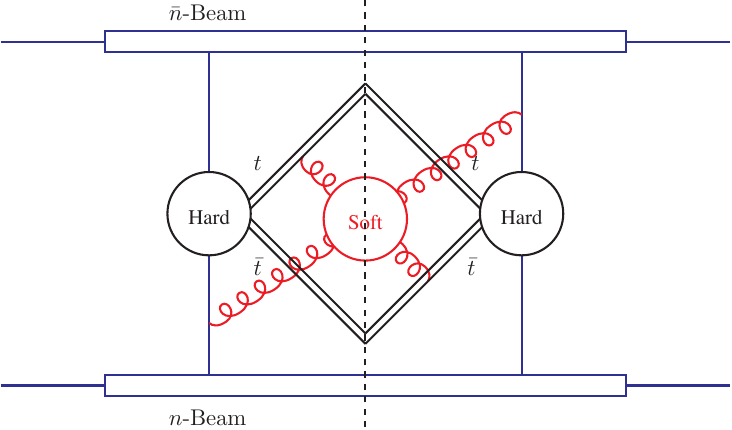}
  \caption{
    Dynamic regions in the $N_{\mathrm{J}}=0$ configuration. Herein, the black bubbles stand for the hard modes.
    The red bubble and springs encode the soft fluctuations connecting the heavy partons represented by the black double lines as well as the collinear modes depicted in  the blue straight lines.
    \label{fig:PSS:NJ0}
  }
\end{figure}

In this part, we will discuss the dynamic regions contributing to
the $N_{\mathrm{J}}=0$ configuration, which comprises the hard,
beam-collinear, and soft regions, as exhibited in
Eqs.~\eqref{eq:regions:hard}-\eqref{eq:regions:beam:nbar}.
As opposed to the beam-collinear and soft regions, which can be
assigned to both the phase space and loop integrals, the hard
mode is only present in the virtual processes.
In this regard, the transverse momentum for the $\ttbar$ system
observes only the isotropic pattern in Eq.~\eqref{eq:def:region:tt:sym},
which thus permits us to perform the expansion on the kinematic
variables in Eq.~\eqref{eq:QCDF} as follows,
\begin{align}
  \label{eq:Pzt:NJ0}
  P^z_t\Bigg|_{N_{\mathrm{J}}=0}
  &=\;
    \frac{x_t\bttbar }{\sqrt{1-\beta^2_{\ttbar}}}\,m_t\,
    +\order(\lamtau)
  \,,
\end{align}
where $\bttbar\equiv\sqrt{1-4m_t^2/\Mttbar^2}$ with $m_t$
denoting the mass of the (anti-)top quark.
$x_t$ is the cosine of the scattering angle of the top quark
in the $\ttbar$ rest frame.
Further,
\begin{align}
  \label{eq:Ettbar:NJ0}
  \MTttbar\Bigg|_{N_{\mathrm{J}}=0}
  &=\;
    \Mttbar\,
    +\order(\lamtau)\,,
\end{align}
and, thus,
\begin{align}
  \label{eq:thetakin:NJ0}
  \Theta_{\mathrm{kin}}\Bigg|_{N_{\mathrm{J}}=0}
  &=\;
    \Theta\left(\sqrt{s}-\Mttbar^2\right)\,
    \Theta\left(\Mttbar-2\sqrt{m_t^2+(\vec{P}^{\perp}_{t})^2}\right)\,
    \Theta\left[\sinh^{-1}\left(\frac{s-\Mttbar^2}{2\Mttbar\sqrt{s}}\right)-|\Yttbar|\right]
    +\order(\lamtau)\nnb\\
  &\equiv\;
    \Theta_{\mathrm{kin}}^{(0)}+\order(\lamtau)\,.
\end{align}
Please note, in the results above only the LP contributions are kept.
Similarly, in the following, we define the LP term in
Eq.~\eqref{eq:thetakin:NJ0} as $\Theta_{\mathrm{kin}}^{(0)}$ hereafter.
 
The remaining task is now to expand the partonic convolution
function $\Sigma_{\ttbar}$ in line with
Eqs.~\eqref{eq:regions:hard}-\eqref{eq:regions:beam:nbar}.
This can be achieved by means of the effective field theories,
i.e.\ \SCETII and HQET.
Therein, the beam-collinear regions are embodied in terms of
the gluon and quark fields, $A_{n(\bar{n})}$ and $\xi_{n(\bar{n})}$~\cite{Beneke:2002ni,Hill:2002vw,Lange:2003pk,Beneke:2003pa} in \SCETII.
The soft fluctuations on the heavy (anti-)quark are encoded by
the field $h_v(\chi_v)$ from HQET, while those from the incoming
gluons and massless quarks are reflected by the fields $A_{s}$
and $q_{s}$~\cite{Lange:2003pk,Beneke:2003pa} in \SCETII.
The hard mode can be taken care of by the effective Hamiltonian
$\mathcal{H}_{\ttbar}^{\mathrm{eff}}$~\cite{Ahrens:2010zv},
which consists of products of Wilson coefficients and the
corresponding field operators.

Up to LP, the interactions in the individual momentum regions are
governed by the effective Lagrangians
\cite{Bauer:2002aj,Lange:2003pk,Beneke:2003pa,Neubert:1993mb},
\begin{align}
  \label{eq:scetII:Ln}
  \mathcal{L}_{n}
  =&\;
    \bar{\xi}_n \Big( i n \cdot D_n +   i \slashed{D}_{n\perp} \frac{1}{i \nbar \cdot D_n} i \slashed{D}_{n\perp} \Big) \frac{\slashed{\nbar}}{2} \xi_n - \frac{1}{2} \mathbf{Tr} \Big[ F^{\mu\nu}_n F_{\mu\nu}^n \Big] \,,
  \\
  \label{eq:scetII:Ls}
  \mathcal{L}_{s}
  =&\;
    - \frac{1}{2} \mathbf{Tr} \Big[ F^{\mu\nu}_{\text{s}} F_{\mu\nu}^{\text{s}}\Big] +\bar{q}_{\text{s}} i \slashed{D}_{\text{s}}q_{\text{s}}\, ,
  \\
  \label{eq:hqet:Lv}
  \mathcal{L}_{v}
  =&\;
    \bar{h}_v \left( i \partial \cdot v\right) h_v + \bar{\chi}_v \left( i \partial \cdot v  \right) \chi_v\,,
\end{align}
where $D^{\mu}_{n(s)}\equiv\partial^{\mu}-ig_sA^{\mu}_{n(s)}$ and
$F_{n(s)}^{\mu\nu}\equiv(i/g_s)[D^{\mu}_{n(s)},D^{\nu}_{n(s)}]$
stand for the covariant derivative and the field strength tensor
for the collinear (soft) fields $A^{\mu}_{n(s)}$, respectively.
The Lagrangian $\mathcal{L}_{\nbar}$ can be obtained by exchanging
$n\leftrightarrow\nbar$ in $\mathcal{L}_{n}$.
In deriving Eqs.~\eqref{eq:scetII:Ln}-\eqref{eq:hqet:Lv},
the decoupling transformations~\cite{Bauer:2001yt, Beneke:2010da}
have already been performed to strip the soft particle of the
collinear and heavy partons.
As a result,  the (anti-)top quark is free of any interactions
at this point, while the collinear partons only communicate
with themselves.
The LP contribution of $\Sigma_{\ttbar}$ can be found through
assembling the amplitudes induced by
$\mathcal{H}_{\ttbar}^{\mathrm{eff}}$ and the effective
Lagrangians in Eqs.~\eqref{eq:scetII:Ln}-\eqref{eq:hqet:Lv},
and we follow the scheme of \cite{Ahrens:2010zv,Li:2013mia} to
construct the hard contributions.
In light of the rapidity divergences arising from the soft
and collinear integrals, the exponential regulator proposed
in \cite{Li:2016axz,Li:2016ctv} is applied throughout.
Finally, the decoupling nature of the Lagrangians in
Eqs.~\eqref{eq:scetII:Ln}-\eqref{eq:hqet:Lv} allows us to
rewrite $\Sigma_{\ttbar}$ as,
\begin{align}\label{eq:M2PS:NJ0}
  \Sigma_{\ttbar}\Bigg|_{N_{\mathrm{J}}=0}
  =&\;
    \frac{8\pi^2}{\Mttbar^2}\,
    \sum_{\kappa}\,
    \int {\done^2\bTvec}\,\exp\left(\mathrm{i}\,\bTvec \cdot\qTvec \right)\,
    \widetilde{\Sigma}_{\ttbar}^{ [\kappa]}
    (\bTvec,\Mttbar,\bttbar,x_t,\Yttbar,\mu,\nu)\,
    \mathcal{W}_{t}
    \,
    \mathcal{W}_{\tbar}\,,
\end{align}
where $\bTvec$ is the impact parameter introduced during the Fourier
transformation.
$\Sigma_{\ttbar}$ includes the contributions from all channels $\kappa\in
  \{g_{n}g_{\nbar},q^i_{n}\bar{q}^j_{\nbar},q^i_{\nbar}\bar{q}^j_{n}\}$,
with $i,j\in\{u,d,c,s,b\}$ indicating the flavour of the quark field.
$\mu$ and $\nu$ are the scales associated with the virtuality and
rapidity renormalisations, respectively.
Finally, $\mathcal{W}_{t}$ and $\mathcal{W}_{{\bar{t}}}$ are the heavy parton
correlation functions,
\begin{align}
  \mathcal{W}_{t}
  =&
    \frac{1}{4m_tN_c }\,
    \mathbf{Tr} \,
    \langle 0|   {h}_{v_t} \left(0\right)  |t \rangle
    \langle t| \bar{h}_{v_t}  \left(0\right)\frac{1+\slashed{v}_{t}}{2}|0 \rangle
    =1\,,
    \label{eq:M2PS:NJ0:Wt}\\
  \mathcal{W}_{\tbar}
  =&
    -\frac{1}{4m_tN_c }\,
    \mathbf{Tr} \,
    \langle 0|  \bar{\chi}_{v_{\bar{t}}}   \left(0\right) |\tbar \rangle
    \langle\tbar|  \frac{1-\slashed{v}_{\tbar}}{2} {\chi}_{v_{\tbar}}   \left(0\right) |0 \rangle
    =1\,,
    \label{eq:M2PS:NJ0:Wtbar}
\end{align}
where $N_c=3$ is the colour factor.
$v_{t(\tbar)}$ is the velocity of the (anti)top quark in the rest
frame of the $\ttbar$ system.
Considering that the (anti-)top quark up to the LP accuracy amounts
to a free particle, the correlation functions $\mathcal{W}_{t(\tbar)}$
will never receive any perturbative corrections.
Therefore, in the second steps of
Eqs.~\eqref{eq:M2PS:NJ0:Wt}-\eqref{eq:M2PS:NJ0:Wtbar}, we evaluate
them using the tree-level expressions.

\changed
{Apart from the correlation function $\mathcal{W}_{t(\bar{t})}$, Eq.~\eqref{eq:M2PS:NJ0} entails the partonic cross sections $\widetilde{\Sigma}_{\ttbar}^{[\kappa]}$ as well, which 
are built by suitably combining beam-collinear, soft and hard functions,
}
\begin{align}
  \widetilde{\Sigma}_{\ttbar}^{[q^i_{n}\bar{q}^j_{\nbar}]}
  (\bTvec,\Mttbar,\bttbar,x_t,\Yttbar,\mu,\nu)
  =&
    \left(\frac{1}{2N_c}\right)^2\,
    \mathcal{B}_{n}^{[q_n^i]}(\eta_n,\bT,\mu,\nu)\,
    \mathcal{B}_{\nbar}^{[\bar{q}_n^j]}(\eta_{\nbar},\bT,\mu,\nu) \,
    \nnb\\
  &
    \times\,
    \sum_{\alpha,\beta}\left\{
    \mathcal{H}_{\alpha\beta}^{[q^i_{n}\bar{q}^j_{\nbar}]}(\Mttbar,\bttbar,x_{t},\mu)\,
    \mathcal{S}^{\alpha\beta}_{[q_{n}\bar{q}_{\nbar}]}(\bTvec,\mu,\nu)
    \right\}\,,
    \label{eq:M2PS:NJ0:qqbar}
    \\[2mm]
  \widetilde{\Sigma}_{\ttbar}^{[g_{n}g_{\nbar}]}
  (\bTvec,\Mttbar,\bttbar,x_t,\Yttbar,\mu,\nu)
  =&
    \left(\frac{1}{N^2_c-1}\right)^2\,
    \sum_{\substack{\alpha,\beta,\hn,\hnp,\hnbar,\hnbarp}}
    \left\{
      \mathcal{B}^{[g_{\nbar}]}_{\nbar,\hnbarp\hnbar}
      (\eta_{\nbar},\bTvec,\mu,\nu)  \,
      \mathcal{B}^{[g_n]}_{n,\hnp\hn}
      (\eta_n,\bTvec,\mu,\nu)\,
    \right.\nnb\\
  &
    \left.\times\,
      \mathcal{H}_{\alpha\beta;{\hnbarp}\hnbar;\hnp\hn}^{[g_ng_{\nbar}]}
      (\Mttbar,\bttbar,x_{t},\mu)\,
      \mathcal{S}^{\alpha\beta}_{[g_ng_{\nbar}]}
      (\bTvec,\mu,\nu)
    \right\}\,,
    \label{eq:M2PS:NJ0:gg}
\end{align}
where the momentum fractions $\eta_{n}$ and $\eta_{\nbar}$ are defined
as $\eta_{n}=\Mttbar\,e^{\Yttbar}/\sqrt{s}$ and
$\eta_{\nbar}=\Mttbar\,e^{-\Yttbar}/\sqrt{s}$, respectively.
The expression for
$\widetilde{\Sigma}_{\ttbar}^{[q^i_{\nbar}\bar{q}^j_{ {n}}]} $ can be
derived by exchanging the labels
$n\leftrightarrow\nbar$.

$\mathcal{H}_{\alpha\beta}^{[q^i_{n}\bar{q}^j_{\nbar}]}$ and
$\mathcal{H}_{\alpha\beta;\hnbarp\hnbar;\hnp\hn}^{[g_ng_{\nbar}]}$
are the hard functions for the quark and gluon initiated processes,
respectively.
The indices $\alpha$ and $\beta$ label the colour states to track
the full colour correlation between the hard and the soft functions
detailed below.
Similarly, the $\hn$, $\hnp$, $\hnbar$, and $\hnbarp$
denote the polarisation states of the incoming gluons to capture
all off-diagonal correlation effects of the beam-collinear and
hard functions.
Please note, that the beam-hard function correlation for
massless quarks is devoid of off-diagonal contributions.
The hard functions now account for the LP contributions from the
deep off-shell region in Eq.~\eqref{eq:regions:hard}.
Their expressions read~\cite{Ahrens:2010zv,Li:2013mia},
\begin{align}
  \mathcal{H}_{\alpha\beta}^{[q^i_{n}\bar{q}^j_{\nbar}]}
  (\Mttbar,\bttbar,x_{t},\mu)
  =&
    \sum_{ \alpha' ,\beta'}\,
    \Big( \mathcal{Z}^{h,{\alpha'\alpha}}_{[{q_n\bar{q}_{\nbar}}]}\Big)^*\,
    \mathcal{Z}^{h,{\beta'\beta}}_{[{q_n\bar{q}_{\nbar}}]}\,
    {\sum_{\{a,a',h\}}}\,
    c_{a'_1a'_2a'_3a'_4}^{qq,(\alpha')}\,
    \left[c_{a_1a_2a_3a_4}^{qq,(\beta')}\right]^*\,
    \nnb
    \\
  &
  \quad\times
    \mathcal{M}^*(\bar{q}^{a'_1h_1^\pp}_{\nbar,j}{q}^{a'_2h_2^\pp}_{n,i}\to t_{a'_3h_3^\pp} \bar{t}_{a'_4h_4^\pp})
    \,
  \mathcal{M}(\bar{q}^{a_1h_1}_{\nbar,j}{q}^{a_2h_2}_{n,i}\to t_{a_3h_3} \bar{t}_{a_4h_4})
    \label{eq:M2PS:NJ0:Hqqbar}
    \,,
    \\[2mm]
  \mathcal{H}_{\alpha\beta;h'_1h_1;h'_2h_2}^{[g_ng_{\nbar}]} (\Mttbar,\bttbar,x_{t},\mu)
  =&
  \sum_{ \alpha', \beta'}\,
  \Big( \mathcal{Z}^{h,{\alpha'\alpha}}_{[{g_ng_{\nbar}}]}\Big)^*\,
  \mathcal{Z}^{h,{\beta'\beta}}_{[{g_ng_{\nbar}}]}\,
  {\sum_{\{a,a'\},h_3,h_4}}\,
  c_{a'_1a'_2a'_3a'_4}^{gg,(\alpha')}\,
  \left[c_{a_1a_2a_3a_4}^{gg,(\beta')} \right]^*\,
  \nnb
  \\
  &
  \quad\times
  \mathcal{M}^*(g^{a'_1h'_1}_{\nbar}g^{a'_2h'_2}_{n}\to t_{a'_3h_3^\pp} \bar{t}_{a'_4h_4^\pp})
  \,
  \mathcal{M}(g^{a_1h_1}_{\nbar}g^{a_2h_2}_{n}\to t_{a_3h_3} \bar{t}_{a_4h_4})
    \label{eq:M2PS:NJ0:Hgg}
\end{align}
where $\mathcal{M}$ denotes the amplitude for the partonic process
$\bar{q} q\to \ttbar$ or $gg\to \ttbar$.
Therein, the variables $a_{i}$$(i=1,\dots,4)$ are introduced for the
colour states of the individual external particles.
In particular, every $a_{i}$ runs over the set $\{1,\dots,3\}$ for
quarks and anti-quarks, and $\{1,\dots,8\}$ for gluons.
Also, to facilitate our calculation, the orthonormal colour bases
$c^{qq}_{\{a\}}$ and $c^{gg}_{\{a\}}$~\cite{Beneke:2009rj} are
exploited in Eqs.~\eqref{eq:M2PS:NJ0:Hqqbar}-\eqref{eq:M2PS:NJ0:Hgg},
more explicitly,
\begin{align}\label{eq:def:color:basis:qq}
  c^{qq,(1)}_{a_1a_2a_3a_4}
  &=
    \frac{1}{3}\,\delta_{a_1a_2}\delta_{a_3a_4}\,,\;\;
  c^{qq,(2)}_{a_1a_2a_3a_4}
  =
    \frac{1}{\sqrt{2}}\sum_c T^{c}_{a_1a_2}T^{c}_{a_3a_4}\,,\\
  \label{eq:def:color:basis:gg}
  c^{gg,(1)}_{a_1a_2a_3a_4}
  &=
    \frac{1}{2\sqrt{6}}\,\delta_{a_1a_2}\delta_{a_3a_4}\,,\;\;
  c^{gg,(2)}_{a_1a_2a_3a_4}
  =
    \frac{i}{2\sqrt{3}}\,\sum_c f^{a_1 c a_2}T^{c}_{a_3a_4}\,,\;\;
  c^{gg,(3)}_{a_1a_2a_3a_4}
  =
    \frac{1}{2}\sqrt{\frac{3}{5}}\sum_cd^{a_1 c a_2}T^{c}_{a_3a_4}\,,
\end{align}
where $T^{c}_{ab}$ stands for the  generator in the fundamental
representation of the SU$(3)$ group. $f_{abc}$ and $d_{abc}$ mark
the antisymmetric and symmetric structure constants for the SU$(3)$
group, respectively.
As alluded to above, in calculating the hard functions, due to the
absence of spin-correlations for external quarks, we take the sum
over all the helicity configurations for the quark channel, while,
to capture the non-diagonal gluon polarisation effects, the
explicit dependence of
$\mathcal{H}_{\alpha\beta;h'_1h_1;h'_2h_2}^{[g_ng_{\nbar}]}$
on the gluon helicities is retained.

To evaluate the amplitudes in
Eqs.~\eqref{eq:M2PS:NJ0:Hqqbar}-\eqref{eq:M2PS:NJ0:Hgg}, we make use
of the on-shell prescription to renormalise the top quark mass and
the $n_f=5$ active flavor scheme to handle the UV divergences from
the strong coupling.
The remaining singularities are of infrared origin, and are
subtracted by means of $\mathcal{Z}^{h }_{[{g_ng_{\nbar}}]}$
and $\mathcal{Z}^{h}_{[{q_n\bar{q}_{\nbar}}]}$ following the
method in \cite{Ferroglia:2009ii}.
Up to NLO, the amplitudes of all the helicity and colour configurations
can be extracted from \Recola~\cite{Actis:2012qn,Actis:2016mpe}.
The NNLO results are more involved, and in consequence we will
only address their logarithmic parts in this work.
For reference, grid-based NNLO results can be found in \cite{Chen:2017jvi},
and the progress towards the full analytic evaluations is discussed
in \cite{DiVita:2019lpl,Badger:2021owl,Mandal:2022vju}.


$\mathcal{B}_{n}^{[q_n^i]}$ and $\mathcal{B}^{[g_n]}_{n,h'h}$ are the
quark and gluon beam functions, respectively, collecting the
contributions from the region in Eq.~\eqref{eq:regions:beam:n}.
Their definitions in the exponential regularisation scheme
are~\cite{Li:2016axz,Luo:2019hmp,Luo:2019bmw},
\begin{align}
  \mathcal{B}_{n}^{[q_n^i]}( \eta_{n},\bT,\mu,\nu)
  =&\;
    \mathcal{Z}^{c}_{[q_n]}
    \Big(\mathcal{Z}^{cs}_{[q_n]}\Big)^{-1}\,
    \lim_{\delta\to0^+}\,
    \int\frac{\done b_n}{4\pi}\,
    \exp\left(\frac{-\mathrm{i}\eta_n b_{n}P_{\nbar}}{2}\right)\,
    \label{eq:M2PS:NJ0:Bnq}
  \\
  &
    \times
    \mathbf{Tr}\,
    \langle N(P)|\overline{\xi}^i_{n} W_n(-{i b_0\delta},-{i b_0\delta}+b_n,\bTvec)
    \frac{\slashed{\nbar}}{2}W^{\dagger}_n{\xi}^i_{n }(0) |N(P) \rangle\bigg|_{\delta=\frac{1}{\nu}} \,,
    \nnb\\[2mm]
  \mathcal{B}^{[g_n]}_{n,h'h}( \eta_{n},\bTvec,\mu,\nu)
  =&\;
    \mathcal{Z}^{c}_{[g_n]} \Big({\mathcal{Z}^{cs}_{[g_n]}}\Big)^{-1}\,
    \lim_{\delta\to0^+}\,
    \int\frac{\done b_n}{4\pi}\,
    \exp\left( \frac{  -\mathrm{i} \eta_n b_{n}P_{\nbar}}{2}\right)\,
    \label{eq:M2PS:NJ0:Bng}
    \\
  &
    \times
    \mathbf{Tr}\,
    \langle N(P)| \mathcal{A}^{\perp}_{n,\rho}(-{i b_0\delta},-{i b_0\delta}+b_n,\bTvec) \,
    \epsilon^{\rho}_{n,h'}\,
    \left(-\eta_nP_{\nbar} \right)\, \epsilon^{\sigma*}_{n,h}\, \mathcal{A}^{\perp}_{n,\sigma}(0) |N(P) \rangle\bigg|_{\delta=\frac{1}{\nu}} \,,
    \nnb
\end{align}
 where $\mathcal{Z}^{c}_{[q_n]([g_n])} $ is the
renormalisation constant for the quark (gluon) beam function in
the $\overline{\mathrm{MS}}$ scheme.
$\mathcal{Z}^{cs}_{[q_n]}$ and $\mathcal{Z}^{cs}_{[g_n]}$ represent
the ensuing zero-bin subtrahend to remove the soft-collinear
overlapping terms.
$\delta$ is the exponential regulator suggested in \cite{Li:2016axz},
accompanied by the constant $b_0\equiv2\exp(-\gamma_{\mathrm{E}})$.
$\mu$ and $\nu$ are the scales associated with the virtuality and
rapidity renormalisations, respectively.
Within the matrix elements, ${\xi}^i_{n }$ denotes the collinear
quark field of the flavour $i$ given in Eq.~\eqref{eq:scetII:Ln}.
$\mathcal{A}_{n,\perp}^{\mu}\equiv\frac{1}{g_s}\,W_n^{\dagger}(iD_{n,\perp}^{\mu}W_n)$
signifies the gauge invariant building block for the gluon field
with $W_n$ denoting the collinear Wilson line~\cite{Hill:2002vw}.
Finally, $P$ is the momentum carried by the initial proton with
$P_{\nbar}=P\cdot\nbar$ being the largest light-cone component.
The anti-quark beam function $\mathcal{B}_{n}^{[\bar{q}_i]}$ and
those for the $\nbar$ direction can be obtained by adjusting the
labels and fields in Eqs.~\eqref{eq:M2PS:NJ0:Bnq}-\eqref{eq:M2PS:NJ0:Bng}
appropriately.
 
Comparing with the quark beam function, the gluon case possesses
extra indices, $\hn,\hnp\in\{+,-\}$, to characterise its intrinsic
polarisation effects~\cite{Catani:2010pd}.
In this work, the following helicity basis is adopted,
\begin{align} \label{eq:Bgg:def:hel:space}
  \epsilon^{\mu}_{n,\pm}\equiv \left\{0, \frac{\mp1}{\sqrt{2}},\frac{-\mathrm{i}}{\sqrt{2}},0\right\}\,,
  \qquad \epsilon^{\mu}_{\nbar,\pm}\equiv \left\{0, \frac{\pm1}{\sqrt{2}},\frac{-\mathrm{i}}{\sqrt{2}},0\right\}\,.
\end{align}
In principle, the representations of the helicity polarisation
states $\epsilon^{\mu}_{n,\pm}$ are not unique.
Nevertheless, in order to avoid the appearance of an unphysical
phase factor in the cross section, the helicity space utilised
in the beam sector must synchronise with the one used for the
hard function in Eq.~\eqref{eq:M2PS:NJ0:gg}.
In our case, since we use \Recola to extract the hard function,
Eq.~\eqref{eq:Bgg:def:hel:space} is subject to the choice adopted
in this program.
The quark beam function is known to \NNNLO accuracy
\cite{Luo:2020epw,Luo:2019szz}, while for the gluon case only
the helicity-conserving components
$\mathcal{B}_{++}^{[g_{n(\nbar)}]}$ and
$\mathcal{B}_{--}^{[g_{n(\nbar)}]}$ are known to this order
\cite{Luo:2020epw}.
The helicity-flip components $\mathcal{B}_{+-}^{[g_{n(\nbar)}]}$
and $\mathcal{B}_{-+}^{[g_{n(\nbar)}]}$ are only known to
\NNLO  \cite{Luo:2019bmw,Gutierrez-Reyes:2019rug,Catani:2022sgr}.

Finally, $\mathcal{S}^{\alpha\beta}_{[q_n\bar{q}_{\nbar}]}$ is the
soft function and covers the wide angle domain in Eq.~\eqref{eq:regions:soft}. Its LP expression is,
\begin{align}
  \lefteqn{\hspace*{-10pt}\mathcal{S}^{\alpha\beta}_{[q_n\bar{q}_{\nbar}]}(\bTvec,\mu,\nu)}\nnb\\
  =&\;
    \sum_{\alpha',\beta'}\,
    \Big(\mathcal{Z}^{s,{\alpha\alpha'}}_{[q_n\bar{q}_{\nbar}]}\Big)^*\,
    \mathcal{Z}^{s,\beta\beta'}_{[q_n\bar{q}_{\nbar}]}\,
    \sum_{\{a,a',b\}} \,
    \lim_{\delta\to0^+} \,
    \left[c_{a'_1a'_2a'_3a'_4}^{qq,(\alpha')}\right]^*\,
    c_{a_1a_2a_3a_4}^{qq,(\beta')}\,
    \label{eq:M2PS:NJ0:Sqq}
    \\
  &
    \times
    \langle 0|\Big[\overline{Y}^{\dagger}_{n,b_2^\pp a'_2}
                   \overline{Y}_{\nbar,a'_1b_1^\pp}
                   Y^{\dagger}_{v_{\tbar},b_4^\pp a'_4}\,
                   Y_{v_{t},a'_3b_3^\pp}\Big]
              (-{i b_0\delta},-{i b_0\delta},\bTvec)\,
              \Big[\overline{Y}^{\dagger}_{\nbar,b_1a_1}
                   \overline{Y}_{n,a_2b_2}
                   Y^{\dagger}_{v_t,b_3a_3}
                   Y_{v_{\tbar},a_4b_4}\Big](0)
    |0\rangle \,
    \bigg|_{\delta=\frac{1}{\nu}} \,,
    \nnb\\[2mm]
  \lefteqn{\hspace*{-10pt}\mathcal{S}^{\alpha\beta}_{[g_ng_{\nbar}]}(\bTvec,\mu,\nu)}\nnb\\
  =&\;
    \sum_{\alpha',\beta'}\,
    \Big(\mathcal{Z}^{s,{\alpha\alpha'}}_{[gg]}\Big)^*\,
    \mathcal{Z}^{s,{\beta\beta'}}_{[gg]}  \,
    \sum_{\{a,a',b\}} \,
     \lim_{\delta\to0^+} \,\,
    \left[c_{a'_1a'_2a'_3a'_4}^{gg,(\alpha')}\right]^*\,
    c_{a_1a_2a_3a_4}^{gg,(\beta')}
    \label{eq:M2PS:NJ0:Sgg}
    \\
  &
    \times
    \langle 0|\Big[\overline{\mathcal{Y}}^{\dagger}_{n,b_2^\pp a'_2}
                   \overline{\mathcal{Y}}_{\nbar,a'_1b_1^\pp}
                   Y^{\dagger}_{v_{\tbar},b_4^\pp a'_4}\,
                   Y_{v_{t},a'_3b_3^\pp}\Big]
                   (-{i b_0\delta},-{i b_0\delta},\bTvec)\,
              \Big[\overline{\mathcal{Y}}^{\dagger}_{\nbar,b_1a_1}
                   \overline{\mathcal{Y}}_{n,a_2b_2}
                   Y^{\dagger}_{v_t,b_3a_3}
                   Y_{v_{\tbar},a_4b_4}\Big](0)
    |0\rangle\,
    \bigg|_{\delta=\frac{1}{\nu}}\;,
    \nnb
\end{align}
where $\mathcal{Z}^{s}_{[q_n\bar{q}_{\nbar}]}$ and $\mathcal{Z}^{s}_{[gg]}$
are the renormalisation constants of the soft function in the
\MSbar scheme.
Again, $\delta$ denotes the rapidity regulator \cite{Li:2016axz,Li:2016ctv},
and the $c_{\{a\}}^{qq(gg)}$ are the colour coefficients defined in
Eqs.~\eqref{eq:def:color:basis:qq}-\eqref{eq:def:color:basis:gg}.
$\overline{Y}_{n(\nbar)}$ and $\overline{\mathcal{Y}}_{n(\nbar)}$
describe the incoming soft Wilson lines for the (anti-)quark and
gluon fields, respectively, while the
$Y_{v_t(v_{\tbar})}$ are the outgoing soft Wilson lines of the
(anti-)top quark.
Their specific expressions have been summarised in \cite{Beneke:2009rj}.
\changed{
Even though the azimuthally averaged soft functions have been computed up to
\NNLO~\cite{Zhu:2012ts,Li:2013mia,Angeles-Martinez:2018mqh} recently,
the azimuthally resolved soft functions, as displayed in
Eqs.~\eqref{eq:M2PS:NJ0:Sqq}-\eqref{eq:M2PS:NJ0:Sgg}, are not yet
available in the context of effective field theories.
The relevant function at this point is the soft correlation
factor defined in \cite{Catani:2014qha,Catani:2021cbl} which is
derived in the generalised $\qT$ resummation framework
of~\cite{Collins:1981uk,Collins:1984kg}.
Nevertheless, in this paper, aiming at a self-consistent and
independent study, we will revisit the soft interactions in
the EFT including the exponential rapidity regulator~\cite{Li:2016axz}
at NLO accuracy.
In Sec.~\ref{sec:soft:func:NLO}, we will explicitly calculate
the rapidity and virtuality associated divergences originating
from the soft sector defined in Eqs.~\eqref{eq:M2PS:NJ0:Sqq}-\eqref{eq:M2PS:NJ0:Sgg} and utilise them to examine the
consistency condition required by the factorisation of
Eqs.~(\ref{eq:M2PS:NJ0:qqbar}-\ref{eq:M2PS:NJ0:gg}), providing
a powerful test of its validity.
Moreover, regarding the finite contributions in the soft
function, we will present a comparison between our results and
those obtained in \cite{Catani:2014qha,Catani:2021cbl}.
}
 
Now we can consider the differential cross section in
Eq.~\eqref{eq:QCDF} with the reduced kinematic variables
of Eqs.~\eqref{eq:Pzt:NJ0}-\eqref{eq:thetakin:NJ0} as
well as the expanded partonic contributions in
Eq.~\eqref{eq:M2PS:NJ0}.
We begin by disentangling the Fourier integrals in
Eq.~\eqref{eq:M2PS:NJ0} with the help of the reference
vector $\vec{\tau}$, using the different scaling behaviours
of the components of the vector $\bTvec$ (or correspondingly,
$\qTvec$),
\begin{align}\label{eq:M2PS:NJ0:dec}
  \Sigma_{\ttbar}\Bigg|_{N_{\mathrm{J}}=0}
  \propto\,
    \sum_\kappa
    \int^{\infty}_{-\infty}\done b_{\tau_{\|}}
    \exp\left(\mathrm{i}\,{q}_{\tau_{\|}}b_{\tau_{\|}}\right)\,
    \int^{\infty}_{-\infty}\done b_{\tau_{\perp}}
    \exp\left(\mathrm{i}\,{q}_{\tau_{\perp}}b_{\tau_{\perp}}\right)\,
    \widetilde{\Sigma}_{\ttbar}^{[\kappa]}(\bTvec,\Mttbar,\bttbar,x_t,\Yttbar,\mu,\nu)
    +\dots\,,
\end{align}
where in analogy to the case in Fig.~\ref{fig:KinConfig}, we apply the relationships,
\begin{align}\label{eq:bT:NJ0:dec}
  \bTvec
  =
    \vec{b}_{\tau_{\perp}}
    +\vec{b}_{\tau_{\|}}
  =
    b_{\tau_{\perp}}\vec{\tau}\times\vec{n}
    +b_{\tau_{\|}}\vec{\tau}\,,
  \quad
  \qTvec
  =
    \vec{q}_{\tau_{\perp}}
    +\vec{q}_{\tau_{\|}}
  =
    q_{\tau_{\perp}}\vec{\tau}\times\vec{n}
    +q_{\tau_{\|}}\vec{\tau}\,.
\end{align} 
Substituting the separated expression in Eq.~\eqref{eq:M2PS:NJ0:dec}
into Eq.~\eqref{eq:QCDF} yields,
\begin{align}\label{eq:eq:QCDF:NJ0:dec}
  \frac{\done^5{\sigma_{\ttbar}}}
       {\done \Mttbar^2\,\done^2\vec{P}_{t}^{\perp}\,\done \Yttbar\,\done q_{\tau}}
  \Bigg|_{N_{\mathrm{J}}=0}
  \propto&\,
    \sum_\kappa
    \int^{\infty}_{-\infty} \done b_{\tau_{\|}}  \,
    \int^{\infty}_{-\infty} \,\done q_{\tau_{\|}}\, \delta(q_{\tau}-|q_{\tau_{\|}}|)\,\exp\left(\mathrm{i}\,{q}_{\tau_{\|}} b_{\tau_{\|}}\right)\,
    \\
  &
    \times
    \int^{\infty}_{-\infty}\done b_{\tau_{\perp}}\,
    \widetilde{\Sigma}_{\ttbar}^{[\kappa]}(\bTvec,\Mttbar,\bttbar,x_t,\Yttbar,\mu,\nu)\,
   \nonumber \\
    &
    \times\int^{\infty}_{-\infty}\,
    \done q_{\tau_{\perp}}\,
    \exp\left(\mathrm{i}\,{q}_{\tau_{\perp}} b_{\tau_{\perp}}\right)
    +\dots\,,
    \nnb
\end{align}
where the variables that are independent of $\bTvec$ or $\qTvec$
are omitted for simplicity.
In comparison to Eq.~\eqref{eq:QCDF}, one of the main differences
in Eq.~\eqref{eq:eq:QCDF:NJ0:dec} resides in the absence of the
boundaries on the $q_{\tau_{\perp}}$ integral, which is a consequence
of the $\qTvec$ independence in the expanded function
$\Theta_{\mathrm{kin}}^{(0)}$ of Eq.~\eqref{eq:thetakin:NJ0}.
In this way, the integral over $q_{\tau_{\perp}}$ in the third
line of Eq.~\eqref{eq:eq:QCDF:NJ0:dec} can be completed before
the inverse Fourier transformation, thereby leading to the Dirac
delta function $\delta(b_{\tau_{\perp}})$. \footnote{
  It is worth emphasizing that this operation is prohibited in the
  original expression of Eq.~\eqref{eq:QCDF} due to the explicit
  dependences on  $\qTvec$ in the boundary function of
  Eq.~\eqref{eq:thetakin:ft}.
}         
The integration over $b_{\tau_{\perp}}$ is then straightforward,
reducing the $\bTvec$ dependence of $\widetilde{\Sigma}_{\ttbar}$
to a dependence on $\vec{b}_{\tau_{\|}}$ only.
This leaves the integrals over $b_{\tau_{\|}}$ and $q_{\tau_{\|}}$.
The evaluation on the latter is immediate via the function
$\delta(q_{\tau}-|q_{\tau_{\|}}|)$ in the first line of
Eq.~\eqref{eq:eq:QCDF:NJ0:dec}, eliminating the imaginary
part of $\exp(\mathrm{i}\,{q}_{\tau_{\|}} b_{\tau_{\|}})$.
The integration over $b_{\tau_{\|}}$, finally, is less straightforward to perform, and we turn to the numeric solutions in
Sec.~\ref{sec:results}.

To summarise, specifying the kinematic factors in
Eq.~\eqref{eq:eq:QCDF:NJ0:dec},
we are able to establish the LP contribution from the
$N_{\mathrm{J}}=0$ configuration,
\begin{align}\label{eq:xs:NJ0:fac}
  \frac{ \done^5{\sigma_{\ttbar}} }
       { \done \Mttbar^2\,\done^2\vec{P}_{t}^{\perp}\,
         \done\Yttbar\,\done q_{\tau} }
  \Bigg|_{N_{\mathrm{J}}=0}
  =
    \sum_{\mathrm{sign}[x_t]}
    \frac{ \Theta_{\mathrm{kin}}^{(0)}}
         {16\pi^3\,\bttbar\,|x_t|\,M^4_{\ttbar}\,s}\,
    \sum_{\kappa}\int^{\infty}_{-\infty}
    &
      \done {b}_{\tau_{\|}}\, \cos\left({b}_{\tau_{\|}} {q}_{\tau}\right)\,
      \\
    &
      \times \widetilde{\Sigma}_{\ttbar}^{[\kappa]}
      (b_{\tau_{\|}}\vec{\tau},\Mttbar,\bttbar,x_t,\Yttbar,\mu,\nu)
      \,,\nnb
\end{align}
where the index $\kappa$ runs over
$\{g_{n}g_{\nbar},q^i_{n}\bar{q}^j_{\nbar},q^i_{\nbar}\bar{q}^j_{n}\}$
as before.
$\Theta_{\mathrm{kin}}^{(0)}$ imposes the kinematic constraints as
defined in Eq.~\eqref{eq:Pzt:NJ0}, and the
$\widetilde{\Sigma}_{\ttbar}$ are the contributions of the
individual partonic processes, as presented in
Eqs.~\eqref{eq:M2PS:NJ0:qqbar}-\eqref{eq:M2PS:NJ0:gg}.

\changed
{
It should be stressed that the result in Eq.~\eqref{eq:xs:NJ0:fac}
and the factorisation in Eqs.~(\ref{eq:M2PS:NJ0:qqbar}-\ref{eq:M2PS:NJ0:gg})
are subject to the absence of other sources of divergent behaviour.
In particular, Coulomb divergences encountered in the threshold
region must be avoided.
In the vicinity of the threshold $\beta_{t\bar{t}}=0$, according to
pNRQCD~\cite{Pineda:1997bj,Brambilla:1999xf,Beneke:1999zr,Beneke:1999qg}
and also the analysis in \cite{Ju:2020otc}, the function
$\widetilde{\Sigma}_{\ttbar}^{[\kappa]}$ can develop the power
like singularities, i.e.,
$$\lim_{\beta_{\ttbar}\to0}\, \widetilde{\Sigma}_{\ttbar}^{[\kappa]}\,\to \,\sum_n\,c_n\,\frac{ \alpha^n_s}{\beta_{\ttbar}^n},$$
as a result of the exchanges of Coulomb gluons between top and
antitop quarks.
As  for lower powers of $\alpha_s$, those singular behaviours
are innocuous, thanks to the kinematic suppressions in the
threshold regime.
However, with increasing perturbative accuracy, the severity of the
Coulomb singularities worsens, such that beyond a given precision,
they can not be regularised by any kinematic factors any longer and,
thereby, develop a Coulomb divergence during the phase space integration.
}

\changed{
The emergence of such Coulomb divergences marks a failure of
the factorisation of Eq.~\eqref{eq:M2PS:NJ0} and
Eqs.~\eqref{eq:M2PS:NJ0:qqbar}-\eqref{eq:M2PS:NJ0:gg} within
the threshold domain and therefore prompts a combined treatment
of the Coulomb, soft, and beam-collinear interactions.
The combination of the former two cases has been addressed in
both the static \cite{Beneke:2010da,Beneke:2011mq,Piclum:2018ndt}
and recoiled \cite{Ju:2019lwp} top-antitop systems in the soft limit.
However, with the participation of the beam-collinear sector
and the ensuing appearance of the rapidity divergences,
novel types of subleading vertices may come into play at a
given logarithmic accuracy along with the insertions of Coulomb
potentials, which inevitably requires additional considerations,
generalising the frameworks of
\cite{Beneke:2010da,Beneke:2011mq,Piclum:2018ndt,Ju:2019lwp}
to the present process.
To this end, we will constrain the investigation in this work
to the domain $\Mttbar\ge 400$~GeV, or, equivalently,
$\bttbar\gtrsim 0.5$, to stay well clear of the Coulomb
divergence, and aim to address the relevant subtleties arising
from Coulomb interactions in a future work.
}      
 
\changed{
Finally, we make use of  Eq.~\eqref{eq:xs:NJ0:fac}  to assess the
power accuracy of the $N_{\mathrm{J}}=0$ configuration. 
}%
First of all, since the kinematic constraint $\Mttbar>400$\,GeV
(or equivalently, $\bttbar\gtrsim 0.5$) has been imposed,
the prefactor in front of the integral does not induce any
power-like behaviour for the bulk of phase space, and is thus
of $\order{(1)}$.
Next, in order to determine the powers of the impact parameter
${b}_{\tau_{\|}}$ and the Fourier basis
$\cos\left({b}_{\tau_{\|}}{q}_{\tau}\right)$, note that the
${b}_{\tau_{\|}}$ integral serves in part as the momentum
conservation condition on the beam and soft radiations.
Considering that the transverse momenta for the real emissions
are all of $\order(\lamtau)$ in the $N_{\mathrm{J}}=0$
configuration, we arrive at~\cite{Beneke:2002ph},
\begin{align}  \label{eq:scaling:bTvec}
  {b}_{\tau_{\|}}\sim\order(\lambda^{-1}_{\tau})\,,\quad\quad
  \cos\left(  {b}_{\tau_{\|}} {q}_{\tau} \right)\sim\order{(1)}\,.
\end{align}
The remaining task is to ascertain the power of
$\widetilde{\Sigma}_{\ttbar}$.
As defined in Eqs.~\eqref{eq:M2PS:NJ0:qqbar}-\eqref{eq:M2PS:NJ0:gg},
$\widetilde{\Sigma}_{\ttbar}$ entails the hard, soft, and beam
functions from the various partonic transitions.
The hard sector consists of nothing but the Wilson coefficients
multiplied by the colour (helicity) bases, which invokes no
power-like behaviour and is, hence, of $\order{(1)}$.
The soft functions are defined in
Eqs.~\eqref{eq:M2PS:NJ0:Sqq}-\eqref{eq:M2PS:NJ0:Sgg} as the
products of the soft Wilson lines sandwiched by the vacuum states.
From the scaling prescriptions in \SCETII
\cite{Bauer:2002aj,Lange:2003pk,Beneke:2003pa}, the soft Wilson
lines are also of $\order{(1)}$, regardless of the quark or gluon
channels, which in turn yields,
\begin{align} 
  \mathcal{S}^{\alpha\beta}_{[q_{n}\bar{q}_{\nbar}]}
  \sim\mathcal{S}^{\alpha\beta}_{[g_{n}{g}_{\nbar}]}
  \sim\order{(1)}\,.
\end{align}
The last piece to examine is the beam function, see
Eqs.~\eqref{eq:M2PS:NJ0:Bnq}-\eqref{eq:M2PS:NJ0:Bng}.
It contains the integral of the collinear building blocks,
$(W^{\dagger}_n\xi_n)$ and $\mathcal{A}_{n}^{\perp}$
sandwiched between the proton states, with respect to $b_n$.
Akin to the case of ${b}_{\tau_{\|}}$, the power of $b_n$ is
now related to $P_{\nbar}$, giving $b_n\sim P_{\nbar}\sim\order{(1)}$.
For the integrand, the scaling rules of those operators
and the corresponding external states are
$(W^{\dagger}_n\xi_n)\sim\mathcal{A}_{n}^{\perp}\sim\order(\lamtau)$
and $|q_{n}\rangle\sim|g_n(\pm)\rangle\sim\order(\lambda^{-1}_{\tau})$ \cite{Bauer:2002aj,Lange:2003pk,Beneke:2003pa}, respectively.
We can thus conclude,
\begin{align} 
  \mathcal{B}_{n}^{[q_i]}
  \sim\mathcal{B}_{n,h'h}^{[g_n]}\sim\order{(1)}\,.
\end{align}
Combining the above scaling relationships, we observe that
the only ingredient of Eq.~\eqref{eq:xs:NJ0:fac} that can
bring about a power-like behaviour is the differential
$\done {b}_{\tau_{\|}}$, from which the power of the
$N_{\mathrm{J}}=0$ configuration is lowered by $\lamtau$.
In accordance, the $q_{\tau}$ spectrum behaves as,
\begin{align}\label{eq:xs:NJ0:fac:pc}
  \frac{\done^5{\sigma_{\ttbar}}}
       {\done \Mttbar^2\,\done^2\vec{P}_{t}^{\perp}\,
        \done\Yttbar\,\done q_{\tau} }
  \Bigg|_{N_{\mathrm{J}}=0}
  \sim \order(\lambda^{-1}_{\tau})\,.
\end{align}
Confronting Eq.~\eqref{eq:xs:NJ0:fac:pc} with the series
in Eq.~\eqref{eq:AsyExp}, it illustrates that
Eq.~\eqref{eq:xs:NJ0:fac} can deliver at least in part
the most singular behaviors of Eq.~\eqref{eq:AsyExp}.
In order to assess the existence of other contributions to
the leading asymptotic terms in Eq.~\eqref{eq:AsyExp}, we
devote the following section to investigate all possible
$N_{\mathrm{J}} \ge1 $ configurations.

\subsection{The case of \texorpdfstring{$N_{\mathrm{J}}\ge1$}{NJ>=1}}
\label{sec:fac:NJm}

\begin{figure}[h!]
  \centering
  \includegraphics[width=.5\textwidth]{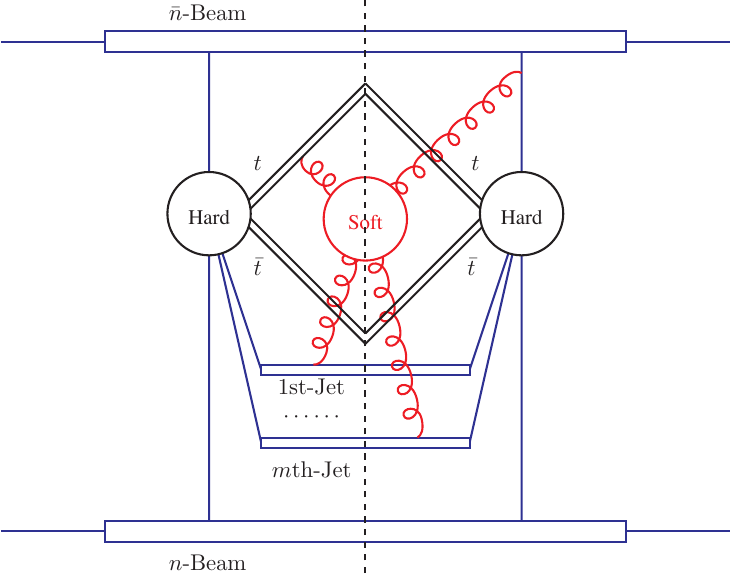}
  \caption{
    Dynamic regions in the $N_{\mathrm{J}}\ge1 $ configuration.
    Herein, the black bubbles stand for the hard modes.
    The red bubble and curly lines encode the soft corrections
    connecting the heavy partons represented by the black double
    lines as well as both the beam- and jet-collinear modes
    depicted in the blue straight lines.
    \label{fig:PSS:NJm}
  }
\end{figure}

This part will discuss the contributions induced
by the hard, beam-collinear, jet-collinear, and soft modes
in the layouts with at least one hard jet.
A representative diagram of the associated dynamic regions is displayed
in Fig.~\ref{fig:PSS:NJm}.

\subsubsection{The \texorpdfstring{$N_{\mathrm{J}}=1$}{NJ=1} configuration}\label{sec:fac:NJ1}

We start our analysis with a single insertion of a jet region.
So far, as the jet momentum is the sole source of energetic
transverse recoil for the $\ttbar$ system, the components
$q_{\tau_{\|}}$ and $q_{\tau_{\perp}}$ must admit the
asymmetric configuration in Eq.~\eqref{eq:def:region:tt:asy}
as a result of momentum conservation.
With this in mind, we can now expand the kinematic variables
and the boundary conditions of Eq.~\eqref{eq:QCDF},
\begin{align}
  \label{eq:Ettbar:NJ1}
  \MTttbar\Bigg|_{N_{\mathrm{J}}=1}
  =&\;
    \sqrt{\Mttbar^2+q^2_{\tau_{\perp}}}+\order(\lamtau)\,
    \equiv\, \MTttbart\,+\order(\lamtau)\,,
    \\[2mm]
  \label{eq:thetakin:NJ1}
  \Theta_{\mathrm{kin}}\Bigg|_{N_{\mathrm{J}}=1}
  =&\;
    \Theta\Big[\sqrt{s}-\MTttbart-|q_{\tau_{\perp}}|\Big]
    \Theta\Big[ \MTttbart-\mTtt-\mTtbart\Big]
    \nnb\\
  &\hspace*{20mm}
    \times
    \Theta\left[\sinh^{-1}
                  \left(\sqrt{\frac{(\Mttbar^2+s)^2}
                                   {4\,s\,{\MTttbart}^{\!\!\!2}}-1}
                  \right)
                                      -|\Yttbar|
          \right]+\order(\lamtau)\nnb\\
  \equiv&\;
    \Theta^{(1)}_{\mathrm{kin}}+\order(\lamtau)\,,
    \\[2mm]
  \label{eq:Pzt:NJ1}
  P^z_t\Bigg|_{N_{\mathrm{J}}=1}
  =&\;
    \pm\,
    \frac{\MTttbart}{2}
    \sqrt{
      1+\frac{\mTtt+\mTtbart}{\MTttbart}
    }\,
    \sqrt{
      1+\frac{\mTtt-\mTtbart}{\MTttbart}
    }\,
    \sqrt{
      1-\frac{\mTtt+\mTtbart}{\MTttbart}
    }\,
    \sqrt{
      1-\frac{\mTtt-\mTtbart}{\MTttbart}
    }\,
    \nnb\\
  &\;{}
    +\order(\lamtau)\,\nnb\\
  \equiv&\;
    \widehat{P}^z_{t}\,+\order(\lamtau)\,,
\end{align}
where the approximate transverse masses of the top and antitop quarks
are defined as $\mTtt\equiv\sqrt{m^2_t+(\vec{q}_{\tau_{\perp}})^2}$ and
$\mTtbart\equiv\sqrt{m^2_t+(\vec{P}_{t}^{\perp}-\vec{q}_{\tau_{\perp}})^2}$,
respectively.
The approximate $\ttbar$ invariant mass $\MTttbart$ is defined in Eq.\
\eqref{eq:Ettbar:NJ1}.
While Eqs.~\eqref{eq:Ettbar:NJ1}-\eqref{eq:thetakin:NJ1} follow
immediately from the definition of the transverse mass ${M}_{\mathrm{T}}$
and the boundary condition of Eq.~\eqref{eq:thetakin:ft},
deriving Eq.~\eqref{eq:Pzt:NJ1} necessitates solving the
energy conservation equation in the zRF,
\begin{align}
  {M}_{\mathrm{T}}
  =
    \sqrt{ {m}_{\mathrm{T}}^2+(P^z_t)^2}
    +\sqrt{ ({m}'_{\mathrm{T}})^2+(P^z_t)^2}\,.
\end{align}
The solution is then expanded using the power counting of
Eq.~\eqref{eq:def:region:tt:asy}, keeping the lowest power
contributions.
Using these results, we can now evaluate the partonic contributions
for $N_{\mathrm{J}}=1$.
We follow the same steps as in the derivation of Eq.~\eqref{eq:M2PS:NJ0}
with the addition of embedding the soft Wilson lines and the jet functions
as appropriate here.
The soft-collinear decomposition as illustrated in
Eqs.~\eqref{eq:scetII:Ln}-\eqref{eq:hqet:Lv} is independent
of the specific configurations and, thus, still holds at present.
After combining all contributions and omitting the unrelated
higher power correction terms, we arrive at the partonic function
$\Sigma_{\ttbar}$ of the $N_{\mathrm{J}}=1$ case,
\begin{align}\label{eq:M2PS:NJ1}
  \Sigma_{\ttbar}\Bigg|_{N_{\mathrm{J}}=1}
  =&\;
    \frac{1}{2\pi s {\eta}'_{n} {\eta}'_{\nbar}}
    \sum_{\kappa,\lambda}
    \int\done^2\vec{k}_{\mathrm{T}}\,\done y_k\,\done^2\bTvec\,
    \exp\left(
      \mathrm{i}\,\bTvec \cdot\qTvec
      +\mathrm{i}\,\bTvec\cdot \vec{k}_{\mathrm{T}}
    \right)
    \widetilde{\Sigma}^{[\kappa]}_{\ttbar,[\lambda]}
    (\bTvec, \vec{P}_{t} ,\vec{k}_{\mathrm{T}},\Yttbar,y_k,\mu,\nu)\;,
\end{align}
where now the indeces $\kappa\in
\{
  g_{n}g_{\nbar},
  q^i_{n}\bar{q}^j_{\nbar},
  \bar{q}^j_{n}q^i_{\nbar},
  q^i_{n}g_{\nbar},
  g_{n}q^i_{\nbar},
  \bar{q}^i_{n}g_{\nbar},
  g_{n}\bar{q}^i_{\nbar}
\}$ 
and $\lambda\in\{g_{n_{\mathrm{J}}},{q}^j_{n_{\mathrm{J}}},\bar{q}^j_{n_{\mathrm{J}}}\}$
label the initial and final state light partons, respectively.
$\vec{P}_{t}\equiv(\vec{P}_{t}^{\perp},\widehat{P}_t^z)$
denotes the spatial momentum of the top quark as measured
in the zRF.
$\vec{k}_{\mathrm{T}}$ and $y_k$ stand for the transverse momentum
and rapidity of the jet in the LF, respectively.
The variables ${\eta}'_{n}$ and ${\eta}'_{\nbar}$ are the
momentum fractions of the active partons in the beam functions
with respect to the colliding protons, namely,
\begin{align}
  {\eta}'_{n}
  &=
    ( \MTttbart\,e^{\Yttbar}+k_{\mathrm{T}}\,e^{y_k})/\sqrt{s}\,,\\
  {\eta}'_{\nbar}
  &=
    ( \MTttbart\,e^{-\Yttbar}+k_{\mathrm{T}}\,e^{-y_k})/\sqrt{s}\,.
\end{align}
In the following discussion, we will particularly focus on the
$q^i_{n}\bar{q}^j_{\nbar}\to\ttbar\,g_{n_{\mathrm{J}}}$ process.
All other partonic processes can be accessed through exchanging
the labels $n\leftrightarrow\nbar$ or the active partons herein,
as appropriate.
The expression of $\widetilde{\Sigma}^{[q^i_{n}\bar{q}^j_{\nbar}]}_{\ttbar,[g_{n_{\mathrm{J}}}]}$ reads,
\begin{align}
  \widetilde{\Sigma}^{[q^i_{n}\bar{q}^j_{\nbar}]}_{\ttbar,[g_{n_{\mathrm{J}}}]}
  (\bTvec,\vec{P}_{t}, \vec{k}_{\mathrm{T}},\Yttbar,y_k,\mu,\nu)
  =&\;
    \left(\frac{1}{2N_c}\right)^2\,
    \mathcal{B}_{n}^{[q_i]}( {\eta}'_n,\bT,\mu,\nu)\,
    \mathcal{B}_{\nbar}^{[\bar{q}_j]}({\eta}'_{\nbar},\bT,\mu,\nu) \,
    \mathcal{J}^{[g]}_{{n_{\mathrm{J}}}}(\vec{k}_{\mathrm{T}},y_k)\,
    \nnb\\
  &\;\;\times
    \sum_{\alpha,\beta}\left\{
    \mathcal{H}_{\alpha\beta,[g_{n_{\mathrm{J}}}]}^{[q^i_{n}\bar{q}^j_{\nbar} ]}(\vec{P}_{t} ,\vec{k}_{\mathrm{T}},\Yttbar,y_k,\mu)\,
    \mathcal{S}^{\alpha\beta,[g_{n_{\mathrm{J}}}]}_{[q_{n}\bar{q}_{\nbar}]}(\bTvec,\mu,\nu)\,
    \right\}\,,
    \label{eq:M2PS:NJ1:qqbar}
\end{align}
where the beam functions $\mathcal{B}_{n}^{[q_i]}$ and
$\mathcal{B}_{\nbar}^{[\bar{q}_j]}$ are defined in the
same way as those in Eqs.~\eqref{eq:M2PS:NJ0:Bnq}-\eqref{eq:M2PS:NJ0:Bng}.
The soft sector
$\mathcal{S}^{\alpha\beta,[g_{n_{\mathrm{J}}}]}_{[q_{n}\bar{q}_{\nbar}]}$
takes the similar appearance to that in eq.~\eqref{eq:M2PS:NJ0:Sqq},
except for the necessary adaptation in the colour bases and the inclusion
of the jet Wilson lines.
The jet function $\mathcal{J}^{[g]}_{{n_{\mathrm{J}}}}$ is the novel
ingredient in the $N_{\mathrm{J}}=1$ configuration,
\begin{align}
  \label{eq:M2PS:NJ1:Jg}
  \mathcal{J}^{[g]}_{{n_{\mathrm{J}}}}(\vec{k}_{\mathrm{T}},y_k)
  =
  -\frac{1}{4\pi(N_c^2-1) }
  \,
  \int
  \done m^2_{k}\,
  \done^4 x\,
  \exp\left(i k\cdot x \right)\,
 \mathbf{Tr}\,  \langle0|  \mathcal{A}^{\bot}_{n_{\mathrm{J}}}(x)  \mathcal{A}_{n_{\mathrm{J}}}^{\bot  }(0)   |0\rangle
  \,
  =\,
  1\,,
\end{align}
where $m^2_{k}=k^2$ measures the offshellness of the jet radiations.
$\mathcal{A}_{n_{\mathrm{J}} }^{\bot}$ stands for the gauge-invariant
collinear fields in \SCETII, see Eq.~\eqref{eq:M2PS:NJ0:Bng}.
In calculating Eq.~\eqref{eq:M2PS:NJ1:Jg}, after using the dimensional
regulator to regulate the UV divergences, completing the coordinate
space integral in Eq.~\eqref{eq:M2PS:NJ1:Jg} always results in
contributions of the form $(m_k^2/\mu^2)^{\epsilon}$
\cite{Fleming:2002rv,Fleming:2002sr,Becher:2010pd,Banerjee:2018ozf,Becher:2006qw,Bruser:2018rad}.
The following integration over the complete $m_k$ range, however, turns
out to be scaleless and, thus, the unmeasured jet functions involved in
this paper will never receive any perturbative corrections in $\alpha_s$.
Therefore, we equate $\mathcal{J}^{[g]}_{{n_{\mathrm{J}}}}$ in
Eq.~\eqref{eq:M2PS:NJ1:Jg} with its tree-level result.
Further, it is worth pointing out that, since we are not observing the
jet itself, but only its recoil on the $\ttbar$ system, the relative
transverse momenta amongst the collinear emissions as well as their
helicity dependence have been integrated out.

The quark jet function shows the same behaviour
\cite{Becher:2006qw,Bruser:2018rad},
\begin{align}
  \label{eq:M2PS:NJ1:Jq}
  \mathcal{J}^{[q]}_{{n_{\mathrm{J}}}}(\vec{k}_{\mathrm{T}},y_k)
  =\,
    \frac{1}{8\pi N_c\nbar_{\mathrm{J}}\cdot p}\,
    \int\done m^2_{k}\,\done^4 x\,
    \exp\left(i k\cdot x \right)\,
    \mathbf{Tr}\,
    \langle0|  \xi_{n_{\mathrm{J}}}(x)  \bar{\xi}_{n_{\mathrm{J}}}(0)\slashed{\nbar}_{\mathrm{J}}   |0\rangle
    \,
  =\,
    1\,.
\end{align}
In addition to the above functions capturing the low-offshellness
effects, $\widetilde{\Sigma}^{[q^i_{n}\bar{q}^j_{\nbar}]}_{\ttbar,[g_{n_{\mathrm{J}}}]}$
also needs the hard function
$\mathcal{H}_{\alpha\beta,[g_{n_{\mathrm{J}}}]}^{[q^i_{n}\bar{q}^j_{\nbar} ]}$,
consisting of the UV-renormalised and IRC-subtracted partonic amplitude
$\mathcal{M}(q\bar{q}\to\ttbar\,g)$ constructed in analogy to
Eqs.~\eqref{eq:M2PS:NJ0:Hqqbar}-\eqref{eq:M2PS:NJ0:Hgg}.
The presence of the jet mode, however, complicates the IRC-subtraction
procedure as singularities arising from the region
$|\vec{k}_{\mathrm{T}}|\to 0$ will need to be treated.
This necessitates the inclusion of lower-multiplicity partonic processes,
e.g.,\ by using 
\begin{align}
\mathcal{Z}_3\otimes\mathcal{M}(q\bar{q}\to\ttbar\,g)+\mathcal{Z}_{2}\otimes\mathcal{M}(q\bar{q}\to\ttbar)
\end{align} 
as the IRC-finite quantity~\cite{Becher:2016mmh,Balsiger:2020ogy}.
Its details, however, are unimportant in the following as
the only quantity of interest in this paper is the scaling
behaviour of the hard function
$\mathcal{H}_{\alpha\beta,[g_{n_{\mathrm{J}}}]}^{[q^i_{n}\bar{q}^j_{\nbar} ]}$
itself.


Substituting Eq.~\eqref{eq:M2PS:NJ1} into Eq.~\eqref{eq:QCDF},
we obtain the master formula for the $N_{\mathrm{J}}=1$ configuration,
\begin{align}\label{eq:xs:NJ1:fac}
  \frac{\done^5{\sigma_{\ttbar}}}
       {\done \Mttbar^2\,\done^2\vec{P}_{t}^{\perp}\,
        \done \Yttbar\,\done q_{\tau}}
  \Bigg|_{N_{\mathrm{J}}=1}
  =&\;
    \sum_{\mathrm{sign}[\widehat{P}^z_{t}]}
    \sum_{\kappa,\lambda}
    \int\done q_{\tau_{\perp}}\,\done y_k\;
    \frac{\Theta^{(1)}_{\mathrm{kin}}}{8 (2\pi)^5}\,
    \frac{\widetilde{\Sigma}_{\ttbar,[\lambda]}^{[\kappa]}
          (\vec{0},\vec{P}_{t},-\vec{q}_{\tau_{\perp}},\Yttbar,y_k,\mu,\nu)}
          {\MTttbart|\widehat{P}^z_{t}|{\eta}'_{n} {\eta}'_{\nbar}s^2}
    \,.
\end{align} 
In deriving this formula,  the argument $\bTvec$ of the function
$\widetilde{\Sigma}_{\ttbar,[\lambda]}^{[\kappa]}$ has been
integrated out following the multipole expansion.
To see this, we apply the decomposition in Eq.~\eqref{eq:bT:NJ0:dec}
again onto the impact parameter $\bTvec$ and the jet transverse
momentum $\kTvec$ in eq.~\eqref{eq:M2PS:NJ1},
\begin{align}\label{eq:bT:NJ1:dec}
  \bTvec
  =
    \vec{b}_{\tau_{\perp}}+ \vec{b}_{\tau_{\|}}
  =
    b_{\tau_{\perp}}\vec{\tau}\times\vec{n}+ b_{\tau_{\|}}\vec{\tau}\;,
  \qquad
  \kTvec
  =
    \vec{k}_{\tau_{\perp}}+ \vec{k}_{\tau_{\|}}
  =
    k_{\tau_{\perp}}\vec{\tau}\times\vec{n}+ k_{\tau_{\|}}\vec{\tau}\;.
\end{align} 
The parallel component ${k}_{\tau_{\|}}$ drops out of the lowest
power hard sector $\mathcal{H}_{\alpha\beta,[\lambda]}^{[\kappa]}$
during the asymptotic expansion, due to the scaling hierarchy
\begin{align}\label{eq:def:hierarchy:NJ1}
  {k}_{\tau_{\|}}\sim\order(\lamtau)
  \ll {k}_{\tau_{\perp}}\sim\order(1)\,.
\end{align}
Therefore, the ${k}_{\tau_{\|}}$ integral therein can be calculated
immediately, resulting in a $\delta(b_{\tau_{\|}})$ distribution in
the integral.
The following ${b}_{\tau_{\|}}$ integral in
$\widetilde{\Sigma}_{\ttbar,[\lambda]}^{[\kappa]}$ can then also
be carried out without further complications.
On the other hand, in the perpendicular direction, the expansion in $\lamtau$ is subject to the relationship
\begin{align}
  q_{\tau_{\perp}}\sim k_{\tau_{\perp}}\sim\order(1)
  \gg k^{\perp}_s\sim k^{\perp}_c\sim k^{\perp}_{\bar{c}}\sim\order(\lamtau) \,,
\end{align} 
where $k^{\perp}_s$ and $k^{\perp}_{c(\bar{c})}$ represent the
transverse momenta of the soft and beam-collinear regions, as
given in Eqs.~\eqref{eq:regions:soft}-\eqref{eq:regions:beam:nbar}.
Thus, the multipole expansion at the hard vertices can
eliminate the argument $\vec{b}_{\tau_{\perp}}$ of the beam-collinear
and soft functions up to $\order(\lambda_\tau)$~\cite{Beneke:2002ph},
such that the inverse Fourier transformation in Eq.~\eqref{eq:M2PS:NJ1}
can be completed prior to the integral over $k_{\tau_{\perp}}$,
giving rise to $\delta(k_{\tau_{\perp}}+q_{\tau_{\perp}})$ and, thus,
sets $\vec{k}_{\mathrm{T}}=-\vec{q}_{\tau_{\perp}}$ upon integration.

Using Eq.~\eqref{eq:xs:NJ1:fac}, we are now ready to determine the
power accuracy of the $N_{\mathrm{J}}=1$ configuration.
Following the assessment of Eq.~\eqref{eq:xs:NJ0:fac}, for the bulk
of the phase space, the kinematic factors in Eq.~\eqref{eq:xs:NJ1:fac},
such as $\MTttbart$ and $\widehat{P}^z_{t}$, invoke no power-like
behaviour, and thus all belong to $\order(1)$.
The hard, beam-collinear, and soft functions are of $\order(1)$ by
construction, and so are the jet functions of
Eqs.~\eqref{eq:M2PS:NJ1:Jg}-\eqref{eq:M2PS:NJ1:Jq}.
Then, the remaining factors that matter to the power counting are
the differentials $\done q_{\tau_{\perp}}$  and $\done y_k$, which
are of $\order(1)$ as well due to the scaling laws in
Eq.~\eqref{eq:def:region:tt:asy} and the definition of
Eq.~\eqref{eq:regions:jets}.
Hence, all the ingredients from Eq.~\eqref{eq:xs:NJ1:fac} are
characterised by the $\order(1)$ behaviour, which then allows us
to establish,
\begin{align}\label{eq:xs:NJ1:fac:pc}
  \frac{\done{\sigma_{\ttbar}}}
  {\done\Mttbar^2\,\done^2\vec{P}_{t}^{\perp}\,  \done\Yttbar\,\done q_{\tau}}\Bigg|_{N_{\mathrm{J}}=1}
  \sim\;
    \order(1)\,.
\end{align}
Comparing Eq.~\eqref{eq:xs:NJ1:fac:pc} with the asymptotic series
in Eq.~\eqref{eq:AsyExp} and the ${N_{\mathrm{J}}=0}$ scaling rule
in Eq.~\eqref{eq:xs:NJ0:fac:pc}, it is noted that the $N_{\mathrm{J}}=1$
configuration here gives rise to the regular behaviors of
Eq.~\eqref{eq:AsyExp}, which pertains to  the NLP corrections and
is one power higher than the ${N_{\mathrm{J}}=0}$ influences.

\subsubsection{The \texorpdfstring{$N_{\mathrm{J}}\ge2$}{NJ>=2} configuration}
\label{sec:fac:NJ2}

The case with (at least) two hard jet insertions differs from
the pervious case where the scaling laws for $q_{\tau_{\|}}$
and $q_{\tau_{\perp}}$ are uniquely determined.
The variety of the jet transverse momenta in the
$N_{\mathrm{J}}\ge2$ configuration can accommodate both the
isotropic and asymmetric recoil configurations in
Eqs.~\eqref{eq:def:region:tt:sym}-\eqref{eq:def:region:tt:asy}.
In the following paragraphs, we will discuss them individually.

\paragraph*{Isotropic recoil.}
We start with the isotropic recoil configuration.
In this case, the transverse components $q_{\tau_{\|}}$ and
$q_{\tau_{\perp}}$ respect the power prescription in
Eq.~\eqref{eq:def:region:tt:sym}, from which the reduced
kinematics variables have been presented in
Eqs.~\eqref{eq:thetakin:NJ0}-\eqref{eq:Pzt:NJ0}.
As demonstrated in Eq.~\eqref{eq:xs:NJ0:fac:pc}, none of them
bears any power-like behaviour.
Then, the problem reduces to the scaling behaviour of the
partonic function $\Sigma_{\ttbar}$ in
Eq.~\eqref{eq:def:QCD:parton} in the $N_{\mathrm{J}}\ge2$
configuration.
%
The calculation of $\Sigma_{\ttbar}$ now follows similarly
to Eq.~\eqref{eq:M2PS:NJ1} in the $N_{\mathrm{J}}=1$ case, aside from
duplicating the jet function and generalizing the hard and soft
sectors in accordance, giving
\begin{align}\label{eq:M2PS:NJm}
  \Sigma_{\ttbar}\Bigg|_{N_{\mathrm{J}}\ge2}\,
  \sim\,
  \int
  \,
  \bigg[  \prod^{N_{\mathrm{J}}}_{i=1} \done y_{i}\, \done^2 \vec{k}^{\perp}_{i}\bigg]\,
  \done^2 \bTvec\,
  \exp\bigg[\mathrm{i}\;\bTvec\cdot
  \Big(\qTvec+\sum^{N_{\mathrm{J}}}_{j=1} \vec{k}^{\perp}_{j} \Big)\bigg]
  \,
  \mathcal{H}
  \,
  \otimes
  \mathcal{S}
  \,
  \otimes
  \mathcal{B}_n
  \,
  \otimes
    \mathcal{B}_{\nbar}
  \,
  \otimes \prod^{N_{\mathrm{J}}}_{l=1}  J_{n_{l}}\,,
\end{align}
where $y_{i}$ and $\vec{k}^{\perp}_{i}$ denote the pseudo-rapidity
and transverse momentum of the $i$-th jet.
For simplicity, the indices specifying the partonic channels have
been omitted here.
Akin to the $N_{\mathrm{J}}=1$ case, the hard, soft, beam, and jet
functions in Eq.~\eqref{eq:M2PS:NJm} are all of $\order(1)$.
To appraise the power accuracy of the pseudo-rapidity $y_{i}$, it
merits noting that due to the scaling rules in
Eq.~\eqref{eq:def:region:jet:parton} and Eq.~\eqref{eq:regions:jets},
the differential $\done y_{i}$ always evaluates to $\order(1)$ and
thus does not influence the scaling behaviour of $\Sigma_{\ttbar}$.

Now the remaining task is to determine the scaling behaviour of
$\vec{k}^{\perp}_{i}$ and $\bTvec$.
It is noted that the scaling behaviour of $\bTvec$ depends on the
regions of $\vec{k}^{\perp}_{i}$.
To exhaust all the possibilities, we regroup the jets here
according to the scaling behaviour of the transverse components,
\begin{align}
  \mathcal{P}_{k}
  \,\equiv&\;
    \{\vec{k}_{i}^{\perp}\in\mathcal{R}_{k};\;
    \order(\lamtau)\sim q_{\tau_{\perp}}\sim q_{\tau_{\|}}\sim
    \vec{k}_{i,\tau_{\perp}} \ll  \vec{k}_{i,\tau_{\|}} \sim
    \order(\lambda^0_{\tau}) \}\,,\label{eq:def:Pset:Iso}\\
  \mathcal{O}_{k}
  \,\equiv&\;
    \{\vec{k}_{i}^{\perp}\in\mathcal{R}_{k};\;
    \order(\lamtau)\sim q_{\tau_{\perp}}\sim q_{\tau_{\|}}\sim
    \vec{k}_{i,\tau_{\|}} \ll  \vec{k}_{i,\tau_{\perp}} \sim
    \order(\lambda^0_{\tau}) \}\,,\label{eq:def:Oset:Iso} \\
  \mathcal{I}_{k}
  \,\equiv&\;
    \{\vec{k}_{i}^{\perp}\in\mathcal{R}_{k};\;
    \order(\lamtau)\sim q_{\tau_{\perp}}\sim q_{\tau_{\|}}\ll
    \vec{k}_{i,\tau_{\perp}}\sim\vec{k}_{i,\tau_{\|}} \sim
    \order(\lambda^0_{\tau}) \}\,,\label{eq:def:Iset:Iso}
\end{align}
where the full set $\mathcal{R}_k$ collects all transverse momenta
for the $N_{\mathrm{J}}$ jets, more explicitly,
\begin{align}
  \mathbf{card}[\mathcal{R}_{k}]=N_{\mathrm{J}}\,.
\end{align}
Here the operator $\mathbf{card}$ evaluates the cardinality of a set.
$\mathcal{P}_{k}$, $\mathcal{O}_{k}$, and $\mathcal{I}_{k}$ are
three non-intersecting subsets of $\mathcal{R}_k$ consisting of
different types of the jet directions.
For instance, $\mathcal{P}_{k}$ contains $\vec{k}_i^{\perp}$
parallel or antiparallel to the reference vector $\vec{\tau}$,
while $\mathcal{O}_{k}$ encompasses the orthogonal ones.
$\mathcal{I}_{k}$ comprises all the other configurations.

We are now ready to investigate the scalings for $\bTvec$.
As required by the label momentum conservation~\cite{Beneke:2002ph},
the scaling power of $b_{\tau_{\|}}$ ($b_{\tau_{\perp}}$) is subject
to the strongest momenta in the $\vec{\tau}$ ($\vec{n}\times\vec{\tau}$)
direction.
To this end, if there are label momenta dictating both sides, namely,
$\mathbf{card}[\mathcal{P}_{k}] \,\mathbf{card}[\order_{k}]+\mathbf{card}[\mathcal{I}_{k}] \ge1$, we have $\done\bTvec\sim\order(1)$.
Otherwise, either the $\vec{\tau}$ or the $\vec{n}\times\vec{\tau}$
direction will be governed by $\order(\lamtau)$ fluctuations,
which gives rise to $\done\bTvec\sim\order(\lambda^{-1}_{\tau})$.
Summarising these relationships, we are capable of establishing the
power counting for $\Sigma_{\ttbar}$,
\begin{align}\label{eq:M2PS:NJm:Sigma:Iso}
  \Sigma_{\ttbar}\Bigg|_{N_{\mathrm{J}}\ge2,\mathbf{Iso}}\,
  \sim\,
    \left\{
      \begin{array}{lr}
        \order{\left(\lamtau^{\mathbf{card}[\mathcal{P}_{k}]+\mathbf{card}[\order_{k}]}\right)},
        &
        \mathbf{card}[\mathcal{P}_{k}]\,\mathbf{card}[\mathcal{O}_{k}]
        +\mathbf{card}[\mathcal{I}_{k}] \ge 1\,,\\
        \order{(\lamtau^{N_{\mathrm{J}}-1})},
        &
        \mathbf{card}[\mathcal{P}_{k}]\,\mathbf{card}[\mathcal{O}_{k}]
        +\mathbf{card}[\mathcal{I}_{k}] =0\,,
      \end{array}
    \right.
\end{align}
where the extra powers of $\lamtau$ come from the differentials
$\done k_{i,\tau_{\perp}}\sim\order(\lamtau)$ of $\mathcal{P}_{k}$
and $\done k_{i,\tau_{\|}}\sim\order(\lamtau)$ of $\mathcal{O}_{k}$.
In light of the non-negative nature of the cardinality, the lowest
power Eq.~\eqref{eq:M2PS:NJm:Sigma:Iso} can reach is $\order(1)$,
where the sets $\mathcal{P}_{k}$ and $\mathcal{O}_{k}$ are both
empty and thus $\mathcal{I}_{k}=\mathcal{R}_{k}$.
It should be emphasised that this finding is not dependent on
the number of the embedded jet modes, or, more specifically, the
result of $\mathbf{card}[\mathcal{I}_{k}]$.
This differs from the na\"ive expectation from the scaling behaviour
of the effective Hamiltonian $\mathcal{H}^{\mathrm{eff}}_{\ttbar}$,
where it appears that the power accuracy of
$\mathcal{H}^{\mathrm{eff}}_{\ttbar}$ grows along with the increase
in the number of jets.
The reason is that every jet in our calculation is unmeasured and
thus participates in the $q_{\tau}$ spectrum such that, when calculating
the contributions in Eqs.~\eqref{eq:M2PS:NJ1:Jg}-\eqref{eq:M2PS:NJ1:Jq},
the 4-dimensional coordinate space integral
$\done^4x\sim\order(\lambda^{-4}_{\tau})$ balances the power suppression
from the collinear field operators and the differential $\done m_k^2$.
Hence, the insertion of the jet modes invokes no power-like behaviour
unless kinematical constraints are imposed on the jet directions,
such as those in $\mathcal{P}_{k}$ or $\mathcal{O}_{k}$.
  
Substituting the result of Eq.~\eqref{eq:M2PS:NJm:Sigma:Iso} into
Eq.~\eqref{eq:QCDF}, we arrive  at
\begin{align}\label{eq:M2PS:NJm:Iso:min}
  \mathrm{Min}
  \left[
    \frac{\done{\sigma_{\ttbar}}}{\done \Mttbar^2\,\done^2\vec{P}_{t}^{\perp}\,
          \done\Yttbar\,\done q_{\tau} }\Bigg|_{N_{\mathrm{J}}\ge2,\mathbf{Iso}}
  \right]
  \,\sim\,\order(\lamtau)\,,
\end{align}
where the integral over $q_{\tau_{\perp}}$ has increased the power
of Eq.~\eqref{eq:M2PS:NJm:Sigma:Iso} by one order of $\lamtau$.
In previous investigations on the $\qT$~spectrum, the scaling
of the isotropic pattern was also addressed in \cite{Collins:1980ui}
and the outcome in Eq.~\eqref{eq:M2PS:NJm:Iso:min} is in agreement
with their findings.
Comparing Eq.~\eqref{eq:M2PS:NJm:Iso:min} with the $N_{\mathrm{J}}=0$
and $N_{\mathrm{J}}=1$ configurations, it is found that the result
here is one order higher than $N_{\mathrm{J}}=1$ case in
Eq.~\eqref{eq:xs:NJ1:fac:pc}, and two orders higher with respect
to the $N_{\mathrm{J}}=0$ one from Eq.~\eqref{eq:xs:NJ0:fac:pc}.

\paragraph*{Asymmetric recoil.}
Turning now to the asymmetric recoil configuration, the transverse
components $q_{\tau_{\|}}$ and $q_{\tau_{\perp}}$ observe the scaling
rules in Eq.~\eqref{eq:def:region:tt:asy}.
The accordingly expanded boundary conditions can be found in
Eqs.~\eqref{eq:thetakin:NJ1}-\eqref{eq:Pzt:NJ1}.
As demonstrated in Eq.~\eqref{eq:xs:NJ1:fac:pc}, those kinematic
factors are all of $\order(1)$.
Thus, the investigation of the scaling behaviour driven by the
asymmetric recoil configuration relies again on the analysis of
the convolution function in Eq.~\eqref{eq:M2PS:NJm}.
As before, to denominate all the possible configurations of the
jet transverse momenta, we introduce the sets,
\begin{align}
  \mathcal{P}_{k}
  \,\equiv&\;
    \{\vec{k}_{i}^{\perp}\in\mathcal{R}_{k};\;
    \order(\lamtau)\sim q_{\tau_{\|}}\sim\vec{k}_{i,\tau_{\perp}}\ll
    q_{\tau_{\perp}}\sim \vec{k}_{i,\tau_{\|}}\sim
    \order(\lambda^0_{\tau}) \}\,,\label{eq:def:Pset:Asy}\\
  \mathcal{O}_{k}
  \,\equiv&\;
    \{\vec{k}_{i}^{\perp}\in\mathcal{R}_{k};\;
    \order(\lamtau)\sim q_{\tau_{\|}}\sim\vec{k}_{i,\tau_{\|}}\ll
    q_{\tau_{\perp}}\sim \vec{k}_{i,\tau_{\perp}}\sim
    \order(\lambda^0_{\tau}) \}\,,\label{eq:def:Oset:Asy}\\
  \mathcal{I}_{k}
  \,\equiv&\;
    \{\vec{k}_{i}^{\perp}\in\mathcal{R}_{k};\;
    \order(\lamtau)\sim q_{\tau_{\|}}\ll q_{\tau_{\perp}}\sim
    \vec{k}_{i,\tau_{\perp}}\sim\vec{k}_{i,\tau_{\|}}\sim
    \order(\lambda^0_{\tau}) \}\,.\label{eq:def:Iset:Asy}
\end{align}
Since there are at least one pair of label momenta presiding
over the direction $\vec{n}\times \vec{\tau}$ now,
the differential $\done b_{\tau_{\perp}}$ is always of
$\order(\lambda^0_{\tau})$.
As for the $\vec{\tau}$ direction, if there also appear any
label momenta, namely
$\mathbf{card}[\mathcal{O}_{k}]+\mathbf{card}[\mathcal{I}_{k}]\ge1$,
the power scaling of $b_{\tau_{\perp}}$ will give the same
result as the perpendicular piece, i.e.\
$\done b_{\tau_{\perp}}\sim\order(\lambda^0_{\tau})$.
Otherwise, this direction will still be occupied by the soft
and beam fluctuations, which leads to
$\done b_{\tau_{\perp}}\sim\order(\lambda^{-1}_{\tau})$.
Summarizing those observations, it follows,
\begin{align}\label{eq:M2PS:NJm:Sigma:Asy}
  \Sigma_{\ttbar}\Bigg|_{N_{\mathrm{J}}\ge2,\mathbf{Asy}}\,\sim\,
  \left\{
    \begin{array}{lr}
      \order{\left(\lamtau^{\mathbf{card}[\mathcal{P}_{k}]+\mathbf{card}[\order_{k}]   }\right)},
      &
      \mathbf{card}[\order_{k}]+\mathbf{card}[\mathcal{I}_{k}]\ge1\,,\\
      \order{(\lamtau^{N_{\mathrm{J}}-1})},
      &
      \mathbf{card}[\order_{k}]+\mathbf{card}[\mathcal{I}_{k}]  =0\,.
    \end{array}
  \right.
\end{align}
Combining  this result with Eq.~\eqref{eq:QCDF} and exploiting
the non-negative nature of the cardinality, the minimal power
of the cross section in the asymmetric recoil configuration
can be derived,
\begin{align}\label{eq:M2PS:NJm:Ays:min}
  \mathrm{Min}
  \left[
    \frac{\done{\sigma_{\ttbar}}}{\done\Mttbar^2\,\done^2\vec{P}_{t}^{\perp}\,
          \done\Yttbar\,\done q_{\tau} }\Bigg|_{N_{\mathrm{J}}\ge2,\mathbf{Asy}}
  \right]
  \,\sim\,\order(1)\,,
\end{align}
where the scaling rule $q_{\tau_{\perp}}\sim\order(\lambda^0_{\tau})$
has been utilised in line with Eqs.~\eqref{eq:def:region:tt:asy}.
Comparing the outcome with Eq.~\eqref{eq:xs:NJ1:fac:pc} of the
$N_{\mathrm{J}}=1$ case, it is
interesting to note that the lowest power behaviour for the
asymmetric configuration is not impacted by the increase in the
number of jets $N_{\mathrm{J}}$, both are of $\order(1)$.
However, in light of the asymptotic series in Eq.~\eqref{eq:AsyExp}
and the power accuracy of the $N_{\mathrm{J}}=0$ configuration
in Eq.~\eqref{eq:xs:NJ0:fac:pc}, it is noted that
Eq.~\eqref{eq:M2PS:NJm:Ays:min} is only able to account for the
regular terms in part, which belongs to at least NLP and thus will
not be the main concern in the latter numeric evaluations.

\subsubsection{Summary and discussion}\label{sec:fac:recap}
 
\changed{In Sec.~\ref{sec:fac:NJ0}, \ref{sec:fac:NJ1}, \ref{sec:fac:NJ2},
 we have enumerated all  configurations that are
relevant to the regime $q_{\tau}\to0$ and determined the corresponding  power
precision  with the help of the effective
field theories \SCETII and HQET.} We find that the leading-power
contribution of $ \order(\lambda^{-1}_{\tau})$ is given by the
$N_{\mathrm{J}}=0$ configuration in Eq.~\eqref{eq:xs:NJ0:fac:pc}.
It is followed at next-to-leading power at $\order(\lambda^{0}_{\tau})$
by the asymmetric recoil configurations regardless of the total
number of the embedded jet modes, see Eq.~\eqref{eq:xs:NJ1:fac:pc}
for $N_\mathrm{J}=1$ and Eq.~\eqref{eq:M2PS:NJm:Ays:min} for
$N_\mathrm{J}\ge 2$, respectively.
The highest order contribution at $\order(\lamtau)$ is produced
by the isotropic recoil configuration in Eq.~\eqref{eq:M2PS:NJm:Iso:min}
with the $N_{\mathrm{J}}\ge2$ jets having been incorporated.

Comparing those power rules with the asymptotic series in
Eq.~\eqref{eq:AsyExp}, we observe that the leading singular
terms are solely governed by the $N_{\mathrm{J}}=0$ configuration.
As will be illustrated below, this finding imposes non-trivial
constraints on the ingredients of
Eqs.~\eqref{eq:M2PS:NJ0:qqbar}-\eqref{eq:M2PS:NJ0:gg}.
Firstly, due to the IRC-safe nature of the observable
$\done\sigma_{\ttbar}/\done q_{\tau}$, the asymptotic terms in
Eq.~\eqref{eq:AsyExp} are finite at each power in $\lamtau$.
In terms of the EFT ingredients, this means the IRC subtraction
factor of the hard sector in
Eqs.~\eqref{eq:M2PS:NJ0:Hqqbar}-\eqref{eq:M2PS:NJ0:Hgg} and
the renormalisation constants of the beam and soft functions
in Eqs.~\eqref{eq:M2PS:NJ0:Bnq}-\eqref{eq:M2PS:NJ0:Bng} and Eqs.~\eqref{eq:M2PS:NJ0:Sqq}-\eqref{eq:M2PS:NJ0:Sgg} must
cancel after the combination,
\begin{align}\label{eq:M2PS:NJ0:Zuv}
  \sum_{\alpha,\beta}\,
  \Big( \mathcal{Z}^{h,{\alpha'\alpha}}_{[\kappa]}\mathcal{Z}^{s,{\alpha\alpha''}}_{[\kappa]}\Big)^*
  \,
  \mathcal{Z}^{h,{\beta'\beta}}_{[\kappa]}
  \,
  \mathcal{Z}^{s,\beta\beta''}_{[\kappa]}
  \,
  \mathcal{Z}^{c}_{[\kappa_n]}\,   \mathcal{Z}^{c}_{[\kappa_{\nbar}]}
  \,
  =
  \delta_{\alpha'\alpha''}   \delta_{\beta'\beta''} \,,
\end{align}
where the superscript $\kappa$ here runs again over the partonic
channels
$\{g_{n}g_{\nbar},q^i_{n}\bar{q}^j_{\nbar},q^i_{\nbar}\bar{q}^j_{n}\}$
as in Eq.~\eqref{eq:M2PS:NJ0}.
$\kappa_n$ and $\kappa_{\nbar}$ mark the incoming active partons
along the $n$ and $\nbar$ directions, respectively.
We have omitted the contributions from
$\mathcal{Z}^{cs}_{[\kappa_{n(\nbar)}]}$ of Eq.~\eqref{eq:M2PS:NJ0:Bnq}
as it was solely introduced to remove the soft-collinear overlap
\cite{Luo:2019hmp,Luo:2019bmw} and thus is not involved with the
renormalisation of the beam function.
The relationship in Eq.~\eqref{eq:M2PS:NJ0:Zuv} will be examined
in Sec.~\ref{sec:soft:func:NLO} through an explicit NLO calculation.

From Eq.~\eqref{eq:M2PS:NJ0:Zuv}, we can infer the scale evolution
relationship amongst the elements of
Eqs.~\eqref{eq:M2PS:NJ0:qqbar}-\eqref{eq:M2PS:NJ0:gg}.
It is worth reminding that the scale dependences in the hard, beam,
and soft functions are all brought about through the IRC subtraction
or the UV/rapidity renormalisation.
To this end, the cancellation of those subtraction factors and
the renormalisation constants renders the partonic convolution
functions $\widetilde{\Sigma}_{\ttbar}^{[\kappa]}$ in
Eqs.~\eqref{eq:M2PS:NJ0:Hqqbar}-\eqref{eq:M2PS:NJ0:Hgg} independent
of the scale $\mu$ or $\nu$, more specifically,
\begin{align}\label{eq:M2PS:NJ0:MuNu:indep}
  \frac{\done}{\done\ln\mu}
  \widetilde{\Sigma}_{\ttbar}^{[\kappa]}
  (\bTvec,\Mttbar,\bttbar,x_t,\Yttbar,\mu,\nu)
  =\frac{\done}{\done \ln\nu}
   \widetilde{\Sigma}_{\ttbar}^{[\kappa]}
   (\bTvec,\Mttbar,\bttbar,x_t,\Yttbar,\mu,\nu)=0\,.
\end{align}
This result correlates the RGEs and RaGEs of the relevant ingredients
therein, and also permits us to leave out the scales $\mu$ and $\nu$
from the arguments of $\widetilde{\Sigma}_{\ttbar}^{[\kappa]}$ hereafter.
In Sec.~\ref{sec:res}, we will utilise Eq.~\eqref{eq:M2PS:NJ0:MuNu:indep}
to derive the evolution equations for the soft interactions.
 
Apart from the leading singularities in Eq.~\eqref{eq:AsyExp},
our analyses in Sec.~\ref{sec:fac:NJ1} and Sec.~\ref{sec:fac:NJ2}
demonstrate that the following subleading terms entail the
participations of the jet modes.
In the previous investigations, a similar point was first addressed
in \cite{Collins:1980ui}.
There, the power scaling of the jet contributions is determined
by assessing the relevant pinch singularity surfaces in the low
$\qT$~domain, from which the asymptotic behaviors of the $\qT$
spectra are related to the power laws of the hard
vertices\footnote{%
  The focus of \cite{Collins:1980ui} is only on the Drell-Yan process.
  However, in absence of the super renormalisable vertices, such as the
  Coulomb exchanges, the findings of \cite{Collins:1980ui} are, however,
  generalisable to top quark pair production at LHC.
},
\begin{align}\label{eq:collins:qT}
  \frac{\done\sigma_{\ttbar}}{\done^2\qTvec}
  \propto\delta^2(\qTvec-\vec{K}_{1}^{\perp}-\vec{K}_{2}^{\perp}),
\end{align}
where $\vec{K}_{1,2}^{\perp}$ are the transverse momenta of two
active partons connecting the hard vertex producing the top quark
pair.
If $\vec{K}_{1,2}^{\perp}$ and $\qTvec$, including all their components,
are of $\order(\lamtau)$, it yields
$\done\sigma_{\ttbar}/\done^2\qTvec\sim \order(\lambda^{-2}_{\tau})$,
which is in agreement with our result in Eq.~\eqref{eq:xs:NJ0:fac:pc}
from the $N_{\mathrm{J}}=0$ configuration after integrating out
$q_{\tau_{\perp}}$.
When the jet modes are taken into account in the isotropic recoil
configuration in Eq.~\eqref{eq:def:region:tt:sym}, it follows that
$\qTvec\sim\order(\lamtau)$ and $\vec{K}_{1,2}^{\perp}\sim\order(1)$,
which gives rise to the scaling behaviour
$\done\sigma_{\ttbar}/\done^2\qTvec\sim \order(1)$ from
Eq.~\eqref{eq:collins:qT}, also coinciding with the expression in
Eq.~\eqref{eq:M2PS:NJm:Iso:min} once the integral over $q_{\tau_{\perp}}$
is performed.
This congruence of findings is a consequence of the equivalence of
Eq.~\eqref{eq:collins:qT} and our approach in regards to the isotropic
configuration in Eq.~\eqref{eq:def:region:tt:sym}.
During our derivations, the power behaviour of the $q_{\tau}$ spectrum
is extracted in part from the scaling laws of the impact parameter
$\bTvec$, which is in practice correlated to the delta function in
Eq.~\eqref{eq:collins:qT} by means of the inverse Fourier transformation,
as illustrated in \eqref{eq:scaling:bTvec}.

Nevertheless, once the asymmetric recoil configuration of Eq.~\eqref{eq:def:region:tt:asy} is encountered, it is not
straightforward to apply Eq.~\eqref{eq:collins:qT}.
The power rules for the transverse components are
$q_{\tau_{\|}}\sim\order(\lamtau)$ and $q_{\tau_{\perp}}\sim\order(1)$
here, from which the right-hand side of Eq.~\eqref{eq:collins:qT}
is of $\order(\lambda^{-1}_{\tau})$ in the $N_{\mathrm{J}}=1$
situation and $\order(1)$ in the $N_{\mathrm{J}}\ge2$ case.
Even though this accidentally agrees with our EFT-based derivation
in the $N_{\mathrm{J}}\ge2$ case, see Eq.~\eqref{eq:M2PS:NJm:Ays:min},
its prediction in the $N_{\mathrm{J}}=1$ case is one power lower than
our EFT-based outcome in Eq.~\eqref{eq:xs:NJ1:fac:pc}.
This mismatch originates in part in the fact that in \cite{Collins:1980ui}
all the jet transverse momenta possess homogenous components, as is the
case our $\mathcal{I}_{k}$ classification in Eq.~\eqref{eq:def:Iset:Iso}
and Eq.~\eqref{eq:def:Iset:Asy}.
This arrangement works well in the isotropic configuration of
Eq.~\eqref{eq:def:region:tt:sym}.
However, if the jet orientations are of particular concern, such as in
the  $N_{\mathrm{J}}=1$ case, the extra power suppression will come
into play by means of the integrals over the transverse components,
e.g.\ the projected element $k_{\tau_{\|}}$ detailed in
Eq.~\eqref{eq:def:hierarchy:NJ1}.

\subsection{The soft function with the exponential regulator}\label{sec:soft:func:NLO}

\begin{figure} [h!]
  \centering
  \begin{subfigure}{.28\textwidth}
    \centering
    \includegraphics[width=0.8\textwidth]{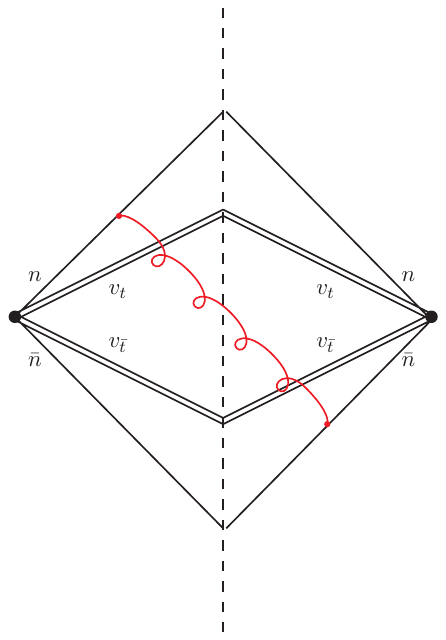}
    \caption{Light-light correlation\\\phantom{H}}
    \label{fig:FeynDiags:Soft:Func:nn}
  \end{subfigure}
  \begin{subfigure}{.28\textwidth}
    \centering
    \includegraphics[width=0.8\textwidth]{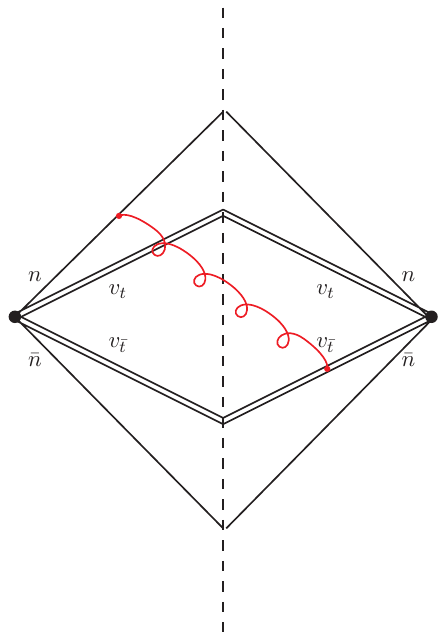}
    \caption{Light-heavy correlation\\\phantom{H}}
    \label{fig:FeynDiags:Soft:Func:nv}
  \end{subfigure}
  \begin{subfigure}{.28\textwidth}
    \centering
    \includegraphics[width=0.8\textwidth]{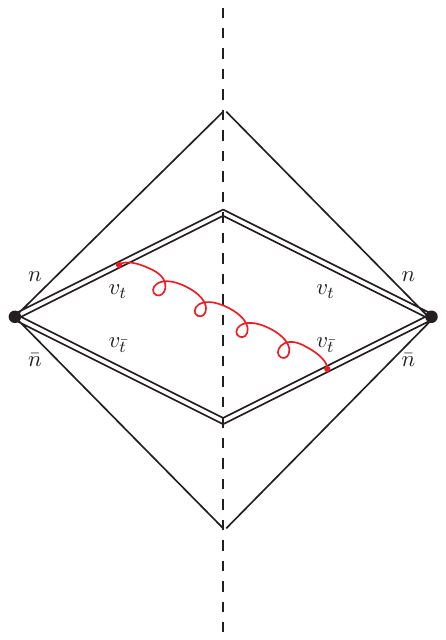}
    \caption{heavy-heavy correlation\\\phantom{H}}
    \label{fig:FeynDiags:Soft:Func:vv}
  \end{subfigure}
  \caption{
    Representative Feynman diagrams for the NLO soft function.
    The double line represents the gauge links attached to the
    top or antitop quark.
    The single line stands for that from the incoming massless
    parton, such as gluon, quark, and antiquark.
  }
  \label{fig:FeynDiags:Soft:Func}
\end{figure}

In the previous parts, we have derived the factorisation formula
for the leading singular behaviour of the $q_{\tau}$ spectrum in
Eqs.~\eqref{eq:M2PS:NJ0:qqbar}-\eqref{eq:M2PS:NJ0:gg}, which entails
the soft functions $\mathcal{S}^{\alpha\beta}_{[\dots]}$ to
accommodate the wide angle correlations amongst the active partons.
The field-operator definitions of the soft elements have been
presented in Eqs.~\eqref{eq:M2PS:NJ0:Sqq}-\eqref{eq:M2PS:NJ0:Sgg}
in terms of the soft Wilson lines, see \cite{Beneke:2009rj},
sandwiched by the orthonormal colour basis in
Eqs.~\eqref{eq:def:color:basis:qq}-\eqref{eq:def:color:basis:gg}.
In this part, we will calculate the soft function to NLO accuracy.
Their three typical contributions are depicted in
Fig.~\ref{fig:FeynDiags:Soft:Func}.

Without any loss of generality, the  fixed-order results can be parameterised as
\begin{align}
  \mathcal{S}^{\alpha\beta}_{[\kappa]}(\bTvec,\mu,\nu)
  \equiv
    \sum_{m=0}^\infty\,
    \left[\frac{\alpha_s(\mu)}{4\pi}\right]^m\,
    \mathcal{S}^{\alpha\beta,(m)}_{[\kappa]}(\bTvec,\mu,\nu)\,,
\end{align}
where $\kappa\in\{g_{n}g_{\nbar},q^i_{n}\bar{q}^j_{\nbar},q^i_{\nbar}\bar{q}^j_{n}\}$
as that in Eq.~\eqref{eq:M2PS:NJ0}.
The coefficients for the first two orders in $\alpha_s$ read
\begin{align} 
  \label{eq:def:SF:LO}
  \mathcal{S}^{\alpha\beta,(0)}_{[\kappa]}(\bTvec,\mu,\nu)
  \,=&\;
    \delta_{\alpha\beta}\,, \\
  \label{eq:def:SF:NLO}
  \mathcal{S}^{\alpha\beta,(1)}_{[\kappa]}(\bTvec,\mu,\nu)
  \,=&\;
    \sum_{a,b}\,
    \langle c^{\alpha}_{\kappa}|\,
            \mathbf{T}_a\!\cdot\mathbf{T}_b\,|c^{\beta}_{\kappa}
    \rangle\;
    \mathcal{I}_{ab}(\bTvec,\mu,\nu)\,.
\end{align}
In absence of any soft interactions, the LO result
$\mathcal{S}^{\alpha\beta,(0)}_{[\kappa]}$ is equal to the
identity matrix as a consequence of the orthonormality of our
colour basis in
Eqs.~\eqref{eq:def:color:basis:qq}-\eqref{eq:def:color:basis:gg}.
For the NLO expression, we employ the colour algebra formalism
suggested in \cite{Catani:1996vz} to illustrate the outcome.
Therein, the flavour subscripts $a,b\in\{q,\bar{q},g,t,\bar{t}\}$
denote the active partons participating in the hard kernel.
$\mathbf{T}_a$ signifies the colour charge operator for the parton
$a$, and $\mathbf{T}_b$ likewise.
$|c_{\kappa}\rangle$ is the vector representation of
$c^{\kappa}_{\{a\}}$ in
Eqs.~\eqref{eq:def:color:basis:qq}-\eqref{eq:def:color:basis:gg}
in colour space.
The coefficient function $\mathcal{I}_{ab}$ contains the
contributions from the squared soft amplitudes induced by the
Wilson lines in Eqs.~\eqref{eq:M2PS:NJ0:Sqq}-\eqref{eq:M2PS:NJ0:Sgg}.
Its expression in the exponential regularisation scheme
\cite{Li:2016axz,Li:2016ctv} is
\begin{align} \label{eq:def:MI:soft}
  \mathcal{I}_{ab}(\bTvec,\mu,\nu)
  \equiv\,
    -\frac{\left(\mu^2 e^{\gamma_{\mathrm{E}}}\pi\right)^{\epsilon}}{\pi}\,
    \lim_{\delta\to0}\int\done k_n\, \done k_{\nbar}\,\done^{2-2\epsilon}\;
    \vec{k}_{\mathrm{T}}\,\delta^+(k^2)\,
    \frac{v_a\cdot v_b}{(k\cdot v_a)(k\cdot v_b)}\,
    e^{\mathrm{i}\vec{k}_{\mathrm{T}}\cdot\bTvec- b_0 E_k\delta}
    \Bigg|_{\delta\to\frac{1}{\nu}}\,
    +\mathcal{Z}_{\mathrm{c}.\mathrm{t}.}^{ab}\,,
\end{align}
where $b_0=2\exp(-\gamma_{\mathrm{E}})$ with $\gamma_{\mathrm{E}}$
being the Euler constant.
$\epsilon$ and $\delta$ are the regulators for the virtuality and
rapidity divergences, respectively.
$v_a$ denotes the four-velocity of parton $a$, for instance
$n^\mu(\nbar^{\mu})$ for $a=q_{n(\nbar)}$ and
$v^{\mu}_{t(\bar{t})}=P^{\mu}_{t(\bar{t})}/m_t$ for the $a=t(\tbar)$.
$\mathcal{Z}_{\mathrm{c}.\mathrm{t}.}$ stems from the perturbative
expansion of the renormalisation constant $\mathcal{Z}_{[\kappa]}^{s}$
and its complex conjugate in
Eqs.~(\ref{eq:M2PS:NJ0:Sqq}-\ref{eq:M2PS:NJ0:Sgg}), defined in
the \MSbar scheme throughout this paper.
  
The result of Eq.~\eqref{eq:def:MI:soft} depends on the partons
$a$ and $b$.
We will first examine the case of $a$ and $b$ being light flavours,
$a,b=l\in\{q,\bar{q},g\}$, depicted in Fig.\ \ref{fig:FeynDiags:Soft:Func:nn}.
Starting with the case $a=b=l$, the light-like nature of the
vectors $n$ and $\nbar$ trivialises the calculation and we have,
\begin{align} \label{eq:def:MI:soft:ll}
  \mathcal{I}_{ll}(\bTvec,\mu,\nu)=0\,,\qquad
  \mathcal{Z}_{\mathrm{c}.\mathrm{t}.}^{ll}(\bTvec,\mu,\nu,\epsilon)=0\,.
\end{align}
However, if the participants consist of different light particles,
$a=l\ne b=l'$, $\mathcal{I}_{ll'} $ does not vanish in general.
To calculate its contribution, we follow \cite{Li:2016axz} and
first integrate out the longitudinal components $k_{n}$ and
$k_{\nbar}$ in line with the exponential regularisation.
Then, we expand the result in $\delta$ and truncate the series to
$\order(\delta^0)$.
Finally, we carry out the integral over $\vec{k}_{\mathrm{T}}$
with the aid of the dimensional regulator, giving
\begin{align}\label{eq:def:MI:soft:llp}
  \mathcal{I}_{ll'}(\bTvec,\mu,\nu)
  =
    L^2_{\mathrm{T}}-2\,L_{\mathrm{T}}\,L_{\nu} +\frac{\pi^2}{6}\,,
    \qquad
  \mathcal{Z}_{\mathrm{c}.\mathrm{t}.}^{ll'}(\bTvec,\mu,\nu,\epsilon)
  =
    \frac{2}{\epsilon^2}+\frac{2\, }{\epsilon}L_{\nu}\,,
\end{align}
where $L_{\nu}=\ln[\mu^2/\nu^2]$ and $L_{\mathrm{T}}=\ln[\bT^2\mu^2/b_0^2]$.
We have compared Eq.~\eqref{eq:def:MI:soft:llp} to the expressions
in \cite{Li:2016axz} and find the agreement after synchronising the
overall colour factors.
   
In addition to those correlations between the incoming particles,
Eq.~\eqref{eq:def:MI:soft} also involves the contributions from the
heavy top quark, $h\in\{t,\bar{t}\}$, see
Fig.\ \ref{fig:FeynDiags:Soft:Func:nv}.
The presence of the massive partons complicates the calculation
substantially.
Since the denominators at this moment are not homogeneous in
$k_{n}$ and $k_{\nbar}$, performing the integral over those
longitudinal components is not straightforward, especially
when involving the exponent $E_k\delta$.
To this end, we resort to the Mellin-Barnes (MB) representation
\cite{Smirnov:1999gc,Tausk:1999vh} to recast the inhomogeneous
propagators in a first step.
For $\mathcal{I}_{lh}$, we thus apply the following
substitution~\cite{Smirnov:2012gma,Bejdakic:2009zz},
\begin{align}
  \frac{1}{(k\cdot v_{h})^{\lambda}}  \,
  \xRightarrow[\text{ }]{\textbf{M.B. }}\,
  -\frac{1}{\Gamma(\lambda)}\frac{1}{4\pi^2}
  &
    \int^{c_i+\mathrm{i}\infty}_{c_i-\mathrm{i}\infty}\done z_1\,\done z_2\;
    \Gamma(-z_1)\,\Gamma(-z_2)\,\Gamma(\lambda+z_1+z_2)\\
  &
    \qquad
    \times\,
    \left[\frac{(k\cdot n)(v_h\cdot\nbar)}{2}\right]^{z_1}\,
    \left[\frac{(k\cdot{\nbar})(v_h\cdot{n})}{2}\right]^{z_2}\,
    \left(-\vec{k}_{\mathrm{T}}\cdot \vec{v}^{\perp}_{h}
    \right)^{-\lambda-z_1-z_2}\,,\nnb
\end{align}
where the contours (or the values $c_i$) are chosen such that
the poles from  $\Gamma (\lambda+ z_1+z_2)$ are to the left of
the path, while those of $\Gamma( -z_1)\Gamma( -z_2)$ are to
the right.
Integrating the MB-transformed propagators over $k_{n{(\nbar)}}$
and $\vec{k}_{\mathrm{T}}$ now follows the similar pattern to
that in deriving Eq.~\eqref{eq:def:MI:soft:llp}.
Then, to perform the $\delta$-expansion, we first make use of
the package \texttt{MB} \cite{Czakon:2005rk} to determine the
contours for $z_1$ and $z_2$, and then feed the outputs to
\texttt{MBasymptotics}~\cite{Czakon:Hepforge}
for deriving the asymptotic series in $\delta$.
The remaining integrals are those over $z_1$ and $z_2$, for
which we utilise \texttt{MBsums} \cite{Ochman:2015fho} to
implement Cauchy's residue theorem and sum up the ensuing
residues from the software \texttt{Mathematica}.
The final expression is,
\begin{align}
  \label{eq:def:MI:soft:lh}
  \mathcal{I}_{lh}(\bTvec,\mu,\nu)
  \,=&\;
    \frac{L^2_{\mathrm{T}}}{2}
    -L_{\mathrm{T}}\,L_{\nu}
    +2\,L_{\mathrm{T}}\,\ln\left(v_l\cdot v_h\right)\nnb\\
  &\;{}
    +2\,\mathrm{arcsinh}^2\Big[|\vec{v}_{h}^{\perp}|\sin(\varphi_{hb})\Big]
    -(2\pi\mathrm{i})\,\mathrm{arcsinh}
      \Big[|\vec{v}_{h}^{\perp}|\cos(\varphi_{hb})\Big]\,
    \nnb\\
  &\;{}
    +\frac{\pi^2}{12}
    +\mathcal{D}_{lh}(\varphi_{hb})
    +\mathcal{D}_{lh}(-\varphi_{hb})\,,\\
  \label{eq:def:MI:soft:lh:Z}
  \mathcal{Z}_{\mathrm{c}.\mathrm{t}.}^{lh}(\bTvec,\mu,\nu,\epsilon)
  \,=&\;
    \frac{1}{\epsilon^2}
    +\frac{1}{\epsilon}\Big[L_{\nu}-2\ln(v_l\cdot v_h)\Big]\,,
\end{align}
where $\varphi_{hb}$ denotes the azimuthal opening angle between
the vectors $\vec{v}_{h}^{\perp}$ and $\bTvec$.
The function $\mathcal{D}_{lh}$ is defined as
\begin{align}
  \mathcal{D}_{lh}(\theta)
  =
    \frac{4|\vec{v}_{h}^{\perp}|}{\pi}\,
    \int_0^{\frac{\pi}{2}}\done{\phi}\,
    \ln\left[\cot(\phi)\right]\,
    \Bigg\{
      \frac{\cos(\theta+\phi)}
           {\sqrt{1+|\vec{v}_{h}^{\perp}|^2\sin^2(\theta+\phi)}}
      \arcsin\bigg[\frac{|\vec{v}_{h}^{\perp}|\cos(\theta+\phi)}
                        {\sqrt{1+|\vec{v}_{h}^{\perp}|^2}}\bigg]
    \Bigg\}\,.
\end{align}
An analogous strategy is also applicable for the evaluations of the
$a=b=h$ and $a=h\neq b=h'$ cases, $\mathcal{I}_{hh}$ and
$\mathcal{I}_{hh'}$, of Fig.\ \ref{fig:FeynDiags:Soft:Func:vv}.
Since rapidity divergences do not emerge from those time-like gauge
links, we can set the regulator $\delta=0$ from the beginning and
solve the $k^{\mu}$-integrals in the conventional dimensional
regularisation.
The results  exhibit explicit $z_{1}$ and $z_{2}$ dependences,
which are treated with the above MB-Tools~\cite{Czakon:2005rk,Ochman:2015fho}
to complete the inverse MB transformations.
Eventually, they yield,
\begin{align}
  \label{eq:def:MI:soft:hh}
  \mathcal{I}_{hh}(\bTvec,\mu,\nu)
  \,=&\;
    2\,L_{\mathrm{T}}\,
    +\frac{{r}^{h}_b}{\sqrt{1+({r}^{h}_b)^2}}
     \Bigg[4\,\mathrm{arcsinh}({r}^{h}_b)-(2\pi\mathrm{i})\Bigg]\,,
    \\
  \label{eq:def:MI:soft:hh:Z}
  \mathcal{Z}^{hh}_{\mathrm{c.t.}}(\bTvec,\mu,\nu,\epsilon)
  \,=&\;
    -\frac{2}{\epsilon}\,,\\
  \label{eq:def:MI:soft:hhp}
  \mathcal{I}_{hh'}(\bTvec,\mu,\nu)
  \,=&\;
    L_{\mathrm{T}}\,
    \frac{4\,v_{h}\cdot v_{h'}}{\sqrt{(v_{h}\cdot v_{h'})^2-1}}\,
    \mathrm{arctanh}\bigg[\sqrt{\frac{v_{h}\cdot v_{h'}-1}
                                     {v_{h}\cdot v_{h'}+1}}\bigg]\nnb\\
  &\;
    +\int_{0}^1\done\zeta\,\frac{{r}^{hh'}_b(\zeta)}{\sqrt{1+\left({r}^{hh'}_b(\zeta)\right)^2}}\,
     \left(\frac{v_{h}\cdot v_{h'}}{{v}^2_{hh'}(\zeta)}\right)\,
     \Bigg[4\,\mathrm{arcsinh}\left({r}^{hh'}_b(\zeta)\right)-(2\pi\mathrm{i})\Bigg]\,,\\
  \label{eq:def:MI:soft:hhp:Z}
  \mathcal{Z}^{hh'}_{\mathrm{c.t.}}(\bTvec,\mu,\nu,\epsilon)
  \,=&\;
    -\frac{1}{\epsilon}\frac{4\,v_{h}\cdot v_{h'} }
                            {\sqrt{(v_{h}\cdot v_{h'})^2-1}}\,
     \mathrm{arctanh}\bigg[\sqrt{\frac{v_{h}\cdot v_{h'}-1}
                                      {v_{h}\cdot v_{h'}+1}}\bigg]\,,
\end{align}
where
\begin{align}
  {v}^{\mu}_{hh'}(\zeta)
  \,=\;
    \zeta\,v_h^{\mu}+(1-\zeta)\,v_{h'}^{\mu}\,,
  \qquad
  {r}^{h}_b
  \,=\;
    \widehat{b}_{\mathrm{T}}\cdot\vec{v}^{\perp}_{h}\,,
  \qquad
  {r}^{hh'}_b(\zeta)
  \,=\;
    \widehat{b}_{\mathrm{T}}\cdot\frac{\vec{v}_{hh'}^{\perp}(\zeta)}
                                      {\sqrt{  v^{2}_{hh'}(\zeta)}}\,.
\end{align}
During the calculations, we have exploited the on-shell conditions
$v_t^2=v_{\bar{t}}^2=1$ and introduced the definition
$\widehat{b}_{\mathrm{T}}\equiv \bTvec/\bT$.

With all relevant coefficient functions $\mathcal{I}_{ab}$ at hand,
together with the colour factors of Eq.~\eqref{eq:def:SF:NLO},
we can now establish the NLO soft function together with the
renormalisation constants $\mathcal{Z}^{s}_{[\kappa]}$.
For the latter case, we have checked that the results in
Eqs.~\eqref{eq:def:MI:soft:llp}-\eqref{eq:def:MI:soft:hhp:Z}
indeed satisfy the identity in Eq.~\eqref{eq:M2PS:NJ0:Zuv},
where the hard and beam renormalisation constants are extracted
from \cite{Ferroglia:2009ii} and \cite{Luo:2019hmp,Luo:2019bmw},
respectively.
Regarding the renormalised finite parts $\mathcal{I}_{ab}$,
we compare our expressions with those from the CSS framework
\cite{Catani:2014qha} at the scale $\mu=\nu=b_0^2/\bT^2$ and
find full agreement.
Furthermore, since during the derivation we did not utilise the
relationship $\vec{v}_{t}^{\perp}+\vec{v}_{\bar{t}}^{\perp}=\vec{0}$
for simplification, the results in
Eqs.~(\ref{eq:def:MI:soft:llp}-\ref{eq:def:MI:soft:hhp})
are also comparable with the soft function in the process
$pp\to \ttbar H$~\cite{Catani:2021cbl}.
However, at this moment, although the real parts of
$\mathcal{I}_{ab}$ still coincide with those in \cite{Catani:2021cbl},
the signs in front of the terms $2\pi\mathrm{i}$ in
Eq.~\eqref{eq:def:MI:soft:lh}, Eq.~\eqref{eq:def:MI:soft:hh},
and Eq.~\eqref{eq:def:MI:soft:hhp} are inverted.
This purely imaginary term does not enter the present calculation
on the $\ttbar$ production or the comparison with \cite{Catani:2014qha},
as these imaginary contributions cancel through the momentum
conservation $\vec{v}_{t}^{\perp}+\vec{v}_{\bar{t}}^{\perp}=\vec{0}$.
Nonetheless, this difference can influence the transverse
momentum spectrum in $\ttbar H$ production, or similar processes,
especially within the domain $M_{\ttbar H}\gtrsim \Mttbar\sim 2m_t$.
We leave it to a forthcoming publication to
elaborate on those processes and, in particular, deliver a
numeric comparison of our results, to those from
\cite{Catani:2021cbl}, and a fixed-order QCD calculation from
\Sherpa~\cite{Gleisberg:2003xi,Gleisberg:2008ta,Sherpa:2019gpd}.

\section {Resummation}\label{sec:res}

In the previous section we have analysed the dynamic regions
contributing to the leading singularities in Eq.~\eqref{eq:AsyExp}
and established the corresponding factorisation formula in impact
parameter space.
This section will now be devoted to the resummation of these
asymptotic behaviours.

\subsection{Asymptotic behavior}\label{sec:res:log:sing}

We start with identifying the singular terms in
Eq.~\eqref{eq:xs:NJ0:fac}.
Therefore, without loss of generality, we parametrise the
perturbative expansion of $\widetilde{\Sigma}_{\ttbar}$
as follows,
\begin{align}\label{eq:xs:NJ0:log}
  \widetilde{\Sigma}_{\ttbar}\,
  \sim\,
  \sum_{m,n}\,\alpha^m_s(\Mttbar)\;
  L^n_{\mathrm{M}}\,
  \Big\{
    s_{m,n}(\bttbar,x_t,\Yttbar)
    +a_{m,n}(\mathrm{sign}[b_{\tau_{\|}}], \bttbar,x_t,\Yttbar)
  \Big\}\;,
\end{align}
where $L_{\mathrm{M}}\equiv\log\left[{{b}^2_{\tau_{\|}}\Mttbar^2}/{b_0^2}\right]$
collects all the relevant dimensionful quantities.
Order by order, there are two dimensionless coefficients,
$s_{m,n}$ and $a_{m,n}$.
While both depend on $\bttbar$, $x_t$, and $\Yttbar$,
$s_{m,n}$ is independent of the magnitude and orientation
of the impact parameter.
Thus, we will refer them to the azimuthal \emph{symmetric}
terms (AST) hereafter.
On the other hand, in presence of the helicity-flipping
beam radiation in Eq.~\eqref{eq:M2PS:NJ0:Bng} and the
wide angle soft correlations from
Eqs.~\eqref{eq:M2PS:NJ0:Sqq}-\eqref{eq:M2PS:NJ0:Sgg},
$\widetilde{\Sigma}_{\ttbar}$ also contains contributions
sensitive to the orientation of the impact parameter, i.e.\
the azimuthal {\it asymmetric} term (AAT)
$a_{m,n}$~\cite{Catani:2010pd,Catani:2017tuc}.
Since the rejection component $b_{\tau_{\perp}}$ has been
integrated out in Eq.~\eqref{eq:eq:QCDF:NJ0:dec}, the
orientation dependence is reduced to a dependence on
the sign of $b_{\tau_{\|}}$ in Eq.~\eqref{eq:xs:NJ0:log}.
Please note, in deriving eq.~\eqref{eq:xs:NJ0:log}, we
have set the scales $\mu=\nu=\Mttbar$ in
Eqs.~\eqref{eq:M2PS:NJ0:qqbar}-\eqref{eq:M2PS:NJ0:gg} and
then extracted the $L_{\mathrm{M}}$ terms through
dimensional analysis.
Other scales choices $\{\mu,\nu\}$ lead to the same result
after regrouping the dimensional quantities appropriately,
on account of the (rapidity) renormalisation group discussed
in Sec.~\ref{sec:fac:recap}.

From Eq.~\eqref{eq:xs:NJ0:log}, we can evaluate the
$q_{\tau}$ spectrum by completing the inverse Fourier
transformation in Eq.~\eqref{eq:xs:NJ0:fac}.
Exploiting the fact that the function
$\cos(b_{\tau_{\|}}q_{\tau})$ is even under the integral
over $b_{\tau_{\|}}$, we have
\begin{align}\label{eq:xs:NJ0:log:invFT}
  \frac{\done{\sigma_{\ttbar}}}{\done\Mttbar^2\,\done^2\vec{P}_{t}^{\perp}\,
        \done \Yttbar\,\done q_{\tau} }
  \,\sim&\;
    \sum\limits_{m,n}
    \int^{\infty}_{-\infty}\done b_{\tau_{\|}}\,
    \cos(b_{\tau_{\|}}q_{\tau})\,L^n_{\mathrm{M}}\,
    \Big\{
      s_{m,n}( \bttbar,x_t,\Yttbar)
      +
      a_{m,n}(\mathrm{sign}[b_{\tau_{\|}}], \bttbar,x_t,\Yttbar)
    \Big\}\\
  \,=&\;
    \sum\limits_{m,n}
    \Big\{
      2s_{m,n}(\bttbar,x_t,\Yttbar)
      +a_{m,n}^+(\bttbar,x_t,\Yttbar)
      +a_{m,n}^-(\bttbar,x_t,\Yttbar)
    \Big\}\;
    \mathcal{F}^{(n)}_{\tau}(q_{\tau},\Mttbar)\,,\nnb
\end{align}
where $a_{m,n}^\pm$ are the decomposition of the AAT components
according to $\mathrm{sign}[b_{\tau_{\|}}]$ and the function
$\mathcal{F}^{(n)}_{\tau}$ is defined as,
\begin{align}
  \mathcal{F}^{(n)}_{\tau}(q_{\tau},\Mttbar)
  =\int^{\infty}_{0}\done b_{\tau_{\|}}\,
  \cos(b_{\tau_{\|}}q_{\tau})\,
  L^n_{\mathrm{M}}\,.
\end{align}
To appraise $\mathcal{F}^{(n)}_{\tau}(q_{\tau},\Mttbar)$, we
follow the strategy of \cite{Bozzi:2005wk} and introduce the
generating function,
\begin{align}\label{eq:def:gen:Ftau}
  \mathbf{F}_{\tau}(\eta,q_{\tau},\Mttbar)
  \,=&\;
    \sum_{n=0}^{\infty}\,\frac{\eta^n}{n!}\,
    \mathcal{F}^{(n)}_{\tau}(q_{\tau},\Mttbar)
  \,=\;
    \int^{\infty}_{0} \done {b}_{\tau_{\|}}\,
    \cos\left(  {b}_{\tau_{\|}} {q}_{\tau} \right)\,
    \left(\frac{{b}^2_{\tau_{\|}}\Mttbar^2}{b_0^2}\right)^{\eta}\nnb\\
  =&\;
    -\frac{4^{-\eta}e^{2\gamma_{\mathrm{E}}\eta}\,\sin(\pi\eta)\,\Gamma[2\eta+1]}
          {{q}_{\tau}}\,
     \left(\frac{\Mttbar^2}{q^2_{\tau}}\right)^{\eta}\,.
\end{align}
Hence, $\mathcal{F}^{(n)}_{\tau}(q_{\tau},\Mttbar)$ corresponds
to the $n$-th derivative of $\mathbf{F}_{\tau}(\eta,q_{\tau},\Mttbar)$
with respect to $\eta$ at the point $\eta=0$.
The results of the first few ranks read
\begin{equation}
  \begin{split}\label{eq:InvFT:F}
  \mathcal{F}^{(0)}_{\tau}(q_{\tau},\Mttbar)
  \,=&\;\;
    0\,,
    \\
  \mathcal{F}^{(1)}_{\tau}(q_{\tau},\Mttbar)
  \,=&\;
    -\frac{\pi}{q_{\tau}}\,,
    \\
  \mathcal{F}^{(2)}_{\tau}(q_{\tau},\Mttbar)
  \,=&\;
    -\frac{2\pi}{q_{\tau}}\ln\left[\frac{\Mttbar^2}{4q_{\tau}^2}\right]\,,
    \\
  \mathcal{F}^{(3)}_{\tau}(q_{\tau},\Mttbar)
  \,=&\;
    -\frac{3\pi}{q_{\tau}}\ln^2\left[\frac{\Mttbar^2}{4q_{\tau}^2}\right]\,
    -\frac{\pi^3}{q_{\tau}},
    \\
  \mathcal{F}^{(4)}_{\tau}(q_{\tau},\Mttbar)
  \,=&\;
    -\frac{4\pi}{q_{\tau}}\ln^3\left[\frac{\Mttbar^2}{4q_{\tau}^2}\right]\,
    -\frac{4\pi^3}{q_{\tau}}\ln\left[\frac{\Mttbar^2}{4q_{\tau}^2}\right]\,
    +\frac{64\pi \zeta_3}{q_{\tau}},
    \\
  \vdots\;&\\
  \mathcal{F}^{(k)}_{\tau}(q_{\tau},\Mttbar)
  \,=&\;
    -\frac{k\pi}{q_{\tau}}\ln^{k-1}\left[\frac{\Mttbar^2}{4q_{\tau}^2}\right]
    +\ldots\,.
  \end{split}
\end{equation}
Here we have focussed on the regime where $q_{\tau}$ is small, but
always larger than 0.
Otherwise, contributions such as $\delta[q_{\tau}]$ will enter the
expressions above.
From these results, it is observed that the coefficients $s_{m,n}$
and $a_{m,n}^\pm$ cannot induce any divergent behaviour by themselves
as exhibited in Eq.~\eqref{eq:InvFT:F}, whilst the logarithmic terms
$L_{\mathrm{M}}^k$ produce the singular series up to
$\ln^{k-1}[{\Mttbar^2}/{(4q_{\tau}^2)}]/q_{\tau}$, i.e.\  the entire
LP asymptotic behaviors of Eq.~\eqref{eq:AsyExp}.
As a consequence, the resummation of the $q_{\tau}$ spectrum can now be
expressed as an exponentiation of the $L_{\mathrm{M}}$, resembling
the corresponding resummation in the Drell-Yan processes and Higgs
production~\cite{Collins:1984kg,Catani:2000vq,Bozzi:2005wk,Bozzi:2007pn,
  Becher:2010tm,GarciaEchevarria:2011rb,Becher:2011dz,Chiu:2011qc,
  Chiu:2012ir,Li:2016axz,Li:2016ctv}.
In Sec.~\ref{sec:rge:rage}, we will use the solutions of RaGE and RGE
to accomplish this exponentiation.

 It is interesting to note that, by analogy to the $q_{\tau}$ resummation, the leading singular behaviour of the azimuthally averaged $\qT$ distribution is also governed by the characteristic logarithmic terms from the impact space \cite{Zhu:2012ts,Li:2013mia}, such that the R(a)GE framework is in principle applicable therein as well to accomplish the resummation.  However, aside from those two observables,   the transverse momentum resummation is generally  more involved on the $pp\to \ttbar+X$ process. 
For instance, the singular behaviour of the double differential observable
$\done\sigma_{\ttbar}/\done \qTvec$ is not only contained in the
logarithmic terms, but the AATs  can also make up in part the
asymptotic series~\cite{Catani:2017tuc}.
The appearance of these asymmetric divergent terms can have
non-trivial impacts on the pattern of the LP singularities
and the choice of resummation scheme.
In App.~\ref{app:phib:qTvec}, we will deliver a comparative study on
this issue.

\subsection{Evolution equations}\label{sec:rge:rage}

We will now elaborate on the evolution equations of the hard,
beam, and soft functions introduced in
Eqs.~\eqref{eq:M2PS:NJ0:qqbar}-\eqref{eq:M2PS:NJ0:gg}.

The hard function, see Eqs.~\eqref{eq:M2PS:NJ0:Hqqbar}-\eqref{eq:M2PS:NJ0:Hgg}, contains the squared UV-renormalised amplitudes multiplied by
the IRC regulator.
Considering that the scale dependences induced by the UV
renormaliation cancel fully within the amplitudes,
the evolution equation of the hard sector is governed solely
by the IRC counterterm $\mathcal{Z}^{h}_{[\kappa]}$.
According to the parametrisation of
\cite{Ferroglia:2009ii,Ferroglia:2009ep,Ahrens:2010zv}, we have,
\begin{align} \label{eq:methods:res_rge_H}
  \frac{\done}{\done\ln \mu^2}\,
  \mathcal{H}^{[\kappa]}_{\alpha\beta}\big(\Mttbar^2,\bttbar,x_t,\mu\big)
  \,=&\;
    -C_{[\kappa]}\,
    \Gamma_{\mathrm{cusp}}\,
    \ln\left[\frac{\mu^2}{\Mttbar^2}\right]\,
    \mathcal{H}^{[\kappa]}_{\alpha\beta}\big(\Mttbar^2,\bttbar,x_t,\mu\big)\,
    \nnb\\
  &\;{}
    +\sum_{\gamma}
     \left[
       \frac{\mathbf{\gamma}^{[\kappa]}_{h,\beta\gamma}}{2}\,
       \mathcal{H}^{[\kappa]}_{\alpha\gamma}\big(\Mttbar^2,\bttbar,x_t,\mu\big)
       +\left(
          \frac{\mathbf{\gamma}^{[\kappa]}_{h,\alpha\gamma}}{2}
        \right)^{\!\!\!*}
      \mathcal{H}^{[\kappa]}_{\gamma\beta}\big(\Mttbar^2,\bttbar,x_t,\mu\big)
     \right]\,,
\end{align}
where the subscripts  $\{\alpha,\beta,\gamma\}$ represent the colour
indices in the set of colour basis in
Eqs.~\eqref{eq:def:color:basis:qq}-\eqref{eq:def:color:basis:gg}.
The helicity indices involved in the gluonic channel have been omitted
for brevity.
$\kappa$ runs over
$\{g_{n}g_{\nbar},q^i_{n}\bar{q}^j_{\nbar},q^i_{\nbar}\bar{q}^j_{n}\}$,
indicating the partonic channel.
$C_{[\kappa]}$ denotes the colour factor in QCD with
\begin{align}
  \kappa\in\{g_{n}g_{\nbar}\}\,:\;
  C_{[\kappa]}=C_A\;,\qquad\qquad
  \kappa\in\{q^i_{n}\bar{q}^j_{\nbar},q^i_{\nbar}\bar{q}^j_{n}\}\,:\;
  C_{[\kappa]}=C_F\,.
\end{align}
$\Gamma_{\mathrm{cusp}}$ is the cusp anomalous dimension.
It is needed to three-loop accuracy \cite{Moch:2004pa} in
this paper, but is available to four-loop precision in the
literature \cite{Henn:2019swt,vonManteuffel:2020vjv}.
A numeric estimation of the five-loop contribution is
addressed in \cite{Herzog:2018kwj}.
Analogously, $\gamma^{[\kappa]}_{h}$ is the non-cusp anomalous
dimension for the hard contribution.
Their analytic expressions up to \NNLO~can be found in
\cite{Ferroglia:2009ii,Ferroglia:2009ep} and progress towards
the three-loop result has been made in \cite{Liu:2022elt}.

The quark and gluon beam functions are given in
Eqs.~\eqref{eq:M2PS:NJ0:Bnq}-\eqref{eq:M2PS:NJ0:Bng} in
terms of the \SCETII field operators.
They are same as those participating into the Drell-Yan
processes and Higgs production.
They admit the RGEs~\cite{Li:2016axz},
\begin{align}
  \frac{\partial}{\partial\ln\mu^2}
  \ln\mathcal{B}_{n}^{[q_n^i]}(\eta_{n},\bT,\mu,\nu)
  \,=&\;
    C_F\,\Gamma_{\mathrm{cusp}}\,
    \ln\bigg[\frac{\nu}{\eta_{n}\,\sqrt{s}}\bigg]
    +\gamma^{[q]}_b\;,\label{eq:methods:res_rge_Bq}
    \\
  \frac{\partial}{\partial\ln\mu^2}
  \ln\mathcal{B}^{[g_n]}_{n,h'h}(\eta_{n},\bTvec,\mu,\nu)
  \,=&\;
    C_A\,\Gamma_{\mathrm{cusp}}\,
    \ln\bigg[\frac{\nu}{\eta_n\,\sqrt{s}}\bigg]
    +\gamma^{[g]}_b\;,\label{eq:methods:res_rge_Bg}
\end{align}
as well as the RaGEs~\cite{Li:2016axz},
\begin{align}
  -\frac{2}{C_F}\,\frac{\partial}{\partial\ln\nu^2}
  \ln\mathcal{B}_{n}^{[q_i]}(\eta_{n},\bT,\mu,\nu)
  \,=&\;
    -\frac{2}{C_A}\,\frac{\partial}{\partial\ln\nu^2}
    \ln\mathcal{B}^{[g_n]}_{n,h'h}(\eta_{n},\bTvec,\mu,\nu)
    \nnb\\
  \,=&\;
    \gamma_r
    \left[\alpha_s\left(\frac{b_0}{\bT}\right)\right]\,
    +\int_{\mu^2}^\frac{b_0^2}{\bT^2}\,
     \frac{\done\bar{\mu}^2}{\bar{\mu}^2}\,
     \Gamma_{\mathrm{cusp}}\left[\alpha_s(\bar{\mu})\right]\,.
     \label{eq:methods:res_rage_B}
\end{align}
Here, $\eta_{n}$ stands for the momentum fraction along the
$n$ direction.
$\gamma^{[q,g]}_b$ and $\gamma_r$ are the non-cusp anomalous
dimensions brought about in the virtuality and rapidity
renormalisations, respectively.
Their specific expressions are dependent on the choice of
regularisation prescription, and this work will use
those that correspond with our choice of using the exponential
rapidity regulator~\cite{Li:2016axz}.
They are known at \NNLO accuracy \cite{Luo:2019bmw,Luo:2019hmp},
which we use in the following.
\NNNLO results are also available in the literature
\cite{Li:2016ctv,Vladimirov:2016dll,Luo:2020epw,Ebert:2020yqt,Luo:2019szz},
and N$^4$LO corrections
\cite{Das:2019btv,Duhr:2022cob,Duhr:2022yyp,Moult:2022xzt} have appeared
recently.
   
In order to derive the evolution equations for the soft sector,
we utilise the scale invariance condition in
Eq.~\eqref{eq:M2PS:NJ0:MuNu:indep} and the R(a)GEs above.
It follows that
\begin{align}
  \frac{\partial}{\partial\ln\mu^2}
  \mathcal{S}_{[\kappa]}^{\alpha\beta}\big(\bTvec,\mu,\nu\big)
  \,=&\;
    -\bigg\{
       C_{[\kappa]}\,\Gamma_{\mathrm{cusp}}\,
       \ln\left[\frac{\nu^2}{\mu^2}\right]
       +2\gamma_b^{[\kappa]}
     \bigg\}\,
     \mathcal{S}_{[\kappa]}^{\alpha\beta}\big(\bTvec,\mu,\nu\big)
  \nnb\\
  &\;
    -\sum_{\gamma}\,
     \left[
       \mathcal{S}_{[\kappa]}^{\alpha\gamma}\big(\bTvec,\mu,\nu\big)\,
       \frac{\mathbf{\gamma}^{[\kappa]}_{h,\gamma\beta} }{2}
       +\mathcal{S}_{[\kappa]}^{\gamma\beta}\big(\bTvec,\mu,\nu\big)\,
        \left(\frac{\mathbf{\gamma}^{[\kappa]}_{h,\gamma\alpha} }{2}\right)^{\!\!\!*}\,
     \right] \,,
  \label{eq:methods:res_rge_S}
\end{align}
and 
\begin{align}
  \frac{1}{C_{[\kappa]}}\,\frac{\partial}{\partial\ln \nu^2}
  \ln\mathcal{S}^{\alpha\beta}_{[\kappa]}\big(\bTvec,\mu,\nu\big)
  \,=&\;
    \gamma_r
    \left[\alpha_s\left(\frac{b_0}{\bT}\right)\right]\,
    +\int_{\mu^2}^\frac{b_0^2}{\bT^2}\,
     \frac{\done\bar{\mu}^2}{\bar{\mu}^2}\,
     \Gamma_{\mathrm{cusp}}\left[\alpha_s(\bar{\mu}) \right]\,   .
  \label{eq:methods:res_rage_S}
\end{align}
For brevity, we make use of $\gamma_b^{[\kappa]}$ here to
represent the virtuality anomalous dimensions in Eqs.~\eqref{eq:methods:res_rge_Bq}-\eqref{eq:methods:res_rge_Bg},
more specifically,
\begin{equation}
  \kappa\in\{g_{n}g_{\nbar}\}\,:\;
  \gamma_b^{[\kappa]}=\gamma_b^{[g]}\,,
  \qquad\qquad
  \kappa\in\{q^i_{n}\bar{q}^j_{\nbar},q^i_{\nbar}\bar{q}^j_{n}\}\,:\;
  \gamma_b^{[\kappa]}=\gamma_b^{[q]}\,.
\end{equation}
As observed in Eqs.~\eqref{eq:methods:res_rge_S}-\eqref{eq:methods:res_rage_S},
since the anomalous dimensions herein are all independently
extracted from the hard and beam functions, these evolution
equations provide a non-trivial opportunity to examine our
soft function of Sec.~\ref{sec:soft:func:NLO} and in turn
the factorisation in Eqs.~\eqref{eq:M2PS:NJ0:qqbar}-\eqref{eq:M2PS:NJ0:gg}.
Substituting the expressions of Eq.~\eqref{eq:def:SF:NLO}
into Eqs.~\eqref{eq:methods:res_rge_S}-\eqref{eq:methods:res_rage_S},
we have checked that our results indeed satisfy the criteria
above on the scale dependences.
 
Solving those RGEs and RaGEs permits us to bridge the intrinsic
scales of the hard, beam, and soft ingredients, thereby
exponentiating the characteristic logarithmic terms in
Eq.~\eqref{eq:xs:NJ0:log}.
Substituting these solutions into the master formula of
Eq.~\eqref{eq:xs:NJ0:fac}, we arrive at the resummed
$q_{\tau}$ spectrum,
\begin{align}\label{eq:xs:NJ0:res}
  \frac{\done{\sigma^{\mathrm{res}}_{\ttbar}}}
       {\done \Mttbar^2\,\done^2\vec{P}_{t}^{\perp}\,\done \Yttbar\,
        \done q_{\tau}}
  \,=&\;
    \sum_{\mathrm{sign}[x_t]}
    \frac{\Theta_{\mathrm{kin}}^{(0)}}
         {16\pi^3\,\bttbar\,|x_t|\,M^4_{\ttbar}\,s}
    \sum_{\kappa}\int^{\infty}_{-\infty}
    \done {b}_{\tau_{\|}}\,\cos\left({b}_{\tau_{\|}} {q}_{\tau} \right)\,
    \widetilde{\Sigma}_{\ttbar}^{\mathrm{res},[\kappa]}(b_{\tau_{\|}}\vec{\tau},
    \Mttbar,\bttbar,x_t,\Yttbar)\,,
    \end{align}
where
\begin{align}
  \label{eq:M2PS:NJ0:qqbar:res}
  &\hspace*{-5mm}
  \widetilde{\Sigma}_{\ttbar}^{\mathrm{res},[q^i_{n}\bar{q}^j_{\nbar}]}
  (\bTvec,\Mttbar,\bttbar,x_t,\Yttbar)\\
  \,=&\;
  \left(\frac{1}{2N_c}\right)^2\,
  \mathcal{D}^{\mathrm{res}}_{[q^i_{n}\bar{q}^j_{\nbar}]}
  (\bT,\Mttbar,\mu_h,\mu_b,\mu_s,\nu_b,\nu_s)\,
  \times
  \mathcal{B}_{n}^{[q_n^i]}(\eta_n,\bT,\mu_b,\nu_b)\,
  \mathcal{B}_{\nbar}^{[\bar{q}_\nbar^j]}(\eta_{\nbar},\bT,\mu_b,\nu_b) \,
  \nnb\\
  &
  \sum_{\{\alpha,\beta\}}\Bigg\{
  \mathcal{S}^{\alpha_1\beta_1}_{[q_{n}\bar{q}_{\nbar}]}(\bTvec,\mu_s,\nu_s)
  \left[\mathcal{V}^{[q_{n}\bar{q}_{\nbar}]}_{\alpha_1\alpha_2}(\bttbar,x_{t},\mu_s,\mu_h)\right]^*\,
  \mathcal{V}^{[q_{n}\bar{q}_{\nbar}]}_{\beta_1\beta_2}(\bttbar,x_{t},\mu_s,\mu_h)\,
  \mathcal{H}_{\alpha_2\beta_2}^{[q^i_{n}\bar{q}^j_{\nbar}]}(\Mttbar,\bttbar,x_{t},\mu_h)\,
  \Bigg\}\,,
  \nnb\\[2mm]
  \label{eq:M2PS:NJ0:gg:res}
  &\hspace*{-5mm}
  \widetilde{\Sigma}_{\ttbar}^{\mathrm{res},[g_{n}g_{\nbar}]}
  (\bTvec,\Mttbar,\bttbar,x_t,\Yttbar)\\
  \,=&\;
  \left(\frac{1}{N^2_c-1}\right)^2\,
  \mathcal{D}^{\mathrm{res}}_{[g_{n}g_{\nbar}]}
  (\bT,\Mttbar,\mu_h,\mu_b,\mu_s,\nu_b,\nu_s)\,
  \sum_{\{\alpha,\beta, h,h'\}}
  \Bigg\{
      \mathcal{S}^{\alpha_1\beta_1}_{[g_ng_{\nbar}]}(\bTvec,\mu_s,\nu_s)\,
    \nnb\\
  &\times\,
    \mathcal{B}^{[g_n]}_{n,h_n'{h_n}}(\eta_n,\bTvec,\mu_b,\nu_b)\,
    \mathcal{B}^{[g_{\nbar}]}_{\nbar,h_{\nbar}'{h_{\nbar}}}(\eta_{\nbar},\bTvec,\mu_b,\nu_b)  \,
      \left[\mathcal{V}^{[g_{n}g_{\nbar}]}_{\alpha_1\alpha_2}(\bttbar,x_{t},\mu_s,\mu_h)\right]^*\,
  \mathcal{V}^{[g_{n}g_{\nbar}]}_{\beta_1\beta_2}(\bttbar,x_{t},\mu_s,\mu_h)\,
  \nnb\\
  &\times
    \mathcal{H}_{\alpha_2\beta_2;{h_{\nbar}'}h_{\nbar};{h_n'}h_n}^{[g_ng_{\nbar}]}(\Mttbar,\bttbar,x_{t},\mu_h)\,
  \Bigg\}\,.\nnb
\end{align}
The expression of
$\widetilde{\Sigma}_{\ttbar}^{\mathrm{res},[q^i_{\nbar}\bar{q}^j_{n}]}$
can be obtained from Eq.~\eqref{eq:M2PS:NJ0:qqbar:res} by adjusting the
label momenta of the beam-collinear modes as appropriate.
As is apparent in Eqs.~(\ref{eq:M2PS:NJ0:qqbar:res}-\ref{eq:M2PS:NJ0:gg:res}),
two sets of auxiliary scales $\{\mu_h,\mu_b,\mu_s\}$ and
$\{\nu_b,\nu_s\}$ have been introduced to define the initial
conditions utilised in solving the RaGEs and RGEs above.
An appropriate choice of their values minimises the missing
higher-order corrections, and in this paper, in a bid to
minimise the logarithmic dependences on the respective
sectors, the following values will be taken as defaults
in this paper~\cite{Chiu:2012ir,Neill:2015roa},
\begin{align} \label{eq:scale:nat}
  \mu^{\mathrm{def}}_h=\nu^{\mathrm{def}}_{b}=\Mttbar\,,\quad
  \mu^{\mathrm{def}}_b= \mu^{\mathrm{def}}_s=\nu^{\mathrm{def}}_s= b_0/|b_{\tau_{\|}}|\,.
\end{align}
With the choices in Eq.~\eqref{eq:scale:nat}, the impact
space integral in Eq.~\eqref{eq:xs:NJ0:res} may approach
or cross the Landau singularity in the large $|b_{\tau_{\|}}|$
regime.
In order to avoid the divergence, we impose upper and lower boundaries
$|b_{\tau_{\|}}|\le b^{\mathrm{cut}}_{\tau_{\|}}=2\,\text{GeV}^{-1}$
in Eq.~\eqref{eq:xs:NJ0:res} \cite{Neill:2015roa}.
Alternative schemes have been discussed in \cite{Becher:2010tm,Kang:2017cjk}.

Further, Eqs.~\eqref{eq:M2PS:NJ0:qqbar:res}-\eqref{eq:M2PS:NJ0:gg:res}
also include the kernels $\mathcal{D}^{\mathrm{res}}_{[\kappa]}$
and $\mathcal{V}^{[\kappa]}_{\alpha\beta}$ to evolve the
intrinsic scales amongst the fixed-order functions.
$\mathcal{D}^{\mathrm{res}}_{[\kappa]}$ is induced
by the diagonal anomalous dimensions, such as
$\gamma_{\mathrm{cusp}}$, $\gamma_b^{[\kappa]}$,
and $\gamma_r$ in
Eqs.~\eqref{eq:methods:res_rge_H}-\eqref{eq:methods:res_rage_B}.
Its definition reads
\begin{align} \label{eq:def:Dres}
  &\ln\mathcal{D}^{\mathrm{res}}_{[\kappa]}
    (\bT,\Mttbar,\mu_h,\mu_b,\mu_s,\nu_b,\nu_s)
    \nnb\\
  &\,=\;
    \int^{\mu^2_s}_{\mu^2_b} \,
    \frac{\done\bar{\mu}^2}{\bar{\mu}^2}\,
    \Bigg\{
      C_{[\kappa]}\, \Gamma_{\mathrm{cusp}}\big[\alpha_s(\bar{\mu}) \big]\,
      \ln\bigg[ \frac{\nu_b^2}{\Mttbar^2}\bigg]
      +2\,\gamma^{[\kappa]}_b\big[\alpha_s(\bar{\mu}) \big]\,
    \Bigg\}
    -\int^{\mu_s^2}_{\mu_h^2}\frac{\done \bar{\mu}^2}{\bar{\mu}^2}\,
    \bigg\{
    {C_{[\kappa]}}\,
    \Gamma_{\mathrm{cusp}}\big[\alpha_s(\bar{\mu}) \big]\,
    \ln\left[\frac{\bar{\mu}^2}{\Mttbar^2}\right]\,\bigg\}\nnb
    \\
  &\,\phantom{=}\;{}
    +C_{[\kappa]}\,
    \ln\bigg[\frac{\nu_s^2}{\nu_b^2}\bigg]\,
    \int_\frac{b_0^2}{\bT^2}^{\mu_s^2}\,
    \frac{\done\bar{\mu}^2}{\bar{\mu}^2}\,
    \Gamma_{\mathrm{cusp}}\big[\alpha_s(\bar{\mu}) \big]\,
    -C_{[\kappa]}\,
     \ln\bigg[\frac{\nu_s^2}{\nu_b^2}\bigg]\,
     \gamma_r
     \bigg[\alpha_s\left(\frac{b_0}{\bT}\right)\bigg]\,.
\end{align}
$\mathcal{V}^{[\kappa]}_{\alpha\beta}$ accounts for the
contributions from the non-diagonal anomalous dimension
$\gamma_{h,\alpha\beta}^{[\kappa]}$, which, in principle,
can be extracted from the solutions of the RGE of the
hard function in Eq.~\eqref{eq:methods:res_rge_H}.
However, in the presence of the non-diagonal elements
in $\gamma_h^{[\kappa]}$, it is somewhat challenging to
achieve a closed expression.
Hence, in this work we adopt the perturbative approaches
suggested in \cite{Buras:1991jm,Buchalla:1995vs,Ahrens:2010zv}.
The details on the implementation are collected in
App.~\ref{app:non:dia:AD}.

\begin{table}[h!]
  \centering
  \begin{tabular}{|c|c|c|c|c|c|} \hline
    Logarithmic accuracy\hl & $\mathcal{H}$, $\mathcal{S}$, $\mathcal{B}$ & $\Gamma_{\text{cusp}}$& $\gamma_{h,s,b,r}$ \\ \hline
    \NLLp\hl  & $\order(\alpha^0_s)$& $\order(\alpha^2_s)$& $\order(\alpha_s)$ \\ \hline
    \NNLL\hl & $\order(\alpha_s)$& $\order(\alpha^3_s)$& $\order(\alpha^2_s)$ \\ \hline
    \NNLLp\hl & $\order(\alpha^2_s)$& $\order(\alpha^3_s)$& $\order(\alpha^2_s)$ \\ \hline
  \end{tabular}
  \caption{
    Precision prerequisites on the anomalous dimensions and the fixed-order functions for a given 
    logarithmic accuracy.
    \label{tab:methods:res_accuracy}
  }
\end{table}

Equipped with Eqs.~\eqref{eq:M2PS:NJ0:qqbar:res}-\eqref{eq:M2PS:NJ0:gg:res},
we can calculate the resummed projected transverse momentum distributions.
In this work, we will present the results in particular at \NLLp, \NNLL,
and approximate \NNLLp~levels.
For \NLLp, \NNLL, and strict \NNLLp accurate calculations,
the required precisions for the various anomalous dimensions
and fixed-order functions have been summarised in
Table.~\ref{tab:methods:res_accuracy}.
In particular, for the \NNLLp~result, the hard and soft sectors
need to be known at full \NNLO accuracy.
However, only the logarithmic terms, which are derived from Eq.~\eqref{eq:methods:res_rge_H} and
Eqs.~\eqref{eq:methods:res_rge_S}-\eqref{eq:methods:res_rage_S},
are included in this work.
We thus label our results with this approximation as \aNNLLp
in the rest of this paper.

\subsection{Observables}
\label{sec:methods:obs}
 
In Eq.~\eqref{eq:xs:NJ0:res}, we have presented the master
formula of the resummed $q_{\tau}$ spectrum with a general
choice of $\vec{\tau}$.
In Sec.~\ref{sec:results}, we will investigate three
observables, $q_{\mathrm{T,out}}$, $ q_{\mathrm{T,in}}$,
and $\dPhittbar$.
The calculations of the first two observables are immediate from
Eq.~\eqref{eq:xs:NJ0:res} by choosing the reference vector
$\vec{\tau}$ to be perpendicular or parallel with respect
to the top-quark transverse momentum $\vec{P}_t^{\perp}$,
\begin{align}
  \label{eq:def:qTout}
  q_{\tau}
  \,=\,~&
    q_{\mathrm{T,out}}\,,
    \qquad
    \mathrm{if}~\vec{\tau}
    =
      \pm\;\vec{n}\times\frac{\vec{P}_t^{\perp}}{|\vec{P}_t^{\perp}|}\,,\\
  \label{eq:def:qTin}
  q_{\tau}
  \,=\,~&
    q_{\mathrm{T,in}}\,,
    \qquad~~\!
    \mathrm{if}~\vec{\tau}
    =
      \pm\;\frac{\vec{P}_t^{\perp}}{|\vec{P}_t^{\perp}|}\,,
\end{align}
where the unit vector $\vec{n}$ characterises the flight
direction of one of the colliding protons.
Since the value of $q_{\tau}$ only concerns the magnitude
of the projected component, the calculation with either
$+$ or $-$ sign in Eqs.~\eqref{eq:def:qTout}-\eqref{eq:def:qTin}
gives the same result.
In order to determine the azimuthal distribution $\dPhittbar$,
it is worth noting that in the vicinity of
$\dPhittbar\to\pi $, we are able to perform the
expansion in $\dphittbar\equiv(\pi-\dPhittbar)\sim\lamtau$,
\begin{align}
  \dPhittbar
  \equiv
    \arccos\left[\frac{\vec{P}_t^{\perp}\cdot\vec{P}_{\bar{t}}^{\perp}}
                      {|\vec{P}_t^{\perp}||\vec{P}_{\bar{t}}^{\perp}|}\right]
  \sim
    \pi-\frac{q_{\mathrm{T,out}}}{|\vec{P}_t^{\perp}|}
    +\order(\lambda^2_{\tau})\,.
\end{align}
In this work, only the leading kinematical effects will
be taken into account, such that the azimuthal spectrum
can be calculated from the results on $q_{\mathrm{T,out}}$,
\begin{align}\label{eq:def:dphi:qTout:cor}
  \frac{\done{\sigma^{\mathrm{res}}_{\ttbar}}}
       {\done\Mttbar^2\,\done^2\vec{P}_{t}^{\perp}\,
        \done\Yttbar\,\done\dphittbar}
  \,=&\;
    \left|\vec{P}_{t}^{\perp}\right|\,
    \frac{\done{\sigma^{\mathrm{res}}_{\ttbar}}}
         {\done\Mttbar^2\,\done^2\vec{P}_{t}^{\perp}\,
          \done\Yttbar\,\done q_{\mathrm{T,out}}}
    +\order(\lambda^0_{\tau})\,.
\end{align}
\changed{
As demonstrated in Sec.~\ref{sec:res:log:sing}, all $q_{\tau}$ associated observables are free of azimuthal asymmetric divergences in momentum space and so are the spectra of $q_{\mathrm{T,out}}$, $ q_{\mathrm{T,in}}$,
and $\dphittbar$. 
This  justifies our application of the resummation
schemes in Tab.~\ref{tab:methods:res_accuracy} and also the
R(a)GE framework in Eqs.~\eqref{eq:xs:NJ0:res}-\eqref{eq:M2PS:NJ0:gg:res} during the calculation.
}

\subsection{Matching to fixed-order QCD}
\label{sec:methods:mat}

We are now ready to match the resummed predictions derived
in the previous sections to exact fixed-order QCD calculations.
As the expansion in $\lamtau$ has been applied in the derivation
of the resummed results of Eq.~\eqref{eq:xs:NJ0:res}, its
validation is maintained only within the asymptotic domain.
To continue the resummed spectra to the entire phase space,
we match Eq.~\eqref{eq:xs:NJ0:res} onto the fixed-order result
using a multiplicative scheme~\cite{Banfi:2012jm,Banfi:2012yh,Bizon:2018foh},
\begin{equation}\label{eq:def:mat}
  \begin{split}
    \cfrac{\done \sigma_{\ttbar}^{\rm{mat}}}{\done \mathcal{Q}}
    \,\equiv&\;
      \Bigg\{
        \left[
          \cfrac{\done\sigma_{\ttbar}^{\rm{res}}}{\done \mathcal{Q}}
          -\cfrac{\done \sigma_{\ttbar}^{\rm{s}}(\mu_{\mathrm{f.o.}})}
                 {\done \mathcal{Q}}
        \right]\,
        f_{\rm{tran}}({\mathcal{Q}},c_{\mathrm{m}},r_{\mathrm{m}})
        +\cfrac{\done \sigma_{\ttbar}^{\rm{s}}(\mu_{\mathrm{f.o.}})}
               {\done \mathcal{Q}}
      \Bigg\}\;
      \mathcal{R}_{\mathrm{fs}}(\mu_{\mathrm{f.o.}})\\
    =&\;
      f_{\rm{tran}}({\mathcal{Q}},c_{\mathrm{m}} ,r_{\mathrm{m}})\,
      \left(\cfrac{\done\sigma_{\ttbar}^{\rm{res}}}{\done \mathcal{Q}}\right)\,
      \mathcal{R}_{\mathrm{fs}}(\mu_{\mathrm{f.o.}})\Bigg|_{\mathrm{exp}}
      +\Big\{
         1-{f}_{\rm{tran}}({\mathcal{Q}},c_{\mathrm{m}} ,r_{\mathrm{m}})
       \Big\}\,
       \cfrac{\done \sigma_{\ttbar}^{\rm{f.o.}}(\mu_{\mathrm{f.o.}})}
             {\done \mathcal{Q}}
      +\dots\,,
\end{split}
\end{equation}
where
$\mathcal{Q}\in\{q_{\mathrm{T,out}},q_{\mathrm{T,in}},\Delta\phi_{\ttbar}\}$
represents a general observable of our concern.
${\done\sigma_{\ttbar}^{\rm{res}}}/{\done\mathcal{Q}}$ is
the resummed differential distribution calculated from
Eq.~\eqref{eq:xs:NJ0:res}.
${\done\sigma_{\ttbar}^{\rm{s}}}/{\done\mathcal{Q}}$
stands for the perturbative expansion of
${\done\sigma_{\ttbar}^{\rm{res}}}/{\done\mathcal{Q}}$
at the scale $\mu_{\mathrm{f.o.}}$ and also corresponds to the leading singular terms in Eq.~\eqref{eq:AsyExp}.
$\mu_{\mathrm{f.o.}}$ is the scale of the fixed-order expansion,
and typically identified with $\mu_R$ and $\mu_F$ in an exact QCD
fixed-order calculation.
In the numerical study presented in the next section of this paper,
we will take as default choice
\begin{equation}\label{eq:scale:mat}
  \mu_{\mathrm{f.o.}}^{\mathrm{def}}=\Mttbar\,.
\end{equation}
Taking the difference between $\sigma_{\ttbar}^{\rm{res}}$
and $\sigma_{\ttbar}^{\rm{s}}$ yields the pure resummation
corrections beyond the fixed-order accuracy of the calculation
that our result is matched to, as shown in the square brackets
of Eq.~\eqref{eq:def:mat}.
Multiplying this difference by the transition function $f_{\rm{tran}}$
permits us to regulate the active range of the resummation
through the shape of $f_{\rm{tran}}$, and avoid double counting
at the same time.
To accomplish a continuous and progressive transition
towards the exact result, this paper will employ the
following piecewise form of $f_{\rm{tran}}$,
\begin{equation}
  f_{\rm{tran}}({\mathcal{Q}},c_{\mathrm{m}},r_{\mathrm{m}}) =
    \begin{dcases}
      1\,, & \mathcal{Q}\le c_{\rm{m}}-r_{\mathrm{m}}\,;\\
      1-\frac{(\mathcal{Q}-c_{\rm{m}}+r_{\mathrm{m}})^2}{2r_{\mathrm{m}}^2} \,,& c_{\rm{m}}-r_{\mathrm{m}}<\mathcal{Q}\le c_{\rm{m}}\,;\\
         \frac{ (\mathcal{Q}-c_{\rm{m}}-r_{\mathrm{m}})^2 }{2r_{\mathrm{m}}^2}\,,& c_{\rm{m}}<\mathcal{Q}\le c_{\rm{m}}+r_{\mathrm{m}}\,;\\
      0 \,,&  c_{\rm{m}}+r_{\mathrm{m}}\le \mathcal{Q} \,,
    \end{dcases}
    \hspace*{0.04\textwidth}
   \begin{minipage}{0.3\textwidth}
     \includegraphics{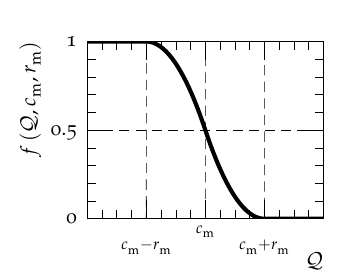}
   \end{minipage}
\end{equation}
where the parameters $c_{\rm{m}}$ and $r_{\mathrm{m}}$
are introduced to measure the focal point and the
transition radius, respectively.
In our calculation, the following parameters will
be taken by default,
\begin{align}\label{eq:scale:ftran:qTin}
  c^{\mathrm{def}}_{\rm{m}}
  &=30~\mathrm{GeV}\,,
  r^{\mathrm{def}}_{\mathrm{m}}
  =20~\mathrm{GeV}\,,
  \qquad
  \mathrm{if}~\mathcal{Q}\in\{q_{\mathrm{T,out}},q_{\mathrm{T,in}}\}\,,\\
  \label{eq:scale:ftran:dphi}
  c^{\mathrm{def}}_{\rm{m}}
  &=0.3\,,
  r^{\mathrm{def}}_{\mathrm{m}}=0.2,~\qquad\qquad\qquad
  \mathrm{if}~\mathcal{Q}\in\{\dphittbar\}\,.
\end{align}
Please note, that the focal point of our transition
function, $c_\mathrm{m}$, plays the role of a traditional
matching scale for the $q_{\mathrm{T,out}}$ and
$q_{\mathrm{T,in}}$ spectra.
As the \dphittbar~spectrum, however, is dimensionless,
$c_\mathrm{m}$ is dimensionless, and can thus not be
directly connected to an ``intrinsic'' scale of the
scattering process.
Instead, we choose it solely on the basis of the quality
of the approximation of the fixed-order expansion of
Eq.\ \eqref{eq:xs:NJ0:res} wrt.\ the exact QCD result,
see Sec.\ \ref{sec:validations}.
Besides the resummed differential cross sections
and its perturbative expansions aforementioned, to
compensate for the power corrections having been
truncated during the asymptotic expansions,
Eq.~\eqref{eq:def:mat} also includes the ratio of the
exact spectra to the leading singular terms derived
from \SCETII and HQET,
\begin{align}\label{eq:def:mat:R}
  \mathcal{R}_{\mathrm{fs}}(\mu_{\mathrm{f.o.}})
  \equiv
    \cfrac{{\done\sigma_{\ttbar}^{\mathrm{f.o.}}(\mu_{\mathrm{f.o.}})}/
           {\done\mathcal{Q}}}
          {{\done\sigma_{\ttbar}^{\mathrm{s}}(\mu_{\mathrm{f.o.}})}/
           {\done\mathcal{Q}}}\,.
\end{align}
Herein, $\sigma_{\ttbar}^{\mathrm{f.o.}}$ denotes the
fixed-order QCD results assessed at the identical
scale to that of $\sigma_{\ttbar}^{\rm{s}}$, and
will be evaluated by the program
\Sherpa~\cite{Gleisberg:2003xi,Gleisberg:2008ta,Sherpa:2019gpd}.
Starting from \NNLO, the fixed-order predictions are not
positive definite on the whole range of $\mathcal{Q}$, and
indeed can turn negative in the asymptotic domain that
will be improved by our resummation.
This invariably leads to the vanishing denominators in
Eq.~\eqref{eq:def:mat:R}.
We thus expand $\mathcal{R}_{\mathrm{fs}}$ in
$\alpha_s(\mu_{\mathrm{f.o.}})$ in the second step
of Eq.~\eqref{eq:def:mat} following the spirit of
\cite{Bizon:2018foh}.

\section{Numerical Results}
\label{sec:results}

\subsection{Parameters and uncertainty estimates}
\label{sec:results:params}

In this part, we will present numeric results for the
observables $q_{\mathrm{T,out}}$, $q_{\mathrm{T,in}}$,
and $\dphittbar$, which are calculated using the master
formulae in Eq.~\eqref{eq:xs:NJ0:res} and Eq.~\eqref{eq:def:mat}.
The resummed result of
Eq.~\eqref{eq:xs:NJ0:res} comprises a convolution of
the hard, beam, soft functions.
We calculate the NLO amplitudes of all the helicity and
color configurations of the hard sector  using program 
\Recola~\cite{Actis:2012qn,Actis:2016mpe}, and then
evolve them by means of the RGE in
Eq.~\eqref{eq:methods:res_rge_H} to derive the logarithmic
contributions at \NNLO.
We strictly adhere to the on-shell prescription in
renormalizing the top quark mass, and take its value
from the Particle Data Group (PDG) \cite{Workman:2022ynf}.
For computing the beam functions, the package
\texttt{HPOLY}~\cite{Ablinger:2018sat} is embedded to
calculate the harmonic poly-logarithms participating in
the hard-collinear coefficients in
\cite{Luo:2019bmw,Luo:2020epw,Luo:2019szz}.
The partonic content of the proton is parametrised
using the \texttt{NNPDF31\_nnlo\_as\_0118}~\cite{NNPDF:2017mvq}
parton distribution function, interfaced through
\LHAPDF~\cite{Buckley:2014ana,Bothmann:2022thx}.
To be consistent, we use the corresponding value of
the strong coupling with $\alpha_s(m_Z)=0.118$.
The evaluation of the soft ingredients at NLO accuracy is
straightforward from the analytic expressions presented
in Sec.~\ref{sec:soft:func:NLO}.
To access the \NNLO logarithmic terms, we expand the
solutions of Eqs.~\eqref{eq:methods:res_rge_S}-\eqref{eq:methods:res_rage_S}
up to $\mathcal{O}(\alpha^2_s)$.
In addition to those fixed-order constituents,
Eq.~\eqref{eq:xs:NJ0:res} also requires the evolution
kernels~$\mathcal{D}^{\mathrm{res}}_{[\kappa]}$ and
$\mathcal{V}^{[\kappa]}_{\alpha\beta}$.
The analytic result of $\mathcal{D}^{\mathrm{res}}_{[\kappa]}$
can be found with the approach of \cite{Becher:2006mr}.
To appraise $\mathcal{V}^{[\kappa]}_{\alpha\beta}$,
we first diagonalise the one-loop anomalous dimension
$\mathbf{\gamma}^{[\kappa]}_{h}$ (see Eq.~\eqref{eq:methods:res_rge_H})
by means of \texttt{Diag}~\cite{Hahn:2006hr} to reach
NLL accuracy.
Then, based on the perturbative scheme proposed in
\cite{Buras:1991jm,Buchalla:1995vs,Ahrens:2010zv},
we reinstate the higher order corrections from
$\mathbf{\gamma}^{[\kappa]}_{h}$ to address the
\NNLL requirements and beyond.

Combining these evolution kernels and fixed-order
functions permits us to evaluate the differential
cross sections.
To assess the phase-space and impact-space integrals
therein, the package \texttt{Cuba} is employed to
manage the relevant multidimensional numerical integrations.
Over the course, to circumvent the threshold regime
$\beta_{\ttbar}\sim0$, where the Coulomb singularity
manifests itself and sabotages the factorisation
formula established in Eq.~\eqref{eq:xs:NJ0:fac},
the constraint $M_{\ttbar}\ge400~\mathrm{GeV}$,
resulting in $\beta_{\ttbar}\gtrsim 0.5$, is
imposed in the phase integral.

With the resummed spectra in hand, we can proceed with
the matching procedure formulated in Eq.~\eqref{eq:def:mat}.
The leading singular contribution
$\done\sigma_{\ttbar}^{\mathrm{s}}/\done\mathcal{Q}$
is obtained by expanding
$\done\sigma_{\ttbar}^{\mathrm{res}}/\done\mathcal{Q}$
in $\alpha_s$.
To calculate the exact fixed-order QCD differential cross
sections, we use \Sherpa~\cite{Gleisberg:2003xi,
  Gleisberg:2008ta,Sherpa:2019gpd} together with
\Recola~\cite{Actis:2012qn,Actis:2016mpe}
and \Rivet~\cite{Buckley:2010ar,Bierlich:2019rhm}.
In particular, we will restrict the fixed-order
calculations to the domains
\begin{align}\label{eq:cuts:lower}
  \mathcal{Q}&\ge10^{-1}~\mathrm{GeV}\,,
  \qquad\mathrm{if}~\mathcal{Q}\in\{q_{\mathrm{T,out}},q_{\mathrm{T,in}}\}\,,\\
  \mathcal{Q}&\ge10^{-2}\,,~\qquad\qquad\mathrm{if}~\mathcal{Q}
  \in\{\dphittbar\}
\end{align}
to avoid numerical inaccuracies.
The NLO calculations involve only the tree
level amplitudes\footnote{
  In this work, the perturbative accuracy is counted with
  respect to the Born cross section.
  Thus, the NLO contributions here correspond to the tree-level
  amplitudes of the process $pp\to\ttbar+\mathrm{jet}$.
}, which can thus be generated by the built-in tree-level
matrix element generator \Amegic~\cite{Krauss:2001iv} and then processed
by \Rivet to extract the observables $q_{\mathrm{T,out}}$,
$q_{\mathrm{T,in}}$, and $\dphittbar$.
To access the \NNLO results, \Recola is used to compute the
renormalised one-loop virtual amplitudes of the relevant
subprocesses, while the program \Amegic  calculates the
real emission corrections and performs the dipole subtraction
in the Catani-Seymour scheme~\cite{Catani:1996vz,Catani:2002hc,
Gleisberg:2007md,Schonherr:2017qcj}.
The subsequent event analysis procedures again proceed
through \Rivet as in the NLO case.
 
Our calculations, Eq.~\eqref{eq:xs:NJ0:res} and Eq.~\eqref{eq:def:mat},
involve a set of auxiliary scales,
$\{\mu_i,\nu_i\}\equiv\{\mu_h,\mu_b,\mu_s,\mu_{\mathrm{f.o.}},\nu_b,\nu_s\}$
and two shape parameters $\{ c_m,r_m\}$ of the transition
function in the matching procedure.
To estimate the theoretical uncertainties of choosing the
default values of those scales, as presented in
Eqs.~\eqref{eq:scale:nat} and~\eqref{eq:scale:mat},
we appraise the differential cross sections with the scales
varied to twice or half their default values independently.
The deviations from the calculation using the default scales
are then combined in the quadrature.
The so estimated error is referred to as
$\delta_{\mathrm{scale}}$ hereafter.
Moreover, to investigate the sensitivity to the shape parameters
of the transition function of
Eqs.~\eqref{eq:scale:ftran:qTin}-\eqref{eq:scale:ftran:dphi},
the differential spectra are also calculated with the combinations,
\begin{align}\label{eq:scale:ftran:qTin:var}
  \{c_{\rm{m}},r_{\mathrm{m}}\}&=\{25~\mathrm{GeV},15~\mathrm{GeV}\},\{35~\mathrm{GeV},25~\mathrm{GeV}\}\,, \qquad\mathrm{if}~\mathcal{Q}\in\{q_{\mathrm{T,out}},q_{\mathrm{T,in}}\}\,,\\
  \label{eq:scale:ftran:dphi:var}
  \{c_{\rm{m}},r_{\mathrm{m}}\}&=\{0.25,0.15\} ,\{0.35,0.25\} \,,\,\qquad\qquad\qquad\qquad\mathrm{if}~\mathcal{Q}\in\{\dphittbar\}\,,
\end{align}
which amount to fixing the lower boundaries of
$f_{\mathrm{tran}}$ but adjusting the descending
gradients around the central choice in
Eqs.~\eqref{eq:scale:ftran:qTin}-\eqref{eq:scale:ftran:dphi}.
Again, the deviation from the central value using the
default choices defines the uncertainty, which we denote
by $\delta_{\mathrm{tran}}$.
The total theoretical error is then obtained through
$\delta_{\mathrm{tot}}=\sqrt{\delta^2_{\mathrm{scale}}+\delta^2_{\mathrm{tran}}}$.

\subsection{Validation}\label{sec:validations}

\begin{figure}[t!]
  \centering
  \begin{subfigure}{0.49\textwidth}
    \centering
    \includegraphics[width=.9\linewidth, height=0.98\linewidth]{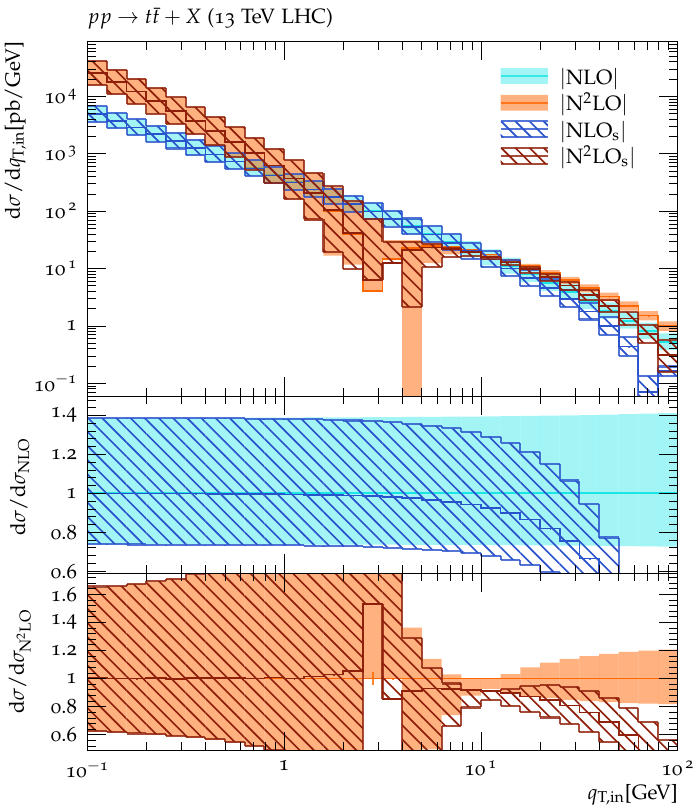}
    \caption{ Numeric results of $\done \sigma_{\ttbar}/\done q_{\mathrm{T,in}}$}
    \label{fig:results:val:qTin:a}
  \end{subfigure}
  \begin{subfigure}{0.49\textwidth}
    \centering
    \includegraphics[width=.9\linewidth, height=0.98\linewidth]{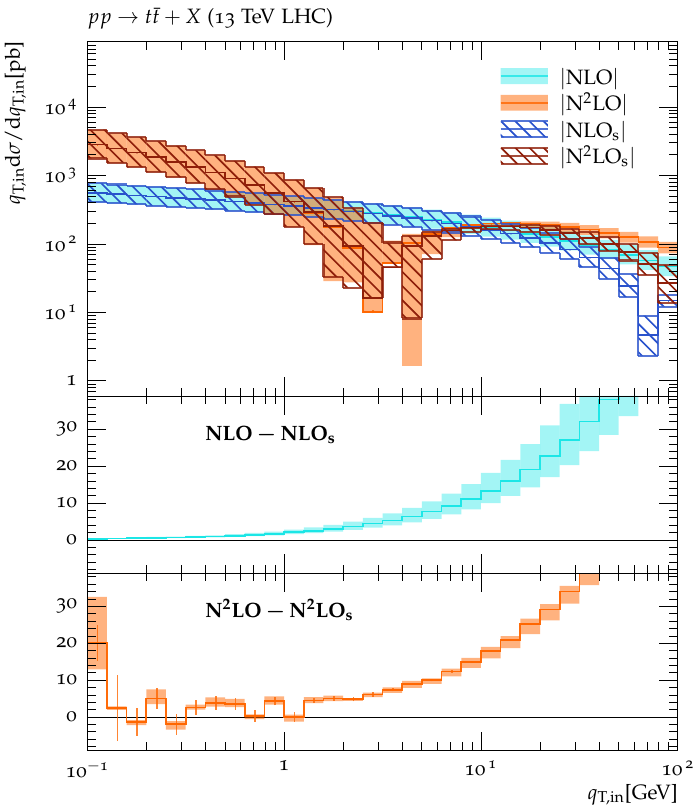}
    \caption{ Numeric results of $\done \sigma_{\ttbar}/\done \ln q_{\mathrm{T,in}}$ }
    \label{fig:results:val:qTin:b}
  \end{subfigure}
  \caption{
    The fixed-order results of the $q_{\mathrm{T,in}}$
    and the weighted $q_{\mathrm{T,in}}$ spectra of the
    process $pp\to\ttbar+X$ at $\sqrt{s}=13$~TeV.
    \NLO/\NNLO represents the differential cross section
    calculated in the full QCD, while \NLOs/\NNLOs
    encodes the leading singular behaviour derived
    from \SCETII$+$HQET.
  }
  \label{fig:results:val:qTin}
\end{figure}

In this part, we will confront the differential cross
sections derived from the factorisation formula of
Eq.~\eqref{eq:xs:NJ0:fac} with those evaluated in the
full theory.
At this point, it merits reminding that in establishing
Eq.~\eqref{eq:xs:NJ0:fac}, the asymptotic expansion has
been carried out in $\lamtau$ by means of the
region expansion of \cite{Beneke:1997zp,Smirnov:2002pj,
  Smirnov:2012gma,Jantzen:2011nz}, and only the leading
singular contributions have been taken into account in
the approximate outputs.
In order to assess the effectiveness of this
approximation, and in turn the resummation scheme of
Eqs.~\eqref{eq:xs:NJ0:res}-\eqref{eq:M2PS:NJ0:gg:res},
it is of the crucial importance for this work to compare
the numeric performances of the exact and approximate
spectra in the asymptotic regime.
  
Fig.~\ref{fig:results:val:qTin} exhibits the calculation
of the $q_{\mathrm{T,in}}$ and the weighted
$q_{\mathrm{T,in}}$ spectra at \NLO and \NNLO precision.
In the left panel, we focus on the distributions
$\done\sigma^{\mathrm{s}}_{\ttbar}/\done q_{\mathrm{T,in}}$
and $\done\sigma^{\mathrm{f.o.}}_{\ttbar}/\done q_{\mathrm{T,in}}$,
in the notation of Sec.\ \ref{sec:methods:mat}.
Please note that while the \NLO fixed-order expansions are
strictly positive throughout, the \NNLO ones are negative
at small $q_{\mathrm{T,in}}$ and only return to the positive
for $q_{\mathrm{T,in}}\gtrsim 3\,\text{GeV}$.
As shown in the top plot, owing to the singular contributions
from the asymptotic expansion of Eq.~\eqref{eq:AsyExp},
the absolute values of the \NLOs and \NNLOs results are
both in excellent agreement with the exact results in
the low $q_{\mathrm{T,in}}$ region.
As the value of $q_{\mathrm{T,in}}$ grows, the acuteness
of the singular terms becomes gradually alleviated, from which the overall cross-section is reduced.
As a result, the \NLOs and \NNLOs approximations show
progressively larger deviations.
The fluctuations visible around $q_{\mathrm{T,in}}=3\,\text{GeV}$
are caused by the slightly different crossing points from the
negative to the positive realm of the exact \NNLO and approximate
\NNLOs results.
At some point, far outside its validity range, the approximate
\NLOs and \NNLOs again become unphysical, e.g.\ at
$ q_{\mathrm{T,in}}\approx80\,\text{GeV}$ for the \NLOs results.

Below the main plot, we show two \emph{ratios} to highlight the
relationship between the exact and approximate spectra.
Independent of the order of the expansion or the choice
on the scale, as indicated by their respective bands,
excellent agreement is observed between
$\done\sigma^{\mathrm{s}}_{\ttbar}/\done q_{\mathrm{T,in}}$ and
$\done\sigma^{\mathrm{f.o.}}_{\ttbar}/\done q_{\mathrm{T,in}}$
below  $q_{\mathrm{T,in}}=1~$GeV.
Within a few percents of deviation, this holds up to
$q_{\mathrm{T,in}}\sim10~$GeV for the \NLOs bands and
$q_{\mathrm{T,in}}\sim20~$GeV in the \NNLOs results.
Further increase in $q_{\mathrm{T,in}}$ will render
the power suppressed terms in Eq.~\eqref{eq:AsyExp}
manifestly important, thereby developing appreciable
discrepancies between the QCD and EFT outcomes.
For instance, it is seen that in the domain
$q_{\mathrm{T,in}}\sim30\,\text{GeV}$, the leading
singular terms only account for about
$75\%$ of the \NLO and $\sim80\%$ of the \NNLO
spectra.

In order to further ascertain the asymptotic properties
in the $q_{\mathrm{T,in}}\to 0$ limit, we investigate
the $q_{\mathrm{T,in}}$-weighted differential distributions
$\done\sigma_{\ttbar}/\done \ln q_{\mathrm{T,in}}$ in
Fig.~\ref{fig:results:val:qTin:b}.
Akin to Eq.~\eqref{eq:AsyExp}, the asymptotic expansion
can also be applied onto the present case,
\begin{align}\label{eq:AsyExp:qTweighted}
  \frac{\done{\sigma_{\ttbar}}}{\done \ln q_{\tau} }
  \,\sim\;
    \sigma^{\ttbar}_{\mathrm{B}}\,
    \sum_{m,n}\,
    \left[\frac{\alpha_s(\Mttbar)}{4\pi}\right]^m\,
    \left[\,
      \underbrace{c^{(0)}_{m,n}\,{\ln^{n}(\lamtau)}}_{\mathrm{LP}}
      +\,\underbrace{c^{(1)}_{m,n}\,\lamtau\ln^{n}(\lamtau)}_{\mathrm{NLP}}
      +\,\underbrace{c^{(2)}_{m,n}\,\lambda^2_{\tau}\ln^{n}(\lamtau)}_{\mathrm{N}^2\mathrm{LP}}
      +\dots
    \right]\,.
\end{align}
Comparing with Eq.~\eqref{eq:AsyExp}, the power series
of Eq.~\eqref{eq:AsyExp:qTweighted} exhibits a less
singular behaviour in each order and power in $\lamtau$.
This characteristic is reflected by the shallower slopes
of the \NLOs and \NNLOs curves in the main plot of
Fig.~\ref{fig:results:val:qTin:b}.
More explicitly, whilst the magnitudes of
$\done\sigma^{\mathrm{s}(\mathrm{f.o.})}_{\ttbar}/\done \ln q_{\mathrm{T,in}}$
are nearly 10 times smaller than those of
$\done\sigma^{\mathrm{s}(\mathrm{f.o.})}_{\ttbar}/\done q_{\mathrm{T,in}}$
in the vicinity of $q_{\mathrm{T,in}}=0$,
they approach each other when entering the area
$q_{\mathrm{T,in}}\sim10~$GeV.
On the other hand, differing from Eq.~\eqref{eq:AsyExp},
the leading power terms of
$\done \sigma_{\ttbar}/\done \ln q_{\mathrm{T,in}}$ are
expected to capture the entire singular behaviour of the
spectrum as all subleading power corrections are finite
in the limit $q_{\mathrm{T,in}}\to 0$.
To verify this property and in turn scrutinise our
calculations of the leading singular contributions,
the \emph{difference} between the full theory and the
EFT results is addressed in the middle and bottom graphs,
at \NLO and \NNLO accuracy, respectively, of
Fig.~\ref{fig:results:val:qTin:b}.
For the \NLOs results, with the decrease in $q_{\mathrm{T,in}}$,
the difference between the full QCD result and our approximation
continuously decreases for all scale choices.
In particular, in the vicinity of $q_{\mathrm{T,in}}\sim 0.1\,\text{GeV}$,
the gaps between the \NLO and \NLOs results shrink to
$\sim0.3~$pb, almost a thousand times smaller than the magnitudes of
$\done\sigma^{\mathrm{s}(\mathrm{f.o.})}_{\ttbar}/\done\ln q_{\mathrm{T,in}}$.
We thus conclude that all the singularities of
$\done\sigma^{\mathrm{f.o.}}_{\ttbar}/\done\ln q_{\mathrm{T,in}}$
at \NLO have been successfully incorporated in
$\done\sigma^{\mathrm{s}}_{\ttbar}/\done \ln q_{\mathrm{T,in}}$.
Analogous behaviors can also be found in the \NNLOs results
from the region $q_{\mathrm{T,in}}\ge2~$GeV, where
the Monte-Carlo integration errors $\epsilon_{_{\mathrm{MC}}}$
for the  difference $\NNLO-\NNLOs$
(indicated by the red bars) are still under control.
Further reducing $q_{\mathrm{T,in}}$ leads to
growing numerical uncertainties in the calculation of
this small residual difference of increasingly large
individual weighted cross sections.
Nonetheless, numeric zero is still within a few standard
deviations $\epsilon_{_{\mathrm{MC}}}$ and, thus, the EFT spectra can
still be regarded as compatible with the exact results.
It should again be stressed that since the absolute values
of the \NNLOs curves are generally above $\mathcal{O}(10^3)$
within the asymptotic domain, mitigating their absolute
uncertainties down to the same level as those in the \NLOs
results necessitates the sub-permille relative accuracy
in running the programs, which is rather demanding in time
and thus has to be postponed to future research.

\begin{figure}[t!]
  \centering
  \begin{subfigure}{0.49\textwidth}
    \centering
    \includegraphics[width=.9\linewidth, height=0.98\linewidth]{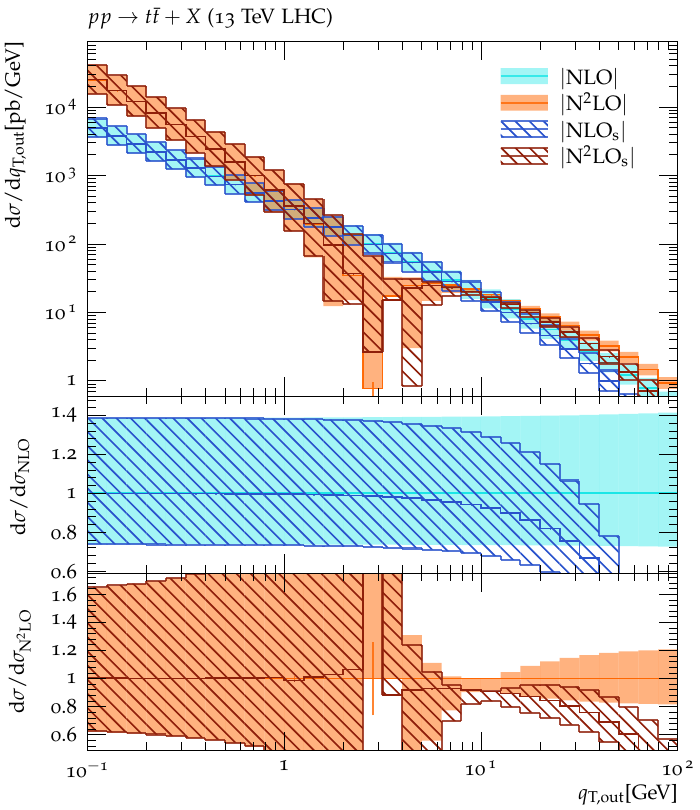}
    \caption{ Numeric results of $\done \sigma_{\ttbar}/\done q_{\mathrm{T,out}}$}
    \label{fig:results:val:qTout:a}
  \end{subfigure}
  \begin{subfigure}{0.49\textwidth}
    \centering
    \includegraphics[width=.9\linewidth, height=0.98\linewidth]{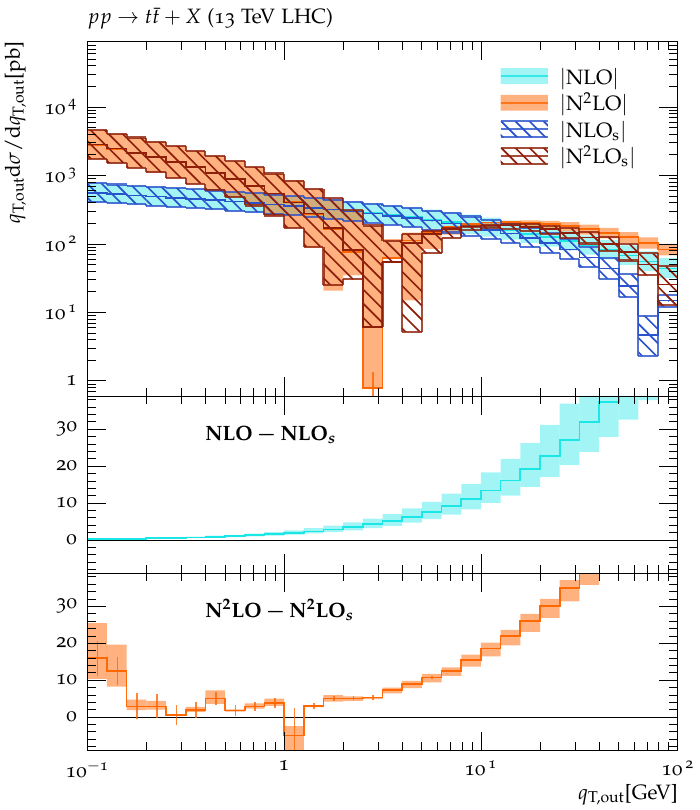}
    \caption{ Numeric results of $\done \sigma_{\ttbar}/\done \ln q_{\mathrm{T,out}}$ }
    \label{fig:results:val:qTout:b}
  \end{subfigure}
  \caption{
    The fixed-order results of the $q_{\mathrm{T,out}}$
    and the weighted $q_{\mathrm{T,out}}$ spectra of the
    process $pp\to\ttbar+X$ at $\sqrt{s}=13$~TeV.
    \NLO/\NNLO represents the differential cross section
    calculated in the full QCD, while \NLOs/\NNLOs
    encodes the leading singular behaviour derived
    from \SCETII$+$HQET.
  }
  \label{fig:results:val:qTout}
\end{figure}

In Fig.~\ref{fig:results:val:qTout}, we illustrate the
fixed-order results of the distributions
$\done\sigma_{\ttbar}/\done q_{\mathrm{T,out}}$ and
$\done\sigma_{\ttbar}/\done \ln q_{\mathrm{T,out}}$.
Within the asymptotic regime, since the $q_{\mathrm{T,out}}$
observable is subject to the same factorisation formula
as that in the $q_{\mathrm{T,in}}$ case,  the general pattern of the $q_{\mathrm{T,out}}$ and
the weighted spectra are similar to those of
Fig.~\ref{fig:results:val:qTin}.
Furthermore, due to the fact that the double-logarithmic
terms, induced by the cusp anomalous dimensions in
Eq.~\eqref{eq:def:Dres} which dominate the leading singular
behaviour, are independent of the choice of the reference
vector $\vec{\tau}$, the absolute values for
$\done\sigma_{\ttbar}/\done q_{\mathrm{T,out}}$ and
$\done\sigma_{\ttbar}/\done\ln q_{\mathrm{T,out}}$ are
close to those of $q_{\mathrm{T,in}}$ in
Fig.~\ref{fig:results:val:qTin}.
In addition to the top diagrams focusing on the magnitudes
of the respective spectra, the lower two panels of
Fig.~\ref{fig:results:val:qTout} are again dedicated
to the comparisons of the exact and approximate results.
It is seen that, up to statistical uncertainties, the
leading singular terms have suitably reproduced the
asymptotic behaviour of the full theory in the small
$q_{\mathrm{T,out}}$ domain.
 
\begin{figure}[t!]
  \centering
  \begin{subfigure}{0.49\textwidth}
    \centering
    \includegraphics[width=.9\linewidth, height=0.98\linewidth]{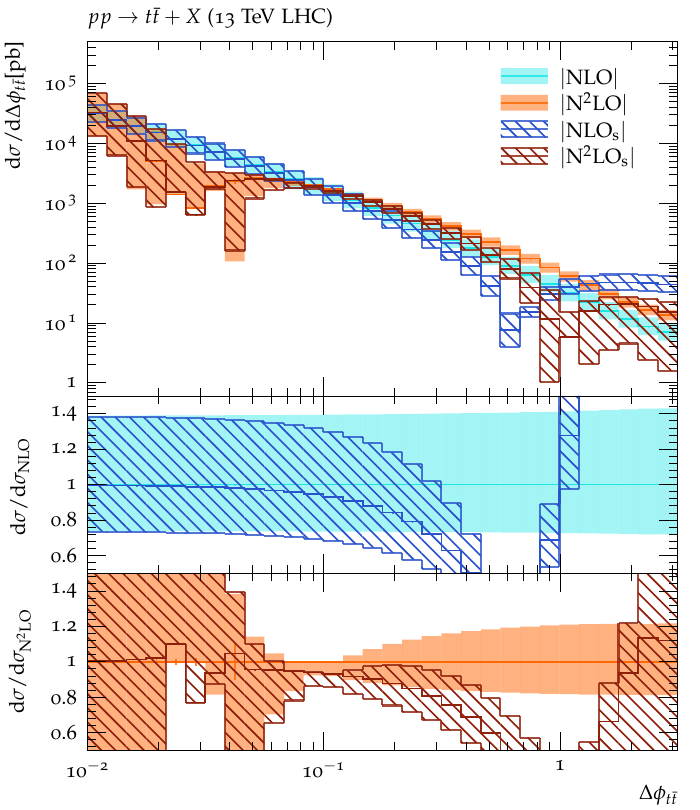}
    \caption{ Numeric results of $\done \sigma_{\ttbar}/\done \dphittbar$}
    \label{fig:results:val:dphi:a}
  \end{subfigure}
  \begin{subfigure}{0.49\textwidth}
    \centering
    \includegraphics[width=.9\linewidth, height=0.98\linewidth]{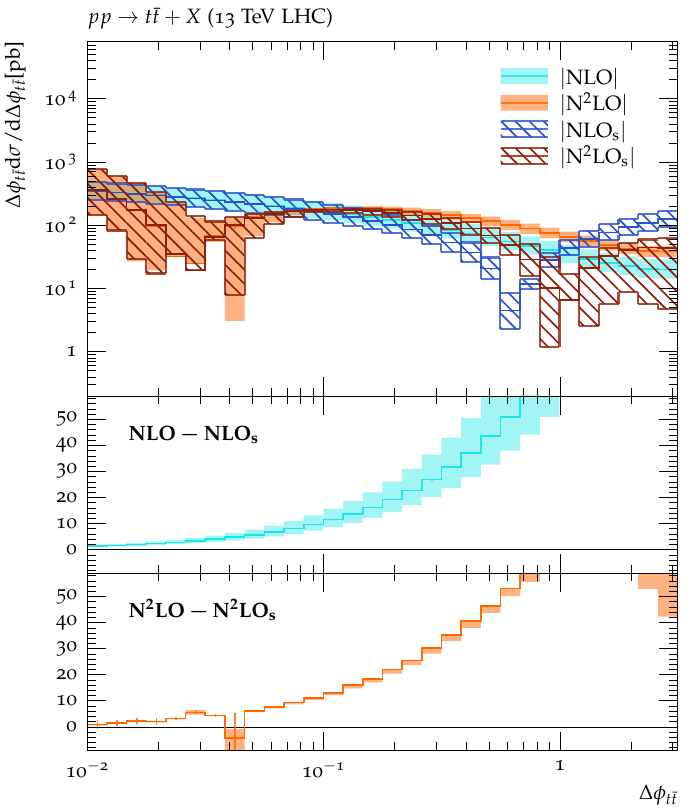}
    \caption{ Numeric results of $\done \sigma_{\ttbar}/\done \ln \dphittbar$ }
    \label{fig:results:val:dphi:b}
  \end{subfigure}
  \caption{
    The fixed-order results of the $\dphittbar$
    and the weighted $\dphittbar$ spectra of the
    process $pp\to\ttbar+X$ at $\sqrt{s}=13$~TeV.
    \NLO/\NNLO represents the differential cross section
    calculated in the full QCD, while \NLOs/\NNLOs
    encodes the leading singular behaviour derived
    from \SCETII$+$HQET.
  }
  \label{fig:results:val:dphi}
\end{figure}

Fig.~\ref{fig:results:val:dphi} finally depicts
the results for the
$\done\sigma_{\ttbar}/\done\dphittbar$ and
$\done\sigma_{\ttbar}/\done\ln\dphittbar$ spectra.
Taking into account the facts that the azimuthal distributions
are in practice derived from the $q_{\mathrm{T,out}}$
spectrum by means of Eq.~\eqref{eq:def:dphi:qTout:cor}
and also that the contributions from the region
$\vec{P}_{t}^{\perp}\sim\mathcal{O}(10^2)\,\text{GeV}$
provide the bulk of the cross section~\cite{Czakon:2015owf,Czakon:2016ckf},
we find that the $\done\sigma_{\ttbar}/\done\dphittbar$
and $\done\sigma_{\ttbar}/\done\ln\dphittbar$ results in
the interval $\dphittbar\in[0.01,1]$ of
Fig.~\ref{fig:results:val:dphi} present a corresponding
behaviour to the $q_{\mathrm{T,out}}$ spectra within the
$q_{\mathrm{T,out}}\in[1,100]\,\text{GeV}$ region of
Fig.~\ref{fig:results:val:qTout}.
For example, the \NLOs curves intercept the \NNLOs ones
around $\dphittbar=0.01$ and $\dphittbar=0.1$ in
Fig.~\ref{fig:results:val:dphi}, while similar intercepts
also take place close to
$q_{\mathrm{T,out}}=1\,\text{GeV}$ and $q_{\mathrm{T,out}}=10\,\text{GeV}$
in Fig.~\ref{fig:results:val:qTout}.
Finally, as before, Fig.~\ref{fig:results:val:dphi} delivers the
ratios of the $\dphittbar$ spectra as well as the differences
of the weighted $\dphittbar$ distributions, to compare the
approximate and exact spectra.
As before, except for the calculations crossing from negative
to positive cross sections in the vicinity of $\dphittbar\sim0.03$
of Fig.~\ref{fig:results:val:dphi:a} and the numerical instability
observed at $\dphittbar\sim0.04$ in Fig.~\ref{fig:results:val:dphi:b},
we observe excellent agreement in the asymptotic domain
$\dphittbar\lesssim 0.1$.

\subsection{Resummation improved results}
\label{sec:res:num}
 
\begin{figure}[t!]
  \centering
  \begin{subfigure}{0.49\textwidth}
    \centering
    \includegraphics[width=.9\linewidth, height=0.9\linewidth]{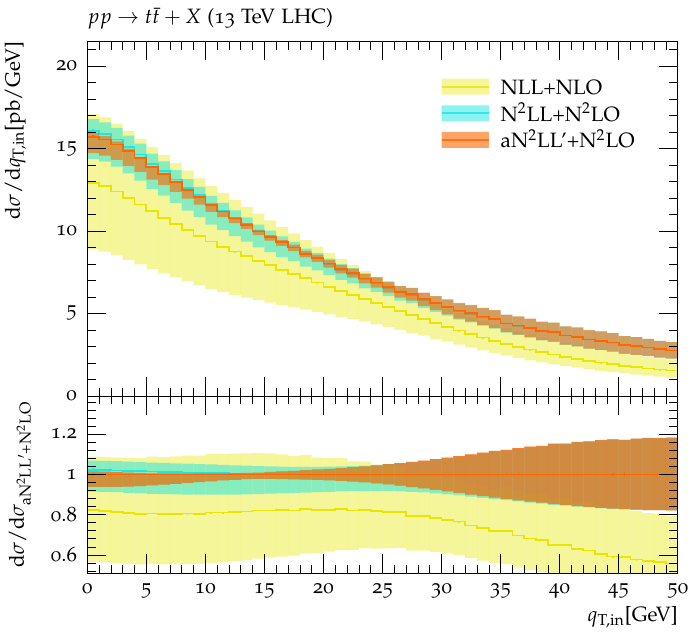}
    \label{fig:results:res:qTin}
  \end{subfigure}
  \begin{subfigure}{0.49\textwidth}
    \centering
    \includegraphics[width=.9\linewidth, height=0.9\linewidth]{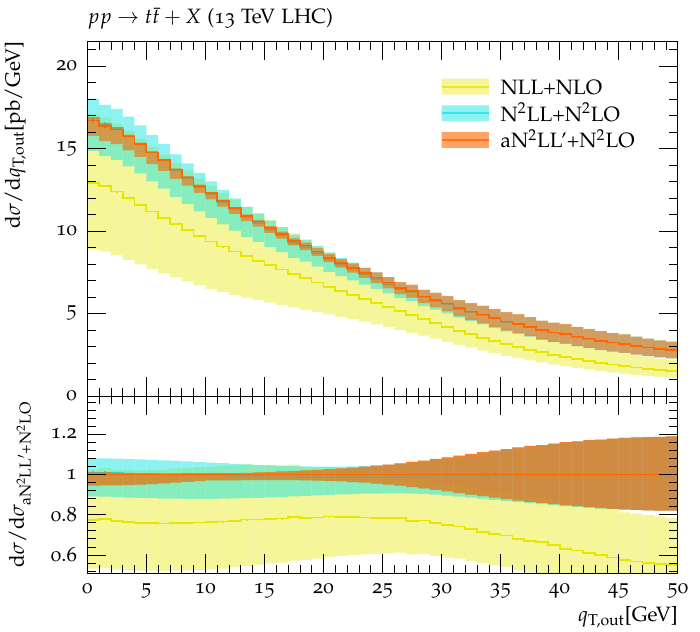}
    \label{fig:results:res:qTout}
  \end{subfigure}
  \caption{
    The resummation improved spectra of $q_{\mathrm{T,in}}$
    and $q_{\mathrm{T,out}}$ at \NLLNLO, \NNLLNNLO, and
    approximate \NNLLpNNLO accuracies (\aNNLLpNNLO).
  }
  \label{fig:results:res:qTin:qTout}
\end{figure}

We now match our resummed calculation, validated in the previous
section, to the exact fixed-order result, according to Eq.~\eqref{eq:def:mat}.
To facilitate our calculations of the exact spectra,
lower cutoffs in the respective observable have been
implemented as shown in Eq.~\eqref{eq:cuts:lower}.
They are justified by the above findings,
namely the agreement of the fixed-order expansion of the
approximation and the full QCD calculation, in the asymptotic domain.
It is important to note that the full theory calculation
participates in the matching procedure through the ratio
$\mathcal{R}_{\mathrm{fs}}$ defined in Eq.~\eqref{eq:def:mat:R}.
As illustrated in Figs.~\ref{fig:results:val:qTin:a},
\ref{fig:results:val:qTout:a}, and \ref{fig:results:val:dphi:a},
this ratio approaches unity in the asymptotic region for all
observables investigated in this paper.
Consequently, for practical evaluations, we are justified
to truncate the impact of the full theory below the
boundaries in Eq.~\eqref{eq:cuts:lower}.
Further lowering these boundaries in principle helps to
suppress the power corrections, which however entails much
longer run time of our programs.

The other major ingredient in our matching procedure is the
transition function $f_{\mathrm{tran}}$ which, together with
the shape parameters $\{c_{\rm{m}},r_{\mathrm{m}}\}$, governs
the active range of the resummation.
In light of the excellent agreement between QCD and EFT,
see Figs.~\ref{fig:results:val:qTin}-\ref{fig:results:val:dphi},
the chosen default values of $\{c_{\rm{m}},r_{\mathrm{m}}\}$
are given in Eq.~\eqref{eq:scale:ftran:qTin} and
Eq.~\eqref{eq:scale:ftran:dphi}.
The transition function ensures that the resummation of
Eq.~\eqref{eq:xs:NJ0:res} is fully operational in the
asymptotic regions.
In the present paper, we choose these regions to be
$q_{\mathrm{T,in(out)}}\le 10\,\text{GeV}$ and
$\dphittbar\le 0.1$.
The contribution of the resummation to the matched result is
then continuously reduced, reaching half its intrinsic value
at the focal point $c_\mathrm{m}$, 30\,GeV in the
$q_{\mathrm{T,in(out)}}$ spectra and 0.3 in the $\dphittbar$
spectrum.
The resummation is eventually fully switched off for
$\mathcal{Q}\ge c_\mathrm{m}+r_\mathrm{m}$, $50\,\text{GeV}$
for $\mathcal{Q}=q_{\mathrm{T,in(out)}}$ and $0.5$ for
$\mathcal{Q}=\dphittbar$.
Alternative settings for $c_\mathrm{m}$ and $r_\mathrm{m}$,
see Eq.~\eqref{eq:scale:ftran:qTin:var} and
Eq.~\eqref{eq:scale:ftran:dphi:var}, are assessed to
estimate the associated theoretical uncertainty of
the matching processes.

In Fig.~\ref{fig:results:res:qTin:qTout}, we present the
resummation improved differential cross sections
$\done\sigma_{\ttbar}/\done q_{\mathrm{T,in}}$ and
$\done\sigma_{\ttbar}/\done q_{\mathrm{T,out}}$.
Within the small $q_{\mathrm{T,in(out)}}$ regimes,
in contrast to the fixed-order results of
Figs.~\ref{fig:results:val:qTin:a}-\ref{fig:results:val:dphi:a},
where substantial \NNLO contributions were observed,
\changed{
the resummed spectra here have been stabilised at the respective accuracies. 
Especially, 
}a reduction in the theoretical
uncertainties, detailed through their respective coloured
bands, can be found with the increase in the logarithmic accuracy.
For example, in the limits $q_{\mathrm{T,in}}\to0$
and $q_{\mathrm{T,out}}\to 0$, the results
on the \aNNLLpNNLO level, that constitute our best prediction,
possess relative uncertainties of about ${+1\%}$ and ${-5\%}$.
They are greatly reduced from the uncertainties of the
\NNLLNNLO calculations, being from $+6\%$ to $-12\%$.
For illustrational purposes we also include the lowest order
\NLLNLO calculation showing uncertainties of $\pm30\%$.
\changed{
The primary driver behind those theoretical uncertainties is the
variation of the beam scale $\mu_b$.
}

\changed{
On the other hand, as exhibited in Fig.~\ref{fig:results:res:qTin:qTout}, 
the central values in higher precision are in general contained
in the uncertainty bands at the lower accuracy and in particular,
the central result of \NNLLNNLO almost coincides with that at
\aNNLLpNNLO.
At this point, it merits reminding that with our
default choice in Eq.~\eqref{eq:scale:nat}, the logarithmic
contributions from the  fixed order functions are all diminished,
such that the discrepancy between \NNLLNNLO and \aNNLLpNNLO
central curves is  determined by the constant terms, which,
in our case, are those from the beam sectors only.
Further inclusion of the \NNLO constant terms from the hard
and soft ingredients may alter this difference, which nevertheless
has to be verified by an exact \NNLLpNNLO calculation.
}

\changed{
Departing from the asymptotic regime, the curves in Fig.~\ref{fig:results:res:qTin:qTout} enter transitional period
ranging from $q_{\mathrm{T,in(out)}}=25~$GeV to
$q_{\mathrm{T,in(out)}}=35~$GeV, where the error bands are
progressively aligning as the different resummations are faded
out and the fixed order contributions, which are the same in the
\NNLLNNLO and \aNNLLpNNLO calculations, gradually take over.
The cause for the uncertainties in this interval is multifaceted,
comprising the variation of the beam scale $\mu_b$, the fixed order
scale $\mu_{\mathrm{f.o.}}$, and the reshaping of the transition
function $f_{\mathrm{tran}}$ through the inputs $\{c_m,r_m\}$.
Further increasing the magnitudes of $q_{\mathrm{T,in(out)}}$
dives into the tail domain, where the central values of the
respective spectra are dominated by the fixed order results
and the uncertainties therein are ruled by the sensitivity
to the scale $\mu_{\mathrm{f.o.}}$.
Improving the theoretical uncertainties at this moment is out
of the scope of resummation and entails a full N$^3$LO QCD
calculation of $pp\to t\bar{t}+X$.
}

Comparing the spectra of $q_{\mathrm{T,in}}$ with the
$q_{\mathrm{T,out}}$ ones, it is interesting to find
that their \NLLNLO results are in close agreement with
each other, whereas those at \NNLLNNLO or \aNNLLpNNLO
accuracy start to exhibit larger differences.
In particular, the $q_{\mathrm{T,out}}$ spectra are
raised by $\approx1~$pb/GeV with regard to the
$q_{\mathrm{T,in}}$ one in the asymptotic regime.
To interpret this, it is beneficial to remind that
according to Tab.~\ref{tab:methods:res_accuracy},
the NLL resummation only involves the evolution kernels
$\mathcal{D}^{\mathrm{res}}_{[\kappa]}$ and
$\mathcal{V}^{[\kappa]}$ in
Eqs.~\eqref{eq:M2PS:NJ0:qqbar:res}-\eqref{eq:M2PS:NJ0:gg:res},
both of which, as defined in Eq.~\eqref{eq:def:Dres} and
Eq.~\eqref{eq:def:Vres:NLL}, concern only the magnitude
of $\vec{b}_{\tau_{\|}}$ and thus fail to differentiate between
the $q_{\mathrm{T,out}}$ and $q_{\mathrm{T,in}}$ observables.
However, starting from \NNLL, the $\vec{\tau}$-dependent
components, such as the soft function $\mathcal{S}_{[\kappa]}$
and the gluon beam sector $\mathcal{B}^{[g_{n(\bar{n})}]}$,
come into play, thus including a non-trivial impact that  the
projected direction $\vec{\tau}$ can make on the results.

Furthermore, contrary to the azimuthally averaged transverse
momentum distributions calculated in \cite{Catani:2018mei,Li:2013mia},
no Sudakov peak is formed in the projected transverse momentum
case, as displayed in Fig.~\ref{fig:results:res:qTin:qTout}.
This originates in part from the kinematic differences of these
two observables.
In the R(a)GE framework introduced in Sec.~\ref{sec:rge:rage},
the resummed $q_{\mathrm{T}}$ distributions can be addressed as,
\begin{align}\label{eq:def:qTav:res}
  \frac{\done\sigma_{\ttbar}}{\done q_{\mathrm{T}}}
  \,\sim\;
    q_{\mathrm{T}}\,\sum_{\kappa}\,
    \int\done\bTvec\,J_{0}(\bT\qT)\,
    \widetilde{\Sigma}^{\mathrm{res},[\kappa]}_{\ttbar}
    (\bTvec,\Mttbar,\bttbar,x_t,\Yttbar)\,,
\end{align}
where $J_{0}$ stands for the zeroth-rank Bessel function.
Taking the limit $\qT\to0$, while the integrand in
Eq.~\eqref{eq:def:qTav:res} approaches a constant value,
the pre-factor $q_{\mathrm{T}}$ linearly suppresses the
resummed spectra, which in turn leads to the formation of
a Sudakov peak.
Such a linear factor is absent for $q_{\mathrm{T,in}}$
and $q_{\mathrm{T,out}}$.
As formulated in Eq.~\eqref{eq:xs:NJ0:res},
the situation of the $q_{\tau}$ resummation is
\begin{align}
  \frac{\done\sigma_{\ttbar}}{\done q_{\tau}}
  \,\sim\;
    \sum_{\kappa}\int\done  b_{\tau_{\|}}\,
    \cos(b_{\tau_{\|}} q_{\tau})\,
    \widetilde{\Sigma}^{\mathrm{res},[\kappa]}_{\ttbar}
    (b_{\tau_{\|}}\vec{\tau},\Mttbar,\bttbar,x_t,\Yttbar)\,.
\end{align}
Thus, the $q_{\tau}$ spectra here receive no kinematical
suppressions in the asymptotic limit, explaining the
absence of the Sudakov peaks in Fig.~\ref{fig:results:res:qTin:qTout}.

\begin{figure}[t!]
  \centering
  \begin{subfigure}{0.49\textwidth}
    \centering
    \includegraphics[width=.9\linewidth, height=0.9\linewidth]{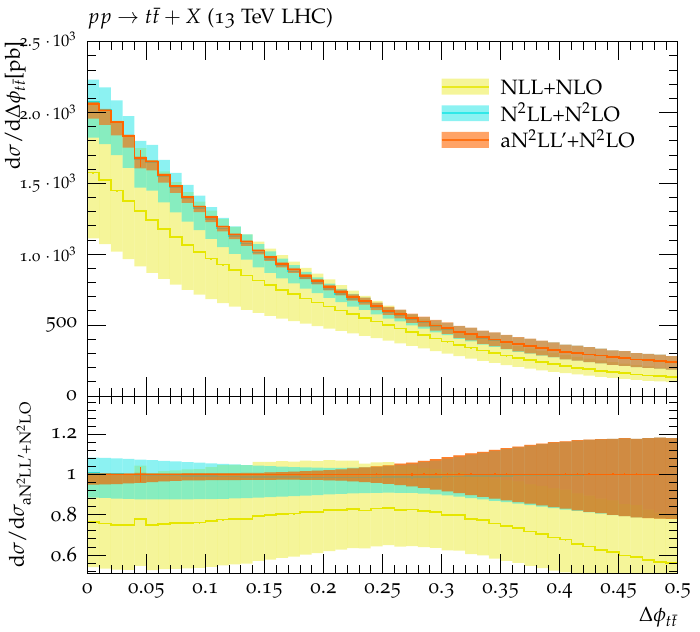}
  \end{subfigure}
  \caption{
    The resummation improved azimuthal spectra
    $\done\sigma_{\ttbar}/\done\dphittbar$ at \NLLNLO,
    \NNLLNNLO, and approximate \NNLLpNNLO accuracies (\aNNLLpNNLO).
  }
  \label{fig:results:res:dphi}
\end{figure}

Finally, in Fig.~\ref{fig:results:res:dphi}, we show the
resummation improved azimuthal separations between the
top and anti-top quarks.
Since at the leading power, the observable $\dphittbar$
exhibits the same factorisation and resummation patterns
as the $q_{\mathrm{T,out}}$ and $q_{\mathrm{T,in}}$ spectra,
the curves from Fig.~\ref{fig:results:res:dphi} present
analogous behaviors to the ones in
Fig.~\ref{fig:results:res:qTin:qTout}, including the
converging theoretical uncertainties with the increasing
of the precision of the calculation.
Due to the same reasons as before, a Sudakov peak cannot
be observed in the asymptotic domain.
In particular, the latter case appears to be in conflict
with the finding in a similar observable in Drell-Yan
production~\cite{Ju:2021lah} where Sudakov peaks were
observed in the azimuthal spectra.
This contradiction is, however, found to be spurious.
The constraints on the transverse momentum that had
been imposed on the final state in \cite{Ju:2021lah}
through fiducial cuts and the projection of the observable
into multiple transverse momentum slices introduced
an effective kinematic damping factor through the shrinking
phase space, thereby restoring
a similar Sudakov peak structure as observed in the
azimuthally averaged transverse momentum distribution of \cite{Ju:2021lah}.
On the contrary, despite requiring $\Mttbar\ge400\,\text{GeV}$
in this paper and the prefactor $\vec{P}^{\perp}_{t}$ having
emerged from the expansions in Eq.~\eqref{eq:def:dphi:qTout:cor},
neither of them can collapse the phase space for
$\dphittbar\to0$ sufficiently for such a structure to emerge.

\section{Conclusions}
\label{sec:conclusions}

In this paper, we investigated the projected transverse
momentum spectrum $q_{\tau}$, which can be evaluated by
integrating out the rejection components of $\qTvec$ with
respect to the reference unit vector $\vec{\tau}$.
We focussed on its singular behaviour in the low $q_{\tau}$
domain in particular.

Using  the region expansion methods \cite{Beneke:1997zp,
    Smirnov:2002pj,Smirnov:2012gma,Jantzen:2011nz} we resolved
the phase space and loop integrals in terms of the dynamic
regions that capture the asymptotic properties of
$\done\sigma_{\ttbar}/\done q_{\tau}$ in the vicinity of
$q_{\tau}=0$.
It comprises the usual hard, beam-collinear, jet-collinear,
and soft domains.
Then, in the context of the effective field theories, \SCETII
and HQET, we enumerated the possible configurations constructed
by those momentum modes, and determined the power accuracy
in each case by probing the corresponding factorisation formulae.
It is observed that the leading singular terms of the
$q_{\tau}$ distribution are governed by the hard, beam-collinear,
and soft regions, akin to the $\qT$ spectrum~\cite{Zhu:2012ts,
  Li:2013mia,Catani:2014qha,Catani:2017tuc,Catani:2018mei,
  Collins:1980ui}, whilst the higher power regular terms entail
the participations of the jet-collinear regime.
 
In the subsequent resummation, we focussed on the leading
asymptotic behaviour of $\done\sigma_{\ttbar}/\done q_{\tau}$.
By inspecting the corresponding factorisation formula in
impact space, we illustrated that the driving factor of
the singular behaviour in momentum space is the characteristic
logarithm $\ln^{k}[{{b}^2_{\tau_{\|}}\Mttbar^2}/{b_0^2}]$,
for $k\ge1$, which therefore permitted us to apply the
framework of RGE and RaGE~\cite{Chiu:2011qc,Chiu:2012ir,
  Li:2016axz,Li:2016ctv} to accomplish the logarithmic
exponentiation and in turn the resummation.
It is worth emphasizing that in presence of the azimuthal
asymmetric terms, this phenomenon is not general on the process $pp\to \ttbar+X$.
The $q_{\tau}$ spectra and the azimuthally averaged case $\done\sigma_{\ttbar}/\done\qT$ are known to be the only two observables which stem from the double-differential distribution $\done\sigma_{\ttbar}/\done\qT$ but are free of the azimuthal asymmetric divergences. 
 \footnote{
  It is interesting to note that in a very recent
  calculation~\cite{Gao:2022bzi} on the semi-inclusive
  deep-inelastic scattering (SIDIS), the observable
  $q_{*}$ is proposed and also demonstrated to be
  promising to cope with the azimuthal asymmetric
  divergences.
  Within the asymptotic regime, the $q_{*}$ spectrum
  actually bears a close resemblance to the
  $q_{\mathrm{T,out}}$ distribution used in this
  work, up to rapidity-dependent factors.  
}

Although the derivation and findings summarised above
hold for an arbitrary choice of reference vector $\vec{\tau}$,
we focussed on three specific observables to demonstrate
the validity and usefulness of our framework, $q_{\mathrm{T,in}}$,
$ q_{\mathrm{T,out}}$, and $\dphittbar$.
The first two cases are obtained by assigning $\vec{\tau}$
to be parallel and perpendicular with respect to the
transverse momentum of the top quark, respectively.
The azimuthal spectrum $\done\sigma_{\ttbar}/\done\dphittbar$
is extracted from $\done\sigma_{\ttbar}/\done q_{\mathrm{T,out}}$
according to the kinematical relationship in between.
Owing to the independence from azimuthally asymmetric
divergences, we were able to calculate those observables
at \NLLNLO, \NNLLNNLO, and approximate \NNLLpNNLO (\aNNLLpNNLO)
accuracy.
From the numerical results, a manifest perturbative convergence
is observed in the asymptotic regime, i.e. the central values
at the higher precisions are generally contained in the
uncertainty bands of the lower ones and the respective uncertainties
are systematically reduced down to the percent level at  
\aNNLLpNNLO.
\changed{
During our calculation, we have focussed on the domain
$M_{t\bar{t}}\ge400~$GeV to evade Coulomb divergences
which inevitably manifest themselves in the higher order
corrections.
It is expected that our results can be compared against
the very recent measurement on the double differential spectra
$\done\sigma_{\ttbar}/(\done\Mttbar\,\done\dphittbar)$~\cite{CMS:2022uae}
within the corresponding $\Mttbar$ slices upon the full publication
of the experimental data.
To expand our results to the full phase space, including
the threshold regime $\Mttbar<400~$GeV, the
emergence of Coulomb singularities at the $\ttbar$ production
threshold necessitates a combined
resummation of the beam-collinear, soft, and Coulomb interactions,
which we seek to address in a future publication.
}

\subsection*{Acknowledgements}

WJ is grateful to Jean-Nicolas Lang for providing detailed
instructions on the usage of \Recola.
WJ also  would like to thank L. Chen,  M. Czakon and R. Poncelet
for the helpful discussions on the \NNLO hard function.
This work is supported by a Royal Society through a University Research Fellowship
(URF\textbackslash{}R1\textbackslash{}180549) and an Enhancement Award 
(RGF\textbackslash{}EA\textbackslash{}181033,
 CEC19\textbackslash{}100349, and RF\textbackslash{}ERE\textbackslash{}210397).

\appendix
\section{The double differential transverse momentum distribution \texorpdfstring{$\done^2\sigma_{\ttbar}/\done^2\qTvec$}{dsigma/dqTvec}}
\label{app:phib:qTvec}
 
As discussed in Sec.~\ref{sec:res:log:sing}, in absence of the
\emph{azimuthal asymmetric term} (AAT), the main task in
resumming the singular behaviors on the $q_{\tau}$ spectra is
to exponentiate the characteristic logarithmic terms in impact
space.
Nevertheless, in light of the non-trivial contributions of the
helicity-flipping beam functions as well as the soft correlations,
the resummation of a general observable is more complicated.
Hence, this appendix will calculate the double differential
spectrum as an example to illustrate the effect of the AATs on
the fixed-order expansion and the organisation of the logarithmic terms.
Alternative cases with active AATs contributions can also be
found in \cite{Nadolsky:2007ba,Catani:2010pd,Catani:2014qha,Catani:2017tuc,Chien:2019gyf,delCastillo:2021znl,Tong:2022zwp,Shao:2022stc}.
 
Considering that the contributions to the $\qTvec$ distributions
from the hard jet regions, $N_\mathrm{J}\ge 1$, have been
demonstrated to be power suppressed \cite{Collins:1980ui,Collins:1981uk},
the factorisation formula is now subject to only the $N_{\mathrm{J}}=0$
configuration at leading power.
Therefore, they can be extracted from Eq.~\eqref{eq:M2PS:NJ0}
after reducing the heavy parton correlation functions with
Eqs.~\eqref{eq:M2PS:NJ0:Wt}-\eqref{eq:M2PS:NJ0:Wtbar}.
It follows that,
\begin{align}\label{eq:xs:qT2D:fac}
  \frac{\done{\sigma_{\ttbar}}}
       {\done\Mttbar^2\,\done^2\vec{P}_{t}^{\perp}\,
        \done\Yttbar\,\done^2 \qTvec}
  =
    \sum_{\mathrm{sign}[x_t]}\sum_{\kappa}\,
    \frac{\Theta_{\mathrm{kin}}^{(0)}}
         {64\pi^4\,\bttbar\,|x_t|\,\Mttbar^4\,s}\,
    \int\done^2\bTvec\,\exp\left(\qTvec\cdot\bTvec\right)\,
    \widetilde{\Sigma}_{\ttbar}^{[\kappa]}
    (\bTvec,\Mttbar,\bttbar,x_t,\Yttbar)\,,
\end{align}
where $\kappa\in\{g_{n}g_{\bar{n}},q^i_{n}\bar{q}^j_{\bar{n}},q^i_{\bar{n}}\bar{q}^j_{n}\}$ again denotes the partonic channel.
The constituents of $\widetilde{\Sigma}_{\ttbar}^{[\kappa]}$
have been specified in Eqs.~\eqref{eq:M2PS:NJ0:qqbar}-\eqref{eq:M2PS:NJ0:gg}
for each partonic channel.
Upon the usage of the identities in
Eqs.~\eqref{eq:M2PS:NJ0:Wt}-\eqref{eq:M2PS:NJ0:Wtbar},
Eq.~\eqref{eq:xs:qT2D:fac} inherits the problems from
Coulomb exchanges as well, akin to Eq.~\eqref{eq:xs:NJ0:fac}.
To this end, we will constrain the following investigation
within the domain $\Mttbar\ge 400\,\text{GeV}$, or equivalently,
$\bttbar\gtrsim 0.5$, to circumvent the threshold regime.

Generically, the expressions of
$\widetilde{\Sigma}^{[\dots]}_{\ttbar}$ at an arbitrary
order in $\alpha_s$ can be parameterized as
\begin{align} \label{eq:def:partonic:qT2D}
  \widetilde{\Sigma}^{[\dots]}_{\ttbar}
  \,\sim\;
    \sum_{m,n}\,\alpha^m_s(\Mttbar)\,L_{\mathrm{T}}^n\,
    \Big\{
      s_{m}(\bttbar,x_t,\Yttbar)
      +
      a_{m}(\widehat{b}_{\mathrm{T}},\bttbar,x_t,\Yttbar)
    \Big\}
    \,,
\end{align}
where $L_{\mathrm{T}}\equiv\log \left[{\bT^2\Mttbar^2}/{b_0^2}\right]$
and $\widehat{b}_{\mathrm{T}}\equiv \bTvec/\bT$.
The dimensionless coefficients $s_{m,n}$ and $a_{m,n}$
encode the azimuthal \emph{symmetric} and \emph{asymmetric}
contributions, respectively.
Substituting Eq.~\eqref{eq:def:partonic:qT2D} into
Eq.~\eqref{eq:xs:qT2D:fac} permits us to evaluate
the impact  space integrals, namely,
\begin{align}\label{eq:xs:qT2D:fac:invFT}
  \frac{\done{\sigma_{\ttbar}}}
       {\done\Mttbar^2\,\done^2\vec{P}_{t}^{\perp}\,
        \done\Yttbar\,\done^2\qTvec}
  \,\sim\;
    \sum_{m,n}\alpha^m_s(\Mttbar)
    \Bigg\{
      s_{m,n}(\bttbar,x_t,\Yttbar)\,
      \mathcal{F}^{(n)}_{\mathrm{T}}(\qT,\Mttbar)
      +
      \mathcal{A}^{(n)}_{\mathrm{T}}\Big[\qTvec,\Mttbar,a_{m,n}\Big]
    \Bigg\}\,.
\end{align}
Herein, since the ASTs are all independent of the impact parameter
$\bTvec$, we can factor them out of the $\bTvec$-integrals and apply
the inverse Fourier transformations only onto the logarithmic terms,
i.e.
\begin{align}
  \mathcal{F}^{(n)}_{\mathrm{T}}&(\qT,\Mttbar)
  \,=\;
    \int\done^2\bT\,\exp\left( \qTvec\cdot\bTvec\right)\,
    L^n_{\mathrm{T}}\,.
\end{align}
The calculation of $\mathcal{F}^{(n)}_{\mathrm{T}}$ follows a
similar pattern to Eq.~\eqref{eq:def:gen:Ftau}.
The corresponding generating function now reads~\cite{Bozzi:2005wk},
\begin{align}\label{eq:def:gen:FT}
  \mathbf{F}_{\mathrm{T}}(\eta,\qT,\Mttbar)
  \,=&\;
    \sum_n\frac{\eta^{n}}{n!}\,
    \mathcal{F}^{(n)}_{\mathrm{T}}(\qT,\Mttbar)
  \,=\;
    \int\done^2\bTvec\,\exp\left(\qTvec\cdot\bTvec\right)\,\left(\frac{\bT^2\Mttbar^2}{b_0^2}\right)^{\eta}\nnb\\
  \,=&\;
    \frac{4\pi e^{2\gamma_{\mathrm{E}}\eta}}{\qT^2}\,
    \frac{\Gamma[1+\eta]}{\Gamma[-\eta]}\,
    \left(\frac{\Mttbar^2}{\qT^2}\right)^{\eta}\,.
\end{align}
$\mathcal{F}^{(n)}_{\mathrm{T}}$ can be extracted by expanding
$\mathbf{F}_{\mathrm{T}}(\eta,\qT,\Mttbar)$ in $\eta$ as appropriate.
The expressions of the first few ranks are
\begin{align}
  \mathcal{F}^{(0)}_{\mathrm{T}}(\qT,\Mttbar)
  &=0\,,
  \label{eq:def:gen:FTn0}\\
  \mathcal{F}^{(1)}_{\mathrm{T}}(\qT,\Mttbar)
  &=-\frac{4\pi}{\qT^2}\,,
  \label{eq:def:gen:FTn1}\\
  \mathcal{F}^{(2)}_{\mathrm{T}}(\qT,\Mttbar)
  &=-\frac{8\pi}{\qT^2}\ln\left[\frac{\Mttbar^2}{\qT^2}\right]\,,
  \label{eq:def:gen:FTn2}\\
  &\;\,\vdots\nnb
\end{align}
From Eq.~\eqref{eq:def:gen:FTn0} it is seen that the ASTs
themselves cannot induce any asymptotic behaviour in the
small $\qT$ domain, whilst the logarithmic term
$L_{\mathrm{T}}^m$ $(m\ge1)$ is able to produce singular
contributions up to $\ln^{m-1}[\Mttbar^2/\qT^2]/\qT^2$
after completing the inverse Fourier transformations.
This observation is found by analogy with
Eq.~\eqref{eq:InvFT:F} and also the
circumstances of the azimuthally averaged observables in
\cite{Catani:2018mei,Zhu:2012ts,Li:2013mia}.
 
On the other hand, Eq.~\eqref{eq:xs:qT2D:fac:invFT} also
includes contributions from the AATs $a_{m,n}$.
In light of their explicit dependence on the orientation
of $\bTvec$, the inverse Fourier transformation now
comprises both $a_{m,n}$ and $L^n_{\mathrm{T}}$, that is,
\begin{align}
  \mathcal{A}^{(n)}_{\mathrm{T}}\Big[\qTvec,\Mttbar,A\Big]
  \,=\;
    \int\done^2\bT\,\exp\left(\qTvec\cdot\bTvec\right)\,
    L^n_{\mathrm{T}}\,
    A(\phi_{qb})\,,
\end{align}
where $A(\phi_{qb})$ represents a generic function of the
azimuthal difference $\phi_{qb}$ between the vector
$\bTvec$ and $\qTvec$, which thereafter participates in
the functional $\mathcal{A}^{(n)}_{\mathrm{T}}$ as one of
its arguments.
Appraising $\mathcal{A}^{(n)}_{\mathrm{T}}$ will benefit
from its generational form as well,
\begin{align}\label{eq:def:gen:AT}
  \mathbf{A}_{\mathrm{T}}\Big[\eta,\qTvec,\Mttbar,A\Big]
  \,=&\;
    \sum_n\,\frac{\eta^{n}}{n!}\,
    \mathcal{A}^{(n)}_{\mathrm{T}}\Big[\qTvec,\Mttbar,A\Big]
  \,=\,
    \int\,\done^2\bTvec\,\exp\left(\qTvec\cdot\bTvec\right)\,
    \left(\frac{\bT^2\Mttbar^2}{b_0^2}\right)^{\eta}\,
    A(\phi_{qb})\,\nnb\\
  \,=&\;
    -\frac{\Gamma[2\eta+2]}{\qT^2}\,
     \frac{e^{2\gamma_{\mathrm{E}}\eta}}{4^{\eta}}
     \left[\frac{\Mttbar^2}{\qT^2}\right]^{\eta}\,
     \int^{\frac{\pi}{2}}_{0}\,\done\phi\,
     \Bigg\{
       c_{\phi}^{-2(1+\eta)}\,e^{\mathrm{i}\pi\eta}\,\bigg[A (\phi )-A\Big(\frac{\pi}{2}\Big)+c_{\phi}A'\Big(\frac{\pi}{2}\Big)\bigg]\,\nnb\\
  &\;{}
    +s_{\phi}^{-2(1+\eta)}\,e^{-\mathrm{i}\pi\eta}\,\bigg[A \Big(\phi +\frac{\pi}{2}\Big)-A\Big(\frac{\pi}{2}\Big)-s_{\phi}A'\Big(\frac{\pi}{2}\Big)
    \bigg]+c_{\phi}^{-2(1+\eta)}\,e^{-\mathrm{i}\pi\eta}\bigg[A \Big(\phi +\pi\Big)\,\nnb\\
  &\;{}
    -A\Big(\frac{3\pi}{2}\Big)+c_{\phi}A'\Big(\frac{3\pi}{2}\Big)
    \bigg]+s_{\phi}^{-2(1+\eta)}\,e^{\mathrm{i}\pi\eta}\,\bigg[A \Big(\phi +\frac{3\pi}{2}\Big)-A\Big(\frac{3\pi}{2}\Big)-s_{\phi}A'\Big(\frac{3\pi}{2}\Big)
    \bigg]\Bigg\}\,\nnb\\
  &\;{}
    -\sqrt{\pi}\, \frac{\Gamma[2\eta+2]}{\qT^2}\frac{ e^{2\gamma_{\mathrm{E} }\eta }}{4^{\eta}}\left[\frac{\Mttbar^2}{\qT^2}\right]^{\eta}\,
    \Bigg\{
    \frac{\cos(\pi\eta)\Gamma[-\frac{1}{2}-\eta]}{\Gamma[-\eta]}\bigg[A\Big(\frac{\pi}{2}\Big)+A\Big(\frac{3\pi}{2}\Big)\bigg]
    \,\nnb\\
  &\;{}
    - \frac{\mathrm{i}\sin(\pi\eta)\Gamma[ -\eta]}{\Gamma[\frac{1}{2}-\eta]}\bigg[A'\Big(\frac{\pi}{2}\Big)-A'\Big(\frac{3\pi}{2}\Big)\bigg]
    \Bigg\}\,,
\end{align}
where $c_{\phi}=\cos(\phi)$ and $s_{\phi}=\sin(\phi)$.
In deriving Eq.~\eqref{eq:def:gen:AT}, the function $A(\phi_{qb})$
is assumed to be analytically continuous in the entire domain
$\phi_{qb}\in[0,2\pi]$, and the result is expected to be
applicable to any azimuthal asymmetric contributions that
stem from the beam functions, the soft sectors, or the
products of these two cases.
In the following discussion, we will focus on scenarios of
the beam constituent.
The other two pieces observe the same scaling manner but
incur lengthier expressions.

As defined in Eq.~\eqref{eq:M2PS:NJ0:Bng}, the gluon beam
function constitutes a $2\times2$ matrix in the helicity space.
With the choice of helicity bases of
Eq.~\eqref{eq:Bgg:def:hel:space}, while the diagonal
entries characterise the ASTs, the off-diagonal ones,
which are always proportional to $e^{\pm\mathrm{i}2\phi_{tb}}$,
encode the azimuthal asymmetric contributions, namely,
$A(\phi_{qb})\propto e^{\pm\mathrm{i}2(\phi_{qb}+\phi_{tq})}$.
Here $\phi_{tb}$ ($\phi_{tq}$) signifies the azimuthal
difference between the top quark with respect to the
vector $\bTvec$ ($\qTvec$).
Plugging this expression into Eq.~\eqref{eq:def:gen:AT}
and then performing the expansions in $\eta$, it yields,
\begin{align}
  \mathcal{A}^{(0)}_{\mathrm{T}}
  \Big[\qTvec,\Mttbar,e^{\pm\mathrm{i}2(\phi_{qb}+\phi_{tq})}\Big]
  \,=&\;
    -\frac{4\pi}{\qT^2}\,e^{\pm\mathrm{i}2\phi_{tq}}\,,
    \label{eq:def:gen:ATn0}\\
  \mathcal{A}^{(1)}_{\mathrm{T}}
  \Big[\qTvec,\Mttbar,e^{\pm\mathrm{i}2(\phi_{qb}+\phi_{tq})}\Big]
  \,=&\;
    -\frac{4\pi}{\qT^2}\,
     \bigg\{\ln\left[\frac{\Mttbar^2}{\qT^2}\right]+1\bigg\}\,
     e^{\pm\mathrm{i}2 \phi_{tq} }\,,\label{eq:def:gen:ATn1}\\
  \mathcal{A}^{(2)}_{\mathrm{T}}
  \Big[\qTvec,\Mttbar,e^{\pm\mathrm{i}2(\phi_{qb}+\phi_{tq})}\Big]
  \,=&\;
    -\frac{4\pi}{\qT^2}\,
     \bigg\{
       \ln^2\left[\frac{\Mttbar^2}{\qT^2}\right]
       +2\ln\left[\frac{\Mttbar^2}{\qT^2}\right]
     \bigg\}\,
     e^{\pm\mathrm{i}2\phi_{tq}}\,,\label{eq:def:gen:ATn2} \\
  \vdots\;&\nnb
\end{align}
It is noted that the situation of
$\mathcal{A}^{(m)}_{\mathrm{T}}$ above is quite different
from that for the $\mathcal{F}^{(m)}_{\mathrm{T}}$ aforementioned.
Firstly, as opposed to
Eqs.~\eqref{eq:def:gen:FTn0}-\eqref{eq:def:gen:FTn2}, where
the logarithmic contribution is the only driver for the
singularities in the momentum space,
Eq.~\eqref{eq:def:gen:ATn0} illustrates that the AAT itself can
serve as a standalone factor to induce an asymptotic behavior.
Secondly, in comparison with
Eqs.~\eqref{eq:def:gen:FTn1}-\eqref{eq:def:gen:FTn2}, the
product of $L^m_{\mathrm{T}}$ $(m\ge1)$ and AATs can induce asymptotic
terms up to $\ln^m\left[{\Mttbar^2}/{\qT^2}\right]/\qT^2$, which
are stronger than the case of ASTs by one power in
$\ln\left[{\Mttbar^2}/{\qT^2}\right]$.
Thirdly, differing from
Eqs.~\eqref{eq:def:gen:FTn0}-\eqref{eq:def:gen:FTn2} concerning
only the absolute value of $\qTvec$, the expressions for
$\mathcal{A}^{(m)}_{\mathrm{T}}$ additionally depend on the
orientation of $\qTvec$ through the phase factor $\exp(\pm2 \phi_{tq})$.
It is worth emphasising that even though the gluon beam function
is utilised here as an example to derive the
$\mathcal{A}^{(m)}_{\mathrm{T}}$s above,
as a matter of fact, the phase factor
$\exp(\pm k \phi_{tq})$ $(k\in\mathbb{Z})$ can also participate
in the soft sectors at an arbitrary perturbative order upon which
the spherical harmonics expansion was performed \cite{Catani:2017tuc}.
Given those non-trivial impacts from the AATs, the fixed-order
results in the low $\qT$ regime and the subsequent resummation
may show a different behaviour from that solely driven
by the ASTs.
The following paragraphs will take the NLO double differential
spectrum $\done\sigma_{\ttbar}/(\done\qT\done\Delta\Phi_{tX})$
and its required ingredients for a \NLLp resummation as examples
to elucidate this.
 Here $\Delta\Phi_{tX}$ stands for the azimuthal separation
between the top quark and the total momentum of the emitted partons. 

\begin{figure}[h!]
  \centering
  \includegraphics[width=0.32\textwidth]{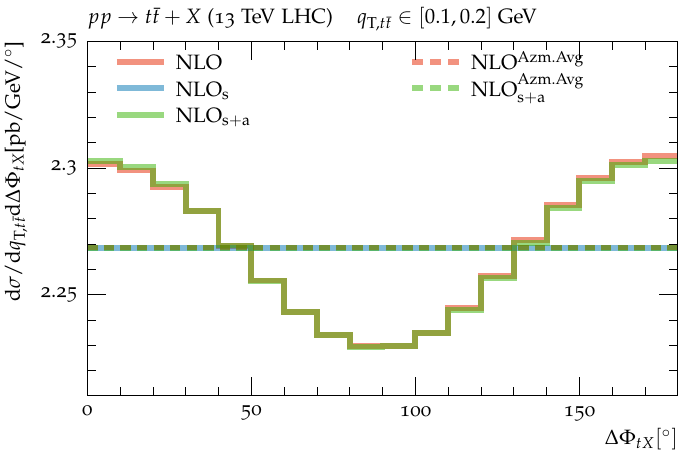}
  \includegraphics[width=0.32\textwidth]{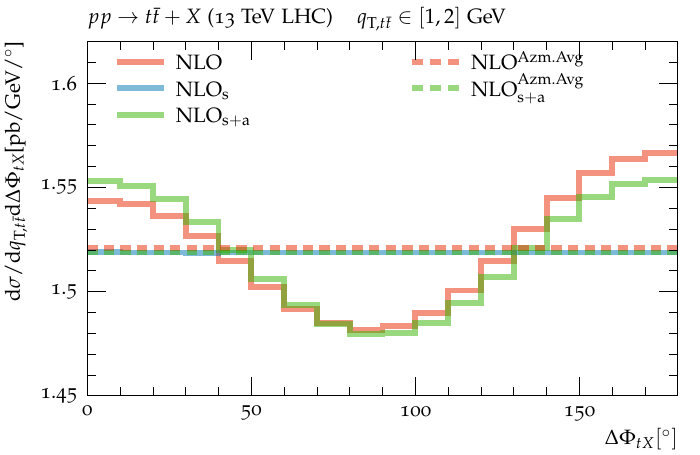}
  \includegraphics[width=0.32\textwidth]{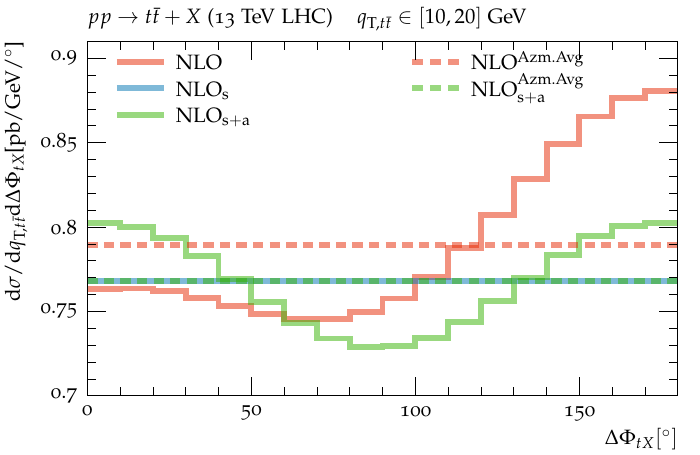}
  \caption{
    Comparison of the NLO double differential spectra
    $\done\sigma_{\ttbar}/(\done\qT\done\Delta\Phi_{tX})$,
    shown as distributions of $\Delta\Phi_{tX}$ in slices
    of $\qT$.
    Herein, NLO represents the QCD result calculated in
    \Sherpa~\cite{Gleisberg:2003xi,Gleisberg:2008ta,Sherpa:2019gpd}.
    $\text{NLO}_\mathrm{s}$ labels the azimuthally \emph{symmetric}
    and $\text{NLO}_\mathrm{s+a}$ marks the full, i.e.\ azimuthally
    \emph{symmetric} and \emph{asymmetric}, contributions
    derived from SCET$_{\mathrm{II}}+$HQET, respectively.
    The superscript ``$\mathrm{Azm.Avg}$'' denotes the
    azimuthally averaging operation on the exact and
    approximate spectra.
  }
  \label{fig:results:val:qT2D:dphi}
\end{figure}

\begin{figure}[h!]
  \centering
  \includegraphics[width=0.32\textwidth]{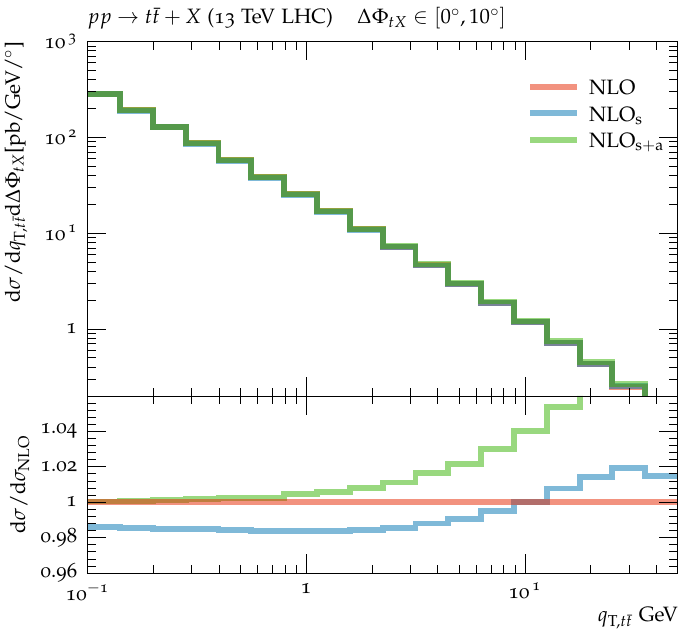}
  \includegraphics[width=0.32\textwidth]{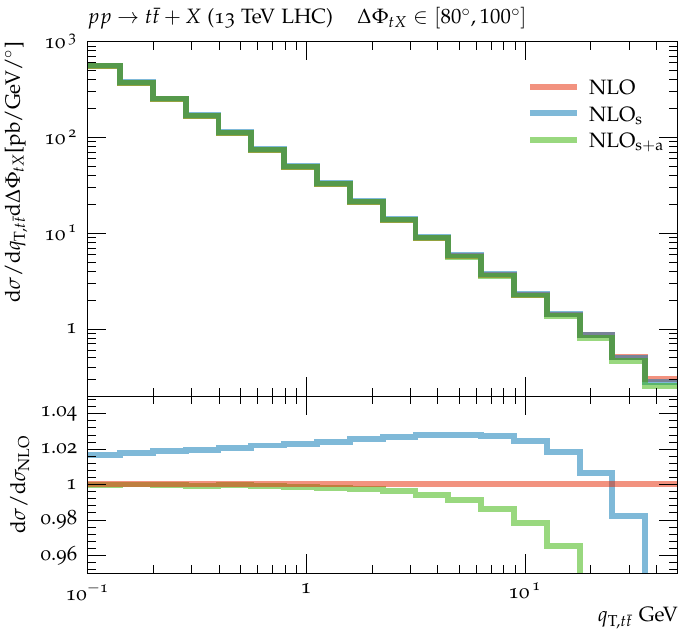}
  \includegraphics[width=0.32\textwidth]{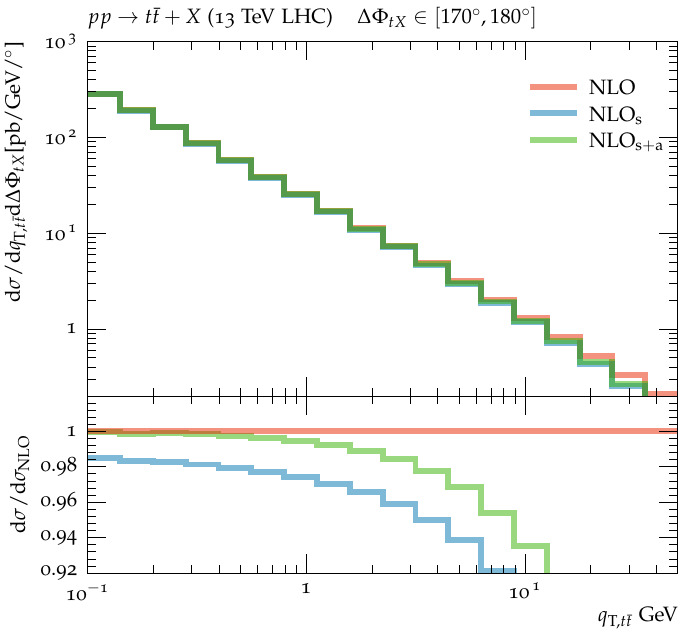}
  \caption{
    Comparison of the NLO double differential spectra
    $\done\sigma_{\ttbar}/(\done\qT\done\Delta\Phi_{tX})$,
    shown as distributions of $\qT$ in slices
    of $\Delta\Phi_{tX}$.
    Herein, NLO represents the QCD result calculated in
    \Sherpa~\cite{Gleisberg:2003xi,Gleisberg:2008ta,Sherpa:2019gpd}.
    $\text{NLO}_\mathrm{s}$ labels the azimuthally \emph{symmetric}
    and $\text{NLO}_\mathrm{s+a}$ marks the full, i.e.\ azimuthally
    \emph{symmetric} and \emph{asymmetric}, contributions
    derived from SCET$_{\mathrm{II}}+$HQET, respectively.    
  }
  \label{fig:results:val:qT2D:qT}
\end{figure}

\begin{figure}[h!]
  \centering
  \includegraphics[width=0.32\textwidth]{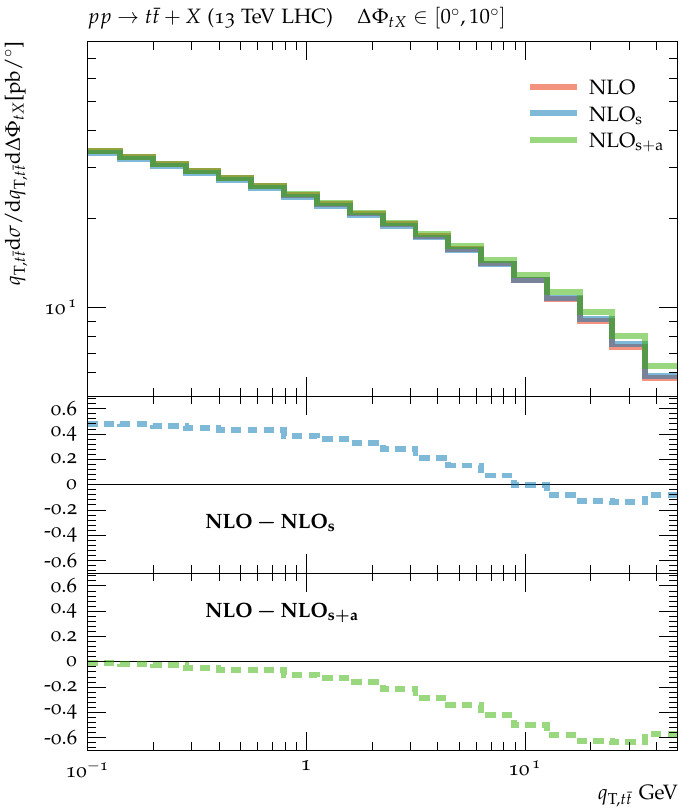}
  \includegraphics[width=0.32\textwidth]{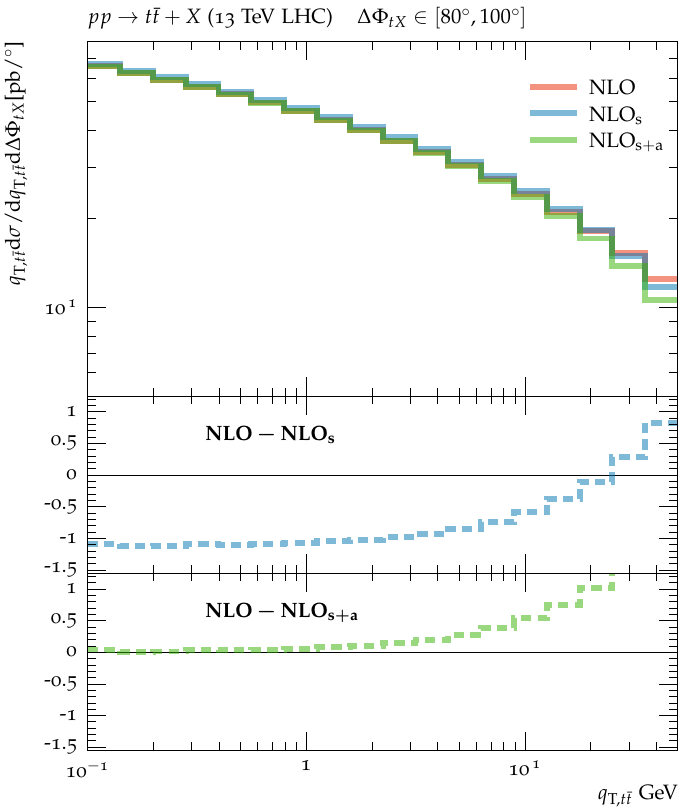}
  \includegraphics[width=0.32\textwidth]{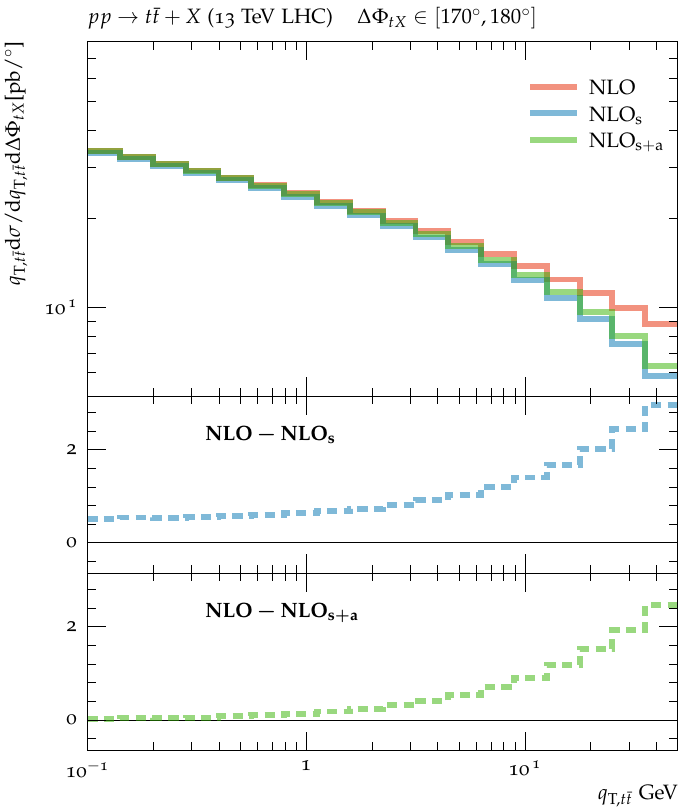}
  \caption{
    Comparison of the \qT-weighted NLO double differential spectra
    $\qT\done\sigma_{\ttbar}/(\done\qT\done\Delta\Phi_{tX})$,
    shown as distributions of $\qT$ in slices
    of $\Delta\Phi_{tX}$.
     Herein, NLO represents the QCD result calculated in
    \Sherpa~\cite{Gleisberg:2003xi,Gleisberg:2008ta,Sherpa:2019gpd}.
    $\text{NLO}_\mathrm{s}$ labels the azimuthally \emph{symmetric}
    and $\text{NLO}_\mathrm{s+a}$ marks the full, i.e.\ azimuthally
    \emph{symmetric} and \emph{asymmetric}, contributions
    derived from SCET$_{\mathrm{II}}+$HQET, respectively.   
  }
  \label{fig:results:val:qT2D:qTwt}
\end{figure}

In Figs.~\ref{fig:results:val:qT2D:dphi}-\ref{fig:results:val:qT2D:qTwt},
we display a comparison of the double differential spectra
$\done\sigma_{\ttbar}/(\done\qT\done\Delta\Phi_{tX})$ among 
the exact NLO result, the approximate spectrum NLO$_{\mathrm{s}}$
comprising only the AST contributions, and the full EFT-derived
spectrum NLO$_{\mathrm{s}+\mathrm{a}}$ including both AST and
AAT ingredients.
As a first observable, Fig.~\ref{fig:results:val:qT2D:dphi}
presents the $\Delta\Phi_{tX}$ distributions, ranging from
the parallel configuration, $\Delta\Phi_{tX}=0^{\circ}$, to
the opposite one, $\Delta\Phi_{tX}=180^{\circ}$, within three
$\qT$ intervals, $\qT\in[0.1,0.2]$\,GeV, $\qT\in[1,2]$\,GeV,
and $\qT\in[10,20]$\,GeV.
As observed in the left panel of Fig.~\ref{fig:results:val:qT2D:dphi},
which focusses on the domain closest to the singularity,
$\qT\in[0.1,0.2]$~GeV, the curves of NLO and NLO$_{\mathrm{s}+\mathrm{a}}$
are in very good agreement and exhibit a nearly cosinusoidal
behaviour on account of the phase factors in
Eqs.~\eqref{eq:def:gen:ATn0}-\eqref{eq:def:gen:ATn2} and
those stemming from the soft functions.
However, if we were to remove the AAT contributions, leaving
only the ASTs as active ingredients of our approximation,
governed now by Eqs.~\eqref{eq:def:gen:FTn0}-\eqref{eq:def:gen:FTn2}
which depend only on the magnitude of $\qTvec$,
the resulting curve NLO$_{\mathrm{s}}$ is independent of
$\Delta\Phi_{tX}$.
After taking the azimuthal averaging operations upon the exact
and approximate spectra, however, it is interesting to note that
the result of NLO$^{\mathrm{Azm.Avg}}_{\mathrm{s}+\mathrm{a}}$
coincides with that of NLO$_{\mathrm{s}}$, and both of them reproduce
the exact NLO$^{\mathrm{Azm.Avg}}$ result well.
This observation demonstrates that the asymptotic terms in
Eqs.~\eqref{eq:def:gen:ATn0}-\eqref{eq:def:gen:ATn2} are indeed
able to be eliminated by completing the $\phi_{tq}$ integral,
which confirms the discussions in \cite{Catani:2017tuc} at least
at the NLO level.
In the centre and right plots of Fig.~\ref{fig:results:val:qT2D:dphi},
we move on to the slices $\qT\in[1,2]$\,GeV and $\qT\in[10,20]$\,GeV.
Even though the general patterns herein follow that in the left panel,
it is seen that the power corrections gradually become more important
with the increase in $\qT$.
They give rise to the growing discrepancies between NLO and
NLO$_{\mathrm{s}+\mathrm{a}}$ results, and also lead to the deviations
amid the azimuthally averaged curves.

In Fig,~\ref{fig:results:val:qT2D:qT}, we exhibit the $\qT$
distributions in slices of $\Delta\Phi_{tX}\in[0^{\circ},10^{\circ}]$,
$\Delta\Phi_{tX}\in[80^{\circ},100^{\circ}]$, and
$\Delta\Phi_{tX}\in[170^{\circ},180^{\circ}]$, which correspond
to the peak and trough regions in Fig.~\ref{fig:results:val:qT2D:dphi}.
We observe that for the majority of the $\qT$ range, the
magnitudes of the exact results are close to the approximate
calculations NLO$_{\mathrm{s}}$ and NLO$_{\mathrm{s}+\mathrm{a}}$,
however, only the NLO$_{\mathrm{s}+\mathrm{a}}$ reproduces
it well  in the region of $\qT\lesssim 1\,$GeV.
To interpret this, it is worth noting that at NLO, the most
singular terms generated by the ASTs and AATs are
$\propto\ln\left[{\Mttbar^2}/{\qT^2}\right]/\qT$ and
$1/\qT$, respectively.
In this regard, the asymptotic behaviors in the low $\qT$
regime are actually dominated by the AST contributions, which
therefore is the cause for the \emph{nearly} coinciding curves
in the main plot of Fig.~\ref{fig:results:val:qT2D:qT}.
However, it should be stressed that in spite of the minor
roles played by AATs, their participations are still of the
crucial importance in reproducing the correct singular manner
of the exact spectra.
As illustrated in the ratio diagrams of
Fig.~\ref{fig:results:val:qT2D:qT}, whereas the AST contributions
either underestimate or overshoot the NLO results by $\sim2\%$
in the vicinity of $\qT=0.1~$GeV, the curves of
NLO$_{\mathrm{s}+\mathrm{a}}$ achieve permille level agreement
with the exact spectra for all three slices.
These phenomena can also be confirmed by
Fig.~\ref{fig:results:val:qT2D:qTwt}, especially in the centre
and bottom diagrams therein.

In light of the non-trivial contributions of the AATs
on the leading singular behaviour, the subsequent task
is to work out a consistent approach to embed their
contributions in the resummation.
In circumstances where the singular behaviour
is governed solely by the ASTs, the resummation can
proceed with exponentiating the characteristic logarithmic
terms in the impact space, as discussed in Sec.~\ref{sec:res}.
Considering that the AAT emergence still preserves
the factorisation formula and so the according R(a)GEs,
we can still apply the logarithmic exponentiation onto
the present situation, which then transforms
Eq.~\eqref{eq:def:partonic:qT2D} into the following expression,
\begin{align} \label{eq:def:partonic:qT2D:expon}
  \widetilde{\Sigma}^{[\dots]}_{\ttbar}
  \,\sim&\;
    \exp\bigg[\;
        L_{\mathrm{T}}\, f_0(\alpha_sL_{\mathrm{T}})
      + f_1(\alpha_sL_{\mathrm{T}})
      + {\alpha_s\, f_2(\alpha_sL_{\mathrm{T}})}
      +{\alpha^2_s\, f_3(\alpha_sL_{\mathrm{T}})}
      +\ldots\;
    \bigg] \nnb\\
  &\times\,
    \sum_{m}\,\alpha^m_s\,
    \Big\{
      \hat{s}_{m}( \bttbar,x_t,\Yttbar)
      +\hat{a}_{m}(\widehat{b}_{\mathrm{T}}, \bttbar,x_t,\Yttbar)
    \Big\}
    \,,
\end{align}
where the coefficient functions $f_{k}(k\ge0)$ are introduced
to collect the logarithmic terms at the respective accuracies.
$\hat{s}_{m}$ and $\hat{a}_{m}$ stem from the coefficients in
Eq.~\eqref{eq:def:partonic:qT2D} capturing the azimuthally
symmetric and asymmetric contributions, respectively.
In spite of the similar appearance of
Eq.~\eqref{eq:def:partonic:qT2D:expon} to the AST-driven result
in Eq.~\eqref{eq:def:partonic:qT2D}, it is worth emphasising
that the perturbative series in
Eq.~\eqref{eq:def:partonic:qT2D:expon} still possesses the
constituents $\hat{a}_{m}$, which can induce a leading asymptotic
behaviour after completing the $\bTvec$-integral.
To this end, it is of essence to clarify the higher order
performance of $\hat{a}_{m}$ so as to determine the subsequent
treatments during the resummation.
If, akin to the logarithmic contributions in
Eqs.~\eqref{eq:def:gen:FTn0}-\eqref{eq:def:gen:FTn2},
$\hat{a}_{m}$ invokes deteriorating singularities with the
increase in $\alpha_s$, a second exponentiating procedure will be
necessary to mitigate the theoretical uncertainty caused by the
perturbative truncation.
Otherwise, after suitable adaptations, we are in principle able
to regard the AAT contributions in the similar manner to the ASTs.

To facilitate the discussion, we note that as for an arbitrary
order in $\alpha_s$, the term $\hat{a}_{m}$ exists as a
function of $\phi_{tb}$ and can thus be expanded in terms of
the spherical harmonics,
\begin{align}
  \hat{a}_{m}(\widehat{b}_{\mathrm{T}}, \bttbar,x_t,\Yttbar)
  \,=\,
    \sum_{k}\,\hat{a}_{m}^{(k)}(\bttbar,x_t,\Yttbar)\;
    e^{\mathrm{i}k\phi_{tb}}\,,
\end{align}
where we have utilised the momentum conservation condition
$\vec{v}_{t}=-\vec{v}_{\bar{t}}$ to reduce the dependence on the
azimuthal angle $\phi_{\bar{t}b}$.
Substituting this harmonic series into Eq.~\eqref{eq:def:gen:AT}
and then repeating the derivation of
Eqs.~\eqref{eq:def:gen:ATn0}-\eqref{eq:def:gen:ATn2} yields
\begin{align}
  e^{\mathrm{i}k\phi_{tb}}
  \xRightarrow[\text{ }]{\textbf{F.T. }}
  \frac{e^{\mathrm{i}k\phi_{tq}}}{\qT^2} \,.
\end{align}
As this consequence holds independent of the rank of the
spherical harmonics or the specific expression of
$\hat{a}_{m}$, we can now conclude that for all perturbative
orders, the singular behaviour induced by $\hat{a}_{m}$ is
always $\propto1/\qT^2$.
Therefore, we can carry out a perturbative truncation
of the fixed-order functions in
Eq.~\eqref{eq:def:partonic:qT2D:expon}\footnote{
  This sort of perturbative truncatablity has also been
  addressed in \cite{Catani:2017tuc} in the context of
  the CSS resummation framework~\cite{Collins:1984kg}.
}.
According to Eq.~\eqref{eq:intro:def_res_order}, there
are two kinds of customary prescriptions in the AST-driven
resummation to truncate the perturbative series, i.e.\
the N$^m$LL and N$^m$LL$'$ schemes.
In the following paragraphs, we will implement both methods
separately for the resummation of the double differential
distribution $\done\sigma_{\ttbar}/(\done\qT\done\Delta\Phi_{tX})$
at NLL and \NLLp accuracy.
Then we will compare their first few perturbative terms
against the solely AST governed case, e.g.\ $q_{\tau}$
resummation.
   
Based on the factorisation in Eq.~\eqref{eq:def:partonic:qT2D}
and the exponentiated impact space cross section in
Eq.~\eqref{eq:def:partonic:qT2D:expon}, the resummed $\qTvec$
spectra at NLL can be organized as follows,
\begin{align}  \label{eq:def:qT2D:NLL:exp}
  \frac{\done\sigma_{\ttbar}}{\done\qT\done\Delta\Phi_{tX}}\Bigg|_{\mathrm{NLL}}
  \,\sim&\;
    \int\done^2\bTvec\,\exp\left( \qTvec\cdot\bTvec \right)\,
    \exp\bigg[
      L_{\mathrm{T}}\, f_0(\alpha_sL_{\mathrm{T}})
      +f_1(\alpha_sL_{\mathrm{T}})
    \bigg] \nnb\\
  \sim&\;
    \alpha_s
    \bigg\{
      \underbrace{\frac{1}{\qT}\ln\left[\frac{\Mttbar^2}{\qT^2}\right]}_{\mathrm{complete}}
      +\underbrace{ \frac{1}{\qT}}_{\mathrm{incomplete}}
    \bigg\}
    +\alpha_s^2
     \bigg\{
       \underbrace{\frac{1}{\qT}\ln^3\left[\frac{\Mttbar^2}{\qT^2}\right]}_{\mathrm{complete}}
       +\underbrace{ \frac{1}{\qT}\ln^2\left[\frac{\Mttbar^2}{\qT^2}\right]}_{\mathrm{incomplete}}
       +\underbrace{ \frac{1}{\qT}\ln\left[\frac{\Mttbar^2}{\qT^2}\right]}_{\mathrm{incomplete}}
          \nnb\\
  &\hspace*{60mm}{}
      +\underbrace{ \frac{1}{\qT}}_{\mathrm{incomplete}}
    \bigg\}
    +\dots\,,
\end{align}
where in the second step, we expand the resummed result
in $\alpha_s$ and keep the first two orders contributing
to the low $\qT$ domain.
For brevity, we omit the specific coefficient in front of
each singular term.
With the underbraces, we manifest the status of each term.
For instance, in presence of the LL anomalous dimensions,
the most singular behaviour at each perturbative order
can be exactly reproduced, whereas, for the lack of the
AAT correction and the N$^m$LL, $m\ge2$, anomalous dimensions,
the terms $1/\qT^2$ at NLO and those of
$\ln^k[\Mttbar^2/\qT^2]/\qT^2$, $k\le2$, at \NNLO are not
complete and thus will receive further correction once
the calculation at the higher accuracy is performed.

A similar analysis can also be applied to the projected
transverse momentum resummation with the help of
Eq.~\eqref{eq:xs:NJ0:res},
\begin{align}  \label{eq:def:qtau:NLL:exp}
  \frac{\done\sigma_{\ttbar}}{\done q_{\tau}}\Bigg|_{\mathrm{NLL}}
  \,\sim&\;
    \int^{\infty}_{-\infty}\done {b}_{\tau_{\|}}\,
    \cos\left(  {b}_{\tau_{\|}} {q}_{\tau}\right)\;
    \exp\bigg[\;
      L_{\mathrm{M}}\, f_0(\alpha_sL_{\mathrm{M}})
      + f_1(\alpha_sL_{\mathrm{M}})
    \bigg] \nnb\\
  \sim&\;
    \alpha_s
    \bigg\{
      \underbrace{\frac{1}{q_{\tau}}\ln\left[\frac{\Mttbar^2}{q_{\tau}^2}\right]}_{\mathrm{complete}}
      +\underbrace{\frac{1}{q_{\tau}}}_{\mathrm{complete}}
    \bigg\}
    +\alpha_s^2
     \bigg\{
       \underbrace{\frac{1}{q_{\tau}}\ln^3\left[\frac{\Mttbar^2}{q_{\tau}^2}\right]}_{\mathrm{complete}}
       +\underbrace{\frac{1}{q_{\tau}}\ln^2\left[\frac{\Mttbar^2}{q_{\tau}^2}\right]}_{\mathrm{complete}}
       +\underbrace{ \frac{1}{q_{\tau}}\ln\left[\frac{\Mttbar^2}{q_{\tau}^2}\right]}_{\mathrm{incomplete}}
          \nnb\\
  &\hspace*{60mm}{}
       +\underbrace{ \frac{1}{q_{\tau}}}_{\mathrm{incomplete}}
     \bigg\}
    +\dots\,,
\end{align}
where $L_{\mathrm{M}}\equiv\log \left[  {{b}^2_{\tau_{\|}}M_{\ttbar}^2} /{b_0^2} \right]$.
Herein, due to the absence of the AAT, the NLO singular
terms of the $q_{\tau}$ spectrum have been completely
reproduced by the NLL expansions, while at \NNLO only
the terms $\ln^k[\Mttbar^2/\qT^2]/\qT^2$, $k\ge2$ can
still be finalised by the occurring Sudakov factor.
This feature permits our matching procedure proceeding
at the \NLLNLO precision as illustrated in
Sec.~\ref{sec:results}.
This, however, is infeasible for the double differential
spectrum $\done\sigma_{\ttbar}/(\done\qT\done\Delta\Phi_{tX})$
as their NLO asymptotic behaviour is incomplete, as
demonstrated in Eq.~\eqref{eq:def:qT2D:NLL:exp}.

We now continue to explore the resummation at the \NLLp level.
In comparison with the NLL case, the present calculation entails
an additional fixed-order constituent.
It follows that
\begin{align}  \label{eq:def:qT2D:NLLp:exp}
  \frac{\done\sigma_{\ttbar}}{\done\qT\done\Delta\Phi_{tX}}\Bigg|_{\mathrm{NLL}'}
  \,\sim&\;
    \int\done^2 \bTvec\,\exp\left(\qTvec\cdot\bTvec\right)\;
    \exp\bigg[
      L_{\mathrm{T}}\, f_0(\alpha_sL_{\mathrm{T}})
      + f_1(\alpha_sL_{\mathrm{T}})
    \bigg]
    \nnb\\
  &\hspace*{40mm}{}\times\;
    \Bigg\{
      1
      +\alpha_s
       \bigg[\hat{s}_{1}(\bttbar,x_t,\Yttbar)
             +\hat{a}_{1}(\widehat{b}_{\mathrm{T}},\bttbar,x_t,\Yttbar)\bigg]
    \Bigg\}
    \nnb\\
  \sim&\;
    \alpha_s
    \bigg\{
      \underbrace{ \frac{1}{\qT}\ln\left[\frac{\Mttbar^2}{\qT^2}\right]}_{\mathrm{complete}}
      +\underbrace{ \frac{1}{\qT}}_{\mathrm{complete}}
    \bigg\}
    +\alpha_s^2
     \bigg\{
       \underbrace{ \frac{1}{\qT}\ln^3\left[\frac{\Mttbar^2}{\qT^2}\right]}_{\mathrm{complete}}
       +\underbrace{ \frac{1}{\qT}\ln^2\left[\frac{\Mttbar^2}{\qT^2}\right]}_{\mathrm{complete}}
       +\underbrace{ \frac{1}{\qT}\ln\left[\frac{\Mttbar^2}{\qT^2}\right]}_{\mathrm{complete}}
          \nnb\\
  &\hspace*{60mm}{}
       +\underbrace{ \frac{1}{\qT}}_{\mathrm{incomplete}}
     \bigg\}
    +\dots\,,
\end{align}
and
\begin{align}  \label{eq:def:qtau:NLLp:exp}
  \frac{\done\sigma_{\ttbar}}{\done q_{\tau}}\Bigg|_{\mathrm{NLL}'}
  \,\sim&\;
    \int^{\infty}_{-\infty}\done {b}_{\tau_{\|}}\,
    \cos\left( {b}_{\tau_{\|}} {q}_{\tau} \right)
    \exp\bigg[
      L_{\mathrm{M}}\, f_0(\alpha_sL_{\mathrm{M}})
      + f_1(\alpha_sL_{\mathrm{M}})
    \bigg]
    \nnb\\
  &\hspace*{40mm}{}\times\;
    \Bigg\{
      1
      +\alpha_s \bigg[\hat{s}_{1}( \bttbar,x_t,\Yttbar)
      +\hat{a}_{1}(\mathrm{sign}[b_{\tau_{\|}}], \bttbar,x_t,\Yttbar)
    \bigg] \Bigg\}\nnb\\
  \sim&\;
    \alpha_s
    \bigg\{
      \underbrace{ \frac{1}{q_{\tau}}\ln\left[\frac{\Mttbar^2}{q_{\tau}^2}\right]}_{\mathrm{complete}}
      +\underbrace{ \frac{1}{q_{\tau}}}_{\mathrm{complete}}
    \bigg\}
    +\alpha_s^2
     \bigg\{
       \underbrace{ \frac{1}{q_{\tau}}\ln^3\left[\frac{\Mttbar^2}{q_{\tau}^2}\right]}_{\mathrm{complete}}
       +\underbrace{ \frac{1}{q_{\tau}}\ln^2\left[\frac{\Mttbar^2}{q_{\tau}^2}\right]}_{\mathrm{complete}}
       +\underbrace{ \frac{1}{q_{\tau}}\ln\left[\frac{\Mttbar^2}{q_{\tau}^2}\right]}_{\mathrm{complete}}
          \nnb\\
  &\hspace*{60mm}{}
       +\underbrace{ \frac{1}{q_{\tau}}}_{\mathrm{incomplete}}
     \bigg\}
    +\dots\,.
\end{align}
Confronting Eqs.~\eqref{eq:def:qT2D:NLLp:exp}-\eqref{eq:def:qtau:NLLp:exp}
with Eqs.~\eqref{eq:def:qT2D:NLL:exp}-\eqref{eq:def:qtau:NLL:exp}, it is
seen that the inclusion of the fixed-order function can make up the
missing singular parts at NLO, and that at \NNLO the asymptotic
performance of the spectrum
$\done\sigma_{\ttbar}/(\done\qT\done\Delta\Phi_{tX})$ is aligned with
the AST-driven case, i.e.\ $\done\sigma_{\ttbar}/\done q_{\tau}$.
It should be stressed that this kind of alignment can also hold
in the perturbative series for other logarithmic accuracies, if
one notes that as to a given precise N$^k$LL$'$, $k\ge1$, the
missing AAT contributions caused by the perturbative truncation
in practice start from $\alpha^{k+1}_s \hat{a}_{k+1}$, which are
mapped onto the singular terms $\alpha^{k+1}_s/\qT^2$ through the
inverse Fourier transformation and thus amount to the N$^{k+1}$LL$'$,
$k\ge1$, corrections.
To this end, differing from the AST-driven observables, which are
able to switch between the unprimed and primed schemes flexibly,
the double differential spectra
$\done\sigma_{\ttbar}/(\done\qT\done\Delta\Phi_{tX})$ prefers the
primed prescriptions in organising the relevant ingredients in
regard to the intact asymptotic series therein and so the improved
matching precision.
    
 \changed{
  Also, it is worth noting that the discussions above are  subject to the condition that  AATs do not participate into the anomalous dimensions, which is true for the process $pp\to t\bar{t}$ as well as the associated productions $pp\to t\bar{t} B(B=H,W^{\pm},Z)$ as required by the consistency condition in Eq.~\eqref{eq:M2PS:NJ0:MuNu:indep}, but can not be guaranteed in the processes with measured jet(s) in the final state, e.g., \cite{Chien:2019gyf,delCastillo:2021znl}, where the azimuthally asymmetric anomalous dimensions may emerge from the soft and coft decomposition and thereby invoke extra divergences in the scale evolution.
  In the recent years, different strategies have been devoted to this issue \cite{Chien:2019gyf,delCastillo:2021znl}, and it is also unveiled that, akin to the top-pair production in this paper, the azimuthal decorrelation in the jet-boson~\cite{Chen:2018fqu,Chien:2020hzh,Bouaziz:2022tik,Chien:2022wiq} and dijet \cite{Banfi:2008qs,Zhang:2022wvs} productions can   be utilized as well to circumvent the azimuthal asymmetric divergences, including those stemming from the resummation exponentials.   }
 
\section{Evolution kernel for the non-diagonal anomalous dimension}
\label{app:non:dia:AD}
  
In this part, we will elaborate on the calculation on the
evolution kernel $\mathcal{V}_{\alpha\beta}^{[\kappa]}$
which is induced by the hard non-diagonal anomalous
dimensions $\gamma_{h,\alpha\beta}^{[\kappa]}$ in
Eq.~\eqref{eq:methods:res_rge_H}.
Here $\kappa$ runs over
$\{g_{n}g_{\bar{n}},q^i_{n}\bar{q}^j_{\bar{n}},q^i_{\bar{n}}\bar{q}^j_{n}\}$,
indicating the partonic channel.

We define the perturbative series of
$\gamma_{h,\alpha\beta}^{[\kappa]}$ below
\begin{align}
  \gamma_{h,\alpha\beta}^{[\kappa]}
  \,=\,
    \sum_{m=0}\,
    \left(\frac{\alpha_s}{4\pi}\right)^{m+1}\,
    \gamma_{h,\alpha\beta}^{[\kappa],(m)}\,,
\end{align}
where the $\gamma_{h,\alpha\beta}^{[\kappa],(m)}$ are the
coefficients in each order.
The \NNLO expressions can be found in
\cite{Ferroglia:2009ii,Ferroglia:2009ep}, while progress
towards \NNNLO precision has been made in \cite{Liu:2022elt}.
Other than that, the resummation procedure also entails
the QCD-$\beta$ functions,
\begin{align}
  \frac{\done\alpha_s}{\done\ln\mu}
  \,=\;
    -2\alpha_s\,\sum_{k=0} \bigg(\frac{\alpha_s}{4\pi}\bigg)^{k+1}\beta_{k}\,
\end{align}
where the $\beta_{k}$ characterise the anomalous dimensions
for the strong coupling $\alpha_s$.
By now, they are known to five-loop accuracy \cite{Baikov:2016tgj}.

Equipped with these expressions, we are now ready to
calculate the kernel $\mathcal{V}_{\alpha\beta}^{[\kappa]}$.
At NLL, according to Tab.~\ref{tab:methods:res_accuracy},
the resummation only concerns the contribution of
$\gamma_{h,\alpha\beta}^{[\kappa],(0)}$ to the kernel
$\mathcal{V}_{\alpha\beta}^{[\kappa]}$, which is
independent of either the scale evolution or the
$\alpha_s$ running, and thus allows us to solve the
hard RG equations in diagonal space.
It yields,
\begin{align}\label{eq:def:Vres:NLL}
  \mathbf{V}_h^{[\kappa]}(\bttbar,x_t,\mu_f,\mu_i)\Bigg|_{\mathrm{NLL}(')}
  \,=\;
  \mathbf{R}^{-1}_{[\kappa]}\,
  \exp\bigg\{
    \frac{\mathbf{r}^{[\kappa],(0)}_{h}}{2\beta_0}
    \ln\bigg[
      \frac{\alpha_s(\mu_i)}{\alpha_s(\mu_f)}
    \bigg]
  \bigg\}\,
  \mathbf{R}_{[\kappa]}\,,
\end{align}
where $\mathbf{V}_h^{[\kappa]}$ stands for the matrix
representation of the evolution kernel
$\mathcal{V}_{\alpha\beta}^{[\kappa]}$.
$\mathbf{r}^{[\kappa],(0)}_{h}$ denotes the resulting
diagonal matrix of $\gamma_{h,\alpha\beta}^{[\kappa],(0)}$
with the invertible transformation matrix
$\mathbf{R}_{[\kappa]}$, namely
\begin{align}\label{eq:def:rh0}
  \mathbf{r}^{[\kappa],(0)}_{h}
  \,=\;
    \mathbf{R}_{[\kappa]}\gamma_{h }^{[\kappa],(0)}\mathbf{R}^{-1}_{[\kappa]}\,.
\end{align} 
However, the situation on the \NNLLp level is different
on the ground that due to the participation of
$\gamma_{h }^{[\kappa],(1)}$, factoring out all the scale
dependent pieces is not straightforward any longer.
To this end, we resort to the perturbative resolution
suggested in \cite{Buras:1991jm,Buchalla:1995vs,Ahrens:2010zv}
to introduce the correction matrices,
\begin{align}
  \mathbf{J}^{[\kappa]}_{ij}
  \,=\;
    \mathbf{r}^{{[\kappa]},(0)}_{h,ii}\;
    \delta_{ij}\,\frac{\beta_1}{2\beta^2_0}
    -\frac{\mathbf{r}^{{[\kappa]},(1)}_{h,ij}}
          {2\beta_0+\mathbf{r}^{{[\kappa]},(0)}_{h,ii}
           -\mathbf{r}^{{[\kappa]},(0)}_{h,jj}}\,,
\end{align}
where $\delta_{ij}$ is the Kronecker delta function with the
indices $i$ and $j$ running over the set $\{1,2\}$ ($\{1,2,3\}$)
for the quark (gluon) channel.
$\mathbf{r}^{{[\kappa]},(1)}_{h}$ is defined in the similar
way of Eq.~\eqref{eq:def:rh0} except for replacing
$\gamma_{h }^{[\kappa],(0)}$ with $\gamma_{h }^{[\kappa],(1)}$.

Combining the $\mathbf{J}^{[\kappa]}$ matrix above with
Eq.~\eqref{eq:def:Vres:NLL}, we then arrive at the result
for \NNLLp precision,
\begin{align}\label{eq:def:Vres:N2LL}
  \mathbf{V}_h^{[\kappa]}(\bttbar,x_t,\mu_f,\mu_i)\Bigg|_{\mathrm{N}^2\mathrm{LL}(')}
  \,=\;
    \mathbf{R}^{-1}_{[\kappa]}\,
    \bigg[\mathbf{I}+\frac{\alpha_s(\mu_f)}{4\pi}\mathbf{J}^{[\kappa]}\bigg]\,
    \exp\bigg\{
      \frac{\mathbf{r}^{[\kappa],(0)}_{h}}{2\beta_0}
      \ln\bigg[
        \frac{\alpha_s(\mu_i)}{\alpha_s(\mu_f)}
      \bigg]
    \bigg\}\,
    \bigg[
      \mathbf{I}-\frac{\alpha_s(\mu_i)}{4\pi}\mathbf{J}^{[\kappa]}
    \bigg]\,
    \mathbf{R}_{[\kappa]}\,,
\end{align}
where  $\mathbf{I}$ signifies  the unit matrix.

\section{Decomposition of the theoretical uncertainty estimate}
\label{app:sv:qTout}

In Sec.~\ref{sec:res:num} the theoretical uncertainties of the
resummation-improved differential cross sections have been estimated
by combining the results of the individual variations of each
auxiliary scale in Eq.~\eqref{eq:scale:nat} or shape parameters
in Eqs.~\eqref{eq:scale:ftran:qTin}-\eqref{eq:scale:ftran:dphi}
in quadrature.
In the following, we will take the
$\done\sigma_{t\bar{t}}/\done q_{\mathrm{T,out}}$ spectrum as an
example to inspect the contributions of the individual variations.
     
\begin{figure}[t!]
  \centering
  \begin{subfigure}{0.32\textwidth}
    \centering
    \includegraphics[width=.9\linewidth, height=0.9\linewidth]{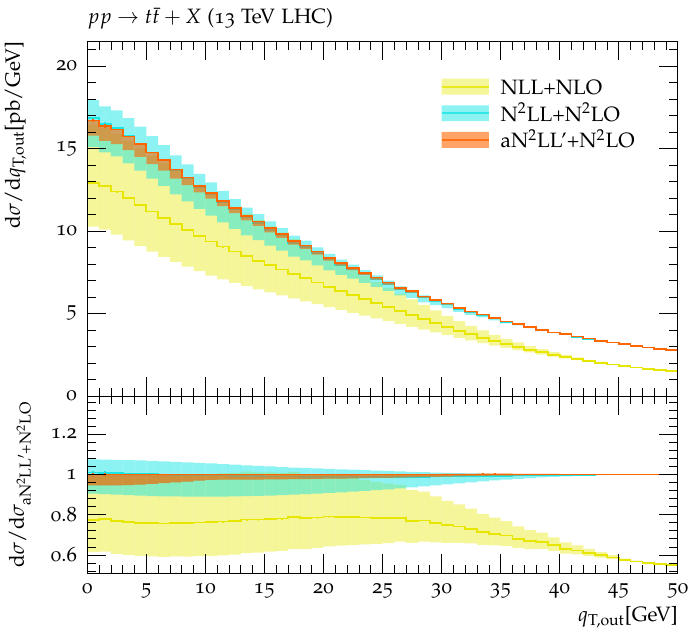}
  \caption{Beam scale $\mu_b$ variation}
   \label{fig:sv:mub:qTout}
  \end{subfigure}
    \begin{subfigure}{0.32\textwidth}
    \centering
    \includegraphics[width=.9\linewidth, height=0.9\linewidth]{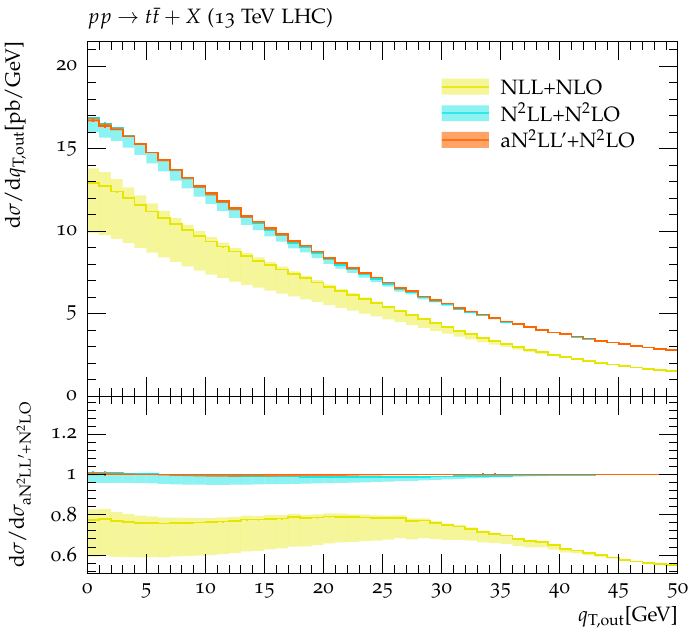}
  \caption{Soft scale $\mu_s$ variation}
    \label{fig:sv:mus:qTout}
  \end{subfigure}
      \begin{subfigure}{0.32\textwidth}
    \centering
    \includegraphics[width=.9\linewidth, height=0.9\linewidth]{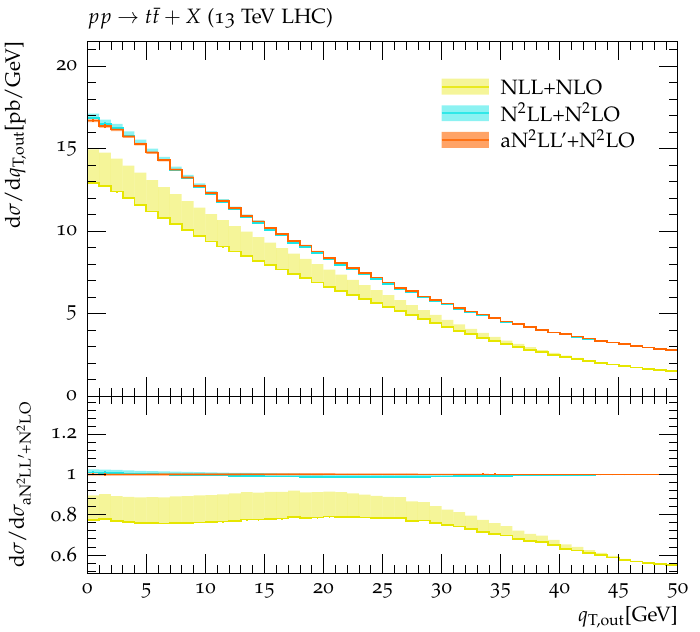}
  \caption{Hard scale $\mu_h$ variation}
    \label{fig:sv:muh:qTout}
  \end{subfigure} 
  \caption{
    Dependence on the virtuality scales associated with the
    beam-collinear, soft, and hard sectors of the differential
    cross section $\done\sigma_{\ttbar}/\done q_{\mathrm{T,out}}$.
  }
  \label{fig:results:sv:vir:qTout}
\end{figure}
 
Fig.~\ref{fig:results:sv:vir:qTout} details the individual variations of
the virtuality scales associated with the beam-collinear, soft, and hard
sectors, $\mu_b$, $\mu_s$, and $\mu_h$, respectively.
They arise from the renormalisation or subtraction of the virtuality
divergences in each regime.
As demonstrated in \cite{Li:2016axz}, the beam scale $\mu_b$ is associated
with both the QCD factorisation and the hard-soft-collinear decomposition,
such that any variations of $\mu_b$ value will induce perturbative
corrections from both the solutions of the DGLAP equation
\cite{Lipatov:1974qm,Gribov:1972ri,Altarelli:1977zs,Dokshitzer:1977sg} and
the RGE in Eqs.~\eqref{eq:methods:res_rge_Bq}-\eqref{eq:methods:res_rge_Bg}.
Therefore, as observed in Fig.~\ref{fig:results:sv:vir:qTout}, a variation
of the beam scale $\mu_b$ generates a larger theoretical uncertainty than
a variation of the soft or hard scales, that are only governed by their
respective RGEs only, as illustrated in Eq.~\eqref{eq:methods:res_rge_H}
and Eq.~\eqref{eq:methods:res_rge_S}.
It is interesting to note that the default choices for $\mu_s$ and $\mu_h$
produce results that lie on one edge of their respective uncertainty interval,
see Fig.~\ref{fig:sv:mus:qTout} and Fig.~\ref{fig:sv:muh:qTout}.
This is particularly noticable at NLL.
This phenomenon can be interpreted through the evolution kernel in
Eq.~\eqref{eq:def:Dres}, more explicitly,
\begin{align}
\label{eq:sv:hard:nll}
\ln\bigg[\frac{\mathcal{D}^{\mathrm{res}}_{[\kappa]}    (\bT,\Mttbar,\mu_h,\mu^{\mathrm{def}}_b,\mu^{\mathrm{def}}_s,\nu^{\mathrm{def}}_b,\nu^{\mathrm{def}}_s)}
{\mathcal{D}^{\mathrm{res}}_{[\kappa]}   (\bT,\Mttbar,\mu^{\mathrm{def}}_h,\mu^{\mathrm{def}}_b,\mu^{\mathrm{def}}_s,\nu^{\mathrm{def}}_b,\nu^{\mathrm{def}}_s)} \bigg] \Bigg|_{\mathrm{NLL}}
\,
&=
\,
C_{[\kappa]}\Gamma^{(0)}_{\mathrm{cusp}}\frac{\alpha_s(\Mttbar)}{4\pi}\frac{1}{2}\ln^2\left[ \frac{\mu^2_h}{(\mu^{\mathrm{def}}_h)^2}\right]+\dots
\,
,
\\
\label{eq:sv:soft:nll}
\ln\bigg[\frac{\mathcal{D}^{\mathrm{res}}_{[\kappa]}    (\bT,\Mttbar,{\mu}^{\mathrm{def}}_h,\mu^{\mathrm{def}}_b,\mu_s,\nu^{\mathrm{def}}_b,\nu^{\mathrm{def}}_s)}
{\mathcal{D}^{\mathrm{res}}_{[\kappa]}   (\bT,\Mttbar,\mu^{\mathrm{def}}_h,\mu^{\mathrm{def}}_b,\mu^{\mathrm{def}}_s,\nu^{\mathrm{def}}_b,\nu^{\mathrm{def}}_s)} \bigg] \Bigg|_{\mathrm{NLL}}
\,
&=
\,
-C_{[\kappa]}\Gamma^{(0)}_{\mathrm{cusp}}\frac{\alpha_s(\Mttbar)}{4\pi}\frac{1}{2}\ln^2\left[ \frac{\mu^2_s}{(\mu^{\mathrm{def}}_s)^2}\right]+\dots\,,
\end{align}
wherein the perturbative expansion of the cusp dimension is given by
\begin{align}
\Gamma_{\mathrm{cusp}}=\sum_{k=0}^{\infty}\,\Gamma_{\mathrm{cusp}}^{(k)}\,\left(\frac{\alpha_s}{4\pi}\right)^k\,.
\end{align}
In Eqs.~\eqref{eq:sv:hard:nll}-\eqref{eq:sv:soft:nll}, we only present
the leading contributions from the cusp dimension, which dictates the
scale dependences at NLL.
We observe that at NLL, the default choice of the hard (soft) scale is
situated close to the saddle point of the evolution function
$\mathcal{D}^{\mathrm{res}}_{[\kappa]} $, such that any changes in the
scale will only enhance (reduce) the cross section, prompting the
behaviour displayed in Fig.~\ref{fig:sv:mus:qTout} and
Fig.~\ref{fig:sv:muh:qTout}.
Further increasing the logarithmic accuracy may bring in the fixed order contributions counterbalancing the alternations of
$\mathcal{D}^{\mathrm{res}}_{[\kappa]} $.
In particular, as for the results at \aNNLLpNNLO, the inclining trends
of the default curves are nearly reversed in comparison to those at
lower accuracies.

\begin{figure}[t!]
  \centering
  \begin{subfigure}{0.32\textwidth}
    \centering
    \includegraphics[width=.9\linewidth, height=0.9\linewidth]{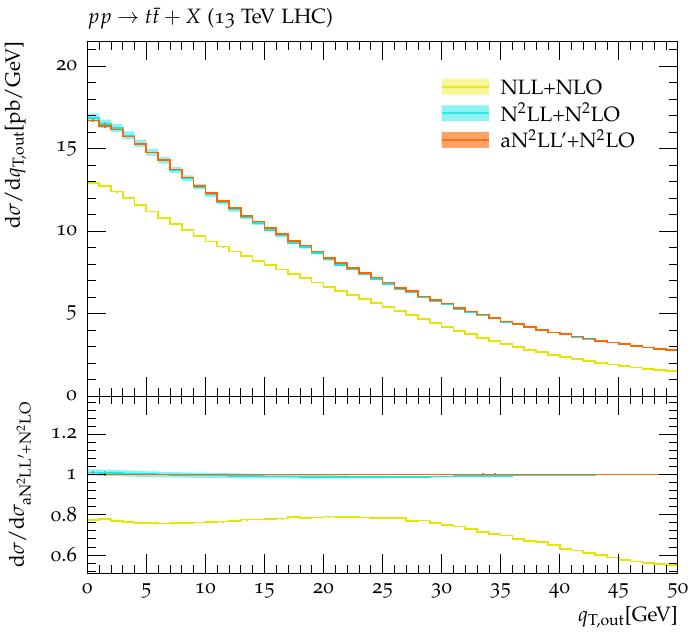}
    \caption{Beam scale $\nu_b$ variation}
  \end{subfigure}
  \begin{subfigure}{0.32\textwidth}
    \centering
    \includegraphics[width=.9\linewidth, height=0.9\linewidth]{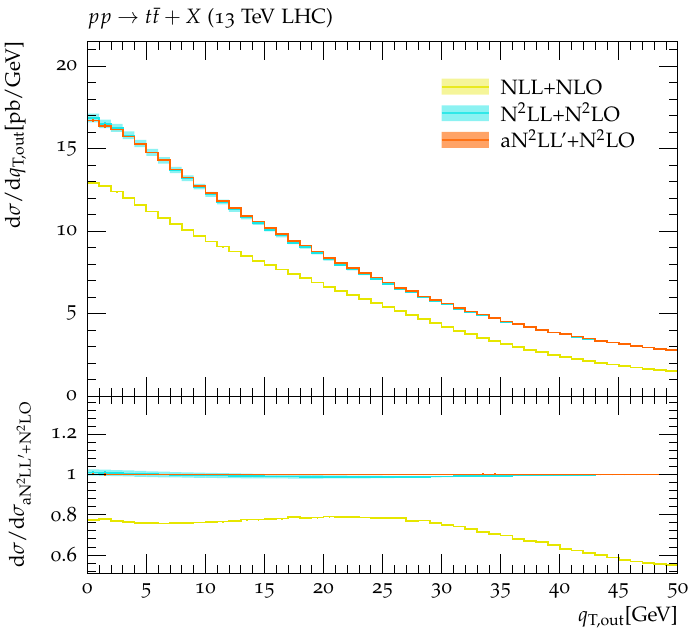}
    \caption{Soft scale $\nu_s$ variation}
  \end{subfigure}  
  \caption{
    Dependence on the rapidity scales associated with the
    beam-collinear and soft sectors of the differential
    cross section $\done\sigma_{\ttbar}/\done q_{\mathrm{T,out}}$.
  }
  \label{fig:results:sv:rap:qTout}
\end{figure}
  
Fig.~\ref{fig:results:sv:rap:qTout} illustrates the variations induced by
scales introduced by regulating the rapidity divergences, $\nu_b$ and
$\nu_s$, respectively.
Differing from the converging behaviour in
Fig.~\ref{fig:results:sv:vir:qTout}, the theoretical uncertainty here
grows from \NLL to \NNLL.
As exhibited in Tab.~\ref{tab:methods:res_accuracy}, \NLL curves only
comprise tree-level fixed order functions and thus the $\nu_s$ and $\nu_b$ dependences are entirely determined by Eq.~\eqref{eq:def:Dres}, or,
more specifically, the rapidity anomalous dimension $\gamma_r$ in Eq.~\eqref{eq:methods:res_rage_B}.
However, as presented in \cite{Li:2016axz}, $\gamma_r$ vanishes in the exponential regularisation prescription at LO.
Subsequently, there is no rapidity scale dependence at \NLL, as observed
in Fig.~\ref{fig:results:sv:vir:qTout}, and the first non-trivial
variation induced by $\nu_s$ and $\nu_b$ occurs at \NNLL.
Confronting \NNLLNNLO and \aNNLLpNNLO shows that the manifestly convergent
behaviour is also found for the rapidity scale depedences.
  
\begin{figure}[t!]
  \centering
  \begin{subfigure}{0.32\textwidth}
    \centering
    \includegraphics[width=.9\linewidth, height=0.9\linewidth]{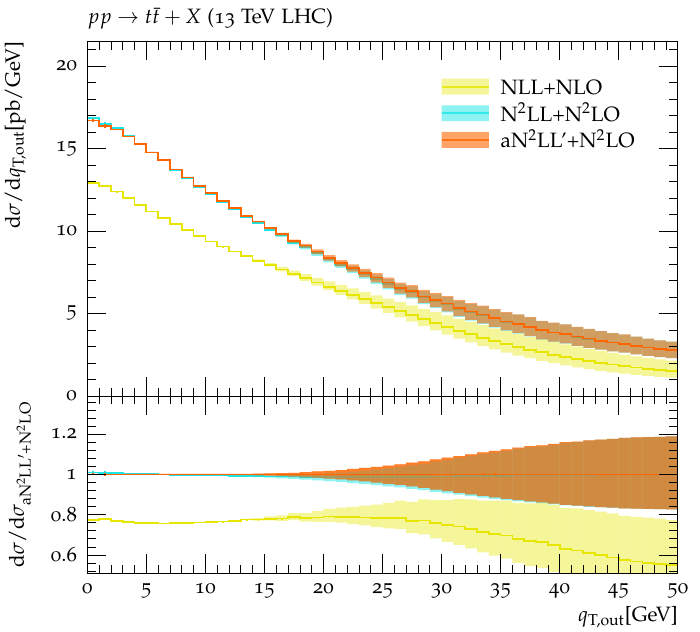}
    \caption{FO scale $\mu_{\mathrm{f.o.}}$ variation}
    \label{fig:sv:mufo:qTout}
  \end{subfigure}
  \begin{subfigure}{0.32\textwidth}
    \centering
    \includegraphics[width=.9\linewidth, height=0.9\linewidth]{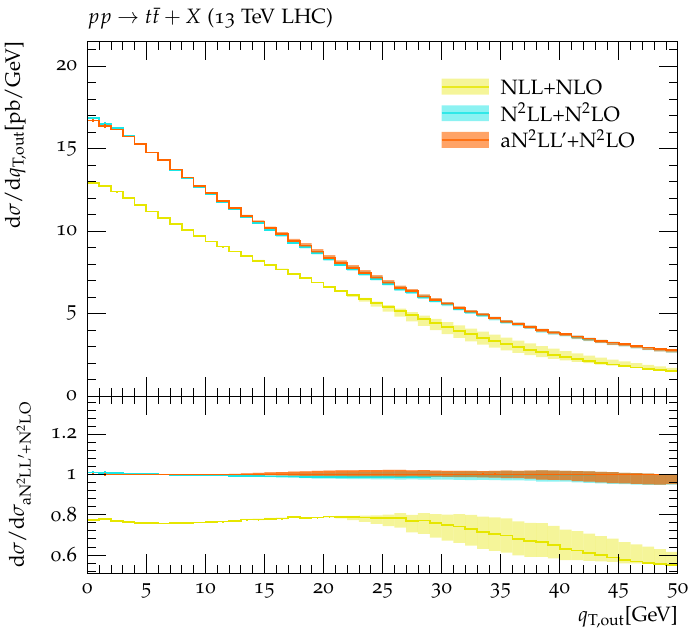}
  \caption{Focal point $c_m$ and the transition radius $r_m$ variation}
  \label{fig:sv:rm:cm:qTout}
  \end{subfigure}  
  \caption{
    Dependence on the fixed-order scale and the matching parameters
    of the differential
    cross section $\done\sigma_{\ttbar}/\done q_{\mathrm{T,out}}$.
  }
  \label{fig:results:sv:mat:qTout}
\end{figure}

At last, we present the variations of the auxiliary inputs pertinent
to the matching procedure, such as the fixed-order scale
$\mu_{\mathrm{f.o.}}$ and the shape parameters $\{c_m, r_m\}$ in the
transition function.
In Fig.~\ref{fig:sv:mufo:qTout}, the $\mu_{\mathrm{f.o.}}$ variation
has been assessed.
It is seen that the theoretical uncertainties in the asymptotic regime
are highly suppressed due to the excellent agreement between QCD and
EFT outputs, as demonstrated in
Figs.~\ref{fig:results:val:qTin}-\ref{fig:results:val:dphi}, but
experience continuous growth with the increase in
$q_{\mathrm{T},\mathrm{out}}$, as a result of the manifesting power
corrections, until the stabilising resummation is switched off and the
pure fixed-order behaviour is recovered.
The sensitivity of the $q_{\mathrm{T},\mathrm{out}}$ spectrum on
the matching transition function parameters $\{c_m, r_m\}$ is
illustrated in Fig.~\ref{fig:sv:rm:cm:qTout}.
It, in essence, reflects both the size of the higher power corrections and fixed-order corrections.
As expected, the reduction in error bands is discovered along with the improvements on the logarithmic and fixed-order precisions.

\bibliographystyle{amsunsrt_mod}
\bibliography{ref_generator3}

\begin{thebibliography}{100}

\bibitem{ATLAS:2021xhc}
ATLAS collaboration, \emph{{Measurement of the $t\bar{t}$ production
  cross-section using dilepton events in $pp$ collisions at $\sqrt{s}=5.02$ TeV
  with the ATLAS detector}}. \relax
 \relax
\bibitem{ATLAS:2022jbj}
\href{https://inspirehep.net/literature?q=2207.01354}{}ATLAS collaboration,
  \emph{{Measurement of the $t\bar{t}$ production cross-section in $pp$
  collisions at $\sqrt{s}=5.02$ TeV with the ATLAS detector}},
  \href{http://arXiv.org/pdf/2207.01354}{{\tt arXiv:2207.01354}} [hep-ex].
  \relax
 \relax
\bibitem{CMS:2017zpm}
A.~M. Sirunyan et~al., CMS collaboration, \emph{{Measurement of the inclusive $
  \mathrm{t}\overline{\mathrm{t}} $ cross section in pp collisions at $
  \sqrt{s}=5.02 $ TeV using final states with at least one charged lepton}},
  JHEP \textbf{03} (2018),
  \href{https://inspirehep.net/literature?q=1711.03143}{115},
  [\href{http://arXiv.org/pdf/1711.03143}{{\tt arXiv:1711.03143}} [hep-ex]].
  \relax
 \relax
\bibitem{CMS:2021gwv}
A.~Tumasyan et~al., CMS collaboration, \emph{{Measurement of the inclusive $
  \mathrm{t}\overline{\mathrm{t}} $ production cross section in proton-proton
  collisions at $ \sqrt{s} $ = 5.02 TeV}}, JHEP \textbf{04} (2022),
  \href{https://inspirehep.net/literature?q=2112.09114}{144},
  [\href{http://arXiv.org/pdf/2112.09114}{{\tt arXiv:2112.09114}} [hep-ex]].
  \relax
 \relax
\bibitem{CMS:2013yjt}
S.~Chatrchyan et~al., CMS collaboration, \emph{{Measurement of the $t\bar{t}$
  Production Cross Section in the All-Jet Final State in pp Collisions at
  $\sqrt{s}$ = 7 TeV}}, JHEP \textbf{05} (2013),
  \href{https://inspirehep.net/literature?q=1302.0508}{065},
  [\href{http://arXiv.org/pdf/1302.0508}{{\tt arXiv:1302.0508}} [hep-ex]].
  \relax
 \relax
\bibitem{CMS:2016yys}
V.~Khachatryan et~al., CMS collaboration, \emph{{Measurement of the t-tbar
  production cross section in the e-mu channel in proton-proton collisions at
  sqrt(s) = 7 and 8 TeV}}, JHEP \textbf{08} (2016),
  \href{https://inspirehep.net/literature?q=1603.02303}{029},
  [\href{http://arXiv.org/pdf/1603.02303}{{\tt arXiv:1603.02303}} [hep-ex]].
  \relax
 \relax
\bibitem{ATLAS:2022aof}
\href{https://inspirehep.net/literature?q=2205.13830}{}ATLAS, CMS
  collaboration, \emph{{Combination of inclusive top-quark pair production
  cross-section measurements using ATLAS and CMS data at $\sqrt{s}= 7$ and 8
  TeV}},  \href{http://arXiv.org/pdf/2205.13830}{{\tt arXiv:2205.13830}}
  [hep-ex]. \relax
 \relax
\bibitem{CMS:2015auz}
V.~Khachatryan et~al., CMS collaboration, \emph{{Measurement of the
  $\mathrm{t}\overline{{\mathrm{t}}}$ production cross section in the all-jets
  final state in pp collisions at $\sqrt{s}=8$ $\,\text {TeV}$}}, Eur. Phys. J.
  C \textbf{76} (2016), no.~3,
  \href{https://inspirehep.net/literature?q=1509.06076}{128},
  [\href{http://arXiv.org/pdf/1509.06076}{{\tt arXiv:1509.06076}} [hep-ex]].
  \relax
 \relax
\bibitem{CMS:2016csa}
V.~Khachatryan et~al., CMS collaboration, \emph{{Measurements of the
  $\mathrm{t}\overline{\mathrm{t}}$ production cross section in lepton+jets
  final states in pp collisions at 8 $\,\text {TeV}$ and ratio of 8 to 7
  $\,\text {TeV}$ cross sections}}, Eur. Phys. J. C \textbf{77} (2017), no.~1,
  \href{https://inspirehep.net/literature?q=1602.09024}{15},
  [\href{http://arXiv.org/pdf/1602.09024}{{\tt arXiv:1602.09024}} [hep-ex]].
  \relax
 \relax
\bibitem{ATLAS:2019hau}
G.~Aad et~al., ATLAS collaboration, \emph{{Measurement of the $t\bar{t}$
  production cross-section and lepton differential distributions in $e\mu $
  dilepton events from $pp$ collisions at $\sqrt{s}=13\,\text {TeV}$ with the
  ATLAS detector}}, Eur. Phys. J. C \textbf{80} (2020), no.~6,
  \href{https://inspirehep.net/literature?q=1910.08819}{528},
  [\href{http://arXiv.org/pdf/1910.08819}{{\tt arXiv:1910.08819}} [hep-ex]].
  \relax
 \relax
\bibitem{ATLAS:2020aln}
G.~Aad et~al., ATLAS collaboration, \emph{{Measurement of the $t\bar{t}$
  production cross-section in the lepton+jets channel at $\sqrt{s}=13$ TeV with
  the ATLAS experiment}}, Phys. Lett. B \textbf{810} (2020),
  \href{https://inspirehep.net/literature?q=2006.13076}{135797},
  [\href{http://arXiv.org/pdf/2006.13076}{{\tt arXiv:2006.13076}} [hep-ex]].
  \relax
 \relax
\bibitem{ATLAS:2020ccu}
G.~Aad et~al., ATLAS collaboration, \emph{{Measurements of top-quark pair
  single- and double-differential cross-sections in the all-hadronic channel in
  $pp$ collisions at $\sqrt{s}=13~\textrm{TeV}$ using the ATLAS detector}},
  JHEP \textbf{01} (2021),
  \href{https://inspirehep.net/literature?q=2006.09274}{033},
  [\href{http://arXiv.org/pdf/2006.09274}{{\tt arXiv:2006.09274}} [hep-ex]].
  \relax
 \relax
\bibitem{CMS:2019snc}
A.~M. Sirunyan et~al., CMS collaboration, \emph{{Measurement of the top quark
  pair production cross section in dilepton final states containing one $\tau$
  lepton in pp collisions at $\sqrt{s}=$ 13 TeV}}, JHEP \textbf{02} (2020),
  \href{https://inspirehep.net/literature?q=1911.13204}{191},
  [\href{http://arXiv.org/pdf/1911.13204}{{\tt arXiv:1911.13204}} [hep-ex]].
  \relax
 \relax
\bibitem{CMS:2016rtp}
CMS collaboration, \emph{{Measurement of the ${\rm t}{\rm \bar{t}}$ production
  cross section at 13 TeV in the all-jets final state}}. \relax
 \relax
\bibitem{CMS:2018fks}
A.~M. Sirunyan et~al., CMS collaboration, \emph{{Measurement of the
  $\mathrm{t}\overline{\mathrm{t}}$ production cross section, the top quark
  mass, and the strong coupling constant using dilepton events in pp collisions
  at $\sqrt{s} =$ 13 TeV}}, Eur. Phys. J. C \textbf{79} (2019), no.~5,
  \href{https://inspirehep.net/literature?q=1812.10505}{368},
  [\href{http://arXiv.org/pdf/1812.10505}{{\tt arXiv:1812.10505}} [hep-ex]].
  \relax
 \relax
\bibitem{CMS:2021vhb}
A.~Tumasyan et~al., CMS collaboration, \emph{{Measurement of differential $t
  \bar t$ production cross sections in the full kinematic range using
  lepton+jets events from proton-proton collisions at $\sqrt {s}$ =
  13\,\,TeV}}, Phys. Rev. D \textbf{104} (2021), no.~9,
  \href{https://inspirehep.net/literature?q=2108.02803}{092013},
  [\href{http://arXiv.org/pdf/2108.02803}{{\tt arXiv:2108.02803}} [hep-ex]].
  \relax
 \relax
\bibitem{CMS:2022elr}
CMS collaboration, \emph{{First measurement of the top quark pair production
  cross section in proton-proton collisions at $\sqrt{s}=13.6\,\mathrm{TeV}$}}.
  \relax
 \relax
\bibitem{CMS:2019esx}
A.~M. Sirunyan et~al., CMS collaboration, \emph{{Measurement of $\mathrm{t\bar
  t}$ normalised multi-differential cross sections in pp collisions at $\sqrt
  s=13$ TeV, and simultaneous determination of the strong coupling strength,
  top quark pole mass, and parton distribution functions}}, Eur. Phys. J. C
  \textbf{80} (2020), no.~7,
  \href{https://inspirehep.net/literature?q=1904.05237}{658},
  [\href{http://arXiv.org/pdf/1904.05237}{{\tt arXiv:1904.05237}} [hep-ex]].
  \relax
 \relax
\bibitem{ATLAS:2022xfj}
G.~Aad et~al., ATLAS collaboration, \emph{{Measurements of differential
  cross-sections in top-quark pair events with a high transverse momentum top
  quark and limits on beyond the Standard Model contributions to top-quark pair
  production with the ATLAS detector at $ \sqrt{s} $ = 13 TeV}}, JHEP
  \textbf{06} (2022),
  \href{https://inspirehep.net/literature?q=2202.12134}{063},
  [\href{http://arXiv.org/pdf/2202.12134}{{\tt arXiv:2202.12134}} [hep-ex]].
  \relax
 \relax
\bibitem{CMS:2020tvq}
A.~M. Sirunyan et~al., CMS collaboration, \emph{{Measurement of differential
  $\mathrm{t\bar{t}}$ production cross sections using top quarks at large
  transverse momenta in $pp$ collisions at $\sqrt{s} =$ 13 TeV}}, Phys. Rev. D
  \textbf{103} (2021), no.~5,
  \href{https://inspirehep.net/literature?q=2008.07860}{052008},
  [\href{http://arXiv.org/pdf/2008.07860}{{\tt arXiv:2008.07860}} [hep-ex]].
  \relax
 \relax
\bibitem{CMS:2022uae}
CMS collaboration, \emph{{Measurement of differential cross sections for the
  production of top quark pairs and of additional jets in pp collisions at
  $\sqrt{s}=13~\mathrm{TeV}$}}. \relax
 \relax
\bibitem{Nason:1987xz}
P.~Nason, S.~Dawson and R.~K. Ellis, \emph{{The Total Cross-Section for the
  Production of Heavy Quarks in Hadronic Collisions}}, Nucl. Phys. B
  \textbf{303} (1988),
  \href{https://inspirehep.net/literature?q=Nucl%20Phys%20B,303,607}{607--633}.
  \relax
 \relax
\bibitem{Beenakker:1988bq}
W.~Beenakker, H.~Kuijf, W.~L. van Neerven and J.~Smith, \emph{{QCD Corrections
  to Heavy Quark Production in p anti-p Collisions}}, Phys. Rev. D \textbf{40}
  (1989),
  \href{https://inspirehep.net/literature?q=Phys%20Rev%20D,40,54}{54--82}.
  \relax
 \relax
\bibitem{Beenakker:1990maa}
W.~Beenakker, W.~L. van Neerven, R.~Meng, G.~A. Schuler and J.~Smith,
  \emph{{QCD corrections to heavy quark production in hadron hadron
  collisions}}, Nucl. Phys. B \textbf{351} (1991),
  \href{https://inspirehep.net/literature?q=Nucl%20Phys%20B,351,507}{507--560}.
  \relax
 \relax
\bibitem{Mangano:1991jk}
M.~L. Mangano, P.~Nason and G.~Ridolfi, \emph{{Heavy quark correlations in
  hadron collisions at next-to-leading order}}, Nucl. Phys. B \textbf{373}
  (1992),
  \href{https://inspirehep.net/literature?q=Nucl%20Phys%20B,373,295}{295--345}.
  \relax
 \relax
\bibitem{Czakon:2013goa}
M.~Czakon, P.~Fiedler and A.~Mitov, \emph{{Total Top-Quark Pair-Production
  Cross Section at Hadron Colliders Through $O(\alpha^4_S)$}}, Phys. Rev. Lett.
  \textbf{110} (2013),
  \href{https://inspirehep.net/literature?q=1303.6254}{252004},
  [\href{http://arXiv.org/pdf/1303.6254}{{\tt arXiv:1303.6254}} [hep-ph]].
  \relax
 \relax
\bibitem{Czakon:2015owf}
M.~Czakon, D.~Heymes and A.~Mitov, \emph{{High-precision differential
  predictions for top-quark pairs at the LHC}}, Phys. Rev. Lett. \textbf{116}
  (2016), no.~8, \href{https://inspirehep.net/literature?q=1511.00549}{082003},
   [\href{http://arXiv.org/pdf/1511.00549}{{\tt arXiv:1511.00549}} [hep-ph]].
  \relax
 \relax
\bibitem{Czakon:2016ckf}
M.~Czakon, P.~Fiedler, D.~Heymes and A.~Mitov, \emph{{NNLO QCD predictions for
  fully-differential top-quark pair production at the Tevatron}}, JHEP
  \textbf{05} (2016),
  \href{https://inspirehep.net/literature?q=1601.05375}{034},
  [\href{http://arXiv.org/pdf/1601.05375}{{\tt arXiv:1601.05375}} [hep-ph]].
  \relax
 \relax
\bibitem{Czakon:2017dip}
\href{https://inspirehep.net/literature?q=1704.08551}{M.~Czakon, D.~Heymes and
  A.~Mitov}, \emph{{fastNLO tables for NNLO top-quark pair differential
  distributions}},  \href{http://arXiv.org/pdf/1704.08551}{{\tt
  arXiv:1704.08551}} [hep-ph]. \relax
 \relax
\bibitem{Czakon:2017wor}
M.~Czakon, D.~Heymes, A.~Mitov, D.~Pagani, I.~Tsinikos and M.~Zaro,
  \emph{{Top-pair production at the LHC through NNLO QCD and NLO EW}}, JHEP
  \textbf{10} (2017),
  \href{https://inspirehep.net/literature?q=1705.04105}{186},
  [\href{http://arXiv.org/pdf/1705.04105}{{\tt arXiv:1705.04105}} [hep-ph]].
  \relax
 \relax
\bibitem{Catani:2019iny}
S.~Catani, S.~Devoto, M.~Grazzini, S.~Kallweit, J.~Mazzitelli and H.~Sargsyan,
  \emph{{Top-quark pair hadroproduction at next-to-next-to-leading order in
  QCD}}, Phys. Rev. D \textbf{99} (2019), no.~5,
  \href{https://inspirehep.net/literature?q=1901.04005}{051501},
  [\href{http://arXiv.org/pdf/1901.04005}{{\tt arXiv:1901.04005}} [hep-ph]].
  \relax
 \relax
\bibitem{Catani:2020tko}
S.~Catani, S.~Devoto, M.~Grazzini, S.~Kallweit and J.~Mazzitelli,
  \emph{{Top-quark pair hadroproduction at NNLO: differential predictions with
  the $\overline{MS}$ mass}}, JHEP \textbf{08} (2020), no.~08,
  \href{https://inspirehep.net/literature?q=2005.00557}{027},
  [\href{http://arXiv.org/pdf/2005.00557}{{\tt arXiv:2005.00557}} [hep-ph]].
  \relax
 \relax
\bibitem{Gao:2012ja}
J.~Gao, C.~S. Li and H.~X. Zhu, \emph{{Top Quark Decay at Next-to-Next-to
  Leading Order in QCD}}, Phys. Rev. Lett. \textbf{110} (2013), no.~4,
  \href{https://inspirehep.net/literature?q=1210.2808}{042001},
  [\href{http://arXiv.org/pdf/1210.2808}{{\tt arXiv:1210.2808}} [hep-ph]].
  \relax
 \relax
\bibitem{Brucherseifer:2013iv}
M.~Brucherseifer, F.~Caola and K.~Melnikov, \emph{{$\mathcal O(\alpha_s^2)$
  corrections to fully-differential top quark decays}}, JHEP \textbf{04}
  (2013), \href{https://inspirehep.net/literature?q=1301.7133}{059},
  [\href{http://arXiv.org/pdf/1301.7133}{{\tt arXiv:1301.7133}} [hep-ph]].
  \relax
 \relax
\bibitem{Catani:2019hip}
S.~Catani, S.~Devoto, M.~Grazzini, S.~Kallweit and J.~Mazzitelli,
  \emph{{Top-quark pair production at the LHC: Fully differential QCD
  predictions at NNLO}}, JHEP \textbf{07} (2019),
  \href{https://inspirehep.net/literature?q=1906.06535}{100},
  [\href{http://arXiv.org/pdf/1906.06535}{{\tt arXiv:1906.06535}} [hep-ph]].
  \relax
 \relax
\bibitem{Behring:2019iiv}
A.~Behring, M.~Czakon, A.~Mitov, A.~S. Papanastasiou and R.~Poncelet,
  \emph{{Higher order corrections to spin correlations in top quark pair
  production at the LHC}}, Phys. Rev. Lett. \textbf{123} (2019), no.~8,
  \href{https://inspirehep.net/literature?q=1901.05407}{082001},
  [\href{http://arXiv.org/pdf/1901.05407}{{\tt arXiv:1901.05407}} [hep-ph]].
  \relax
 \relax
\bibitem{Czakon:2020qbd}
M.~Czakon, A.~Mitov and R.~Poncelet, \emph{{NNLO QCD corrections to leptonic
  observables in top-quark pair production and decay}}, JHEP \textbf{05}
  (2021), \href{https://inspirehep.net/literature?q=2008.11133}{212},
  [\href{http://arXiv.org/pdf/2008.11133}{{\tt arXiv:2008.11133}} [hep-ph]].
  \relax
 \relax
\bibitem{Bernreuther:2010ny}
W.~Bernreuther and Z.-G. Si, \emph{{Distributions and correlations for top
  quark pair production and decay at the Tevatron and LHC.}}, Nucl. Phys. B
  \textbf{837} (2010),
  \href{https://inspirehep.net/literature?q=1003.3926}{90--121},
  [\href{http://arXiv.org/pdf/1003.3926}{{\tt arXiv:1003.3926}} [hep-ph]].
  \relax
 \relax
\bibitem{Kuhn:2006vh}
J.~H. Kuhn, A.~Scharf and P.~Uwer, \emph{{Electroweak effects in top-quark pair
  production at hadron colliders}}, Eur. Phys. J. C \textbf{51} (2007),
  \href{https://inspirehep.net/literature?q=hep-ph/0610335}{37--53},
  [\href{http://arXiv.org/pdf/hep-ph/0610335}{{\tt hep-ph/0610335}}]. \relax
 \relax
\bibitem{Bernreuther:2006vg}
W.~Bernreuther, M.~Fuecker and Z.-G. Si, \emph{{Weak interaction corrections to
  hadronic top quark pair production}}, Phys. Rev. D \textbf{74} (2006),
  \href{https://inspirehep.net/literature?q=hep-ph/0610334}{113005},
  [\href{http://arXiv.org/pdf/hep-ph/0610334}{{\tt hep-ph/0610334}}]. \relax
 \relax
\bibitem{Kuhn:2013zoa}
J.~H. K\"uhn, A.~Scharf and P.~Uwer, \emph{{Weak Interactions in Top-Quark Pair
  Production at Hadron Colliders: An Update}}, Phys. Rev. D \textbf{91} (2015),
  no.~1, \href{https://inspirehep.net/literature?q=1305.5773}{014020},
  [\href{http://arXiv.org/pdf/1305.5773}{{\tt arXiv:1305.5773}} [hep-ph]].
  \relax
 \relax
\bibitem{Hollik:2011ps}
W.~Hollik and D.~Pagani, \emph{{The electroweak contribution to the top quark
  forward-backward asymmetry at the Tevatron}}, Phys. Rev. D \textbf{84}
  (2011), \href{https://inspirehep.net/literature?q=1107.2606}{093003},
  [\href{http://arXiv.org/pdf/1107.2606}{{\tt arXiv:1107.2606}} [hep-ph]].
  \relax
 \relax
\bibitem{Pagani:2016caq}
D.~Pagani, I.~Tsinikos and M.~Zaro, \emph{{The impact of the photon PDF and
  electroweak corrections on $t \bar{t}$ distributions}}, Eur. Phys. J. C
  \textbf{76} (2016), no.~9,
  \href{https://inspirehep.net/literature?q=1606.01915}{479},
  [\href{http://arXiv.org/pdf/1606.01915}{{\tt arXiv:1606.01915}} [hep-ph]].
  \relax
 \relax
\bibitem{Gutschow:2018tuk}
C.~G\"utschow, J.~M. Lindert and M.~Sch\"onherr, \emph{{Multi-jet merged
  top-pair production including electroweak corrections}}, Eur. Phys. J. C
  \textbf{78} (2018), no.~4,
  \href{https://inspirehep.net/literature?q=1803.00950}{317},
  [\href{http://arXiv.org/pdf/1803.00950}{{\tt arXiv:1803.00950}} [hep-ph]].
  \relax
 \relax
\bibitem{Denner:2016jyo}
A.~Denner and M.~Pellen, \emph{{NLO electroweak corrections to off-shell
  top-antitop production with leptonic decays at the LHC}}, JHEP \textbf{08}
  (2016), \href{https://inspirehep.net/literature?q=1607.05571}{155},
  [\href{http://arXiv.org/pdf/1607.05571}{{\tt arXiv:1607.05571}} [hep-ph]].
  \relax
 \relax
\bibitem{Kidonakis:2014pja}
N.~Kidonakis, \emph{{NNNLO soft-gluon corrections for the top-quark $p_T$ and
  rapidity distributions}}, Phys. Rev. D \textbf{91} (2015), no.~3,
  \href{https://inspirehep.net/literature?q=1411.2633}{031501},
  [\href{http://arXiv.org/pdf/1411.2633}{{\tt arXiv:1411.2633}} [hep-ph]].
  \relax
 \relax
\bibitem{Kidonakis:2010dk}
N.~Kidonakis, \emph{{Next-to-next-to-leading soft-gluon corrections for the top
  quark cross section and transverse momentum distribution}}, Phys. Rev. D
  \textbf{82} (2010),
  \href{https://inspirehep.net/literature?q=1009.4935}{114030},
  [\href{http://arXiv.org/pdf/1009.4935}{{\tt arXiv:1009.4935}} [hep-ph]].
  \relax
 \relax
\bibitem{Kidonakis:2009ev}
N.~Kidonakis, \emph{{Two-loop soft anomalous dimensions and NNLL resummation
  for heavy quark production}}, Phys. Rev. Lett. \textbf{102} (2009),
  \href{https://inspirehep.net/literature?q=0903.2561}{232003},
  [\href{http://arXiv.org/pdf/0903.2561}{{\tt arXiv:0903.2561}} [hep-ph]].
  \relax
 \relax
\bibitem{Kidonakis:2014isa}
N.~Kidonakis, \emph{{NNNLO soft-gluon corrections for the top-antitop pair
  production cross section}}, Phys. Rev. D \textbf{90} (2014), no.~1,
  \href{https://inspirehep.net/literature?q=1405.7046}{014006},
  [\href{http://arXiv.org/pdf/1405.7046}{{\tt arXiv:1405.7046}} [hep-ph]].
  \relax
 \relax
\bibitem{Kidonakis:2019yji}
N.~Kidonakis, \emph{{Top-quark double-differential distributions at approximate
  N$^3$LO}}, Phys. Rev. D \textbf{101} (2020), no.~7,
  \href{https://inspirehep.net/literature?q=1912.10362}{074006},
  [\href{http://arXiv.org/pdf/1912.10362}{{\tt arXiv:1912.10362}} [hep-ph]].
  \relax
 \relax
\bibitem{Ahrens:2010zv}
V.~Ahrens, A.~Ferroglia, M.~Neubert, B.~D. Pecjak and L.~L. Yang,
  \emph{{Renormalization-Group Improved Predictions for Top-Quark Pair
  Production at Hadron Colliders}}, JHEP \textbf{09} (2010),
  \href{https://inspirehep.net/literature?q=1003.5827}{097},
  [\href{http://arXiv.org/pdf/1003.5827}{{\tt arXiv:1003.5827}} [hep-ph]].
  \relax
 \relax
\bibitem{Ferroglia:2012ku}
A.~Ferroglia, B.~D. Pecjak and L.~L. Yang, \emph{{Soft-gluon resummation for
  boosted top-quark production at hadron colliders}}, Phys. Rev. D \textbf{86}
  (2012), \href{https://inspirehep.net/literature?q=1205.3662}{034010},
  [\href{http://arXiv.org/pdf/1205.3662}{{\tt arXiv:1205.3662}} [hep-ph]].
  \relax
 \relax
\bibitem{Ferroglia:2013awa}
A.~Ferroglia, S.~Marzani, B.~D. Pecjak and L.~L. Yang, \emph{{Boosted top
  production: factorization and resummation for single-particle inclusive
  distributions}}, JHEP \textbf{01} (2014),
  \href{https://inspirehep.net/literature?q=1310.3836}{028},
  [\href{http://arXiv.org/pdf/1310.3836}{{\tt arXiv:1310.3836}} [hep-ph]].
  \relax
 \relax
\bibitem{Pecjak:2016nee}
B.~D. Pecjak, D.~J. Scott, X.~Wang and L.~L. Yang, \emph{{Resummed differential
  cross sections for top-quark pairs at the LHC}}, Phys. Rev. Lett.
  \textbf{116} (2016), no.~20,
  \href{https://inspirehep.net/literature?q=1601.07020}{202001},
  [\href{http://arXiv.org/pdf/1601.07020}{{\tt arXiv:1601.07020}} [hep-ph]].
  \relax
 \relax
\bibitem{Czakon:2018nun}
M.~Czakon, A.~Ferroglia, D.~Heymes, A.~Mitov, B.~D. Pecjak, D.~J. Scott,
  X.~Wang and L.~L. Yang, \emph{{Resummation for (boosted) top-quark pair
  production at NNLO+NNLL' in QCD}}, JHEP \textbf{05} (2018),
  \href{https://inspirehep.net/literature?q=1803.07623}{149},
  [\href{http://arXiv.org/pdf/1803.07623}{{\tt arXiv:1803.07623}} [hep-ph]].
  \relax
 \relax
\bibitem{Hinderer:2014qta}
P.~Hinderer, F.~Ringer, G.~F. Sterman and W.~Vogelsang, \emph{{Toward NNLL
  Threshold Resummation for Hadron Pair Production in Hadronic Collisions}},
  Phys. Rev. D \textbf{91} (2015), no.~1,
  \href{https://inspirehep.net/literature?q=1411.3149}{014016},
  [\href{http://arXiv.org/pdf/1411.3149}{{\tt arXiv:1411.3149}} [hep-ph]].
  \relax
 \relax
\bibitem{Beneke:2009ye}
M.~Beneke, M.~Czakon, P.~Falgari, A.~Mitov and C.~Schwinn, \emph{{Threshold
  expansion of the $gg(q \bar q)$ $\to \overline {QQ} + X$ cross section at
  $O(\alpha^4_s)$}}, Phys. Lett. B \textbf{690} (2010),
  \href{https://inspirehep.net/literature?q=0911.5166}{483--490},
  [\href{http://arXiv.org/pdf/0911.5166}{{\tt arXiv:0911.5166}} [hep-ph]],
  [Erratum: Phys.Lett.B 778, 464--464 (2018)]. \relax
 \relax
\bibitem{Beneke:2010da}
M.~Beneke, P.~Falgari and C.~Schwinn, \emph{{Threshold resummation for pair
  production of coloured heavy (s)particles at hadron colliders}}, Nucl. Phys.
  B \textbf{842} (2011),
  \href{https://inspirehep.net/literature?q=1007.5414}{414--474},
  [\href{http://arXiv.org/pdf/1007.5414}{{\tt arXiv:1007.5414}} [hep-ph]].
  \relax
 \relax
\bibitem{Beneke:2011mq}
M.~Beneke, P.~Falgari, S.~Klein and C.~Schwinn, \emph{{Hadronic top-quark pair
  production with NNLL threshold resummation}}, Nucl. Phys. B \textbf{855}
  (2012), \href{https://inspirehep.net/literature?q=1109.1536}{695--741},
  [\href{http://arXiv.org/pdf/1109.1536}{{\tt arXiv:1109.1536}} [hep-ph]].
  \relax
 \relax
\bibitem{Cacciari:2011hy}
M.~Cacciari, M.~Czakon, M.~Mangano, A.~Mitov and P.~Nason, \emph{{Top-pair
  production at hadron colliders with next-to-next-to-leading logarithmic
  soft-gluon resummation}}, Phys. Lett. B \textbf{710} (2012),
  \href{https://inspirehep.net/literature?q=1111.5869}{612--622},
  [\href{http://arXiv.org/pdf/1111.5869}{{\tt arXiv:1111.5869}} [hep-ph]].
  \relax
 \relax
\bibitem{Piclum:2018ndt}
J.~Piclum and C.~Schwinn, \emph{{Soft-gluon and Coulomb corrections to hadronic
  top-quark pair production beyond NNLO}}, JHEP \textbf{03} (2018),
  \href{https://inspirehep.net/literature?q=1801.05788}{164},
  [\href{http://arXiv.org/pdf/1801.05788}{{\tt arXiv:1801.05788}} [hep-ph]].
  \relax
 \relax
\bibitem{Ju:2020otc}
W.-L. Ju, G.~Wang, X.~Wang, X.~Xu, Y.~Xu and L.~L. Yang, \emph{{Top quark pair
  production near threshold: single/double distributions and mass
  determination}}, JHEP \textbf{06} (2020),
  \href{https://inspirehep.net/literature?q=2004.03088}{158},
  [\href{http://arXiv.org/pdf/2004.03088}{{\tt arXiv:2004.03088}} [hep-ph]].
  \relax
 \relax
\bibitem{Ju:2019mqc}
W.-L. Ju, G.~Wang, X.~Wang, X.~Xu, Y.~Xu and L.~L. Yang, \emph{{Invariant-mass
  distribution of top-quark pairs and top-quark mass determination}}, Chin.
  Phys. C \textbf{44} (2020), no.~9,
  \href{https://inspirehep.net/literature?q=1908.02179}{091001},
  [\href{http://arXiv.org/pdf/1908.02179}{{\tt arXiv:1908.02179}} [hep-ph]].
  \relax
 \relax
\bibitem{Zhu:2012ts}
H.~X. Zhu, C.~S. Li, H.~T. Li, D.~Y. Shao and L.~L. Yang,
  \emph{{Transverse-momentum resummation for top-quark pairs at hadron
  colliders}}, Phys. Rev. Lett. \textbf{110} (2013), no.~8,
  \href{https://inspirehep.net/literature?q=1208.5774}{082001},
  [\href{http://arXiv.org/pdf/1208.5774}{{\tt arXiv:1208.5774}} [hep-ph]].
  \relax
 \relax
\bibitem{Li:2013mia}
H.~T. Li, C.~S. Li, D.~Y. Shao, L.~L. Yang and H.~X. Zhu, \emph{{Top quark pair
  production at small transverse momentum in hadronic collisions}}, Phys. Rev.
  D \textbf{88} (2013),
  \href{https://inspirehep.net/literature?q=1307.2464}{074004},
  [\href{http://arXiv.org/pdf/1307.2464}{{\tt arXiv:1307.2464}} [hep-ph]].
  \relax
 \relax
\bibitem{Catani:2014qha}
S.~Catani, M.~Grazzini and A.~Torre, \emph{{Transverse-momentum resummation for
  heavy-quark hadroproduction}}, Nucl. Phys. B \textbf{890} (2014),
  \href{https://inspirehep.net/literature?q=1408.4564}{518--538},
  [\href{http://arXiv.org/pdf/1408.4564}{{\tt arXiv:1408.4564}} [hep-ph]].
  \relax
 \relax
\bibitem{Catani:2017tuc}
S.~Catani, M.~Grazzini and H.~Sargsyan, \emph{{Azimuthal asymmetries in QCD
  hard scattering: infrared safe but divergent}}, JHEP \textbf{06} (2017),
  \href{https://inspirehep.net/literature?q=1703.08468}{017},
  [\href{http://arXiv.org/pdf/1703.08468}{{\tt arXiv:1703.08468}} [hep-ph]].
  \relax
 \relax
\bibitem{Catani:2018mei}
S.~Catani, M.~Grazzini and H.~Sargsyan, \emph{{Transverse-momentum resummation
  for top-quark pair production at the LHC}}, JHEP \textbf{11} (2018),
  \href{https://inspirehep.net/literature?q=1806.01601}{061},
  [\href{http://arXiv.org/pdf/1806.01601}{{\tt arXiv:1806.01601}} [hep-ph]].
  \relax
 \relax
\bibitem{Alioli:2021ggd}
S.~Alioli, A.~Broggio and M.~A. Lim, \emph{{Zero-jettiness resummation for
  top-quark pair production at the LHC}}, JHEP \textbf{01} (2022),
  \href{https://inspirehep.net/literature?q=2111.03632}{066},
  [\href{http://arXiv.org/pdf/2111.03632}{{\tt arXiv:2111.03632}} [hep-ph]].
  \relax
 \relax
\bibitem{Mazzitelli:2020jio}
J.~Mazzitelli, P.~F. Monni, P.~Nason, E.~Re, M.~Wiesemann and G.~Zanderighi,
  \emph{{Next-to-Next-to-Leading Order Event Generation for Top-Quark Pair
  Production}}, Phys. Rev. Lett. \textbf{127} (2021), no.~6,
  \href{https://inspirehep.net/literature?q=2012.14267}{062001},
  [\href{http://arXiv.org/pdf/2012.14267}{{\tt arXiv:2012.14267}} [hep-ph]].
  \relax
 \relax
\bibitem{Mazzitelli:2021mmm}
J.~Mazzitelli, P.~F. Monni, P.~Nason, E.~Re, M.~Wiesemann and G.~Zanderighi,
  \emph{{Top-pair production at the LHC with MINNLO$_{PS}$}}, JHEP \textbf{04}
  (2022), \href{https://inspirehep.net/literature?q=2112.12135}{079},
  [\href{http://arXiv.org/pdf/2112.12135}{{\tt arXiv:2112.12135}} [hep-ph]].
  \relax
 \relax
\bibitem{Nadolsky:2007ba}
P.~M. Nadolsky, C.~Balazs, E.~L. Berger and C.~P. Yuan, \emph{{Gluon-gluon
  contributions to the production of continuum diphoton pairs at hadron
  colliders}}, Phys. Rev. D \textbf{76} (2007),
  \href{https://inspirehep.net/literature?q=hep-ph/0702003}{013008},
  [\href{http://arXiv.org/pdf/hep-ph/0702003}{{\tt hep-ph/0702003}}]. \relax
 \relax
\bibitem{Catani:2010pd}
S.~Catani and M.~Grazzini, \emph{{QCD transverse-momentum resummation in gluon
  fusion processes}}, Nucl. Phys. B \textbf{845} (2011),
  \href{https://inspirehep.net/literature?q=1011.3918}{297--323},
  [\href{http://arXiv.org/pdf/1011.3918}{{\tt arXiv:1011.3918}} [hep-ph]].
  \relax
 \relax
\bibitem{Collins:1980ui}
J.~C. Collins, \emph{{INTRINSIC TRANSVERSE MOMENTUM. 1. NONGAUGE THEORIES}},
  Phys. Rev. D \textbf{21} (1980),
  \href{https://inspirehep.net/literature?q=Phys%20Rev%20D,21,2962}{2962}.
  \relax
 \relax
\bibitem{Sterman:1978bi}
G.~F. Sterman, \emph{{Mass Divergences in Annihilation Processes. 1. Origin and
  Nature of Divergences in Cut Vacuum Polarization Diagrams}}, Phys. Rev. D
  \textbf{17} (1978),
  \href{https://inspirehep.net/literature?q=Phys%20Rev%20D,17,2773}{2773}.
  \relax
 \relax
\bibitem{Landau:1959fi}
L.~D. Landau, \emph{{On analytic properties of vertex parts in quantum field
  theory}}, Nucl. Phys. \textbf{13} (1959), no.~1,
  \href{https://inspirehep.net/literature?q=Nucl%20Phys,13,181}{181--192}.
  \relax
 \relax
\bibitem{Coleman:1965xm}
S.~Coleman and R.~E. Norton, \emph{{Singularities in the physical region}},
  Nuovo Cim. \textbf{38} (1965),
  \href{https://inspirehep.net/literature?q=Nuovo%20Cim,38,438}{438--442}.
  \relax
 \relax
\bibitem{Beneke:1997zp}
M.~Beneke and V.~A. Smirnov, \emph{{Asymptotic expansion of Feynman integrals
  near threshold}}, Nucl. Phys. B \textbf{522} (1998),
  \href{https://inspirehep.net/literature?q=hep-ph/9711391}{321--344},
  [\href{http://arXiv.org/pdf/hep-ph/9711391}{{\tt hep-ph/9711391}}]. \relax
 \relax
\bibitem{Smirnov:2002pj}
V.~A. Smirnov, \emph{{Applied asymptotic expansions in momenta and masses}},
  Springer Tracts Mod. Phys. \textbf{177} (2002),
  \href{https://inspirehep.net/literature?q=Springer%20Tracts%20Mod%20Phys,177,1}{1--262}.
  \relax
 \relax
\bibitem{Smirnov:2012gma}
V.~A. Smirnov, \emph{{Analytic tools for Feynman integrals}}, vol. 250, 2012.
  \relax
 \relax
\bibitem{Jantzen:2011nz}
B.~Jantzen, \emph{{Foundation and generalization of the expansion by regions}},
  JHEP \textbf{12} (2011),
  \href{https://inspirehep.net/literature?q=1111.2589}{076},
  [\href{http://arXiv.org/pdf/1111.2589}{{\tt arXiv:1111.2589}} [hep-ph]].
  \relax
 \relax
\bibitem{Bauer:2001yt}
C.~W. Bauer, D.~Pirjol and I.~W. Stewart, \emph{{Soft collinear factorization
  in effective field theory}}, Phys. Rev. D \textbf{65} (2002),
  \href{https://inspirehep.net/literature?q=hep-ph/0109045}{054022},
  [\href{http://arXiv.org/pdf/hep-ph/0109045}{{\tt hep-ph/0109045}}]. \relax
 \relax
\bibitem{Bauer:2001ct}
C.~W. Bauer and I.~W. Stewart, \emph{{Invariant operators in collinear
  effective theory}}, Phys. Lett. B \textbf{516} (2001),
  \href{https://inspirehep.net/literature?q=hep-ph/0107001}{134--142},
  [\href{http://arXiv.org/pdf/hep-ph/0107001}{{\tt hep-ph/0107001}}]. \relax
 \relax
\bibitem{Bauer:2000yr}
C.~W. Bauer, S.~Fleming, D.~Pirjol and I.~W. Stewart, \emph{{An Effective field
  theory for collinear and soft gluons: Heavy to light decays}}, Phys. Rev. D
  \textbf{63} (2001),
  \href{https://inspirehep.net/literature?q=hep-ph/0011336}{114020},
  [\href{http://arXiv.org/pdf/hep-ph/0011336}{{\tt hep-ph/0011336}}]. \relax
 \relax
\bibitem{Bauer:2000ew}
C.~W. Bauer, S.~Fleming and M.~E. Luke, \emph{{Summing Sudakov logarithms in $B
  \to  X_s \gamma $in effective field theory.}}, Phys. Rev. D \textbf{63}
  (2000), \href{https://inspirehep.net/literature?q=hep-ph/0005275}{014006},
  [\href{http://arXiv.org/pdf/hep-ph/0005275}{{\tt hep-ph/0005275}}]. \relax
 \relax
\bibitem{Bauer:2002nz}
C.~W. Bauer, S.~Fleming, D.~Pirjol, I.~Z. Rothstein and I.~W. Stewart,
  \emph{{Hard scattering factorization from effective field theory}}, Phys.
  Rev. D \textbf{66} (2002),
  \href{https://inspirehep.net/literature?q=hep-ph/0202088}{014017},
  [\href{http://arXiv.org/pdf/hep-ph/0202088}{{\tt hep-ph/0202088}}]. \relax
 \relax
\bibitem{Beneke:2002ph}
M.~Beneke, A.~P. Chapovsky, M.~Diehl and T.~Feldmann, \emph{{Soft collinear
  effective theory and heavy to light currents beyond leading power}}, Nucl.
  Phys. B \textbf{643} (2002),
  \href{https://inspirehep.net/literature?q=hep-ph/0206152}{431--476},
  [\href{http://arXiv.org/pdf/hep-ph/0206152}{{\tt hep-ph/0206152}}]. \relax
 \relax
\bibitem{Beneke:2002ni}
M.~Beneke and T.~Feldmann, \emph{{Multipole expanded soft collinear effective
  theory with nonAbelian gauge symmetry}}, Phys. Lett. B \textbf{553} (2003),
  \href{https://inspirehep.net/literature?q=hep-ph/0211358}{267--276},
  [\href{http://arXiv.org/pdf/hep-ph/0211358}{{\tt hep-ph/0211358}}]. \relax
 \relax
\bibitem{Bauer:2002aj}
C.~W. Bauer, D.~Pirjol and I.~W. Stewart, \emph{{Factorization and endpoint
  singularities in heavy to light decays}}, Phys. Rev. D \textbf{67} (2003),
  \href{https://inspirehep.net/literature?q=hep-ph/0211069}{071502},
  [\href{http://arXiv.org/pdf/hep-ph/0211069}{{\tt hep-ph/0211069}}]. \relax
 \relax
\bibitem{Lange:2003pk}
B.~O. Lange and M.~Neubert, \emph{{Factorization and the soft overlap
  contribution to heavy to light form-factors}}, Nucl. Phys. B \textbf{690}
  (2004), \href{https://inspirehep.net/literature?q=hep-ph/0311345}{249--278},
  [\href{http://arXiv.org/pdf/hep-ph/0311345}{{\tt hep-ph/0311345}}], [Erratum:
  Nucl.Phys.B 723, 201--202 (2005)]. \relax
 \relax
\bibitem{Beneke:2003pa}
M.~Beneke and T.~Feldmann, \emph{{Factorization of heavy to light form-factors
  in soft collinear effective theory}}, Nucl. Phys. B \textbf{685} (2004),
  \href{https://inspirehep.net/literature?q=hep-ph/0311335}{249--296},
  [\href{http://arXiv.org/pdf/hep-ph/0311335}{{\tt hep-ph/0311335}}]. \relax
 \relax
\bibitem{Eichten:1989zv}
E.~Eichten and B.~R. Hill, \emph{{An Effective Field Theory for the Calculation
  of Matrix Elements Involving Heavy Quarks}}, Phys. Lett. B \textbf{234}
  (1990),
  \href{https://inspirehep.net/literature?q=Phys%20Lett%20B,234,511}{511--516}.
  \relax
 \relax
\bibitem{Georgi:1990um}
H.~Georgi, \emph{{An Effective Field Theory for Heavy Quarks at Low-energies}},
  Phys. Lett. B \textbf{240} (1990),
  \href{https://inspirehep.net/literature?q=Phys%20Lett%20B,240,447}{447--450}.
  \relax
 \relax
\bibitem{Grinstein:1990mj}
B.~Grinstein, \emph{{The Static Quark Effective Theory}}, Nucl. Phys. B
  \textbf{339} (1990),
  \href{https://inspirehep.net/literature?q=Nucl%20Phys%20B,339,253}{253--268}.
  \relax
 \relax
\bibitem{Neubert:1993mb}
M.~Neubert, \emph{{Heavy quark symmetry}}, Phys. Rept. \textbf{245} (1994),
  \href{https://inspirehep.net/literature?q=hep-ph/9306320}{259--396},
  [\href{http://arXiv.org/pdf/hep-ph/9306320}{{\tt hep-ph/9306320}}]. \relax
 \relax
\bibitem{Chiu:2011qc}
J.-y. Chiu, A.~Jain, D.~Neill and I.~Z. Rothstein, \emph{{The Rapidity
  Renormalization Group}}, Phys. Rev. Lett. \textbf{108} (2012),
  \href{https://inspirehep.net/literature?q=1104.0881}{151601},
  [\href{http://arXiv.org/pdf/1104.0881}{{\tt arXiv:1104.0881}} [hep-ph]].
  \relax
 \relax
\bibitem{Chiu:2012ir}
J.-Y. Chiu, A.~Jain, D.~Neill and I.~Z. Rothstein, \emph{{A Formalism for the
  Systematic Treatment of Rapidity Logarithms in Quantum Field Theory}}, JHEP
  \textbf{05} (2012),
  \href{https://inspirehep.net/literature?q=1202.0814}{084},
  [\href{http://arXiv.org/pdf/1202.0814}{{\tt arXiv:1202.0814}} [hep-ph]].
  \relax
 \relax
\bibitem{Li:2016axz}
Y.~Li, D.~Neill and H.~X. Zhu, \emph{{An exponential regulator for rapidity
  divergences}}, Nucl. Phys. B \textbf{960} (2020),
  \href{https://inspirehep.net/literature?q=1604.00392}{115193},
  [\href{http://arXiv.org/pdf/1604.00392}{{\tt arXiv:1604.00392}} [hep-ph]].
  \relax
 \relax
\bibitem{Li:2016ctv}
Y.~Li and H.~X. Zhu, \emph{{Bootstrapping Rapidity Anomalous Dimensions for
  Transverse-Momentum Resummation}}, Phys. Rev. Lett. \textbf{118} (2017),
  no.~2, \href{https://inspirehep.net/literature?q=1604.01404}{022004},
  [\href{http://arXiv.org/pdf/1604.01404}{{\tt arXiv:1604.01404}} [hep-ph]].
  \relax
 \relax
\bibitem{Collins:1984kg}
J.~C. Collins, D.~E. Soper and G.~F. Sterman, \emph{{Transverse Momentum
  Distribution in Drell-Yan Pair and W and Z Boson Production}}, Nucl. Phys. B
  \textbf{250} (1985),
  \href{https://inspirehep.net/literature?q=Nucl%20Phys%20B,250,199}{199--224}.
  \relax
 \relax
\bibitem{Catani:2000vq}
S.~Catani, D.~de~Florian and M.~Grazzini, \emph{{Universality of nonleading
  logarithmic contributions in transverse momentum distributions}}, Nucl. Phys.
  B \textbf{596} (2001),
  \href{https://inspirehep.net/literature?q=hep-ph/0008184}{299--312},
  [\href{http://arXiv.org/pdf/hep-ph/0008184}{{\tt hep-ph/0008184}}]. \relax
 \relax
\bibitem{Bozzi:2005wk}
G.~Bozzi, S.~Catani, D.~de~Florian and M.~Grazzini, \emph{{Transverse-momentum
  resummation and the spectrum of the Higgs boson at the LHC}}, Nucl. Phys. B
  \textbf{737} (2006),
  \href{https://inspirehep.net/literature?q=hep-ph/0508068}{73--120},
  [\href{http://arXiv.org/pdf/hep-ph/0508068}{{\tt hep-ph/0508068}}]. \relax
 \relax
\bibitem{Bozzi:2007pn}
G.~Bozzi, S.~Catani, D.~de~Florian and M.~Grazzini, \emph{{Higgs boson
  production at the LHC: Transverse-momentum resummation and rapidity
  dependence}}, Nucl. Phys. B \textbf{791} (2008),
  \href{https://inspirehep.net/literature?q=0705.3887}{1--19},
  [\href{http://arXiv.org/pdf/0705.3887}{{\tt arXiv:0705.3887}} [hep-ph]].
  \relax
 \relax
\bibitem{Ebert:2016gcn}
M.~A. Ebert and F.~J. Tackmann, \emph{{Resummation of Transverse Momentum
  Distributions in Distribution Space}}, JHEP \textbf{02} (2017),
  \href{https://inspirehep.net/literature?q=1611.08610}{110},
  [\href{http://arXiv.org/pdf/1611.08610}{{\tt arXiv:1611.08610}} [hep-ph]].
  \relax
 \relax
\bibitem{Monni:2016ktx}
P.~F. Monni, E.~Re and P.~Torrielli, \emph{{Higgs Transverse-Momentum
  Resummation in Direct Space}}, Phys. Rev. Lett. \textbf{116} (2016), no.~24,
  \href{https://inspirehep.net/literature?q=1604.02191}{242001},
  [\href{http://arXiv.org/pdf/1604.02191}{{\tt arXiv:1604.02191}} [hep-ph]].
  \relax
 \relax
\bibitem{Bizon:2017rah}
W.~Bizon, P.~F. Monni, E.~Re, L.~Rottoli and P.~Torrielli,
  \emph{{Momentum-space resummation for transverse observables and the Higgs
  p$_{\perp}$ at N$^{3}$LL+NNLO}}, JHEP \textbf{02} (2018),
  \href{https://inspirehep.net/literature?q=1705.09127}{108},
  [\href{http://arXiv.org/pdf/1705.09127}{{\tt arXiv:1705.09127}} [hep-ph]].
  \relax
 \relax
\bibitem{Becher:2010tm}
T.~Becher and M.~Neubert, \emph{{Drell-Yan Production at Small $q_T$,
  Transverse Parton Distributions and the Collinear Anomaly}}, Eur. Phys. J. C
  \textbf{71} (2011),
  \href{https://inspirehep.net/literature?q=1007.4005}{1665},
  [\href{http://arXiv.org/pdf/1007.4005}{{\tt arXiv:1007.4005}} [hep-ph]].
  \relax
 \relax
\bibitem{GarciaEchevarria:2011rb}
M.~G. Echevarria, A.~Idilbi and I.~Scimemi, \emph{{Factorization Theorem For
  Drell-Yan At Low $q_T$ And Transverse Momentum Distributions
  On-The-Light-Cone}}, JHEP \textbf{07} (2012),
  \href{https://inspirehep.net/literature?q=1111.4996}{002},
  [\href{http://arXiv.org/pdf/1111.4996}{{\tt arXiv:1111.4996}} [hep-ph]].
  \relax
 \relax
\bibitem{Becher:2011dz}
T.~Becher and G.~Bell, \emph{{Analytic Regularization in Soft-Collinear
  Effective Theory}}, Phys. Lett. B \textbf{713} (2012),
  \href{https://inspirehep.net/literature?q=1112.3907}{41--46},
  [\href{http://arXiv.org/pdf/1112.3907}{{\tt arXiv:1112.3907}} [hep-ph]].
  \relax
 \relax
\bibitem{Collins:1989gx}
J.~C. Collins, D.~E. Soper and G.~F. Sterman, \emph{{Factorization of Hard
  Processes in QCD}}, Adv. Ser. Direct. High Energy Phys. \textbf{5} (1989),
  \href{https://inspirehep.net/literature?q=hep-ph/0409313}{1--91},
  [\href{http://arXiv.org/pdf/hep-ph/0409313}{{\tt hep-ph/0409313}}]. \relax
 \relax
\bibitem{Manohar:2006nz}
A.~V. Manohar and I.~W. Stewart, \emph{{The Zero-Bin and Mode Factorization in
  Quantum Field Theory}}, Phys. Rev. D \textbf{76} (2007),
  \href{https://inspirehep.net/literature?q=hep-ph/0605001}{074002},
  [\href{http://arXiv.org/pdf/hep-ph/0605001}{{\tt hep-ph/0605001}}]. \relax
 \relax
\bibitem{Ananthanarayan:2018tog}
B.~Ananthanarayan, A.~Pal, S.~Ramanan and R.~Sarkar, \emph{{Unveiling Regions
  in multi-scale Feynman Integrals using Singularities and Power Geometry}},
  Eur. Phys. J. C \textbf{79} (2019), no.~1,
  \href{https://inspirehep.net/literature?q=1810.06270}{57},
  [\href{http://arXiv.org/pdf/1810.06270}{{\tt arXiv:1810.06270}} [hep-ph]].
  \relax
 \relax
\bibitem{Plenter:2020lop}
J.~Plenter and G.~Rodrigo, \emph{{Asymptotic expansions through the loop-tree
  duality}}, Eur. Phys. J. C \textbf{81} (2021), no.~4,
  \href{https://inspirehep.net/literature?q=2005.02119}{320},
  [\href{http://arXiv.org/pdf/2005.02119}{{\tt arXiv:2005.02119}} [hep-ph]].
  \relax
 \relax
\bibitem{Heinrich:2021dbf}
G.~Heinrich, S.~Jahn, S.~P. Jones, M.~Kerner, F.~Langer, V.~Magerya,
  A.~P\"oldaru, J.~Schlenk and E.~Villa, \emph{{Expansion by regions with
  pySecDec}}, Comput. Phys. Commun. \textbf{273} (2022),
  \href{https://inspirehep.net/literature?q=2108.10807}{108267},
  [\href{http://arXiv.org/pdf/2108.10807}{{\tt arXiv:2108.10807}} [hep-ph]].
  \relax
 \relax
\bibitem{Collins:1981ta}
J.~C. Collins and G.~F. Sterman, \emph{{Soft Partons in {QCD}}}, Nucl. Phys. B
  \textbf{185} (1981),
  \href{https://inspirehep.net/literature?q=Nucl%20Phys%20B,185,172}{172--188}.
  \relax
 \relax
\bibitem{Collins:1985ue}
J.~C. Collins, D.~E. Soper and G.~F. Sterman, \emph{{Factorization for Short
  Distance Hadron - Hadron Scattering}}, Nucl. Phys. B \textbf{261} (1985),
  \href{https://inspirehep.net/literature?q=Nucl%20Phys%20B,261,104}{104--142}.
  \relax
 \relax
\bibitem{Collins:1997sr}
J.~C. Collins, \emph{{Proof of factorization for diffractive hard scattering}},
  Phys. Rev. D \textbf{57} (1998),
  \href{https://inspirehep.net/literature?q=hep-ph/9709499}{3051--3056},
  [\href{http://arXiv.org/pdf/hep-ph/9709499}{{\tt hep-ph/9709499}}], [Erratum:
  Phys.Rev.D 61, 019902 (2000)]. \relax
 \relax
\bibitem{Collins:2004nx}
J.~C. Collins and A.~Metz, \emph{{Universality of soft and collinear factors in
  hard-scattering factorization}}, Phys. Rev. Lett. \textbf{93} (2004),
  \href{https://inspirehep.net/literature?q=hep-ph/0408249}{252001},
  [\href{http://arXiv.org/pdf/hep-ph/0408249}{{\tt hep-ph/0408249}}]. \relax
 \relax
\bibitem{Gaunt:2014ska}
J.~R. Gaunt, \emph{{Glauber Gluons and Multiple Parton Interactions}}, JHEP
  \textbf{07} (2014),
  \href{https://inspirehep.net/literature?q=1405.2080}{110},
  [\href{http://arXiv.org/pdf/1405.2080}{{\tt arXiv:1405.2080}} [hep-ph]].
  \relax
 \relax
\bibitem{Schwartz:2018obd}
M.~D. Schwartz, K.~Yan and H.~X. Zhu, \emph{{Factorization Violation and Scale
  Invariance}}, Phys. Rev. D \textbf{97} (2018), no.~9,
  \href{https://inspirehep.net/literature?q=1801.01138}{096017},
  [\href{http://arXiv.org/pdf/1801.01138}{{\tt arXiv:1801.01138}} [hep-ph]].
  \relax
 \relax
\bibitem{Chang:1968bh}
S.-J. Chang and S.-K. Ma, \emph{{Feynman rules and quantum electrodynamics at
  infinite momentum}}, Phys. Rev. \textbf{180} (1969),
  \href{https://inspirehep.net/literature?q=Phys%20Rev,180,1506}{1506--1513}.
  \relax
 \relax
\bibitem{Sterman:1978bj}
G.~F. Sterman, \emph{{Mass Divergences in Annihilation Processes. 2.
  Cancellation of Divergences in Cut Vacuum Polarization Diagrams}}, Phys. Rev.
  D \textbf{17} (1978),
  \href{https://inspirehep.net/literature?q=Phys%20Rev%20D,17,2789}{2789}.
  \relax
 \relax
\bibitem{Mitov:2012gt}
A.~Mitov and G.~Sterman, \emph{{Final state interactions in single- and
  multi-particle inclusive cross sections for hadronic collisions}}, Phys. Rev.
  D \textbf{86} (2012),
  \href{https://inspirehep.net/literature?q=1209.5798}{114038},
  [\href{http://arXiv.org/pdf/1209.5798}{{\tt arXiv:1209.5798}} [hep-ph]].
  \relax
 \relax
\bibitem{Rothstein:2016bsq}
I.~Z. Rothstein and I.~W. Stewart, \emph{{An Effective Field Theory for Forward
  Scattering and Factorization Violation}}, JHEP \textbf{08} (2016),
  \href{https://inspirehep.net/literature?q=1601.04695}{025},
  [\href{http://arXiv.org/pdf/1601.04695}{{\tt arXiv:1601.04695}} [hep-ph]].
  \relax
 \relax
\bibitem{Rogers:2010dm}
T.~C. Rogers and P.~J. Mulders, \emph{{No Generalized TMD-Factorization in
  Hadro-Production of High Transverse Momentum Hadrons}}, Phys. Rev. D
  \textbf{81} (2010),
  \href{https://inspirehep.net/literature?q=1001.2977}{094006},
  [\href{http://arXiv.org/pdf/1001.2977}{{\tt arXiv:1001.2977}} [hep-ph]].
  \relax
 \relax
\bibitem{Donoghue:2009cq}
J.~F. Donoghue and D.~Wyler, \emph{{On Regge kinematics in SCET}}, Phys. Rev. D
  \textbf{81} (2010),
  \href{https://inspirehep.net/literature?q=0908.4559}{114023},
  [\href{http://arXiv.org/pdf/0908.4559}{{\tt arXiv:0908.4559}} [hep-ph]].
  \relax
 \relax
\bibitem{Bauer:2010cc}
C.~W. Bauer, B.~O. Lange and G.~Ovanesyan, \emph{{On Glauber modes in
  Soft-Collinear Effective Theory}}, JHEP \textbf{07} (2011),
  \href{https://inspirehep.net/literature?q=1010.1027}{077},
  [\href{http://arXiv.org/pdf/1010.1027}{{\tt arXiv:1010.1027}} [hep-ph]].
  \relax
 \relax
\bibitem{Fleming:2014rea}
S.~Fleming, \emph{{The role of Glauber exchange in soft collinear effective
  theory and the
  Balitsky\textendash{}Fadin\textendash{}Kuraev\textendash{}Lipatov Equation}},
  Phys. Lett. B \textbf{735} (2014),
  \href{https://inspirehep.net/literature?q=1404.5672}{266--271},
  [\href{http://arXiv.org/pdf/1404.5672}{{\tt arXiv:1404.5672}} [hep-ph]].
  \relax
 \relax
\bibitem{Moult:2022lfy}
\href{https://inspirehep.net/literature?q=2207.02859}{I.~Moult, S.~Raman,
  G.~Ridgway and I.~W. Stewart}, \emph{{Anomalous Dimensions from Soft Regge
  Constants}},  \href{http://arXiv.org/pdf/2207.02859}{{\tt arXiv:2207.02859}}
  [hep-ph]. \relax
 \relax
\bibitem{Bonciani:2015sha}
R.~Bonciani, S.~Catani, M.~Grazzini, H.~Sargsyan and A.~Torre, \emph{{The $q_T$
  subtraction method for top quark production at hadron colliders}}, Eur. Phys.
  J. C \textbf{75} (2015), no.~12,
  \href{https://inspirehep.net/literature?q=1508.03585}{581},
  [\href{http://arXiv.org/pdf/1508.03585}{{\tt arXiv:1508.03585}} [hep-ph]].
  \relax
 \relax
\bibitem{Collins:1981uk}
J.~C. Collins and D.~E. Soper, \emph{{Back-To-Back Jets in QCD}}, Nucl. Phys. B
  \textbf{193} (1981),
  \href{https://inspirehep.net/literature?q=Nucl%20Phys%20B,193,381}{381},
  [Erratum: Nucl.Phys.B 213, 545 (1983)]. \relax
 \relax
\bibitem{Hill:2002vw}
R.~J. Hill and M.~Neubert, \emph{{Spectator interactions in soft collinear
  effective theory}}, Nucl. Phys. B \textbf{657} (2003),
  \href{https://inspirehep.net/literature?q=hep-ph/0211018}{229--256},
  [\href{http://arXiv.org/pdf/hep-ph/0211018}{{\tt hep-ph/0211018}}]. \relax
 \relax
\bibitem{Beneke:2009rj}
M.~Beneke, P.~Falgari and C.~Schwinn, \emph{{Soft radiation in heavy-particle
  pair production: All-order colour structure and two-loop anomalous
  dimension}}, Nucl. Phys. B \textbf{828} (2010),
  \href{https://inspirehep.net/literature?q=0907.1443}{69--101},
  [\href{http://arXiv.org/pdf/0907.1443}{{\tt arXiv:0907.1443}} [hep-ph]].
  \relax
 \relax
\bibitem{Ferroglia:2009ii}
A.~Ferroglia, M.~Neubert, B.~D. Pecjak and L.~L. Yang, \emph{{Two-loop
  divergences of massive scattering amplitudes in non-abelian gauge theories}},
  JHEP \textbf{11} (2009),
  \href{https://inspirehep.net/literature?q=0908.3676}{062},
  [\href{http://arXiv.org/pdf/0908.3676}{{\tt arXiv:0908.3676}} [hep-ph]].
  \relax
 \relax
\bibitem{Actis:2012qn}
S.~Actis, A.~Denner, L.~Hofer, A.~Scharf and S.~Uccirati, \emph{{Recursive
  generation of one-loop amplitudes in the Standard Model}}, JHEP \textbf{04}
  (2013), \href{https://inspirehep.net/literature?q=1211.6316}{037},
  [\href{http://arXiv.org/pdf/1211.6316}{{\tt arXiv:1211.6316}} [hep-ph]].
  \relax
 \relax
\bibitem{Actis:2016mpe}
S.~Actis, A.~Denner, L.~Hofer, J.-N. Lang, A.~Scharf and S.~Uccirati,
  \emph{{RECOLA: REcursive Computation of One-Loop Amplitudes}}, Comput. Phys.
  Commun. \textbf{214} (2017),
  \href{https://inspirehep.net/literature?q=1605.01090}{140--173},
  [\href{http://arXiv.org/pdf/1605.01090}{{\tt arXiv:1605.01090}} [hep-ph]].
  \relax
 \relax
\bibitem{Chen:2017jvi}
L.~Chen, M.~Czakon and R.~Poncelet, \emph{{Polarized double-virtual amplitudes
  for heavy-quark pair production}}, JHEP \textbf{03} (2018),
  \href{https://inspirehep.net/literature?q=1712.08075}{085},
  [\href{http://arXiv.org/pdf/1712.08075}{{\tt arXiv:1712.08075}} [hep-ph]].
  \relax
 \relax
\bibitem{DiVita:2019lpl}
S.~Di~Vita, T.~Gehrmann, S.~Laporta, P.~Mastrolia, A.~Primo and U.~Schubert,
  \emph{{Master integrals for the NNLO virtual corrections to $
  q\overline{q}\to t\overline{t} $ scattering in QCD: the non-planar graphs}},
  JHEP \textbf{06} (2019),
  \href{https://inspirehep.net/literature?q=1904.10964}{117},
  [\href{http://arXiv.org/pdf/1904.10964}{{\tt arXiv:1904.10964}} [hep-ph]].
  \relax
 \relax
\bibitem{Badger:2021owl}
S.~Badger, E.~Chaubey, H.~B. Hartanto and R.~Marzucca, \emph{{Two-loop leading
  colour QCD helicity amplitudes for top quark pair production in the gluon
  fusion channel}}, JHEP \textbf{06} (2021),
  \href{https://inspirehep.net/literature?q=2102.13450}{163},
  [\href{http://arXiv.org/pdf/2102.13450}{{\tt arXiv:2102.13450}} [hep-ph]].
  \relax
 \relax
\bibitem{Mandal:2022vju}
M.~K. Mandal, P.~Mastrolia, J.~Ronca and W.~J. Bobadilla~Torres,
  \emph{{Two-loop scattering amplitude for heavy-quark pair production through
  light-quark annihilation in QCD}}, JHEP \textbf{09} (2022),
  \href{https://inspirehep.net/literature?q=2204.03466}{129},
  [\href{http://arXiv.org/pdf/2204.03466}{{\tt arXiv:2204.03466}} [hep-ph]].
  \relax
 \relax
\bibitem{Luo:2019hmp}
M.-X. Luo, X.~Wang, X.~Xu, L.~L. Yang, T.-Z. Yang and H.~X. Zhu,
  \emph{{Transverse Parton Distribution and Fragmentation Functions at NNLO:
  the Quark Case}}, JHEP \textbf{10} (2019),
  \href{https://inspirehep.net/literature?q=1908.03831}{083},
  [\href{http://arXiv.org/pdf/1908.03831}{{\tt arXiv:1908.03831}} [hep-ph]].
  \relax
 \relax
\bibitem{Luo:2019bmw}
M.-X. Luo, T.-Z. Yang, H.~X. Zhu and Y.~J. Zhu, \emph{{Transverse Parton
  Distribution and Fragmentation Functions at NNLO: the Gluon Case}}, JHEP
  \textbf{01} (2020),
  \href{https://inspirehep.net/literature?q=1909.13820}{040},
  [\href{http://arXiv.org/pdf/1909.13820}{{\tt arXiv:1909.13820}} [hep-ph]].
  \relax
 \relax
\bibitem{Luo:2020epw}
M.-x. Luo, T.-Z. Yang, H.~X. Zhu and Y.~J. Zhu, \emph{{Unpolarized quark and
  gluon TMD PDFs and FFs at N$^{3}$LO}}, JHEP \textbf{06} (2021),
  \href{https://inspirehep.net/literature?q=2012.03256}{115},
  [\href{http://arXiv.org/pdf/2012.03256}{{\tt arXiv:2012.03256}} [hep-ph]].
  \relax
 \relax
\bibitem{Luo:2019szz}
M.-x. Luo, T.-Z. Yang, H.~X. Zhu and Y.~J. Zhu, \emph{{Quark Transverse Parton
  Distribution at the Next-to-Next-to-Next-to-Leading Order}}, Phys. Rev. Lett.
  \textbf{124} (2020), no.~9,
  \href{https://inspirehep.net/literature?q=1912.05778}{092001},
  [\href{http://arXiv.org/pdf/1912.05778}{{\tt arXiv:1912.05778}} [hep-ph]].
  \relax
 \relax
\bibitem{Gutierrez-Reyes:2019rug}
D.~Gutierrez-Reyes, S.~Leal-Gomez, I.~Scimemi and A.~Vladimirov,
  \emph{{Linearly polarized gluons at next-to-next-to leading order and the
  Higgs transverse momentum distribution}}, JHEP \textbf{11} (2019),
  \href{https://inspirehep.net/literature?q=1907.03780}{121},
  [\href{http://arXiv.org/pdf/1907.03780}{{\tt arXiv:1907.03780}} [hep-ph]].
  \relax
 \relax
\bibitem{Catani:2022sgr}
\href{https://inspirehep.net/literature?q=2208.05840}{S.~Catani and P.~K.
  Dhani}, \emph{{Collinear functions for QCD resummations}},
  \href{http://arXiv.org/pdf/2208.05840}{{\tt arXiv:2208.05840}} [hep-ph].
  \relax
 \relax
\bibitem{Angeles-Martinez:2018mqh}
R.~Angeles-Martinez, M.~Czakon and S.~Sapeta, \emph{{NNLO soft function for top
  quark pair production at small transverse momentum}}, JHEP \textbf{10}
  (2018), \href{https://inspirehep.net/literature?q=1809.01459}{201},
  [\href{http://arXiv.org/pdf/1809.01459}{{\tt arXiv:1809.01459}} [hep-ph]].
  \relax
 \relax
\bibitem{Catani:2021cbl}
S.~Catani, I.~Fabre, M.~Grazzini and S.~Kallweit, \emph{{${t {{\bar{t}}}H}$
  production at NNLO: the flavour off-diagonal channels}}, Eur. Phys. J. C
  \textbf{81} (2021), no.~6,
  \href{https://inspirehep.net/literature?q=2102.03256}{491},
  [\href{http://arXiv.org/pdf/2102.03256}{{\tt arXiv:2102.03256}} [hep-ph]].
  \relax
 \relax
\bibitem{Pineda:1997bj}
A.~Pineda and J.~Soto, \emph{{Effective field theory for ultrasoft momenta in
  NRQCD and NRQED}}, Nucl. Phys. B Proc. Suppl. \textbf{64} (1998),
  \href{https://inspirehep.net/literature?q=hep-ph/9707481}{428--432},
  [\href{http://arXiv.org/pdf/hep-ph/9707481}{{\tt hep-ph/9707481}}]. \relax
 \relax
\bibitem{Brambilla:1999xf}
N.~Brambilla, A.~Pineda, J.~Soto and A.~Vairo, \emph{{Potential NRQCD: An
  Effective theory for heavy quarkonium}}, Nucl. Phys. B \textbf{566} (2000),
  \href{https://inspirehep.net/literature?q=hep-ph/9907240}{275},
  [\href{http://arXiv.org/pdf/hep-ph/9907240}{{\tt hep-ph/9907240}}]. \relax
 \relax
\bibitem{Beneke:1999zr}
M.~Beneke, \emph{{Perturbative heavy quark - anti-quark systems}}, PoS
  \textbf{hf8} (1999),
  \href{https://inspirehep.net/literature?q=hep-ph/9911490}{009},
  [\href{http://arXiv.org/pdf/hep-ph/9911490}{{\tt hep-ph/9911490}}]. \relax
 \relax
\bibitem{Beneke:1999qg}
M.~Beneke, A.~Signer and V.~A. Smirnov, \emph{{Top quark production near
  threshold and the top quark mass}}, Phys. Lett. B \textbf{454} (1999),
  \href{https://inspirehep.net/literature?q=hep-ph/9903260}{137--146},
  [\href{http://arXiv.org/pdf/hep-ph/9903260}{{\tt hep-ph/9903260}}]. \relax
 \relax
\bibitem{Ju:2019lwp}
W.-L. Ju and L.~L. Yang, \emph{{Resummation of soft and Coulomb corrections for
  $ t\overline{t}h $ production at the LHC}}, JHEP \textbf{06} (2019),
  \href{https://inspirehep.net/literature?q=1904.08744}{050},
  [\href{http://arXiv.org/pdf/1904.08744}{{\tt arXiv:1904.08744}} [hep-ph]].
  \relax
 \relax
\bibitem{Fleming:2002rv}
S.~Fleming and A.~K. Leibovich, \emph{{The Resummed Photon Spectrum in
  Radiative Upsilon Decays}}, Phys. Rev. Lett. \textbf{90} (2003),
  \href{https://inspirehep.net/literature?q=hep-ph/0211303}{032001},
  [\href{http://arXiv.org/pdf/hep-ph/0211303}{{\tt hep-ph/0211303}}]. \relax
 \relax
\bibitem{Fleming:2002sr}
S.~Fleming and A.~K. Leibovich, \emph{{The Photon Spectrum in Upsilon Decays}},
  Phys. Rev. D \textbf{67} (2003),
  \href{https://inspirehep.net/literature?q=hep-ph/0212094}{074035},
  [\href{http://arXiv.org/pdf/hep-ph/0212094}{{\tt hep-ph/0212094}}]. \relax
 \relax
\bibitem{Becher:2010pd}
T.~Becher and G.~Bell, \emph{{The gluon jet function at two-loop order}}, Phys.
  Lett. B \textbf{695} (2011),
  \href{https://inspirehep.net/literature?q=1008.1936}{252--258},
  [\href{http://arXiv.org/pdf/1008.1936}{{\tt arXiv:1008.1936}} [hep-ph]].
  \relax
 \relax
\bibitem{Banerjee:2018ozf}
P.~Banerjee, P.~K. Dhani and V.~Ravindran, \emph{{Gluon jet function at three
  loops in QCD}}, Phys. Rev. D \textbf{98} (2018), no.~9,
  \href{https://inspirehep.net/literature?q=1805.02637}{094016},
  [\href{http://arXiv.org/pdf/1805.02637}{{\tt arXiv:1805.02637}} [hep-ph]].
  \relax
 \relax
\bibitem{Becher:2006qw}
T.~Becher and M.~Neubert, \emph{{Toward a NNLO calculation of the anti-B
  ---\ensuremath{>} X(s) gamma decay rate with a cut on photon energy. II.
  Two-loop result for the jet function}}, Phys. Lett. B \textbf{637} (2006),
  \href{https://inspirehep.net/literature?q=hep-ph/0603140}{251--259},
  [\href{http://arXiv.org/pdf/hep-ph/0603140}{{\tt hep-ph/0603140}}]. \relax
 \relax
\bibitem{Bruser:2018rad}
R.~Br\"user, Z.~L. Liu and M.~Stahlhofen, \emph{{Three-Loop Quark Jet
  Function}}, Phys. Rev. Lett. \textbf{121} (2018), no.~7,
  \href{https://inspirehep.net/literature?q=1804.09722}{072003},
  [\href{http://arXiv.org/pdf/1804.09722}{{\tt arXiv:1804.09722}} [hep-ph]].
  \relax
 \relax
\bibitem{Becher:2016mmh}
T.~Becher, M.~Neubert, L.~Rothen and D.~Y. Shao, \emph{{Factorization and
  Resummation for Jet Processes}}, JHEP \textbf{11} (2016),
  \href{https://inspirehep.net/literature?q=1605.02737}{019},
  [\href{http://arXiv.org/pdf/1605.02737}{{\tt arXiv:1605.02737}} [hep-ph]],
  [Erratum: JHEP 05, 154 (2017)]. \relax
 \relax
\bibitem{Balsiger:2020ogy}
M.~Balsiger, T.~Becher and A.~Ferroglia, \emph{{Resummation of non-global
  logarithms in cross sections with massive particles}}, JHEP \textbf{09}
  (2020), \href{https://inspirehep.net/literature?q=2006.00014}{029},
  [\href{http://arXiv.org/pdf/2006.00014}{{\tt arXiv:2006.00014}} [hep-ph]].
  \relax
 \relax
\bibitem{Catani:1996vz}
S.~Catani and M.~H. Seymour, \emph{{A General algorithm for calculating jet
  cross-sections in NLO QCD}}, Nucl. Phys. B \textbf{485} (1997),
  \href{https://inspirehep.net/literature?q=hep-ph/9605323}{291--419},
  [\href{http://arXiv.org/pdf/hep-ph/9605323}{{\tt hep-ph/9605323}}], [Erratum:
  Nucl.Phys.B 510, 503--504 (1998)]. \relax
 \relax
\bibitem{Smirnov:1999gc}
V.~A. Smirnov, \emph{{Analytical result for dimensionally regularized massless
  on shell double box}}, Phys. Lett. B \textbf{460} (1999),
  \href{https://inspirehep.net/literature?q=hep-ph/9905323}{397--404},
  [\href{http://arXiv.org/pdf/hep-ph/9905323}{{\tt hep-ph/9905323}}]. \relax
 \relax
\bibitem{Tausk:1999vh}
J.~B. Tausk, \emph{{Nonplanar massless two loop Feynman diagrams with four
  on-shell legs}}, Phys. Lett. B \textbf{469} (1999),
  \href{https://inspirehep.net/literature?q=hep-ph/9909506}{225--234},
  [\href{http://arXiv.org/pdf/hep-ph/9909506}{{\tt hep-ph/9909506}}]. \relax
 \relax
\bibitem{Bejdakic:2009zz}
E.~Bejdakic, \emph{{Feynman integrals, hypergeometric functions and nested
  sums}}, Other thesis, 10 2009. \relax
 \relax
\bibitem{Czakon:2005rk}
M.~Czakon, \emph{{Automatized analytic continuation of Mellin-Barnes
  integrals}}, Comput. Phys. Commun. \textbf{175} (2006),
  \href{https://inspirehep.net/literature?q=hep-ph/0511200}{559--571},
  [\href{http://arXiv.org/pdf/hep-ph/0511200}{{\tt hep-ph/0511200}}]. \relax
 \relax
\bibitem{Czakon:Hepforge}
M.~Czakon, \emph{{MBasymptotics}},  \url{https://mbtools.hepforge.org}. \relax
 \relax
\bibitem{Ochman:2015fho}
M.~Ochman and T.~Riemann, \emph{{MBsums - a Mathematica package for the
  representation of Mellin-Barnes integrals by multiple sums}}, Acta Phys.
  Polon. B \textbf{46} (2015), no.~11,
  \href{https://inspirehep.net/literature?q=1511.01323}{2117},
  [\href{http://arXiv.org/pdf/1511.01323}{{\tt arXiv:1511.01323}} [hep-ph]].
  \relax
 \relax
\bibitem{Gleisberg:2003xi}
T.~Gleisberg, S.~Hoeche, F.~Krauss, A.~Schalicke, S.~Schumann and J.-C. Winter,
  \emph{{SHERPA 1. alpha: A Proof of concept version}}, JHEP \textbf{02}
  (2004), \href{https://inspirehep.net/literature?q=hep-ph/0311263}{056},
  [\href{http://arXiv.org/pdf/hep-ph/0311263}{{\tt hep-ph/0311263}}]. \relax
 \relax
\bibitem{Gleisberg:2008ta}
T.~Gleisberg, S.~Hoeche, F.~Krauss, M.~Schonherr, S.~Schumann, F.~Siegert and
  J.~Winter, \emph{{Event generation with SHERPA 1.1}}, JHEP \textbf{02}
  (2009), \href{https://inspirehep.net/literature?q=0811.4622}{007},
  [\href{http://arXiv.org/pdf/0811.4622}{{\tt arXiv:0811.4622}} [hep-ph]].
  \relax
 \relax
\bibitem{Sherpa:2019gpd}
E.~Bothmann et~al., Sherpa collaboration, \emph{{Event Generation with Sherpa
  2.2}}, SciPost Phys. \textbf{7} (2019), no.~3,
  \href{https://inspirehep.net/literature?q=1905.09127}{034},
  [\href{http://arXiv.org/pdf/1905.09127}{{\tt arXiv:1905.09127}} [hep-ph]].
  \relax
 \relax
\bibitem{Ferroglia:2009ep}
A.~Ferroglia, M.~Neubert, B.~D. Pecjak and L.~L. Yang, \emph{{Two-loop
  divergences of scattering amplitudes with massive partons}}, Phys. Rev. Lett.
  \textbf{103} (2009),
  \href{https://inspirehep.net/literature?q=0907.4791}{201601},
  [\href{http://arXiv.org/pdf/0907.4791}{{\tt arXiv:0907.4791}} [hep-ph]].
  \relax
 \relax
\bibitem{Moch:2004pa}
S.~Moch, J.~A.~M. Vermaseren and A.~Vogt, \emph{{The Three loop splitting
  functions in QCD: The Nonsinglet case}}, Nucl. Phys. B \textbf{688} (2004),
  \href{https://inspirehep.net/literature?q=hep-ph/0403192}{101--134},
  [\href{http://arXiv.org/pdf/hep-ph/0403192}{{\tt hep-ph/0403192}}]. \relax
 \relax
\bibitem{Henn:2019swt}
J.~M. Henn, G.~P. Korchemsky and B.~Mistlberger, \emph{{The full four-loop cusp
  anomalous dimension in $\mathcal{N}=4$ super Yang-Mills and QCD}}, JHEP
  \textbf{04} (2020),
  \href{https://inspirehep.net/literature?q=1911.10174}{018},
  [\href{http://arXiv.org/pdf/1911.10174}{{\tt arXiv:1911.10174}} [hep-th]].
  \relax
 \relax
\bibitem{vonManteuffel:2020vjv}
A.~von Manteuffel, E.~Panzer and R.~M. Schabinger, \emph{{Cusp and collinear
  anomalous dimensions in four-loop QCD from form factors}}, Phys. Rev. Lett.
  \textbf{124} (2020), no.~16,
  \href{https://inspirehep.net/literature?q=2002.04617}{162001},
  [\href{http://arXiv.org/pdf/2002.04617}{{\tt arXiv:2002.04617}} [hep-ph]].
  \relax
 \relax
\bibitem{Herzog:2018kwj}
F.~Herzog, S.~Moch, B.~Ruijl, T.~Ueda, J.~A.~M. Vermaseren and A.~Vogt,
  \emph{{Five-loop contributions to low-N non-singlet anomalous dimensions in
  QCD}}, Phys. Lett. B \textbf{790} (2019),
  \href{https://inspirehep.net/literature?q=1812.11818}{436--443},
  [\href{http://arXiv.org/pdf/1812.11818}{{\tt arXiv:1812.11818}} [hep-ph]].
  \relax
 \relax
\bibitem{Liu:2022elt}
Z.~L. Liu and N.~Schalch, \emph{{Infrared Singularities of Multileg QCD
  Amplitudes with a Massive Parton at Three Loops}}, Phys. Rev. Lett.
  \textbf{129} (2022), no.~23,
  \href{https://inspirehep.net/literature?q=2207.02864}{232001},
  [\href{http://arXiv.org/pdf/2207.02864}{{\tt arXiv:2207.02864}} [hep-ph]].
  \relax
 \relax
\bibitem{Vladimirov:2016dll}
A.~A. Vladimirov, \emph{{Correspondence between Soft and Rapidity Anomalous
  Dimensions}}, Phys. Rev. Lett. \textbf{118} (2017), no.~6,
  \href{https://inspirehep.net/literature?q=1610.05791}{062001},
  [\href{http://arXiv.org/pdf/1610.05791}{{\tt arXiv:1610.05791}} [hep-ph]].
  \relax
 \relax
\bibitem{Ebert:2020yqt}
M.~A. Ebert, B.~Mistlberger and G.~Vita, \emph{{Transverse momentum dependent
  PDFs at N$^3$LO}}, JHEP \textbf{09} (2020),
  \href{https://inspirehep.net/literature?q=2006.05329}{146},
  [\href{http://arXiv.org/pdf/2006.05329}{{\tt arXiv:2006.05329}} [hep-ph]].
  \relax
 \relax
\bibitem{Das:2019btv}
G.~Das, S.-O. Moch and A.~Vogt, \emph{{Soft corrections to inclusive
  deep-inelastic scattering at four loops and beyond}}, JHEP \textbf{03}
  (2020), \href{https://inspirehep.net/literature?q=1912.12920}{116},
  [\href{http://arXiv.org/pdf/1912.12920}{{\tt arXiv:1912.12920}} [hep-ph]].
  \relax
 \relax
\bibitem{Duhr:2022cob}
C.~Duhr, B.~Mistlberger and G.~Vita, \emph{{Soft integrals and soft anomalous
  dimensions at N$^{3}$LO and beyond}}, JHEP \textbf{09} (2022),
  \href{https://inspirehep.net/literature?q=2205.04493}{155},
  [\href{http://arXiv.org/pdf/2205.04493}{{\tt arXiv:2205.04493}} [hep-ph]].
  \relax
 \relax
\bibitem{Duhr:2022yyp}
C.~Duhr, B.~Mistlberger and G.~Vita, \emph{{Four-Loop Rapidity Anomalous
  Dimension and Event Shapes to Fourth Logarithmic Order}}, Phys. Rev. Lett.
  \textbf{129} (2022), no.~16,
  \href{https://inspirehep.net/literature?q=2205.02242}{162001},
  [\href{http://arXiv.org/pdf/2205.02242}{{\tt arXiv:2205.02242}} [hep-ph]].
  \relax
 \relax
\bibitem{Moult:2022xzt}
I.~Moult, H.~X. Zhu and Y.~J. Zhu, \emph{{The four loop QCD rapidity anomalous
  dimension}}, JHEP \textbf{08} (2022),
  \href{https://inspirehep.net/literature?q=2205.02249}{280},
  [\href{http://arXiv.org/pdf/2205.02249}{{\tt arXiv:2205.02249}} [hep-ph]].
  \relax
 \relax
\bibitem{Neill:2015roa}
D.~Neill, I.~Z. Rothstein and V.~Vaidya, \emph{{The Higgs Transverse Momentum
  Distribution at NNLL and its Theoretical Errors}}, JHEP \textbf{12} (2015),
  \href{https://inspirehep.net/literature?q=1503.00005}{097},
  [\href{http://arXiv.org/pdf/1503.00005}{{\tt arXiv:1503.00005}} [hep-ph]].
  \relax
 \relax
\bibitem{Kang:2017cjk}
D.~Kang, C.~Lee and V.~Vaidya, \emph{{A fast and accurate method for
  perturbative resummation of transverse momentum-dependent observables}}, JHEP
  \textbf{04} (2018),
  \href{https://inspirehep.net/literature?q=1710.00078}{149},
  [\href{http://arXiv.org/pdf/1710.00078}{{\tt arXiv:1710.00078}} [hep-ph]].
  \relax
 \relax
\bibitem{Buras:1991jm}
A.~J. Buras, M.~Jamin, M.~E. Lautenbacher and P.~H. Weisz, \emph{{Effective
  Hamiltonians for $\Delta S = 1$ and $\Delta B = 1$ nonleptonic decays beyond
  the leading logarithmic approximation}}, Nucl. Phys. B \textbf{370} (1992),
  \href{https://inspirehep.net/literature?q=Nucl%20Phys%20B,370,69}{69--104},
  [Addendum: Nucl.Phys.B 375, 501 (1992)]. \relax
 \relax
\bibitem{Buchalla:1995vs}
G.~Buchalla, A.~J. Buras and M.~E. Lautenbacher, \emph{{Weak decays beyond
  leading logarithms}}, Rev. Mod. Phys. \textbf{68} (1996),
  \href{https://inspirehep.net/literature?q=hep-ph/9512380}{1125--1144},
  [\href{http://arXiv.org/pdf/hep-ph/9512380}{{\tt hep-ph/9512380}}]. \relax
 \relax
\bibitem{Banfi:2012jm}
A.~Banfi, P.~F. Monni, G.~P. Salam and G.~Zanderighi, \emph{{Higgs and Z-boson
  production with a jet veto}}, Phys. Rev. Lett. \textbf{109} (2012),
  \href{https://inspirehep.net/literature?q=1206.4998}{202001},
  [\href{http://arXiv.org/pdf/1206.4998}{{\tt arXiv:1206.4998}} [hep-ph]].
  \relax
 \relax
\bibitem{Banfi:2012yh}
A.~Banfi, G.~P. Salam and G.~Zanderighi, \emph{{NLL+NNLO predictions for
  jet-veto efficiencies in Higgs-boson and Drell-Yan production}}, JHEP
  \textbf{06} (2012),
  \href{https://inspirehep.net/literature?q=1203.5773}{159},
  [\href{http://arXiv.org/pdf/1203.5773}{{\tt arXiv:1203.5773}} [hep-ph]].
  \relax
 \relax
\bibitem{Bizon:2018foh}
W.~Bizo\'n, X.~Chen, A.~Gehrmann-De~Ridder, T.~Gehrmann, N.~Glover, A.~Huss,
  P.~F. Monni, E.~Re, L.~Rottoli and P.~Torrielli, \emph{{Fiducial
  distributions in Higgs and Drell-Yan production at N$^{3}$LL+NNLO}}, JHEP
  \textbf{12} (2018),
  \href{https://inspirehep.net/literature?q=1805.05916}{132},
  [\href{http://arXiv.org/pdf/1805.05916}{{\tt arXiv:1805.05916}} [hep-ph]].
  \relax
 \relax
\bibitem{Workman:2022ynf}
R.~L. Workman et~al., Particle Data Group collaboration, \emph{{Review of
  Particle Physics}}, PTEP \textbf{2022} (2022),
  \href{https://inspirehep.net/literature?q=PTEP,2022,083C01}{083C01}. \relax
 \relax
\bibitem{Ablinger:2018sat}
J.~Ablinger, J.~Bl\"umlein, M.~Round and C.~Schneider, \emph{{Numerical
  Implementation of Harmonic Polylogarithms to Weight w = 8}}, Comput. Phys.
  Commun. \textbf{240} (2019),
  \href{https://inspirehep.net/literature?q=1809.07084}{189--201},
  [\href{http://arXiv.org/pdf/1809.07084}{{\tt arXiv:1809.07084}} [hep-ph]].
  \relax
 \relax
\bibitem{NNPDF:2017mvq}
R.~D. Ball et~al., NNPDF collaboration, \emph{{Parton distributions from
  high-precision collider data}}, Eur. Phys. J. C \textbf{77} (2017), no.~10,
  \href{https://inspirehep.net/literature?q=1706.00428}{663},
  [\href{http://arXiv.org/pdf/1706.00428}{{\tt arXiv:1706.00428}} [hep-ph]].
  \relax
 \relax
\bibitem{Buckley:2014ana}
A.~Buckley, J.~Ferrando, S.~Lloyd, K.~Nordstr\"om, B.~Page, M.~R\"ufenacht,
  M.~Sch\"onherr and G.~Watt, \emph{{LHAPDF6: parton density access in the LHC
  precision era}}, Eur. Phys. J. C \textbf{75} (2015),
  \href{https://inspirehep.net/literature?q=1412.7420}{132},
  [\href{http://arXiv.org/pdf/1412.7420}{{\tt arXiv:1412.7420}} [hep-ph]].
  \relax
 \relax
\bibitem{Bothmann:2022thx}
\href{https://inspirehep.net/literature?q=2209.00843}{E.~Bothmann, A.~Buckley,
  I.~A. Christidi, C.~G\"utschow, S.~H\"oche, M.~Knobbe, T.~Martin and
  M.~Sch\"onherr}, \emph{{Accelerating LHC event generation with simplified
  pilot runs and fast PDFs}},  \href{http://arXiv.org/pdf/2209.00843}{{\tt
  arXiv:2209.00843}} [hep-ph]. \relax
 \relax
\bibitem{Becher:2006mr}
T.~Becher, M.~Neubert and B.~D. Pecjak, \emph{{Factorization and Momentum-Space
  Resummation in Deep-Inelastic Scattering}}, JHEP \textbf{01} (2007),
  \href{https://inspirehep.net/literature?q=hep-ph/0607228}{076},
  [\href{http://arXiv.org/pdf/hep-ph/0607228}{{\tt hep-ph/0607228}}]. \relax
 \relax
\bibitem{Hahn:2006hr}
\href{https://inspirehep.net/literature?q=physics/0607103}{T.~Hahn},
  \emph{{Routines for the diagonalization of complex matrices}},
  \href{http://arXiv.org/pdf/physics/0607103}{{\tt physics/0607103}}. \relax
 \relax
\bibitem{Buckley:2010ar}
A.~Buckley, J.~Butterworth, D.~Grellscheid, H.~Hoeth, L.~Lonnblad, J.~Monk,
  H.~Schulz and F.~Siegert, \emph{{Rivet user manual}}, Comput. Phys. Commun.
  \textbf{184} (2013),
  \href{https://inspirehep.net/literature?q=1003.0694}{2803--2819},
  [\href{http://arXiv.org/pdf/1003.0694}{{\tt arXiv:1003.0694}} [hep-ph]].
  \relax
 \relax
\bibitem{Bierlich:2019rhm}
C.~Bierlich et~al., \emph{{Robust Independent Validation of Experiment and
  Theory: Rivet version 3}}, SciPost Phys. \textbf{8} (2020),
  \href{https://inspirehep.net/literature?q=1912.05451}{026},
  [\href{http://arXiv.org/pdf/1912.05451}{{\tt arXiv:1912.05451}} [hep-ph]].
  \relax
 \relax
\bibitem{Krauss:2001iv}
F.~Krauss, R.~Kuhn and G.~Soff, \emph{{AMEGIC++ 1.0: A Matrix element generator
  in C++}}, JHEP \textbf{02} (2002),
  \href{https://inspirehep.net/literature?q=hep-ph/0109036}{044},
  [\href{http://arXiv.org/pdf/hep-ph/0109036}{{\tt hep-ph/0109036}}]. \relax
 \relax
\bibitem{Catani:2002hc}
S.~Catani, S.~Dittmaier, M.~H. Seymour and Z.~Trocsanyi, \emph{{The Dipole
  formalism for next-to-leading order QCD calculations with massive partons}},
  Nucl. Phys. B \textbf{627} (2002),
  \href{https://inspirehep.net/literature?q=hep-ph/0201036}{189--265},
  [\href{http://arXiv.org/pdf/hep-ph/0201036}{{\tt hep-ph/0201036}}]. \relax
 \relax
\bibitem{Gleisberg:2007md}
T.~Gleisberg and F.~Krauss, \emph{{Automating dipole subtraction for QCD NLO
  calculations}}, Eur. Phys. J. C \textbf{53} (2008),
  \href{https://inspirehep.net/literature?q=0709.2881}{501--523},
  [\href{http://arXiv.org/pdf/0709.2881}{{\tt arXiv:0709.2881}} [hep-ph]].
  \relax
 \relax
\bibitem{Schonherr:2017qcj}
M.~Sch\"onherr, \emph{{An automated subtraction of NLO EW infrared
  divergences}}, Eur. Phys. J. C \textbf{78} (2018), no.~2,
  \href{https://inspirehep.net/literature?q=1712.07975}{119},
  [\href{http://arXiv.org/pdf/1712.07975}{{\tt arXiv:1712.07975}} [hep-ph]].
  \relax
 \relax
\bibitem{Ju:2021lah}
W.-L. Ju and M.~Sch\"onherr, \emph{{The q$_{T}$ and
  \ensuremath{\Delta}\ensuremath{\phi} spectra in W and Z production at the LHC
  at N$^{3}$LL'+N$^{2}$LO}}, JHEP \textbf{10} (2021),
  \href{https://inspirehep.net/literature?q=2106.11260}{088},
  [\href{http://arXiv.org/pdf/2106.11260}{{\tt arXiv:2106.11260}} [hep-ph]].
  \relax
 \relax
\bibitem{Gao:2022bzi}
\href{https://inspirehep.net/literature?q=2209.11211}{A.~Gao, J.~K.~L. Michel,
  I.~W. Stewart and Z.~Sun}, \emph{{A Better Angle on Hadron Transverse
  Momentum Distributions at the EIC}},
  \href{http://arXiv.org/pdf/2209.11211}{{\tt arXiv:2209.11211}} [hep-ph].
  \relax
 \relax
\bibitem{Chien:2019gyf}
Y.-T. Chien, D.~Y. Shao and B.~Wu, \emph{{Resummation of Boson-Jet Correlation
  at Hadron Colliders}}, JHEP \textbf{11} (2019),
  \href{https://inspirehep.net/literature?q=1905.01335}{025},
  [\href{http://arXiv.org/pdf/1905.01335}{{\tt arXiv:1905.01335}} [hep-ph]].
  \relax
 \relax
\bibitem{delCastillo:2021znl}
R.~F. del Castillo, M.~G. Echevarria, Y.~Makris and I.~Scimemi,
  \emph{{Transverse momentum dependent distributions in dijet and heavy hadron
  pair production at EIC}}, JHEP \textbf{03} (2022),
  \href{https://inspirehep.net/literature?q=2111.03703}{047},
  [\href{http://arXiv.org/pdf/2111.03703}{{\tt arXiv:2111.03703}} [hep-ph]].
  \relax
 \relax
\bibitem{Tong:2022zwp}
\href{https://inspirehep.net/literature?q=2211.01647}{X.-B. Tong, B.-W. Xiao
  and Y.-Y. Zhang}, \emph{{Harmonics of Parton Saturation in Lepton-Jet
  Correlations at the EIC}},  \href{http://arXiv.org/pdf/2211.01647}{{\tt
  arXiv:2211.01647}} [hep-ph]. \relax
 \relax
\bibitem{Shao:2022stc}
\href{https://inspirehep.net/literature?q=2212.05775}{D.~Y. Shao, C.~Zhang,
  J.~Zhou and Y.~Zhou}, \emph{{Azimuthal asymmetries of muon pair production in
  ultraperipheral heavy ion collisions}},
  \href{http://arXiv.org/pdf/2212.05775}{{\tt arXiv:2212.05775}} [hep-ph].
  \relax
 \relax
\bibitem{Chen:2018fqu}
L.~Chen, G.-Y. Qin, L.~Wang, S.-Y. Wei, B.-W. Xiao, H.-Z. Zhang and Y.-Q.
  Zhang, \emph{{Study of Isolated-photon and Jet Momentum Imbalance in $pp$ and
  $PbPb$ collisions}}, Nucl. Phys. B \textbf{933} (2018),
  \href{https://inspirehep.net/literature?q=1803.10533}{306--319},
  [\href{http://arXiv.org/pdf/1803.10533}{{\tt arXiv:1803.10533}} [hep-ph]].
  \relax
 \relax
\bibitem{Chien:2020hzh}
Y.-T. Chien, R.~Rahn, S.~Schrijnder~van Velzen, D.~Y. Shao, W.~J. Waalewijn and
  B.~Wu, \emph{{Recoil-free azimuthal angle for precision boson-jet
  correlation}}, Phys. Lett. B \textbf{815} (2021),
  \href{https://inspirehep.net/literature?q=2005.12279}{136124},
  [\href{http://arXiv.org/pdf/2005.12279}{{\tt arXiv:2005.12279}} [hep-ph]].
  \relax
 \relax
\bibitem{Bouaziz:2022tik}
H.~Bouaziz, Y.~Delenda and K.~Khelifa-Kerfa, \emph{{Azimuthal decorrelation
  between a jet and a Z boson at hadron colliders}}, JHEP \textbf{10} (2022),
  \href{https://inspirehep.net/literature?q=2207.10147}{006},
  [\href{http://arXiv.org/pdf/2207.10147}{{\tt arXiv:2207.10147}} [hep-ph]].
  \relax
 \relax
\bibitem{Chien:2022wiq}
\href{https://inspirehep.net/literature?q=2205.05104}{Y.-T. Chien, R.~Rahn,
  D.~Y. Shao, W.~J. Waalewijn and B.~Wu}, \emph{{Precision boson-jet azimuthal
  decorrelation at hadron colliders}},
  \href{http://arXiv.org/pdf/2205.05104}{{\tt arXiv:2205.05104}} [hep-ph].
  \relax
 \relax
\bibitem{Banfi:2008qs}
A.~Banfi, M.~Dasgupta and Y.~Delenda, \emph{{Azimuthal decorrelations between
  QCD jets at all orders}}, Phys. Lett. B \textbf{665} (2008),
  \href{https://inspirehep.net/literature?q=0804.3786}{86--91},
  [\href{http://arXiv.org/pdf/0804.3786}{{\tt arXiv:0804.3786}} [hep-ph]].
  \relax
 \relax
\bibitem{Zhang:2022wvs}
\href{https://inspirehep.net/literature?q=2211.07071}{C.~Zhang, Q.-S. Dai and
  D.~Y. Shao}, \emph{{Azimuthal decorrelation for photon induced dijet
  production in ultra-peripheral collisions of heavy ions}},
  \href{http://arXiv.org/pdf/2211.07071}{{\tt arXiv:2211.07071}} [hep-ph].
  \relax
 \relax
\bibitem{Baikov:2016tgj}
P.~A. Baikov, K.~G. Chetyrkin and J.~H. K\"uhn, \emph{{Five-Loop Running of the
  QCD coupling constant}}, Phys. Rev. Lett. \textbf{118} (2017), no.~8,
  \href{https://inspirehep.net/literature?q=1606.08659}{082002},
  [\href{http://arXiv.org/pdf/1606.08659}{{\tt arXiv:1606.08659}} [hep-ph]].
  \relax
 \relax
\bibitem{Lipatov:1974qm}
L.~N. Lipatov, \emph{{The parton model and perturbation theory}}, Yad. Fiz.
  \textbf{20} (1974),
  \href{https://inspirehep.net/literature?q=Yad%20Fiz,20,181}{181--198}. \relax
 \relax
\bibitem{Gribov:1972ri}
V.~N. Gribov and L.~N. Lipatov, \emph{{Deep inelastic e p scattering in
  perturbation theory}}, Sov. J. Nucl. Phys. \textbf{15} (1972),
  \href{https://inspirehep.net/literature?q=Sov%20J%20Nucl%20Phys,15,438}{438--450}.
  \relax
 \relax
\bibitem{Altarelli:1977zs}
G.~Altarelli and G.~Parisi, \emph{{Asymptotic Freedom in Parton Language}},
  Nucl. Phys. B \textbf{126} (1977),
  \href{https://inspirehep.net/literature?q=Nucl%20Phys%20B,126,298}{298--318}.
  \relax
 \relax
\bibitem{Dokshitzer:1977sg}
Y.~L. Dokshitzer, \emph{{Calculation of the Structure Functions for Deep
  Inelastic Scattering and e+ e- Annihilation by Perturbation Theory in Quantum
  Chromodynamics.}}, Sov. Phys. JETP \textbf{46} (1977),
  \href{https://inspirehep.net/literature?q=Sov%20Phys%20JETP,46,641}{641--653}.
  \relax
 \relax
\end{thebibliography}

  \end{document}